\documentclass[11pt]{book}
\usepackage{fullpage}
\usepackage{latexsym}
\usepackage{eepic}

\usepackage[english,dutch]{babel}    

\usepackage{amsfonts,amsmath,xspace,graphicx,relsize,bm,mathtools,xcolor}
\usepackage{mathrsfs}
\usepackage{hyperref}

\hypersetup{
    colorlinks,
    linkcolor={blue!100!black},
    citecolor={blue!100!black},
}

\def\01{\{0,1\}}
\newcommand{\eps}{\varepsilon}
\newcommand{\ket}[1]{|#1\rangle}
\newcommand{\bra}[1]{\langle#1|}
\newcommand{\ketbra}[2]{|#1\rangle\langle#2|}
\newcommand{\inp}[2]{\langle{#1}|{#2}\rangle} 
\newcommand{\inpc}[2]{\langle{#1},{#2}\rangle} 
\newcommand{\norm}[1]{{\left\|{#1}\right\|}}

\newcommand{\rank}{\mbox{\rm rank}}
\newcommand{\poly}{\mbox{\rm poly}}
\newcommand{\polylog}{\mbox{\rm polylog}}
\newcommand{\Sc}{{\boldsymbol S}}
\newcommand{\Uc}{{\boldsymbol U}}
\newcommand{\Cc}{{\boldsymbol C}}
\newcommand{\Tr}{{\rm Tr}}
\newcommand{\ceil}[1]{\lceil{#1}\rceil}
\newcommand{\floor}[1]{\lfloor{#1}\rfloor}

\newcommand{\CNOT}{\mbox{\rm CNOT}}
\newcommand{\CZ}{\mbox{\rm CZ}}

\newcommand{\OR}{\mbox{\rm OR}}

\newcommand{\Aut}{\mbox{\rm Aut}}
\newcommand{\ADV}{\mbox{\rm ADV}^\pm}

\newcommand{\Pp}{\mbox{\bf P}}
\newcommand{\NP}{\mbox{\bf NP}}
\newcommand{\PSPACE}{\mbox{\bf PSPACE}}
\newcommand{\QPSPACE}{\mbox{\bf QPSPACE}}
\newcommand{\BPP}{\mbox{\bf BPP}}
\newcommand{\EQP}{\mbox{\bf EQP}}
\newcommand{\BQP}{\mbox{\bf BQP}}

\newcommand{\PP}{\mbox{\bf PP}}
\newcommand{\QMA}{\mbox{\bf QMA}}
\newcommand{\IPr}{\mbox{\bf IP}}
\newcommand{\MA}{\mbox{\bf MA}}
\newcommand{\QCMA}{\mbox{\bf QCMA}}
\newcommand{\QIP}{\mbox{\bf QIP}}
\newcommand{\Exp}{\mathbb{E}}
\newcommand{\Var}{\mbox{Var}}

\newcommand{\R}{\mathbb{R}}

\newtheorem{theorem}{Theorem}
\newtheorem{definition}{Definition}
\newtheorem{fact}{Fact}
\newtheorem{claim}{Claim}
\newenvironment{proof}
{\noindent {\bf Proof. }}
{{\hfill $\Box$}\\
 \smallskip}

\begin{document}
\pagestyle{plain}
\pagenumbering{roman}

\selectlanguage{english}

\title{\bf\Huge Quantum Computing:\\[0.5cm] Lecture Notes\\[2cm]}
\author{\huge Ronald de Wolf\\[1cm]
QuSoft, CWI and University of Amsterdam}
\date{}
\maketitle

\newpage

\vspace*{5cm}

\begin{center}
{\it  Dedicated to the memory of my father\\[1mm]
  Abraham de Wolf (1942--2019)}
\end{center}

\newpage

\section*{Preface from 2011}
These lecture notes were formed in small chunks during my ``Quantum computing'' course
at the University of Amsterdam, Feb-May 2011, and compiled into one text thereafter.
Each chapter was covered in a lecture of $2\times 45$ minutes, with an additional 45-minute lecture for exercises and homework.
The first half of the course (Chapters 1--7) covers quantum algorithms,
the second half covers quantum complexity (Chapters 8--9), 
stuff involving Alice and Bob (Chapters 10--13), and error-correction (Chapter~14).
A 15th lecture about physical implementations and general outlook was more sketchy, and I didn't write lecture notes for it.

These chapters may also be read as a general introduction to the area of quantum computation and information
from the perspective of a theoretical computer scientist.
While I made an effort to make the text self-contained and consistent, it may still be somewhat rough around the edges; 
I hope to continue polishing and adding to it.
Comments \&\ constructive criticism are very welcome, and can be sent to {\tt rdewolf@cwi.nl}

Those who want to read more (much more\ldots): see the book by Nielsen and Chuang~\cite{nielsen&chuang:qc} for the general area, the book of John Watrous~\cite{watrous:tqi} for quantum information theory, and the lecture notes of John Preskill~\cite{preskill:faulttol} for the theoretical physics perspective.

\section*{Attribution, acknowledgments, subsequent updates}
Most of the material in Chapters 1--6 [chapter numbers in this paragraph are for the 2011 version] comes from the first chapter of my PhD thesis~\cite{wolf:thesis},
with a number of additions: the lower bound for Simon, the Fourier transform, the geometric explanation of Grover.
Chapter~7 is newly written for these notes, inspired by Santha's survey~\cite{santha:qrwsurvey}.
Chapters~8 and~9 are largely new as well.
Section~3 of Chapter~8, and most of Chapter~10 are taken (with many changes)
from my ``quantum proofs'' survey paper with Andy Drucker~\cite{drucker&wolf:qproofs}.
Chapters~11 and~12 are partly taken from my non-locality survey with Harry Buhrman, Richard Cleve, and Serge Massar~\cite{bcmw:survey}.
Chapters~13 and~14 are new.
Thanks to Giannicola Scarpa (the teaching assistant for the first two editions of this course) for useful comments on some of the chapters.\\[5mm]
\emph{January'13}: Updated and corrected a few things for the Feb-Mar 2013 version of this course, and included exercises for each chapter.  Thanks to Harry Buhrman, Florian Speelman, and Jeroen Zuiddam for spotting some typos in the earlier version.
\\[5mm]
\emph{April'13}: More updates, clarifications and corrections; moved some material from Chapter~2 to~1; changed and added some exercises. Thanks to Jouke Witteveen for useful comments.
\\[5mm]
\emph{April'14}: Fixed and clarified a few more things.  Thanks to Maarten Wegewijs for spotting a typo in Chapter~\ref{chap:ft}.
\\[5mm]
\emph{March'15}: Updated a few small things.
\\[5mm]
\emph{July'15}: Updated and corrected a few small things, added more exercises. Thanks to Srinivasan Arunachalam, Carla Groenland, and Koen Groenland for useful comments.
\\[5mm]
\emph{May'16}: A few more corrections, thanks to Ralph Bottesch for useful comments.
\\[5mm]
\emph{January'18}: Many more corrections, more exercises, a new chapter about the Hidden Subgroup Problem (Chapter~\ref{chap:hsp}; the above-mentioned chapter numbers are for the earlier version of the notes), and moved the hints about exercises to an Appendix for students who want to try the exercises first without hints. Thanks to Joran van Apeldoorn, Srinivasan Arunachalam, Rens Baardman, Alexander Belov, Koen de Boer, Daniel Chernowitz, Andr\'as Gily\'en, Ronald de Haan, Leon Ingelse, Stacey Jeffery, Rafael Kiesel, Jens Klooster, Sam Kuypers, Christian Nesenberend, and Christian Schaffner for useful comments. 
\\[5mm]
\emph{January'19}: More corrections, clarifications and exercises, and new chapters about Hamiltonian simulation (Chapter~\ref{chap:hamsim}) and the HHL algorithm (Chapter~\ref{chap:hhl}). These two chapters can be taught together in two lectures, with the longer Chapter~\ref{chap:hamsim} spilling over into the second lecture if necessary. I marked by `(H)' the exercises having a hint in Appendix~\ref{app:hints}, and removed citations from exercises to prevent students looking up the original papers when doing the exercises (which is neither necessary nor helpful). Those references are
\cite{superdense,dam:oracle,durr&hoyer:minimum,bht:collision,dhhm:graphproblemsj,freivalds:matrixmult,bs:matrix,bbcmw:polynomialsj,bbbv:str&weak,bjk:q1wayj,gkkrw:1wayj,tsirelson80,amtw:pqc}.
Thanks to Arjan Cornelissen, Sven Cornets de Groot, Gerrit Vos, and Harm de Vries for useful comments, and to Andr\'as Gily\'en for much help with Chapters~\ref{chap:hamsim} and~\ref{chap:hhl}. Thanks to my father and Mieke Beer for hosting me for two months while I was recovering from an ankle fracture in a wheelchair, from which much of these two chapters was written.
\\[5mm]
\emph{July'19}: More corrections, clarifications and exercises. Thanks to Joran van Apeldoorn, Andr\'as Gily\'en, Stephanie Gonzalez, Sander Gribling, Jaco ter Hoeve, Arnold Kole, Lotte Mertens, Stefano Pironio, Merel Schalkers, Jim Skulte, Iris Smit, Manuel Van, and Sebastian Zur for useful comments.
Thanks to Barbara Terhal for suggesting the possibility of dedicating these notes.
\\[5mm]
\emph{January'21}: More corrections, clarifications and exercises, and a new chapter about QMA and the local Hamiltonian problem  (Chapter~\ref{ch:QMAlocham}). Thanks to Dorit Aharonov, Tomas Ehrencron, Alex Grilo,  Joris Kattem\"{o}lle, Stan de Lange, Noah Linden, Tessel Majtlis, Nikhil Mande, Andrea Mazzocco, Robert Modderman, Thomas Preu, Philip Verduyn Lunel, Baer Ververgaert, and Carel Wagenaar for useful comments.
\\[5mm]
\emph{January'22}: More corrections, clarifications and exercises. Thanks to Simon Apers, Christiaan van Asperen, Yanlin Chen, Lynn Engelberts, Sevag Gharibian, Andr\'{a}s Gily\'{e}n, Diego Gonz\'{a}lez-S\'{a}nchez, Bruno Jedynak, Joris Kattem\"{o}lle, Julius Krebbekx, Zeph Landau, Noah Linden, Fr\'{e}d\'{e}ric Magniez, Ryan Mann, and Yanelle Stolwijk for useful comments.
\\[5mm]
\emph{August'22}:
More corrections, clarifications and exercises, and a new chapter about some aspects of quantum machine learning (Chapter~\ref{chap:qml}). 
Thanks to Danish Alvi, Srinivasan Arunachalam, Nikolaos Chatzis, Li Chen, Yanlin Chen, Iordanis Kerenidis, Mikhail Kudinov, Casper Loman, Fr\'{e}d\'{e}ric Magniez, Galina Pass, Benjamin Shaffrey, Mario Szegedy, and  Michael Walter for useful comments. 
\\[5mm]
\emph{January'23}:
More corrections and clarifications, and a new chapter about the generalized adversary bound (Chapter~\ref{chap:adversary}).  
Thanks to Srinivasan Arunachalam, Martijn Brehm, Yanlin Chen, Lynn Engelberts, Noah Linden, Fr\'{e}d\'{e}ric Magniez, Mehrdad Tahmasbi, Quinten Tupker, and Jordi Weggemans for useful comments.\\[3mm]
\emph{August'26}:
More corrections,  clarifications, exercises, and a new chapter about stabilizer states (Chapter~\ref{chap:stabilizer}). 
Thanks to Simon Apers, Floor Beks, Raufs Dunamalijevs, Lynn Engelberts, Chinook Felix, Timo Gils, Chris de Hart, Rik Heurter, Milena Kapr\'{a}lov\'{a}, Lorenzo Laneve, Noah Linden, Manideep Mamindlapally, Lois de la Mar, Harold Nieuwboer, Manaswi Paraashar, Subhasree Patro, Oded Regev, Giannicola Scarpa, Daan Schoneveld, Steven Wagter, and Stefan Wolf for useful comments. And ChatGPT for proofreading.

\vfill

\noindent
\textcopyright~\emph{Ronald de Wolf, August 2026, Amsterdam\\
All translations of these notes into other languages are unauthorized and violate this copyright.}

\newpage

\tableofcontents

\cleardoublepage
\pagestyle{plain}
\pagenumbering{arabic}

\chapter{Quantum Computing}\label{chap:intro}

\section{Introduction}

Today's computers---both in theory (Turing machines\index{Turing machine}) and practice (PCs, HPCs, laptops, tablets, smartphones, \ldots)---are based on classical physics. 
They are limited by locality (operations
have only local effects) and by the classical fact that systems
can be in only one state at a time.
However, modern quantum physics tells us that the world behaves
quite differently. A quantum system can be in a \emph{superposition}
of many different states at the same time, and can exhibit 
\emph{interference} effects during the course of its evolution.
Moreover, spatially separated quantum systems may be \emph{entangled} with 
each other and operations may have ``non-local'' effects because of this (though they won't signal faster than the speed of light).

\emph{Quantum computation}\index{quantum computing} 
is the field that investigates the computational
power and other properties of computers based on quantum-mechanical principles.
It combines two of the most important strands of 20th-century science: quantum mechanics (developed by Planck, Einstein, Bohr, Heisenberg, Schr\"odinger and others in the period 1900--1925) and computer science (whose birth may be dated to Turing's 1936 paper~\cite{turing:compnumb}).
An important objective is to find quantum algorithms that are significantly
faster than any classical algorithm solving the same problem.

Quantum computation started in the early 1980s with suggestions for analog
quantum computers by Yuri Manin~\cite{manin:radio} (and appendix of~\cite{manin:qc}),
Richard Feynman~\cite{feynman:simulating,feynman:qmc}\index{Feynman, R.},
and Paul Benioff~\cite{benioff:hamiltonian}\index{Benioff, P.}, 
and reached more digital ground when in 1985 David Deutsch\index{Deutsch, D.} 
defined the universal quantum Turing machine~\cite{deutsch:uqc}.\index{quantum Turing machine} 
See Preskill~\cite{preskill:qc40years} for more on this early history.
The following years saw only sparse activity, notably the development
of the first algorithms by Deutsch\index{Deutsch, D.} 
and Jozsa\index{Jozsa, R.}~\cite{deutsch&jozsa} 
and by Simon~\cite{simon:power}\index{Simon, D.}, and the development of quantum 
complexity theory by Bernstein\index{Bernstein, E.} and 
Vazirani~\cite{bernstein&vazirani:qcomplexity}.\index{Vazirani, U.}
However, interest in the field increased tremendously 
after Peter Shor's\index{Shor, P.} very surprising 
discovery of efficient quantum algorithms for the problems of integer 
factorization\index{factoring} and discrete 
logarithms\index{discrete logarithm} in 1994~\cite{shor:factoring}, which was inspired by Simon's work. 
Since most of classical cryptography\index{cryptography} is based on
the assumption that these two problems are computationally hard,
the ability to actually build and use a quantum computer 
would allow us to break most current classical cryptographic systems, 
notably the RSA\index{RSA} system~\cite{rsa,rivest:crypto}.\index{Rivest, R.}
In contrast, a \emph{quantum} form of cryptography due to Bennett\index{Bennett, C.} 
and Brassard~\cite{bb84}\index{Brassard, G.} 
is unbreakable even by quantum computers: it is information-theoretically secure under some (idealizing) assumptions.

Here are three reasons to study quantum computers, from practical to more philosophical:
\begin{enumerate}
\item The process of miniaturization that has made current classical computers so powerful and cheap,
has already reached micro-levels where quantum effects occur. Chipmakers tend to go to great lengths to suppress
those quantum effects, forcing their bits and logical operations to behave classically, but instead one might also try to work with them, enabling further miniaturization.
\item Making use of quantum effects allows one to speed up certain computations enormously
(sometimes exponentially), and even enables some things that are impossible for classical computers.
The main purpose of these lecture notes is to explain these advantages of quantum computing (algorithms, crypto, etc.) in detail.
\item Finally, one might say that the main goal of theoretical computer science is to 
``study the power and limitations of the strongest-possible computational devices that Nature allows us.''
Since our current understanding of Nature is quantum mechanical, theoretical computer science 
should arguably be studying the power of quantum computers, not classical ones.
\end{enumerate}
Before limiting ourselves to theory, let us say a few words about 
practice: \emph{to what extent will large quantum computers ever be built}?%
\index{quantum computing!implementation of}
At this point in time, it is just too early to tell.
The first small 2-qubit quantum computer was built in 1997
and in 2001 a 5-qubit quantum computer was used to successfully factor the number~15~\cite{nmr:period}.\index{Vandersypen, L.}\index{Steffen, M.}\index{Breyta, G.}\index{Costantino, S.}\index{Cleve, R.}\index{Chuang, I.} 
Since then, experimental progress on a number of different technologies has been steady but slow. 
The most advanced implementations currently use superconducting qubits and trapped-ion qubits, and typically act on 100--200 qubits.
The arguably first actual quantum computation that is beyond current-day classical computers, was Google's ``quantum supremacy'' experiment~\cite{google:supremacy}. This  performs a complicated but rather useless sampling task on 53 superconducting qubits. This task (especially in later versions of the experiment with more qubits) appears to be no longer simulatable in a reasonable amount of the time on even the largest existing classical supercomputer.

The practical problems facing physical realizations of quantum
computers seem formidable. The problems of noise and decoherence
have to some extent been solved in theory by the discovery 
of quantum error-correcting codes and fault-tolerant computing\index{error-correction}
(see, e.g., Chapter~\ref{chap:qec} in these notes).
Experiments with small fault-tolerant quantum computations have also been done in the last few years, and they behave as predicted, but scaling this up to thousands or millions of qubits is an enormous task.

On the other hand, we should realize that the field of physical 
realization of quantum computing is still in its infancy and
that classical computing had to face and solve many formidable
technical problems as well in the course of many decades---interestingly, often these problems
were even of the same nature as those now faced by quantum computing
(e.g., noise-reduction and error-correction). 
Moreover, while the difficulties facing the implementation of a full
quantum computer may seem daunting, more limited applications involving
quantum \emph{communication} have already been implemented with some success, 
for example teleportation (which is the process of sending qubits using 
entanglement and \emph{classical} communication), and 
versions of BB84 quantum key distribution\index{quantum cryptography} are nowadays even commercially available.\index{teleportation} 

Even if the theory of quantum computing
never materializes to a real large-scale physical computer,  
quantum-mechanical computers are still an extremely interesting
idea which will bear fruit in other areas than practical fast computing.
On the physics side, it may improve our understanding of quantum mechanics.
The emerging theories of entanglement and of Hamiltonian complexity 
have already done this to some extent.
On the computer science side, the theory of quantum computation
generalizes and enriches classical complexity theory and may 
help resolve some of its problems (see Section~\ref{sec:ldclowerbound} for an example).

\section{Quantum mechanics}\label{secqm}

Here we give a brief and abstract introduction to quantum mechanics.\index{quantum mechanics}
In short: a quantum state is a \emph{superposition} of classical states, written as a vector of amplitudes, to which we can apply either a \emph{measurement} or a \emph{unitary operation}.
For the required linear algebra we refer to Appendix~\ref{chaplinalg}.

\subsection{Superposition}\index{superposition}
Consider some physical system that can be in $N$ different, mutually
exclusive classical states. Because we will typically start counting from~0 in these notes, we call these states $\ket{0},\ket{1},\ldots,\ket{N-1}$.
Roughly, by a ``classical'' state we mean a state in which the system can
be found if we observe it.
A \emph{pure quantum state}\index{quantum state!pure} (usually just called \emph{state})
$\ket{\phi}$ is a \emph{superposition} of classical states, written
$$
\ket{\phi}=\alpha_0\ket{0}+\alpha_1\ket{1}+\cdots+\alpha_{N-1}\ket{N-1}.
$$
Here $\alpha_i$ is a complex number that is called the \emph{amplitude}\index{amplitude}
of $\ket{i}$ in $\ket{\phi}$. 
Intuitively, a system in quantum state $\ket{\phi}$ is ``in \emph{all} classical states \emph{at the same time},'' each state having a certain amplitude.
It is in state $\ket{0}$ with amplitude~$\alpha_0$, 
in state $\ket{1}$ with amplitude~$\alpha_1$, and so on.
Mathematically, the states $\ket{0},\ldots,\ket{N-1}$ form an orthonormal
basis of an $N$-dimensional \emph{Hilbert space}\index{Hilbert space} 
(i.e., an $N$-dimensional vector space equipped with an inner product).
A quantum state $\ket{\phi}$ is a vector in this space, usually written as an $N$-dimensional column vector of its amplitudes:
$$
\ket{\phi}=\left(\begin{array}{c}
\alpha_0\\
\vdots\\
\alpha_{N-1}
\end{array}\right).
$$
Such a vector is sometimes called a ``ket.''
Its conjugate transpose is the following row vector, sometimes called a ``bra'':
$$
\bra{\phi}=\left(\alpha_0^*,\dots,\alpha_{N-1}^*\right).
$$
The reason for this terminology (often called ``Dirac notation'' after Paul Dirac) is that an inner product $\inp{\phi}{\psi}$ between two states corresponds to the dot product between a bra and a ket vector (``bracket''): $\inp{\phi}{\psi}=\bra{\phi}\cdot\ket{\psi}$.

We can combine different Hilbert spaces using tensor product: if $\ket{0},\ldots,\ket{N-1}$ are an orthonormal basis of space ${\cal H}_A$ and $\ket{0},\ldots,\ket{M-1}$ are an orthonormal basis of space ${\cal H}_B$, then the tensor product space ${\cal H}={\cal H}_A\otimes{\cal H}_B$ is an $NM$-dimensional space spanned by the set of states $\{\ket{i}\otimes\ket{j} \mid i\in\{0,\ldots,N-1\}, j\in\{0,\ldots,M-1\}\}$. An arbitrary state in $\cal H$ is of the form $\sum_{i=0}^{N-1}\sum_{j=0}^{M-1}\alpha_{ij}\ket{i}\otimes\ket{j}$. Such a state is called \emph{bipartite}.
Similarly we can have \emph{tripartite} states that ``live'' in a Hilbert space that is the tensor product of three smaller Hilbert spaces, etc.

There are two things we can do with a quantum state: 
measure it or let it evolve unitarily without measuring it.
We will deal with measurement first.

\subsection{Measurement}\label{ssecmeasurement}\index{measurement}

\subsubsection{Measurement in the computational basis}
Suppose we measure state $\ket{\phi}$.
We cannot ``see'' a superposition itself, but only classical states. 
Accordingly, if we measure state $\ket{\phi}$ we will see one and only 
one classical state $\ket{j}$. Which specific $\ket{j}$ will we see?
This is not determined in advance; the only thing we can say is that we
will see state $\ket{j}$ with probability $|\alpha_j|^2$,
which is the squared norm of the corresponding amplitude $\alpha_j$.
This is known as ``Born's rule.''
Accordingly, observing a quantum state induces a probability distribution 
on the classical states, given by the squared norms of the amplitudes.
This implies $\sum_{j=0}^{N-1}|\alpha_j|^2=1$, so the vector
of amplitudes has (Euclidean) norm~1.
If we measure $\ket{\phi}$ and get outcome $j$ as a result\footnote{Don't use the ambiguous phrase ``we measure $j$'' in this case, since it's not clear in that phrasing whether $\ket{j}$ is the state you're applying the measurement to, or the outcome of the measurement.}, then $\ket{\phi}$ itself has ``disappeared,'' and all that is left is $\ket{j}$.
In other words, observing $\ket{\phi}$ ``collapses'' the quantum 
superposition $\ket{\phi}$ to the classical state $\ket{j}$ that we saw, 
and all ``information'' that might have been contained 
in the amplitudes $\alpha_i$ is gone.
Note that the probabilities of the various measurement outcomes are exactly the same when we measure the state $\ket{\phi}$ or when we measure the state $e^{i\theta}\ket{\phi}$; because of this we sometimes say that the ``global phase'' $e^{i\theta}$ has no physical significance.

\subsubsection{Projective measurement}
For most of the topics in these notes, the above ``measurement in the computational (or standard) basis'' suffices. However, somewhat more general kinds of measurement than the above are possible and sometimes useful.
The remainder of this subsection may be skipped on a first reading, but will become more relevant in the later parts of these notes, starting from Chapter~\ref{chap:qencodings}.

A \emph{projective} measurement\index{measurement!orthogonal} on some space, with $m$ possible outcomes, is 
a collection of projectors $P_1,\ldots,P_m$ that all act on that same space and that sum to identity, $\sum_{j=1}^m P_j=I$.%
\footnote{The $m$ projectors together form \emph{one} measurement; don't use the word ``measurement'' for individual $P_j$s.}
These projectors are then pairwise orthogonal, meaning that $P_iP_j=0$ if $i\neq j$.
The projector~$P_j$ projects on some subspace $V_j$ of
the total Hilbert space $V$, and every state $\ket{\phi}\in V$ can be 
decomposed in a unique way as $\ket{\phi}=\sum_{j=1}^m\ket{\phi_j}$, 
with $\ket{\phi_j}=P_j\ket{\phi}\in V_j$.
Because the projectors are orthogonal, the subspaces $V_j$ are
orthogonal as well, as are the states $\ket{\phi_j}$.
When we apply this measurement to the pure state $\ket{\phi}$, 
then we will get outcome $j$ with probability 
$\norm{\ket{\phi_j}}^2=\Tr(P_j\ketbra{\phi}{\phi})=\bra{\phi}P_j\ket{\phi}$
and the measured state will then ``collapse'' to the new state 
$\ket{\phi_j}/\norm{\ket{\phi_j}}=P_j\ket{\phi}/\norm{P_j\ket{\phi}}$.\footnote{Don't confuse the \emph{outcome} of the measurement, which is the label $j$ of the projector $P_j$ that was applied, and the \emph{post-measurement state}, which is $P_j\ket{\phi}/\norm{P_j\ket{\phi}}$.}
The probabilities sum to~1 thanks to our assumption that $\sum_{j=1}^m P_j=I$ and the fact that trace is a linear function:
\[
\sum_{j=1}^m \Tr(P_j\ketbra{\phi}{\phi})=
 \Tr((\sum_{j=1}^m P_j)\ketbra{\phi}{\phi})=\Tr(\ketbra{\phi}{\phi})=\inp{\phi}{\phi}=1.
\]
Note carefully that we cannot \emph{choose} which $P_j$ will be applied to the state but can only give a probability distribution.
However, if the state $\ket{\phi}$ that we measure lies fully within one of the subspaces~$V_j$, then the measurement outcome will be that~$j$ with certainty.

For example, a measurement in the computational basis on an $N$-dimensional state is the specific
projective measurement where $m=N$ and $P_j=\ketbra{j}{j}$.
That is, $P_j$ projects onto the computational basis state $\ket{j}$
and the corresponding subspace $V_j\subseteq V$ is the 1-dimensional subspace spanned by $\ket{j}$.
Consider the state $\ket{\phi}=\sum_{j=0}^{N-1}\alpha_j\ket{j}$.
Note that $P_j\ket{\phi}=\alpha_j\ket{j}$, so applying our measurement
to $\ket{\phi}$ will give outcome~$j$ with probability 
$\norm{\alpha_j\ket{j}}^2=|\alpha_j|^2$, and in that case the state 
collapses to $\alpha_j\ket{j}/\norm{\alpha_j\ket{j}}=
\frac{\alpha_j}{|\alpha_j|}\ket{j}$.
The norm-1 factor $\frac{\alpha_j}{|\alpha_j|}$ may be disregarded
because it has no physical significance, 
so we end up with the state $\ket{j}$ as we saw before.

Instead of the standard orthonormal basis, with basis states $\ket{0},\ldots,\ket{N-1}$, we may consider any other orthonormal basis $B$ of states $\ket{\psi_0},\ldots,\ket{\psi_{N-1}}$, and consider the projective measurement defined by the projectors $P_j=\ketbra{\psi_j}{\psi_j}$. This is called ``measuring in basis~$B$.'' Applying this measurement to state $\ket{\phi}$ gives outcome $j$ with probability $\bra{\phi}P_j\ket{\phi}=|\inp{\phi}{\psi_j}|^2$. Note that if $\ket{\phi}$ equals one of the basis vectors $\ket{\psi_j}$, then the measurement gives that outcome~$j$ with probability~1.

In the previous two examples the projectors had rank~1 (i.e., project on 1-dimensional subspaces), but this is not necessary. For example, 
a measurement that distinguishes between
$\ket{j}$ with $j<N/2$  and $\ket{j}$ with $j\geq N/2$
corresponds to the two projectors $P_1=\sum_{j<N/2}\ketbra{j}{j}$ 
and $P_2=\sum_{j\geq N/2}\ketbra{j}{j}$, each of rank~$N/2$ (assume $N$ is even). 
Applying this measurement to the state 
$\ket{\phi}=\frac{1}{\sqrt{3}}\ket{0}+\sqrt{\frac{2}{3}}\ket{N-1}$
gives outcome~1 with probability $\norm{P_1\ket{\phi}}^2=1/3$,
in which case the state collapses to $\ket{0}$.
It gives outcome~2 with probability $\norm{P_2\ket{\phi}}^2=2/3$, 
the state then collapses to~$\ket{N-1}$.

\subsubsection{Observables}
A projective measurement with projectors $P_1,\ldots,P_m$
and associated distinct outcomes $\lambda_1,\ldots,\lambda_m\in\mathbb{R}$, can be written as one matrix $M=\sum_{i=1}^m \lambda_i P_i$, which is called an \emph{observable}.  This is a succinct way of writing down the projective measurement
in one matrix, and has the added advantage that the \emph{expected value} of the outcome can be easily calculated:
if we are measuring a state $\ket{\phi}$, then the probability of outcome $\lambda_i$ is 
$\norm{P_i\ket{\phi}}^2=\Tr(P_i\ketbra{\phi}{\phi})$, so the expected value of the outcome
is $\sum_{i=1}^m \lambda_i\Tr(P_i\ketbra{\phi}{\phi})=\Tr(\sum_{i=1}^m \lambda_iP_i\ketbra{\phi}{\phi})=\Tr(M\ketbra{\phi}{\phi})$.
Note that $M$ is Hermitian: $M=M^*$.
Conversely, since every Hermitian $M$ has a spectral decomposition $M=\sum_{i=1}^m \lambda_i P_i$, there is a direct correspondence between observables and Hermitian matrices.

The Pauli matrices $I,X,Y,Z$ (see Appendix~\ref{app:pauli}) are examples of 2-dimensional observables, with eigenvalues~$\pm 1$. For example, $Z=\ketbra{0}{0}-\ketbra{1}{1}$ corresponds to measurement in the computational basis (with measurement outcomes $+1$ and $-1$ for $\ket{0}$ and $\ket{1}$, respectively). 

Suppose we have a bipartite state. An observable $A$ on the first part of the state corresponds to an observable $A\otimes I$ on the bipartite state.
Similarly, an observable $B$ on the second part of the state corresponds to an observable $I\otimes B$ on the bipartite state.
Separately measuring observables $A$ and $B$ on the two parts of a bipartite state is different from measuring the joint observable $A\otimes B$: the separate measurements give one outcome each, while the joint measurement gives only one outcome, and the distribution on the post-measurement state may be different. What is true, however, is that the measurement statistics of the product of outcomes is the same as the measurement statistics of the outcome of the joint measurement.
    For example consider the case when $A=B=Z$ (these correspond to measurement in the computational basis), and the bipartite state is $\ket{\phi}=\frac{1}{\sqrt{2}}(\ket{0}\otimes\ket{0}+\ket{1}\otimes\ket{1})$. With the separate measurements, the outcomes will be $++$ or $--$ (note that in both cases the product of the two outcomes is $+1$) and the state $\ket{\phi}$ will collapse to either $\ket{0}\otimes\ket{0}$ or $\ket{1}\otimes\ket{1}$. Yet $\ket{\phi}$ remains undisturbed by a joint measurement with $\pm 1$-valued observable $Z\otimes Z$, because $\ket{\phi}$ is a $+1$-eigenstate of $Z\otimes Z$.

\subsubsection{POVM measurement}
If we only care about the final probability distribution on the $m$ outcomes, 
not about the resulting post-measurement state, then the most general type of measurement we can do is a so-called \emph{positive-operator-valued measure} (POVM).  
This is specified by $m$ positive semidefinite (psd) matrices
$E_1,\ldots,E_m$ that sum to identity. When measuring a state $\ket{\phi}$, the probability of outcome $i$ is given by $\Tr(E_i\ketbra{\phi}{\phi})$. A projective measurement is the special case of a POVM where the measurement elements $E_i$ are projectors.\footnote{Note that if $E_i$ is a projector, then $\Tr(E_i\ketbra{\phi}{\phi})=\Tr(E_i^2\ketbra{\phi}{\phi})=\Tr(E_i\ketbra{\phi}{\phi}E_i)=\norm{E_i\ket{\phi}}^2$, using the fact that $E_i=E_i^2$ and the cyclic property of the trace. These equalities can fail if $E_i$ is psd but not a projector.}

There are situations where a POVM can do things a projective measurement cannot do.\footnote{Even though POVMs strictly generalize projective measurements, one can show that every POVM can be ``simulated'' by a projective measurement on a slightly larger space that yields the exact same probability distribution over measurement outcomes (this follows from Neumark's theorem).} 
For example, suppose you have a state in a 2-dimensional space, and you know it is either in state $\ket{0}$ or in state $\ket{+}=\frac{1}{\sqrt{2}}(\ket{0}+\ket{1})$. These two states are not orthogonal, so there is no measurement that distinguishes them perfectly. However, there is a POVM measurement that never makes a mistake, but sometimes gives another outcome 2, meaning ``I don't know.'' That is, you would like to do a measurement with three possible outcome: 0, 1, and 2, such that:
\begin{itemize}
\item If the state is $\ket{0}$, then you get correct outcome 0 with probability 1/4, and outcome 2 with probability 3/4, but never get incorrect outcome 1.
\item If the state is $\ket{+}$, then you get correct outcome 1 with probability 1/4, and outcome 2 with probability 3/4, but never get incorrect outcome 0.
\end{itemize}
You cannot achieve this with a projective measurement on the qubit, but the following 3-outcome POVM does the job:
\begin{quote}
$E_0=\frac{1}{2}\ketbra{-}{-}$ (where $\ket{-}=\frac{1}{\sqrt{2}}(\ket{0}-\ket{1})$, which is orthogonal to the $\ket{+}$ state);\\
$E_1=\frac{1}{2}\ketbra{1}{1}$ (note that this is orthogonal to the $\ket{0}$ state);\\
$E_2=I-E_0-E_1.$
\end{quote}
You can check that $E_0,E_1,E_2$ are psd and add up to identity, so they form a valid POVM. None of the 3 matrices is a projector.
The success probability 1/4 can be improved further, see Exercise~\ref{ex:povmvsprojective}.

\subsection{Unitary evolution}\index{matrix!unitary}
Instead of measuring $\ket{\phi}$, we can also apply some operation to it,
i.e., change the state to some
$$
\ket{\psi}=\beta_0\ket{0}+\beta_1\ket{1}+\cdots+\beta_{N-1}\ket{N-1}.
$$
Quantum mechanics only allows \emph{linear} operations to be applied to 
quantum states in a closed system. What this means is: if we view a state like $\ket{\phi}$ as an
$N$-dimensional vector $(\alpha_0,\ldots,\alpha_{N-1})^T$, then applying
an operation that changes $\ket{\phi}$ to $\ket{\psi}$
corresponds to multiplying $\ket{\phi}$ with an $N\times N$ 
complex-valued matrix $U$:
$$
U
\left(\begin{array}{c}
\alpha_0\\
\vdots\\
\alpha_{N-1}
\end{array}\right)
=
\left(\begin{array}{c}
\beta_0\\
\vdots\\
\beta_{N-1}
\end{array}\right).
$$
Note that by linearity we have 
$\ket{\psi}=U\ket{\phi}=U\left(\sum_i\alpha_i\ket{i}\right)=\sum_i\alpha_iU\ket{i}$.

Because measuring $\ket{\psi}$ should also give a probability
distribution, we have the constraint $\sum_{j=0}^{N-1}|\beta_j|^2=1$ on the new state.
This implies that the operation $U$ must preserve the
norm of vectors, and hence must be a \emph{unitary} transformation (often just called ``a unitary'').
A matrix $U$ is \emph{unitary} if its inverse $U^{-1}$
equals its conjugate transpose $U^*$. This is equivalent to saying that
$U$ always maps a vector of norm~1 to a vector of norm~1.
Because a unitary transformation always has an inverse, it follows that
any (non-measuring) operation on quantum states must be reversible:\index{reversible computing}
by applying $U^{-1}$ we can always ``undo'' the action of $U$, 
and nothing is lost in the process.
On the other hand, a measurement is clearly non-reversible, because
we cannot reconstruct $\ket{\phi}$ from the observed classical state $\ket{j}$.

\section{Qubits and quantum memory}\label{secqmem}

In classical computation, the unit of information is a \emph{bit}, 
which can be 0 or 1. In \emph{quantum} computation, this unit is 
a \emph{quantum} bit (\emph{qubit})\index{qubit}, 
which is a superposition of 0 and~1.
Consider a system with 2 basis states, call them $\ket{0}$ and $\ket{1}$.
We identify these basis states with the two orthogonal vectors 
$\left(\begin{array}{c}1\\ 0\end{array}\right)$ and
$\left(\begin{array}{c}0\\ 1\end{array}\right)$, respectively.
A single qubit can be in any superposition
$$
\alpha_0\ket{0}+\alpha_1\ket{1}, \ |\alpha_0|^2+|\alpha_1|^2=1.
$$
Accordingly, a single qubit ``lives'' in the vector space $\mathbb{C}^2$.

Similarly we can think of systems of more than 1 qubit, which ``live''
in the tensor product space\index{tensor product} of several qubit systems.
For instance, a 2-qubit system has 4 basis states: 
$\ket{0}\otimes\ket{0}$, $\ket{0}\otimes\ket{1}$, 
$\ket{1}\otimes\ket{0}$, $\ket{1}\otimes\ket{1}$.
Here for instance $\ket{1}\otimes\ket{0}$ means that the first qubit is in 
its basis state $\ket{1}$ and the second qubit
is in its basis state $\ket{0}$. 
We will often abbreviate this to
$\ket{1}\ket{0}$, $\ket{1,0}$, or even $\ket{10}$.

More generally, a register of $n$ qubits has $2^n$ basis states,
each of the form $\ket{b_1}\otimes\ket{b_2}\otimes\ldots\otimes\ket{b_n}$, 
with $b_i\in\{0,1\}$.
We can abbreviate this to $\ket{b_1b_2\ldots b_n}$.
We will often abbreviate $0\ldots 0$ to $0^n$.
Since bitstrings of length $n$ can be viewed as integers between
0 and $2^n-1$ (see Appendix~\ref{appusefulCS}),
we can also write the basis states as numbers
$\ket{0},\ket{1},\ket{2},\ldots,\ket{2^n-1}$.
Note that the vector corresponding to $n$-qubit basis state $\ket{x}$ is the $2^n$-dimensional vector that has a 1 at the $x$-th position and 0s elsewhere (here we view $x$ as an integer in $\{0,\ldots,2^n-1\}$ and we count the positions in the vector starting from position~0). This implies that two $n$-qubit basis states $\ket{x}$ and $\ket{y}$ are orthogonal iff $x\neq y$. A different way to see this orthogonality is to use the rules of tensor product (Appendix~\ref{app:tensor}):
$$
\inp{x}{y}=\inp{x_1}{y_1}\otimes\cdots\otimes\inp{x_n}{y_n}=\inp{x_1}{y_1}\cdots\inp{x_n}{y_n}.
$$
Since $\inp{x_k}{y_k}=\delta_{x_k,y_k}$, we see that basis states $\ket{x}$ and $\ket{y}$ will be orthogonal as soon as there is at least one position $k$ at which the bits of $x$ and $y$ differ.

A quantum register of $n$ qubits can be in any superposition\footnote{Don't call such a multi-qubit state or register a ``qubit'' or an ``$n$-qubit''---the term ``qubit'' only refers to the state of a 2-dimensional system. You can use ``$n$-qubit'' as an adjective but not as a noun.}
$$
\alpha_0\ket{0}+\alpha_1\ket{1}+\cdots+\alpha_{2^n-1}\ket{2^n-1}, \ 
\sum_{j=0}^{2^n-1}|\alpha_j|^2=1.
$$
Measuring this in the computational basis, we obtain the $n$-bit state $\ket{j}$ with probability $|\alpha_j|^2$.

Measuring just the first qubit of a state would correspond 
to the projective measurement\index{measurement!orthogonal} 
that has the two projectors $P_0=\ketbra{0}{0}\otimes I_{2^{n-1}}$ 
and $P_1=\ketbra{1}{1}\otimes I_{2^{n-1}}$.
For example, applying this measurement to the state
$\frac{1}{\sqrt{3}}\ket{0}\ket{\phi}+\sqrt{\frac{2}{3}}\ket{1}\ket{\psi}$
gives outcome 0 with probability 1/3; the state then becomes
$\ket{0}\ket{\phi}$. We get outcome 1 with probability 2/3; the state
then becomes $\ket{1}\ket{\psi}$.
Similarly, measuring the first $n$ qubits of an $(n+m)$-qubit state 
in the computational basis corresponds to the projective measurement
that has $2^n$ projectors $P_j=\ketbra{j}{j}\otimes I_{2^{m}}$ for $j\in\01^n$.

An important property that deserves to be mentioned is 
\emph{entanglement},\index{entanglement}
which refers to quantum correlations between different qubits. 
For instance, consider a 2-qubit register that is in the state
$$
\frac{1}{\sqrt{2}}\ket{00}+\frac{1}{\sqrt{2}}\ket{11}.
$$
Such 2-qubit states are sometimes called \emph{EPR-pairs}\index{EPR-pair}
in honor of Einstein, Podolsky, and Rosen~\cite{epr},\index{Einstein, A.}\index{Podolsky, B.}\index{Rosen, N.}
who examined such states and their seemingly paradoxical properties.
Initially neither of the two qubits has a classical value
$\ket{0}$ or $\ket{1}$. 
However, if we measure the first qubit and observe, say,
a $\ket{0}$, then the whole state collapses to $\ket{00}$.
Thus observing the first qubit immediately fixes also 
the second, unobserved qubit to a classical value. 
Since the two qubits that make up the register may be far apart, this example 
illustrates some of the non-local effects that quantum systems can exhibit.
In general, a bipartite state $\ket{\phi}$ is called \emph{entangled}
if it cannot be written as a tensor product $\ket{\phi_A}\otimes\ket{\phi_B}$
where $\ket{\phi_A}$ lives in the first space and $\ket{\phi_B}$ lives in 
the second.\footnote{We often omit the tensor product symbol for such unentangled states, abbreviating $\ket{\phi_A}\otimes\ket{\phi_B}$ to $\ket{\phi_A}\ket{\phi_B}$ (you shouldn't abbreviate this further to $\ket{\phi_A\phi_B}$ though, unless both $\ket{\phi_A}$ and $\ket{\phi_B}$ are computational basis states). Note that there cannot be ambiguity between tensor product and the usual matrix product in this abbreviation, because both $\ket{\phi_A}$ and $\ket{\phi_B}$ are column vectors and hence their matrix product wouldn't even be well-defined (the dimensions ``don't fit'').}

At this point, a comparison with classical probability 
distributions\index{probability distribution} may be helpful. 
Suppose we have two probability spaces, $A$ and $B$,
the first with $2^n$ possible outcomes, the second with $2^m$ possible
outcomes. A probability distribution on the first space can be described by $2^n$
numbers (nonnegative reals summing to 1; actually there are only 
$2^n-1$ degrees of freedom here) and a distribution on the
second by $2^m$ numbers. Accordingly, a \emph{product} distribution 
on the joint space can be described by $2^n+2^m$ numbers.
However, an arbitrary (non-product) distribution on the joint space takes 
$2^{n+m}$ real numbers, since there are $2^{n+m}$ possible outcomes in total.
Analogously, an $n$-qubit state $\ket{\phi_A}$ can be described by $2^n$ 
numbers (complex numbers whose squared moduli sum to 1), an $m$-qubit state 
$\ket{\phi_B}$ by $2^m$ numbers, and their tensor product
$\ket{\phi_A}\otimes\ket{\phi_B}$ by $2^n+2^m$ numbers.
However, an arbitrary (possibly entangled) state in the joint space
takes $2^{n+m}$ numbers, since it lives in a $2^{n+m}$-dimensional space.
We see that the number of parameters required to describe quantum 
states is the same as the number of parameters needed 
to describe probability distributions.
Also note the analogy between statistical independence\footnote{Two random variables $A$ and $B$ are \emph{independent} if their joint probability distribution can be written as a product of individual distributions for~$A$ and for~$B$: $\Pr[A=a\wedge B=b]=\Pr[A=a]\cdot\Pr[B=b]$ for all possible values $a,b$.} of 
two random variables $A$ and $B$ and non-entanglement 
of the product state $\ket{\phi_A}\otimes\ket{\phi_B}$.
However, despite the similarities between probabilities and amplitudes, 
quantum states are much more powerful than distributions,
because amplitudes may have negative (or even complex) parts which can lead 
to \emph{interference} effects.\index{interference} 
Amplitudes only become probabilities when we square them.
The art of quantum computing is to use these special
properties for interesting computational purposes.

\section{Elementary gates}
A unitary that acts on a small number of qubits (say, at most~3) is often called a \emph{gate}, in analogy to classical logic gates like AND, OR, and NOT; more about that in the next chapter.
The Pauli matrices $I,X,Y,Z$ (Appendix~\ref{app:pauli}) are examples of 1-qubit gates. For example, the bitflip gate~$X$ (a.k.a.\ NOT-gate) negates the bit in the computational basis, i.e., it swaps $\ket{0}$ and $\ket{1}$. The phaseflip gate $Z$ puts a $-$ in front of $\ket{1}$. Represented as $2\times 2$ unitary matrices, these are
$$
X=\left(
\begin{array}{rr}
0 & 1\\ 
1 & 0
\end{array}
\right)
\mbox{ , \, }
Z=\left(
\begin{array}{rr}
1 & 0\\ 
0 & -1
\end{array}
\right).
$$
Another important 1-qubit gate is the \emph{phase gate} $R_{\phi}$,\index{phase gate}
which merely rotates the phase of the $\ket{1}$-state by an angle $\phi$:
$$
\begin{array}{l}
R_{\phi}\ket{0}=\ket{0}\\
R_{\phi}\ket{1}=e^{i\phi}\ket{1}
\end{array}
$$
This corresponds to the unitary matrix
$$
R_\phi=\left(
\begin{array}{rr}
1 & 0\\ 
0 & e^{i\phi}
\end{array}
\right).
$$
Note that $Z$ is a special case of this: $Z=R_\pi$, because $e^{i\pi}=-1$.
The $R_{\pi/4}$-gate is often just called the $T$-gate.

Possibly the most important 1-qubit gate is the \emph{Hadamard}
transform\index{Hadamard gate}, specified by:
$$
\begin{array}{l}
\displaystyle H\ket{0}=\frac{1}{\sqrt{2}}\ket{0}+\frac{1}{\sqrt{2}}\ket{1}\\
\displaystyle H\ket{1}=\frac{1}{\sqrt{2}}\ket{0}-\frac{1}{\sqrt{2}}\ket{1}
\end{array}
$$
As a unitary matrix, this is represented as
$$
H=\frac{1}{\sqrt{2}}\left(
\begin{array}{rr}
1 & 1\\ 
1 & -1
\end{array}
\right).
$$
If we apply $H$ to initial state $\ket{0}$ and then measure, we have equal
probability of observing $\ket{0}$ or $\ket{1}$.  Similarly, applying $H$
to $\ket{1}$ and observing gives equal probability of $\ket{0}$ or $\ket{1}$.
However, if we apply $H$ to the superposition 
$\frac{1}{\sqrt{2}}\ket{0}+\frac{1}{\sqrt{2}}\ket{1}$
then we obtain
$$
H(\frac{1}{\sqrt{2}}\ket{0}+\frac{1}{\sqrt{2}}\ket{1})=\frac{1}{\sqrt{2}}H\ket{0}+\frac{1}{\sqrt{2}}H\ket{1}=\frac{1}{2}(\ket{0}+\ket{1})+\frac{1}{2}(\ket{0}-\ket{1})=\ket{0}.
$$
The positive and negative amplitudes for $\ket{1}$ have canceled each other out! 
This effect is called \emph{interference},\index{interference}
and is analogous to interference patterns between light or sound waves.

An example of a 2-qubit gate is the \emph{controlled-not} gate $\CNOT$.\index{controlled-not gate}
It negates the second bit of its input if the first bit is 1,
and does nothing if the first bit is~0:
$$
\begin{array}{l}
\CNOT\ket{0}\ket{b}=\ket{0}\ket{b}\\
\CNOT\ket{1}\ket{b}=\ket{1}\ket{1-b}
\end{array}
$$
The first qubit is called the \emph{control} qubit, the second the \emph{target} qubit.
In matrix form, this is
$$
\CNOT=\left(
\begin{array}{rrrr}
1 & 0 & 0 & 0\\
0 & 1 & 0 & 0\\
0 & 0 & 0 & 1\\
0 & 0 & 1 & 0
\end{array}
\right).
$$
More generally, if $U$ is some $n$-qubit unitary matrix, then the controlled-$U$ operation corresponds to the following $2^{n+1}\times 2^{n+1}$ unitary matrix:
$$
\left(
\begin{array}{cc}
I & 0\\
0 & U
\end{array}
\right),
$$
where $I$ is the $2^n$-dimensional identity matrix and the  two $0$s denote $2^n\times 2^n$ all-0 matrices.

\section{Example: quantum teleportation}\label{sec:teleport}

In the next chapter we will look in more detail at how we can use and combine such elementary gates, but as an example we will here already explain \emph{teleportation}~\cite{teleporting}. 
Suppose there are two parties, Alice and Bob.
Alice has a qubit $\alpha_0\ket{0}+\alpha_1\ket{1}$ that 
she wants to send to Bob via a \emph{classical} channel. 
Without further resources this would be impossible, intuitively because the amplitudes $\alpha_0,\alpha_1$ may require an infinite number of bits of precision to write them down exactly.
However, suppose Alice also shares an EPR-pair 
$$
\frac{1}{\sqrt{2}}(\ket{00}+\ket{11})
$$ 
with Bob (say Alice holds the left qubit and Bob the right qubit\footnote{Some students, on first encountering an EPR-pair, think that Alice holds $\frac{1}{\sqrt{2}}\ket{00}$ and Bob holds $\frac{1}{\sqrt{2}}\ket{11}$. This interpretation doesn't really make sense. For the right intuition, think of the analogous classical case where there is a random coin flip (with value 0 or 1), and Alice and Bob each have a copy of this value. Then the probabilistic joint state is 00 with probability 1/2 (Alice and Bob each hold a bit with value 0), and 11 with probability 1/2 (each holds a bit with value 1). The EPR-pair is similar, but now with amplitudes $1/\sqrt{2}$ instead of probabilities 1/2.}). 
Initially, their joint state is
$$
(\alpha_0\ket{0}+\alpha_1\ket{1})\otimes\frac{1}{\sqrt{2}}(\ket{00}+\ket{11}).
$$
The first two qubits belong to Alice, the third to Bob.
Alice performs a CNOT on her two qubits and then a Hadamard transform
on her first qubit. Their joint 3-qubit state can now be written as
$$
\begin{array}{l}
\frac{1}{2} \ \ket{00}(\alpha_0\ket{0}+\alpha_1\ket{1}) \ +\\
\frac{1}{2} \ \ket{01}(\alpha_0\ket{1}+\alpha_1\ket{0}) \ +\\
\frac{1}{2} \ \ket{10}(\alpha_0\ket{0}-\alpha_1\ket{1}) \ +\\
\frac{1}{2}\underbrace{\ket{11}}_{Alice}\underbrace{(\alpha_0\ket{1}-\alpha_1\ket{0})}_{Bob}.
\end{array}
$$
Alice then measures her two qubits in the computational basis 
and sends the result (2 random classical bits $ab$) to Bob over a classical channel. 
Bob now knows which transformation he must do on his qubit 
in order to regain the qubit $\alpha_0\ket{0}+\alpha_1\ket{1}$.
First, if $b=1$ then he applies a bitflip ($X$-gate) on his qubit; second if $a=1$ then he applies a phaseflip ($Z$-gate). For instance, if Alice sent $ab=11$, then Bob knows that his qubit is
$\alpha_0\ket{1}-\alpha_1\ket{0}$. 
A bitflip followed by a phaseflip will give him Alice's original qubit $\alpha_0\ket{0}+\alpha_1\ket{1}$. 
In fact, if Alice's qubit had been \emph{entangled} with some other qubits,
then teleportation preserves this entanglement: Bob then receives
a qubit that is entangled in the same way as Alice's original qubit was.

Note that the qubit on Alice's side has been destroyed: 
teleporting moves a qubit from Alice to Bob, rather than copying it.
In fact, \emph{copying an unknown qubit is impossible}~\cite{nocloning}, see Exercise~\ref{ex:nocloning}.

\section*{Exercises}

\begin{enumerate}

\item
  \begin{enumerate}
\item What is the inner product between the real vectors $(0,1,0,1)$ and $(0,1,1,1)$?
\item What is the inner product between the states $\ket{0101}$ and $\ket{0111}$?
\end{enumerate}

\item
Compute the result of applying a Hadamard transform to both
qubits of $\ket{0}\otimes\ket{1}$ in two ways (the first way using tensor product of vectors, the second using tensor product of matrices), and show that the two results are equal:
$$
H\ket{0}\otimes H\ket{1}=(H\otimes H)(\ket{0}\otimes\ket{1}).
$$

\item
Show that a bitflip operation, preceded and followed by Hadamard transforms, equals a phaseflip operation: $HXH=Z$.

\item
Show that surrounding a CNOT gate with Hadamard gates switches the role of the control-bit and target-bit of the CNOT: $(H\otimes H)\mbox{CNOT}(H\otimes H)$ is the 2-qubit gate where the second bit controls whether the first bit is negated (i.e., flipped).

\item 
Let 
\[
R_\theta=\left(\begin{array}{rr}\cos\theta & -\sin\theta\\
\sin\theta & \cos\theta
\end{array}\right)
\]
be the single-qubit \emph{rotation gate} that rotates the amplitude-vector counterclockwise over an angle~$\theta$. Show that $HR_{\theta}H=R_{-\theta}$, so conjugation by Hadamard reverses the direction.

\item 
\begin{enumerate}
    \item Show that 
$\frac{1}{2}\left(\begin{array}{cc}1+i & 1-i\\
1-i & 1+i
\end{array}\right)$ is unitary.
\item This gate
is sometimes called ``the square root of NOT.'' Explain why.
\end{enumerate}

\item Simplify the following: $(\bra{0}\otimes I)(\alpha_{00}\ket{00}+\alpha_{01}\ket{01}+\alpha_{10}\ket{10}+\alpha_{11}\ket{11})$.

\item Prove that an EPR-pair $\frac{1}{\sqrt{2}}\left(\ket{00}+\ket{11}\right)$ is an \emph{entangled} state,
  i.e., that it cannot be written as the tensor product of two separate qubits.

\item 
Suppose we have the state $\frac{1}{\sqrt{2}} (\ket{0}\ket{\phi}+\ket{1}\ket{\psi})$, where $\ket{\phi}$ and $\ket{\psi}$ are unknown normalized quantum states with the same number of qubits.
Suppose we apply a Hadamard gate to the first qubit and then measure that first qubit in the computational basis.
Give the probability of measurement outcome~1, as a function of the inner product between the states $\ket{\phi}$ and $\ket{\psi}$.

\item Give the 2-outcome projective measurement on a 2-qubit space that measures the parity (i.e., sum modulo 2) of 2-bit basis states. Also give the corresponding observable.
  
\item\label{ex:povmvsprojective}
(H) Show that the success probability of the POVM at the end of Section~\ref{ssecmeasurement} can be increased from 1/4 to $1/(2+\sqrt{2})$.

\item\label{ex:nocloning} 
(H) Prove the \emph{quantum no-cloning theorem}: there does not exist a 2-qubit unitary $U$ that maps
$$
\ket{\phi}\ket{0}\mapsto\ket{\phi}\ket{\phi}
$$
for every qubit $\ket{\phi}$.

\item Show that unitaries cannot ``delete'' information: there is no 1-qubit unitary $U$ that maps $\ket{\phi}\mapsto\ket{0}$ for every 1-qubit state $\ket{\phi}$.

\item\label{ex:superdense}
Suppose Alice and Bob are not entangled.  If Alice sends a qubit to Bob, then this can give Bob at most one bit of information about Alice.\footnote{This is actually a deep statement, a special case of \emph{Holevo's theorem}. More about this may be found in Chapter~\ref{chap:qencodings}.}  However, if they share an EPR-pair,  $\ket{\psi}=\frac{1}{\sqrt{2}}(\ket{00}+\ket{11})$, then they can transmit \emph{two} classical bits by sending one qubit over the channel; this is called \emph{superdense coding}. This exercise will show how this works.
\begin{enumerate}
\item They start with a shared EPR-pair, $\frac{1}{\sqrt{2}}(\ket{00}+\ket{11})$.
Alice has classical bits $a$ and $b$.  Suppose she does an $X$-gate on her half of the EPR-pair if $a=1$, followed by a $Z$-gate if $b=1$ (she does both if $ab=11$, and neither if $ab=00$).
Write the resulting 2-qubit state for the four different cases that $ab$ could take.
\item Suppose Alice sends her half of the state to Bob, who now has two qubits.
Show that Bob can determine both $a$ and $b$ from his state, using Hadamard and CNOT gates, followed by a measurement in the computational basis. After his final measurement, Bob's first qubit should be $\ket{a}$ and his second qubit should be $\ket{b}$.
\end{enumerate}

\item\label{ex:TrC}
Alice and Bob share an EPR-pair, $\ket{\psi}=\frac{1}{\sqrt{2}}(\ket{00}+\ket{11})$.
\begin{enumerate}
\item Let $C$ be a $2\times 2$ matrix. Show that $\Tr((C\otimes I)\ketbra{\psi}{\psi})=\frac{1}{2}\Tr(C)$.
\item  (H) Alice could apply one of the 4 Pauli matrices ($I,X,Y,Z$) to her qubit. Use part (a) to show that the 4 resulting 2-qubit states form an orthonormal set.
\item Suppose Alice applies one of the 4 Pauli matrices to her qubit and then sends that qubit to Bob. Give the 4 projectors of a 4-outcome projective measurement that Bob could do on his 2 qubits to find out which Pauli matrix Alice actually applied.
\end{enumerate}

\item
Let $\theta\in[0,2\pi)$, $U_\theta=\left(
\begin{array}{rr}
\cos\theta & -\sin\theta\\
\sin\theta & \cos\theta
\end{array}
\right)$, $\ket{\phi}=U_\theta\ket{0}$ and $\ket{\phi^\perp}=U_\theta\ket{1}$.
\begin{enumerate}
\item Show that $ZX\ket{\phi^\perp}=\ket{\phi}$.
\item Show that an EPR-pair, $\frac{1}{\sqrt{2}}(\ket{00}+\ket{11})$, can also be written as $\frac{1}{\sqrt{2}}(\ket{\phi}\ket{\phi}+\ket{\phi^\perp}\ket{\phi^\perp})$.
\item Suppose Alice and Bob start with an EPR-pair. Alice applies $U_\theta^{-1}$ to her qubit and then measures it in the computational basis. What pure state does Bob have if her outcome was~0, and what pure state does he have if her outcome was 1?
\item Suppose Alice knows the number $\theta$ but Bob does not. Give a protocol that uses one EPR-pair and 1 classical bit of communication where Bob ends up with the qubit $\ket{\phi}$ (in contrast to general teleportation of an unknown qubit, which uses 1 EPR-pair and \emph{2} bits of communication).
\end{enumerate}

\end{enumerate}

\chapter{The Circuit Model and the Deutsch-Jozsa Algorithm}\label{chap:qircuitdj}

\section{Quantum computation}\label{secqc}

Below we explain how a quantum computer can apply computational
steps to its register of qubits. Two models exist for this:
the quantum Turing machine~\cite{deutsch:uqc,bernstein&vazirani:qcomplexity}\index{quantum Turing machine}\index{Deutsch, D.}\index{Bernstein, E.}\index{Vazirani, U.}
and the quantum circuit\index{quantum circuit} model~\cite{deutsch:networks,yao:qcircuit}.\index{Deutsch, D.}\index{Yao, A.}
These models are equivalent, in the sense that they can simulate 
each other in polynomial time, assuming the circuits are 
appropriately ``uniform.''
We only explain the circuit model here, which is more popular among researchers.

\subsection{Classical circuits}
In classical complexity theory, a \emph{Boolean circuit}\index{Boolean circuit} 
is a finite directed acyclic graph with AND, OR, and NOT gates. 
It has $n$ input nodes, which contain the $n$ input bits ($n\geq 0$).
The internal nodes are AND, OR, and NOT gates, and there are one or more 
designated output nodes.  The initial input bits are fed into 
AND, OR, and NOT gates according to the circuit, and eventually the output 
nodes assume some value.
We say that a circuit \emph{computes} some Boolean function 
$f: \01^n\rightarrow\01^m$ if the output nodes get the right value $f(x)$
for every input $x\in\01^n$.

A \emph{circuit family}\index{circuit family} 
is a set ${\cal C}=\{C_n\}$ of circuits, one
for each input size $n$. Each circuit has one output bit.
Such a family \emph{recognizes} or \emph{decides} a \emph{language}
$L\subseteq\01^*=\cup_{n\geq 0}\01^n$ if, for every $n$ and every input $x\in\01^n$,
the circuit $C_n$ outputs 1 if $x\in L$ and outputs 0 otherwise.\footnote{We can think of a language~$L$ as a sequence of Boolean functions $f_n:\01^n\to\01$, where $f_n$ takes value~1 exactly on the $n$-bit strings that are in~$L$. The circuit $C_n$ then computes the function~$f_n$.}
Such a circuit family is \emph{uniformly polynomial} if there is 
a deterministic Turing machine\footnote{If you don't know what a Turing machine is, you can just think of a computer program in some fixed language, such as C or Python.} that outputs $C_n$ given $n$ as input, 
using space logarithmic in $n$.\footnote{Logarithmic space implies time that's at most polynomial in $n$, because such a machine will have only $\poly(n)$ different internal states, so it either halts after $\poly(n)$ steps or cycles forever.}
Note that the size (number of gates) of the circuits 
$C_n$ can then grow at most polynomially with $n$.
It is known that uniformly polynomial circuit families are equal in power to 
polynomial-time deterministic Turing machines: a language~$L$ can be 
decided by a uniformly polynomial circuit family iff 
$L\in\Pp$~\cite[Theorem~11.5]{papadimitriou:cc},\index{Papadimitriou, C.}\index{P}
where $\Pp$ is the class of languages decidable by polynomial-time
Turing machines.

Similarly we can consider \emph{randomized} circuits.\index{randomized circuit} 
These receive, in addition to the $n$ input bits, 
also some random bits (``coin flips'') as input. 
A randomized circuit computes a function $f$ if it successfully outputs 
the right answer $f(x)$ with probability at least $2/3$ for every $x$ 
(probability taken over the values of the random bits).
Randomized circuits are equal in power to randomized Turing machines: 
a language $L$ can be decided by a uniformly polynomial randomized circuit 
family iff $L\in\BPP$,\index{BPP} where $\BPP$ (``Bounded-error 
Probabilistic Polynomial time'') is the class of languages that can
efficiently be recognized by randomized Turing machines with success
probability at least $2/3$.
Because we can efficiently reduce the error probability of randomized algorithms (see Appendix~\ref{appusefulCS}), the particular value $2/3$ doesn't really matter here and may be replaced by any fixed constant in $(1/2,1)$.

\subsection{Quantum circuits}
A \emph{quantum circuit}\index{quantum circuit} 
(also called quantum network or quantum gate array)
generalizes the idea of classical circuit families, replacing the
AND, OR, and NOT gates by elementary \emph{quantum gates}\index{quantum gate}. 
A quantum gate is a unitary transformation on a small (usually 1, 2, or 3)
number of qubits. We saw a number of examples already in the previous chapter:
the bitflip gate $X$,  the phaseflip gate $Z$, the Hadamard gate~$H$.
The main 2-qubit gate we have seen is the controlled-NOT (CNOT) gate. 
Adding another control qubit, we get the 3-qubit \emph{Toffoli} gate, also called controlled-controlled-not (CCNOT) gate.
This negates the third bit of its input if \emph{both} of the first two bits are~1.
The Toffoli gate is important because it is complete for classical reversible computation:
any classical computation can be implemented by a circuit of Toffoli gates.
This is easy to see: using auxiliary wires with fixed values, Toffoli can implement AND (fix the 3rd ingoing wire to~0) 
and NOT (fix the 1st and 2nd ingoing wire to~1).
It is known that AND and NOT-gates together suffice to implement any classical Boolean circuit, so if we can apply (or simulate) Toffoli gates, we can implement any classical computation in a reversible manner.

Mathematically, such elementary quantum gates can be composed into bigger unitary operations by taking tensor products (if gates are applied \emph{in parallel} to different parts
of the register), and ordinary matrix products (if gates are applied \emph{sequentially}).
We have already seen a simple example of such a circuit of elementary gates in the previous chapter, namely to implement teleportation.

For example, if we apply the Hadamard gate~$H$ to each qubit in a register of $n$ zeroes,
we obtain
$$
\frac{1}{\sqrt{2^n}}\sum_{j\in\01^n}\ket{j},
$$
which is a superposition
of all $n$-bit strings. More generally, if we apply $H^{\otimes n}$
to an initial state $\ket{i}$, with $i\in\01^n$, we obtain
\begin{equation}\label{eqnhadamard}
H^{\otimes n}\ket{i}=\frac{1}{\sqrt{2^n}}\sum_{j\in\01^n}(-1)^{i\cdot j}\ket{j},
\end{equation}
where $i\cdot j=\sum_{k=1}^n i_kj_k$ denotes the inner product of 
the $n$-bit strings $i,j\in\01^n$.
For example: 
$$
H^{\otimes 2}\ket{01}=\frac{1}{\sqrt{2}}(\ket{0}+\ket{1})\otimes
\frac{1}{\sqrt{2}}(\ket{0}-\ket{1})=\frac{1}{2}\sum_{j\in\01^2}(-1)^{01\cdot j}\ket{j}.
$$
Note that Hadamard happens to be its own inverse (it's unitary and Hermitian, hence $H=H^*=H^{-1}$), so applying it once more on the right-hand side of the above equation would give us back $\ket{01}$.
The $n$-fold Hadamard transform will be very useful for 
quantum algorithms.

As in the classical case, a quantum circuit can be viewed as a finite directed 
acyclic graph of input nodes, gates, and output nodes. The graph needs some annotation to tell us the order of the qubits going into a gate.%
\footnote{E.g., what is the first qubit going into a particular CNOT gate and what is the second qubit. Such extra information is not necessary for classical circuits with symmetric gates such as AND, OR, and NOT, since the order of incoming bits doesn't affect the output bit there.}
There are $n$ nodes that contain the input (as classical bits); 
in addition we may have some more input nodes that are 
initially $\ket{0}$ (``workspace'').
The internal nodes of the quantum circuit are quantum gates
that each operate on at most two or three qubits of the state.
The gates in the circuit transform the initial state vector 
into a final state, which will generally be a superposition.  
We measure some or all qubits of this final state in the computational basis in order to (probabilistically) obtain a classical output to the algorithm. We can think of the measurement of one qubit in the computational basis as one special type of gate. We may assume without much loss of generality that such measurements only happen at the very end of the circuit (see Exercise~\ref{ex:intermediatemeas}).

What about the more general kinds of measurements discussed in Section~\ref{ssecmeasurement}? If we want to apply such a measurement in the circuit model, we will have to implement it using a circuit of elementary gates followed by a measurement in the computational basis.
For example, suppose projectors $P_0$ and $P_1$ form a 2-outcome projective measurement on an $n$-qubit space ($P_0+P_1=I_{2^n}$). Assume for simplicity that $P_0$ and $P_1$ both have rank $2^n/2$. Then there exists a unitary $U$ that maps an $n$-qubit state  $\ket{\phi}$ to a state whose first qubit is $\ket{0}$ whenever $P_0\ket{\phi}=\ket{\phi}$, and that maps $n$-qubit $\ket{\psi}$ to a state whose first qubit is $\ket{1}$ whenever $P_1\ket{\psi}=\ket{\psi}$.
We can now implement the projective measurement by first applying a circuit that implements~$U$, and then measuring (in the computational basis) the first qubit of the resulting state.
This recovers the measurement probabilities of the 2-outcome projective measurement; if we also want to recover the same post-measurement \emph{state}, then we have to multiply by $U^{-1}$ at the end.
The minimal-size circuit to implement $U$ could be very large (i.e., expensive) if the projective measurement is complicated, but that is how it should be.

\begin{figure}[htb]
  \centering
\includegraphics{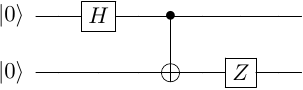}
\caption{Simple circuit for turning $\ket{00}$ into an entangled state}\label{fig:examplecircuit}
\end{figure}

To draw quantum circuits, the convention is to let time progress from left to right: we start with the initial state on the left. Each qubit is pictured as a horizontal wire, and the circuit prescribes which gates are to be applied to which wires. Single-qubit gates like $X$ and $H$ just act on one wire, while multi-qubit gates such as the CNOT act on multiple wires simultaneously.\footnote{Note that the number of wires (qubits) going into a unitary must equal the number of wires going out because a unitary is always invertible (reversible). This differs from the case of classical circuits, where non-reversible gates like AND have more wires going in than out.} 
When one qubit ``controls'' the application of a gate to another qubit, then the controlling wire is drawn with a dot linked vertically to the gate that is applied to the target qubit. This happens for instance with the CNOT, where the applied single-qubit gate is $X$, usually drawn as `$\oplus$' in a circuit picture (similarly, the Toffoli gate is drawn in a circuit with a dot on the two control wires and an `$\oplus$' on the target wire). Figure~\ref{fig:examplecircuit} gives a simple example on two qubits, initially in basis state $\ket{00}$: first apply $H$ to the 1st qubit, then CNOT to both qubits (with the first qubit acting as the control), and then $Z$ to the last qubit. The resulting state is $\frac{1}{\sqrt{2}}(\ket{00}-\ket{11})$.

Note that if we have a circuit for unitary $U$, it is very easy to find a circuit for the inverse $U^{-1}$ with the same complexity: just reverse the order of the gates, and take the inverse of each gate. For example, if $U=U_1U_2U_3$, then $U^{-1}=U_3^{-1}U_2^{-1}U_1^{-1}$.

In analogy to the classical class $\BPP$, we will define $\BQP$ (``Bounded-error 
Quantum Polynomial time'') as the class of languages that can
efficiently be computed  with success probability at least $2/3$ by (a family of) quantum circuits whose size grows at most polynomially with the input-length.
We will study this quantum complexity class and its relation with various classical complexity classes in more detail in Chapter~\ref{chap:qcomplexitytheory}.

\section{Universality of various sets of elementary gates}\label{sec:universalgatesets}

Which set of elementary gates should we allow? 
There are several reasonable choices. 
\begin{quote}
(1) The set of all 1-qubit operations together with the 2-qubit $\CNOT$ gate is universal,
meaning that any other unitary transformation can be built from these gates. 
\end{quote}
Allowing all 1-qubit gates is not very realistic from an implementational 
point of view, as there are continuously many of them, and we cannot expect experimentalists to implement gates to infinite precision. 
However, the model is usually restricted, only allowing 
a small finite set of 1-qubit gates from which
all other 1-qubit gates can be efficiently approximated.
\begin{quote}
(2) The set consisting of CNOT, Hadamard, and the phase-gate $T=R_{\pi/4}$ is universal in the sense of approximation,
meaning that any other unitary can be arbitrarily well approximated using circuits of only these gates.\\
The \emph{Solovay-Kitaev theorem}~\cite[Appendix~3]{nielsen&chuang:qc} says that this approximation is quite efficient:
we can approximate any gate on 1 or~2 qubits up to error $\eps$ using a circuit of only $\polylog(1/\eps)$ gates from our small universal set, i.e., polynomial in the logarithm of $1/\eps$.
In particular, simulating an arbitrary 2-qubit gate up to exponentially small error~$\eps$ costs only a polynomial number of gates from our small universal set. 
\end{quote}
It is often convenient to restrict to real numbers and use an even smaller set of gates:
\begin{quote}
(3) The set of Hadamard and Toffoli (CCNOT) is universal for all unitaries with real entries in the sense of approximation,
meaning that any unitary with only real entries can be arbitrarily well approximated using circuits of only these gates.
\end{quote}

\section{Quantum parallelism}

One uniquely quantum-mechanical effect that we can use 
for building quantum algorithms is \emph{quantum parallelism}.\index{quantum parallelism}
Suppose we have a classical algorithm that computes some function
$f:\01^n\rightarrow\01^m$. Then we can build a quantum circuit $U$ (consisting only of Toffoli gates)
that maps $\ket{z}\ket{0}\rightarrow\ket{z}\ket{f(z)}$ for every $z\in\01^n$.
Now suppose we apply $U$ to a superposition of \emph{all} inputs~$z$ 
(which is easy to build using $n$ Hadamard transforms):
$$
U\left(\frac{1}{\sqrt{2^n}}\sum_{z\in\01^n}\ket{z}\ket{0}\right)=
\frac{1}{\sqrt{2^n}}\sum_{z\in\01^n}\ket{z}\ket{f(z)}.
$$
We applied $U$ just once, but the final superposition contains $f(z)$
for \emph{all} $2^n$ input values~$z$!  However, by itself this is not very
useful and does not give more than classical randomization, 
since observing the final superposition will give just
one uniformly random $\ket{z}\ket{f(z)}$ and all other information will be lost.
As we will see below, quantum parallelism needs to be combined with 
the effects of interference and entanglement in order to get something 
that is better than classical.

\section{The early algorithms}\label{secearlyalgos}

The two best-known successes of quantum algorithms so far are
Shor's factoring algorithm from 1994~\cite{shor:factoring}\index{Grover, L.}
and Grover's search algorithm from 1996~\cite{grover:search},\index{Grover, L.} 
which will be explained in later chapters.
Here we describe some of the earlier quantum 
algorithms that preceded Shor's and Grover's.

Virtually all quantum algorithms work with \emph{queries}\index{query complexity} 
in some form or other. We will explain this model here.
It may look contrived at first, but eventually will lead 
smoothly to Shor's and Grover's algorithm.
We should, however, emphasize that the query complexity model differs 
from the standard model described above, because the input is now given
as a ``black-box'' (also sometimes called an ``oracle'').
This means that the exponential quantum-classical
separations that we describe below do not by themselves
give exponential quantum-classical separations in the standard circuit model (the same applies to Simon's algorithm in the next chapter).

To explain the query setting,
consider an $N$-bit input $x=(x_0,\ldots,x_{N-1})\in\01^N$.
Usually we will have $N=2^n$, so that we can address bit
$x_i$ using an $n$-bit index $i$.
One can think of the input as an $N$-bit memory which we can access
at any point of our choice (a ``Random Access Memory'' or RAM).\index{RAM}
For example, a memory of $N=1024$ bits can be indexed by addresses $i\in\01^{10}$ of $n=10$ bits each.
A memory access is via a so-called ``black-box,''
which is equipped to output the bit $x_i$ on input $i$.
As a quantum operation, this is the following unitary 
mapping on $n+1$ qubits:
$$
O_x:\ket{i,0}\rightarrow\ket{i,x_i}.
$$
The first $n$ qubits of the state are called the \emph{address bits} (or address register), 
while the $(n+1)$st qubit is called the \emph{target bit}.\footnote{It is a common rookie mistake to confuse the $N$ bits of $x$ with the $n$ address bits; don't fall for this!}
Since this mapping must be unitary, we also have to specify
what happens if the initial value of the target bit is 1.
Therefore we actually let $O_x$ be the following unitary transformation:
$$
O_x:\ket{i,b}\rightarrow\ket{i,b\oplus x_i},
$$
here $i\in\01^n$, $b\in\01$, and $\oplus$ denotes exclusive-or (addition modulo 2).
In matrix representation, this $O_x$ is now a permutation matrix and hence unitary.
Note that a quantum computer can apply $O_x$ on a superposition
of various~$i$, something a classical computer cannot do.
One application of this black-box is called a \emph{query},
and counting the required number of queries to compute this 
or that function of $x$ is something we will do a lot in 
the first half of these notes.

Given the ability to make a query of the above type, we can also
make a query of the form $\ket{i}\mapsto(-1)^{x_i}\ket{i}$
by setting the target bit to the state 
$\ket{-}=\frac{1}{\sqrt{2}}(\ket{0}-\ket{1})=H\ket{1}$:
$$
O_x\left(\ket{i}\ket{-}\right)=
\ket{i}\frac{1}{\sqrt{2}}(\ket{x_i}-\ket{1-x_i})=
(-1)^{x_i}\ket{i}\ket{-}.
$$
This $\pm$-kind of query puts the output variable in the phase
of the state: if $x_i$ is 1 then we get a $-1$ in the phase of basis state $\ket{i}$; 
if $x_i=0$ then nothing happens to $\ket{i}$.\footnote{This is sometimes called the ``phase kick-back trick.'' Note that for $\ket{+}=\frac{1}{\sqrt{2}}(\ket{0}+\ket{1})$, we have $O_x\left(\ket{i}\ket{+}\right)=\ket{i}\ket{+}$ irrespective of what $x$ is. This allows us to control on which part of the state a phase-query is applied: we put the control qubit in state $\ket{-}$ for indices $i$ where we want to apply the phase-query, and in state $\ket{+}$ for the indices where we do \emph{not} want to apply a phase-query.} This ``phase-query'' or ``phase-oracle'' is sometimes more convenient than 
the standard type of query. We denote the corresponding $n$-qubit unitary transformation by $O_{x,\pm}$.

\subsection{Deutsch-Jozsa}\label{ssecdeutschjozsa}\index{Deutsch, D.}\index{Jozsa, R.}

{\bf Deutsch-Jozsa problem}~\cite{deutsch&jozsa}:\\
For $N=2^n$, we are given $x\in\01^N$ such that either\\ 
(1) all $x_i$ have the same value (``constant''), or\\ 
(2) $N/2$ of the $x_i$ are 0 and $N/2$ are 1 (``balanced'').\\
The goal is to find out whether $x$ is constant or balanced.

\bigskip

The algorithm of Deutsch and Jozsa is as follows.
We start in the $n$-qubit zero state $\ket{0^n}$, apply 
a Hadamard transform\index{Hadamard gate} 
to each qubit, apply a phase-query,
apply another Hadamard to each qubit, and then measure the final state.
As a unitary transformation, the algorithm would be 
$H^{\otimes n}O_{x,\pm}H^{\otimes n}$.
We have drawn the corresponding quantum circuit 
in Figure~\ref{figdeutschjozsa}, where time again progresses from left to right.
Note that the number of wires going into the query is $n$, not $N$; the basis states on this sequence of wires specify an $n$-bit address (if we use $O_x$ instead of $O_{x,\pm}$ then there is one more wire at the bottom, for the target qubit of $O_x$, initialized to $H\ket{1}=\ket{-}$).

\begin{figure}[hbt]
\centering
\setlength{\unitlength}{0.3mm}
\begin{picture}(200,120)
\put(-10,80){\makebox(20,20){$\ket{0}$}}
\put(-10,40){\makebox(20,20){$\ket{0}$}}
\put(-10,0){\makebox(20,20){$\ket{0}$}}
\put(215,40){\makebox(20,20){\small measure}}
\put(10,90){\line(1,0){25}}
\put(10,50){\line(1,0){25}}
\put(10,10){\line(1,0){25}}
\put(55,90){\line(1,0){25}}
\put(55,50){\line(1,0){25}}
\put(55,10){\line(1,0){25}}
\put(35,80){\framebox(20,20){$H$}}
\put(35,40){\framebox(20,20){$H$}}
\put(35,0){\framebox(20,20){$H$}}
\put(145,80){\framebox(20,20){$H$}}
\put(145,40){\framebox(20,20){$H$}}
\put(145,0){\framebox(20,20){$H$}}
\put(165,90){\line(1,0){25}}
\put(165,50){\line(1,0){25}}
\put(165,10){\line(1,0){25}}
\put(80,0){\framebox(40,100){$O_{x,\pm}$}}
\put(120,90){\line(1,0){25}}
\put(120,50){\line(1,0){25}}
\put(120,10){\line(1,0){25}}
\end{picture}
\caption{The Deutsch-Jozsa algorithm for $n=3$}\label{figdeutschjozsa}
\end{figure}

Let us follow the state through these operations.
Initially we have the state $\ket{0^n}$.
By Equation~(\ref{eqnhadamard}) on page~\pageref{eqnhadamard}, 
after the first Hadamard transforms 
we have obtained the uniform superposition of all $i$:
$$
\frac{1}{\sqrt{2^n}}\sum_{i\in\01^n}\ket{i}.
$$
The $O_{x,\pm}$-query turns this into
$$
\frac{1}{\sqrt{2^n}}\sum_{i\in\01^n}(-1)^{x_i}\ket{i}.
$$
Applying the second batch of Hadamards gives (again by Equation~(\ref{eqnhadamard}))
the final superposition 
$$
\frac{1}{2^n}\sum_{i\in\01^n}(-1)^{x_i}\sum_{j\in\01^n}(-1)^{i\cdot j}\ket{j},
$$
where $i\cdot j=\sum_{k=1}^n i_kj_k$ as before.
Since $i\cdot 0^n=0$ for all $i\in\01^n$, we see that the amplitude 
of the $\ket{0^n}$-state in the final superposition is
$$
\frac{1}{2^n}\sum_{i\in\01^n}(-1)^{x_i}=
\left\{\begin{array}{rl}
1 & \mbox{ if $x_i=0$ for all $i$,}\\
-1 & \mbox{ if $x_i=1$ for all $i$,}\\
0 & \mbox{ if $x$ is balanced.}
\end{array}\right.
$$
Hence the final observation will yield $\ket{0^n}$ if $x$ 
is constant and will yield some other state if $x$ is balanced.
Accordingly, the Deutsch-Jozsa problem can be solved with certainty
using only 1 quantum query and $O(n)$ other operations 
(the original solution of Deutsch and Jozsa used 2 queries, 
the 1-query solution is from~\cite{cemm:revisited}).%
\index{Cleve, R.}\index{Ekert, A.}\index{Macchiavello, C.}\index{Mosca, M.}

In contrast, it is easy to see that any \emph{classical} deterministic
algorithm needs at least $N/2+1$ queries: if it has made only $N/2$
queries and seen only 0s, the correct output is still undetermined.
However, a classical algorithm can solve this problem efficiently
if we allow a small error probability: just query $x$ at two
random positions, output ``constant'' if those bits are the same
and ``balanced'' if they are different.
This algorithm outputs the correct answer with probability 1
if $x$ is constant and outputs the correct answer with probability 1/2 
if $x$ is balanced. Thus the quantum-classical separation of this problem
only holds if we consider algorithms without error probability.

\subsection{Bernstein-Vazirani}\label{ssec:bernsteinvazirani}\index{Bernstein, E.}\index{Vazirani, U.}

{\bf Bernstein-Vazirani problem}~\cite{bernstein&vazirani:qcomplexity}:\\
For $N=2^n$, we are given $x\in\01^N$ with the property that 
there is some unknown $a\in\01^n$ such that $x_i=(i\cdot a)\mod 2$.
The goal is to find $a$.

\bigskip

The Bernstein-Vazirani algorithm is \emph{exactly} the same as
the Deutsch-Jozsa algorithm, but now the final observation
miraculously yields $a$. Since $(-1)^{x_i}=(-1)^{(i\cdot a)\mod 2}=(-1)^{i\cdot a}$,
we can write the state obtained after the query as:
$$
\frac{1}{\sqrt{2^n}}\sum_{i\in\01^n}(-1)^{x_i}\ket{i}=
\frac{1}{\sqrt{2^n}}\sum_{i\in\01^n}(-1)^{i\cdot a}\ket{i}.
$$
Since Hadamard is its own inverse, from Equation~(\ref{eqnhadamard}) we can see that applying a Hadamard to each qubit of the above state will turn it into the classical state $\ket{a}$. This solves the Bernstein-Vazirani problem with 
1 query and $O(n)$ other operations.
In contrast, any classical algorithm (even a randomized one with 
small error probability) needs to ask $n$ queries for 
information-theoretic reasons: the final answer consists 
of $n$ bits and one classical query gives at most 1 bit of information.

Bernstein and Vazirani also defined a recursive version
of this problem, which can be solved exactly by a quantum algorithm
in $\poly(n)$ steps, but for which every classical randomized algorithm needs $n^{\Omega(\log n)}$ steps (queries).
However, we'll see a much better classical-quantum separation in the next chapter.

\section*{Exercises}

\begin{enumerate}

\item Is the controlled-NOT operation $C$ Hermitian?
Determine $C^{-1}$.

\item Construct a CNOT from two Hadamard gates and one controlled-$Z$ (the controlled-$Z$ gate maps $\ket{11}\mapsto -\ket{11}$ and acts like the identity on the other basis states).

\item\label{ex:swapcnot} A SWAP-gate interchanges two qubits: it maps basis state $\ket{a,b}$ to $\ket{b,a}$. Implement a SWAP-gate using a few CNOTs (when using a CNOT, you're allowed to use either of the 2 bits as the control, but be explicit about this).

\item Show that every 1-qubit unitary with \emph{real} entries can be written as a rotation matrix, possibly preceded and followed by $Z$-gates.  More precisely, show that for every $2\times 2$ real unitary $U$, there exist  
signs $s_1,s_2,s_3\in\{1,-1\}$ and angle $\theta\in[0,\pi/2]$ such that
$$
U=s_1\left(\begin{array}{rr}1 & 0\\ 0 & s_2\end{array}\right)\left(\begin{array}{rr}\cos(\theta) & -\sin(\theta)\\ \sin(\theta)& \cos(\theta)\end{array}\right)\left(\begin{array}{rr}1 & 0\\ 0 & s_3\end{array}\right).
$$

\item Show that for every 1-qubit unitary~$U$ there exist angles $\alpha,\beta,\gamma\in[0,2\pi)$ and $\theta\in[0,\pi/2]$ such that
$$
U=e^{i\alpha}\left(\begin{array}{rr}1 & 0\\ 0 & e^{i\beta}\end{array}\right)\left(\begin{array}{rr}\cos(\theta) & -\sin(\theta)\\ \sin(\theta)& \cos(\theta)\end{array}\right)\left(\begin{array}{rr}1 & 0\\ 0 & e^{i\gamma}\end{array}\right).
$$

\item\label{ex:controlledgate} Let $U$ be a 1-qubit unitary of determinant~1 that we would like to implement in a controlled way, i.e., we want to implement a map $\ket{c}\ket{b}\mapsto \ket{c}U^c\ket{b}$ for all $c,b\in\01$ (here $U^0=I$ and $U^1=U$).  One can show there exist 1-qubit unitaries $A$, $B$, and $C$, such that $ABC=I$ and $AXBXC=U$ ($X$ is the NOT-gate); you may assume this without proof. Give a circuit that acts on two qubits and implements a controlled-$U$ gate, using CNOTs and (uncontrolled) $A$, $B$, and $C$ gates.

\item\label{ex:controlledcircuit} (H) Let $C$ be a given quantum circuit consisting of $T$ many gates, which may be CNOTs and single-qubit gates. Show that we can implement $C$ in a controlled way using $O(T)$ Toffoli gates, CNOTs and arbitrary single-qubit gates, and no auxiliary qubits other than the controlling qubit.

\item\label{ex:intermediatemeas} (H) It is possible to avoid doing any intermediate measurements in a quantum circuit, using one auxiliary qubit for each 1-qubit measurement that needs to be delayed until the end of the computation. Show how.

\item\label{ex:negate0circuit}
\begin{enumerate}
\item Give a circuit that maps $\ket{0^n,b}\mapsto\ket{0^n,1-b}$ for $b\in\01$, and that maps $\ket{i,b}\mapsto\ket{i,b}$ whenever $i\in\01^n\backslash\{0^n\}$. You are allowed to use every type of elementary gate mentioned in the lecture notes (incl.~Toffoli gates), as well as auxiliary qubits that are initially $\ket{0}$ and that should be put back to $\ket{0}$ at the end of the computation.
\item Suppose we can make queries of the type $\ket{i,b}\mapsto\ket{i,b\oplus x_i}$ to input $x\in\01^N$, with $N=2^n$. Let $x'$ be the input $x$ with its first bit flipped (e.g., if $x=0110$ then $x'=1110$). Give a circuit that implements a query to $x'$.  Your circuit may use one query to~$x$.
\item Give a circuit that implements a query to an input $x''$ that is obtained from $x$ (analogously to~(b)) by setting its first bit to~0.  Your circuit may use one query to~$x$.
\end{enumerate}

\item In Section~\ref{secearlyalgos} we showed that a standard query, which maps $\ket{i,b}\mapsto\ket{i,b\oplus x_i}$ (where $i\in\{0,\ldots,N-1\}$ and $b\in\{0,1\}$), can be used to implement a phase-query to~$x$, i.e., one of the type $\ket{i}\mapsto(-1)^{x_i}\ket{i}$ (this is an \emph{uncontrolled} phase-query).
\begin{enumerate}
\item  Show that a standard query can be implemented using one \emph{controlled} phase-query to~$x$ (which maps $\ket{c,i}\mapsto(-1)^{cx_i}\ket{c,i}$, so the phase is added only if the control bit is $c=1$), and possibly some auxiliary qubits and other gates.
\item  Can you also implement a standard query using one or more \emph{un}controlled phase-queries to~$x$, and possibly some auxiliary qubits and other gates?
If yes, show how.  If no, prove why not.
\end{enumerate}

\item\label{ex:workspacesinterference} Suppose we have a 2-bit input $x=x_0x_1$ and a phase-query that maps
$$
O_{x,\pm}:\ket{b}\mapsto (-1)^{x_b}\ket{b}\mbox{~~for }b\in\01.
$$
\begin{enumerate}
\item Suppose we run the 1-qubit circuit $HO_{x,\pm}H$ on initial state $\ket{0}$ and then measure (in the computational basis). What is the probability distribution on the output bit, as a function of $x$?
\item Now suppose the query leaves some workspace in a second qubit, which is initially $\ket{0}$:
$$
O'_{x,\pm}:\ket{b,0}\mapsto (-1)^{x_b}\ket{b,b}\mbox{~~for }b\in\01.
$$
Suppose we just ignore the workspace and run the algorithm of (a) on the first qubit with $O'_{x,\pm}$ instead of $O_{x,\pm}$ (and $H\otimes I$ instead of $H$, and initial state $\ket{00}$).
What is now the probability distribution on the output bit (i.e., if we measure the first of the two bits)?\\
{\footnotesize Comment: This exercise illustrates why it's important to ``clean up'' (i.e., set back to $\ket{0}$) workspace qubits of some subroutine before running it on a superposition of inputs: the unintended entanglement between the address and workspace registers can thwart the intended interference effects.}
\end{enumerate}

\item Give a randomized classical algorithm (i.e., one that can flip coins
during its operation) that makes only two queries to $x$, and decides
the Deutsch-Jozsa problem with success probability at least 2/3 on every possible input.
A high-level description is enough, no need to write out the classical circuit.

\item Let $N=2^n$. Suppose our $N$-bit input $x$ satisfies the following promise:\\
either (1) the first $N/2$ bits of $x$ are all 0 and the second $N/2$ bits are all~1; or (2) the number of 1s in the first half of $x$ plus the number of 0s in the second half, equals $N/2$.
Modify the Deutsch-Jozsa algorithm to efficiently distinguish these two cases (1) and (2).

\item\label{ex:parityquery} 
(H) Let $N=2^n$. A \emph{parity query} to input $x\in\01^N$ corresponds to the $(N+1)$-qubit unitary map $Q_x:\ket{y,b}\mapsto \ket{y,b\oplus(x\cdot y)}$, where $x\cdot y=\sum_{i=0}^{N-1}x_iy_i$ mod~2. For a fixed function $f:\01^N\rightarrow\01$, give a quantum algorithm that computes $f(x)$ using only one such query (i.e., one application of $Q_x$), and as many elementary gates as you want. You do not need to give the circuit in full detail, an informal description of the algorithm is good enough.

\end{enumerate}

\chapter{Simon's Algorithm}\label{chap:simon}\index{Simon, D.}

The Deutsch-Jozsa problem showed an exponential quantum improvement over the best \emph{deterministic}
classical algorithms; the Bernstein-Vazirani problem showed a polynomial improvement over
the best \emph{randomized} classical algorithms that have error probability $\leq 1/3$.
In this chapter we will combine these two features: we will see a computational problem due to Simon~\cite{simon:power} where quantum computers
are provably exponentially more efficient (in terms of number of queries) than bounded-error randomized algorithms.
Simon's problem may look rather contrived at first sight, but his quantum algorithm to solve it was the main inspiration for Shor's important quantum algorithm for the very natural problem of integer factorization, which we will see in Chapter~\ref{chap:factoring}. More recently, Simon's algorithm itself was also used to break some classical cryptographic systems~\cite{KLLN:breaking,santoli&schaffner:simon}.

\section{The problem}

Let $N=2^n$, and identify the set $\{0,\ldots,N-1\}$ with $\01^n$.
Let $j\oplus s$ be the $n$-bit string obtained by bitwise 
adding the $n$-bit strings $j$ and $s$ mod 2, so for example $00110\oplus 10101=10011$.

\medskip

\noindent
{\bf Simon's problem}~\cite{simon:power}:\\
For $N=2^n$, we are given $x=(x_0,\ldots,x_{N-1})$, with $x_i\in\01^n$, with the property 
that there is some unknown nonzero $s\in\01^n$ 
such that $x_i=x_j$ iff ($i=j$ or $i=j\oplus s$).
The goal is to find $s$.

\medskip

Note that $x$, viewed as a function from $\{0,\ldots,N-1\}$ to $\{0,\ldots,N-1\}$, is a 2-to-1
function, where the 2-to-1-ness is determined by the unknown \emph{mask}~$s$.
The queries to the input here are slightly different from before: 
the input $x=(x_0,\ldots,x_{N-1})$ now has variables $x_i$ that themselves 
are $n$-bit strings, and one query gives such a string completely 
($\ket{i,0^n}\mapsto\ket{i,x_i}$).
However, we can also view this problem as having $n2^n$ binary variables
that we can query individually. Since we can simulate one $x_i$-query 
using only $n$ binary queries (just query all $n$ bits of $x_i$),
this alternative view will not affect the number of queries very much.

\section{The quantum algorithm}

The main ``subroutine'' of Simon's algorithm starts out very similar to Deutsch-Jozsa:
start in a state of $2n$ zero qubits $\ket{0^n}\ket{0^n}$ and apply
Hadamard transforms to the first $n$ qubits to put them in a uniform superposition, giving
$$
\frac{1}{\sqrt{2^n}}\sum_{i\in\01^n}\ket{i}\ket{0^n}.
$$
At this point, the second $n$-qubit register still holds only zeroes.
A query turns this into
$$
\frac{1}{\sqrt{2^n}}\sum_{i\in\01^n}\ket{i}\ket{x_i}.
$$
Now the algorithm measures the second $n$-qubit register in the computational basis (see Exercise~\ref{ex:2registermeasurement}); this measurement is actually not necessary, but it facilitates analysis.
The measurement outcome will be some value $x_i$ and the first 
register will collapse to the superposition of the two indices 
having that $x_i$-value:
$$
\frac{1}{\sqrt{2}}(\ket{i}+\ket{i\oplus s})\ket{x_i}.
$$
We will now ignore the second register and apply Hadamard 
transforms to the first $n$ qubits. Using Equation~(\ref{eqnhadamard})
and the fact that $(i\oplus s)\cdot j=(i\cdot j)\oplus(s\cdot j)$,
we can write the resulting state as
$$
\frac{1}{\sqrt{2^{n+1}}}\left(\sum_{j\in\01^n}(-1)^{i\cdot j}\ket{j}+
\sum_{j\in\01^n}(-1)^{(i\oplus s)\cdot j}\ket{j}\right)=
$$
$$
\frac{1}{\sqrt{2^{n+1}}}\left(
\sum_{j\in\01^n}(-1)^{i\cdot j}\left(1+(-1)^{s\cdot j}\right)\ket{j}\right).
$$
Note that $\ket{j}$ has nonzero amplitude iff $s\cdot j=0\mod 2$.
Measuring the state gives a uniformly random element from the set $\{j\mid s\cdot j=0\mod 2\}$.  Accordingly, we get a linear equation that gives information about $s$. We repeat this algorithm until we have obtained $n-1$ independent linear equations involving~$s$. The solutions to these equations will be $0^n$ and the correct $s$, which we can compute efficiently by a classical algorithm (Gaussian elimination modulo~2).
This can be done by means of a classical circuit of size roughly $O(n^3)$. 

Note that if the $j$'s you have generated at some point span a space of size $2^k$, for some $k<n-1$, then the probability that your next run of the algorithm produces a $j$ that is linearly independent of the earlier ones, is $(2^{n-1}-2^k)/2^{n-1}\geq 1/2$. Hence an expected number of $O(n)$ runs of the algorithm suffices to find $n-1$ linearly independent $j$'s. Simon's algorithm thus finds $s$ using an expected number of $O(n)$ $x_i$-queries and polynomially many other operations (Exercise~\ref{chap:grover}.\ref{ex:exactsimon} will show how to make the success probability equal to~1).

\section{Classical algorithms for Simon's problem}

\subsection{Upper bound}
Let us first sketch a classical randomized algorithm that solves Simon's problem using $O(\sqrt{2^n})$ queries. The algorithm is based on the so-called ``birthday paradox,'' which is the phenomenon that in a group of only 23 people, there is already a fairly large probability (slightly higher than 1/2) that two people share the same birthday, despite the fact that the number of possible birthdays (365) is much larger than the number of people (23). The intuitive explanation is that the number of \emph{pairs} of people is actually quadratic in the number of people, and each pair has a $1/365$ probability to have the same birthday (assuming dates of birth are distributed uniformly; otherwise that probability is even $>1/365$). Of course, different pairs may overlap and hence are not independent, but the idea still works.

Our algorithm will make $T$ randomly chosen distinct queries $i_1,\ldots,i_T$, for some~$T$ to be determined later.
If there is a \emph{collision} among those queries (i.e., $x_{i_k}=x_{i_{\ell}}$ for some $k\neq\ell$, so $i_k$ and $i_\ell$ happen to have the same ``birthday''), then we are done, because then we know $i_k=i_\ell\oplus s$, equivalently $s=i_k\oplus i_\ell$.
There won't be any collisions if $s=0^n$, but how large should $T$ be such that we are likely to see a collision in case $s\neq 0^n$? 
There are $\binom{T}{2}=\frac{1}{2}T(T-1)$ pairs in our sequence that could be a collision, and since the indices are chosen uniformly at random, the probability for a fixed pair to form a collision is $1/(2^n-1)$.
Hence by linearity of expectation, the \emph{expected} number of collisions in our sequence will be $\binom{T}{2}/2^n\approx T^2/2^{n+1}$.  If we choose $T=\sqrt{2^{n+1}}$, we expect to have roughly 1~collision in our sequence, which is good enough to find~$s$.  Of course, an \emph{expected value} of 1~collision does not mean that we will have at least one collision \emph{with high probability}, but a slightly more involved calculation shows the latter statement as well.

\subsection{Lower bound}
Simon~\cite{simon:power} proved that any classical randomized 
algorithm that finds $s$ with high probability needs to make 
$\Omega(\sqrt{2^n})$ queries, so the above classical algorithm is essentially optimal.
This was the first proven exponential separation between quantum 
algorithms and classical bounded-error algorithms (let us stress again 
that this does not prove an exponential separation in the usual circuit model, 
because we are counting queries rather than ordinary operations here). 
Simon's algorithm inspired Shor to his factoring algorithm, which we describe in Chapter~\ref{chap:factoring}.

We will prove the classical lower bound for a \emph{decision} version of Simon's problem, where we now also allow $s=0^n$ as a possibility:
\begin{quote}
{\bf Given:} input $x=(x_0,\ldots,x_{N-1})$, where $N=2^n$ and $x_i\in\01^n$\\
{\bf Promise:} $\exists s\in\01^n$ such that: $x_i=x_j$ iff ($i=j$ or $i=j\oplus s$)\\
{\bf Task:} decide whether $s=0^n$
\end{quote}
Consider the input distribution $\mu$ that is defined as follows.
With probability 1/2, $x$ is a uniformly random permutation of $\01^n$; 
this corresponds to the case $s=0^n$. With probability 1/2,
we pick a nonzero string $s$ at random, and for each pair $(i,i\oplus s)$, we pick a unique value for $x_i=x_{i\oplus s}$ at random.
If there exists a \emph{randomized} $T$-query algorithm that achieves success probability $\geq 2/3$ under this input distribution $\mu$,
then there also is \emph{deterministic} $T$-query algorithm that achieves success probability $\geq 2/3$ under $\mu$
(because the behavior of the randomized algorithm is an average over a number of deterministic algorithms).
Now consider a deterministic algorithm with error $\leq 1/3$ under $\mu$, 
that makes $T$ queries to $x$.  We want to show that $T=\Omega(\sqrt{2^n})$.

First consider the case $s=0^n$. We can assume the algorithm never queries 
the same point twice. Then the $T$ outcomes of the queries are $T$ distinct
$n$-bit strings, and each sequence of $T$ strings is equally likely.

Now consider the case $s\neq 0^n$. 
Suppose the algorithm queries the indices $i_1,\ldots,i_T$ (this sequence depends on $x$)
and gets outputs $x_{i_1},\ldots,x_{i_T}$. 
Call a sequence of queries $i_1,\ldots,i_T$ 
\emph{good} if it shows a collision (i.e., $x_{i_k}=x_{i_{\ell}}$ for some $k\neq\ell$), and \emph{bad} otherwise.
If the sequence of queries of the algorithm is good, then we
can find $s$, since $i_k\oplus i_\ell=s$.  On the other hand,
if the sequence is bad, then each sequence of $T$ distinct 
outcomes is equally likely---just as in the $s=0^n$ case!
We will now show that the probability of the bad case is very
close to 1 for small $T$. 

If $i_1,\ldots,i_{k-1}$ is bad,
then we have excluded at most $\binom{k-1}{2}$ possible values of $s$ (namely all values $i_j\oplus i_{j'}$ for all distinct $j,j'\in[k-1]$),
and all other values of $s$ are equally likely.
The probability that the next query $i_k$ makes the sequence good,
is the probability that $x_{i_k}=x_{i_j}$ for some $j<k$,
equivalently, that the set $S_k=\{i_k\oplus i_j\mid j<k\}$ happens to 
contain the string $s$. But $S_k$ has only $k-1$ members, while
there are at least $2^n-1-\binom{k-1}{2}$ equally likely remaining possibilities for~$s$.
This means that the probability that the sequence is still bad after 
query $i_k$ is made, is very close to 1.
In formulas:
\begin{eqnarray*}
\Pr[i_1,\ldots,i_T\mbox{ is bad}] & = &
\prod_{k=2}^T \Pr[i_1,\ldots,i_k\mbox{ is bad}\mid i_1,\ldots,i_{k-1}\mbox{ is bad}]\\
& \geq & \prod_{k=2}^T \left(1-\frac{k-1}{2^n-1-\binom{k-1}{2}}\right)\\
& \geq & 1-\sum_{k=2}^T \frac{k-1}{2^n-1-\binom{k-1}{2}}.
\end{eqnarray*}
Here we used the fact that $(1-a)(1-b)\geq 1-(a+b)$ if $a,b\geq 0$. Note that $\sum_{k=2}^T (k-1)=T(T-1)/2\leq T^2/2$, 
and $2^n-1-\binom{k-1}{2}\geq 2^n/2$ as long as $k< \sqrt{2^n}$.
Hence
\begin{eqnarray*}
\Pr[i_1,\ldots,i_T\mbox{ is good}] \leq \sum_{k=2}^T \frac{k-1}{2^n-1-\binom{k-1}{2}}\leq T^2/2^n.
\end{eqnarray*}
Accordingly, if $T\leq c\sqrt{2^n}$ for some small constant~$c$, then with probability nearly~1 (probability taken over the input  distribution $\mu$) 
the algorithm's sequence of queries is bad. 
If it gets a bad sequence, it cannot ``see'' the difference
between the $s=0^n$ case and the $s\neq 0^n$ case, since both cases
result in a uniformly random sequence of $T$ distinct $n$-bit strings as answers to the $T$ queries.
This shows that $T$ has to be $\Omega(\sqrt{2^n})$ in order to enable the algorithm
to get a good sequence of queries with high probability.

\section*{Exercises}

\begin{enumerate}

\item\label{ex:2registermeasurement} Give the projectors of the $2^n$-outcome projective measurement that is applied to the whole $2n$-qubit state in Simon's algorithm right after the query.
  
\item Analyze the different steps of Simon's algorithm if $s=0^n$ (so all $x_i$-values are distinct), and show that the final output $j$ is uniformly distributed over $\01^n$.

\item 
Suppose we run Simon's algorithm on the following input $x$ (with $N=8$ and hence $n=3$):
\begin{quote}
$x_{000}=x_{111}=000$\\
$x_{001}=x_{110}=001$\\
$x_{010}=x_{101}=010$\\
$x_{011}=x_{100}=011$
\end{quote}
Note that $x$ is 2-to-1 and $x_i=x_{i\oplus 111}$ for all $i\in\01^3$, so $s=111$.
\begin{enumerate}
\item Give the starting state of Simon's algorithm.
\item Give the state after the first Hadamard transforms on the first $3$ qubits.
\item Give the state after applying the oracle query.
\item Give the state after measuring the second register
(suppose the measurement gave $\ket{001}$).
\item
Using $H^{\otimes n}\ket{i}=\frac{1}{\sqrt{2^n}}\sum_{j\in\01^n}(-1)^{i\cdot j}\ket{j}$,
give the state after the final Hadamards.
\item Why does a measurement of the first 3 qubits of
the final state give information about~$s$?
\item Suppose the first run
of the algorithm gives $j=011$ and a second run gives $j=101$.
Show that, assuming $s\neq 000$, those two runs of the algorithm already determine $s$.
\end{enumerate}

\item\label{ex:subspacesimon}
Consider the following generalization of Simon's problem:
the input is $x=(x_0,\ldots,x_{N-1})$, with $N=2^n$ and $x_i\in\01^n$, with the property that there is some unknown \emph{subspace} $V\subseteq\01^n$ (where $\01^n$ is the vector space of $n$-bit strings with entrywise addition modulo~2)
such that $x_i=x_j$ iff there exists a $v\in V$ such that $i=j\oplus v$.
The usual definition of Simon's problem corresponds to the case of 1-dimensional subspace $V=\{0,s\}$. 

Show that one run of the circuit of the main subroutine in Simon's algorithm now produces a $j\in\01^n$ that is orthogonal to the whole subspace (i.e., $j\cdot v=0$ mod~2 for every $v\in V$).

\item 
Let $f:\01^n\to\01^{n-1}$ be a 2-to-1 function, meaning that every $y\in\01^{n-1}$ has exactly two distinct pre-images $x,x'\in\01^n$.
Suppose there is an efficient quantum circuit (i.e., with number of elementary gates that's polynomial in $n$) to compute $f$, but no efficient circuit that can produce from given $x\in\01^n$ an $x'\neq x$ such that $f(x)=f(x')$.

Show how a quantum computer can efficiently generate a uniformly random $y\in\01^{n-1}$ and an associated $n$-qubit state $\ket{\phi_y}$ such that:\\
(1) when asked, from $\ket{\phi_y}$ you can efficiently generate an $x$ such that $f(x)=y$;\\
and\\
(2) when asked, you can efficiently sample uniformly from the set\\ 
$\{a\in\01^n : a\cdot (x\oplus x')=0\mbox{ mod }2\}$, where $x$ and $x'$ are the two pre-images of $y$.\\
{\footnotesize Comment: After having generated $y$ and $\ket{\phi_y}$, you should give an algorithm that can do task (1) and another algorithm that can do task (2), but not necessarily an algorithm that can do both task (1) and (2) simultaneously (since doing task (1) might affect the state in a way that makes it impossible to do task (2), and vice versa). This problem may look arbitrary but was recently used to design an efficient protocol through which a \emph{classical} computer can efficiently verify that a quantum computer works as intended~\cite{mahadev:clasveri}.}

\item\label{ex:oracleinter}
\begin{enumerate}
\item  Suppose $x$ is an $N$-bit string.
What happens if we apply a Hadamard transform to each qubit of the $N$-qubit state $\displaystyle\frac{1}{\sqrt{2^N}}\sum_{y\in\01^N}(-1)^{x\cdot y}\ket{y}$?
\item
Give a quantum algorithm that uses $T$ queries to $N$-bit string $x$, and that maps $\ket{y}\mapsto (-1)^{x\cdot y}\ket{y}$
for every $y\in\01^N$ that contains at most $T$ 1s (i.e., for every $y$ of Hamming weight $\leq T$).
You can argue on a high level, no need to write out circuits in detail.
\item (H)
Give a quantum algorithm that with high probability outputs $x$, using at most $N/2+2\sqrt{N}$ queries to $x$.
\item Argue that a classical algorithm needs at least $N$ queries in order to have success probability $>1/2$ of outputting the correct $x$.
\end{enumerate}

\end{enumerate}

\chapter{The Fourier Transform}\label{chap:ft}

\section{The classical discrete Fourier transform}

The Fourier transform occurs in many different versions throughout classical computing,
in areas ranging from signal-processing to data compression to complexity theory.

For our purposes, the Fourier transform is going to be an $N\times N$ unitary matrix, all of whose entries have the same magnitude.
For $N=2$, it's just our familiar Hadamard transform:
$$
F_2=H=\frac{1}{\sqrt{2}}\left(\begin{array}{rr} 1 & 1\\ 1 & -1\end{array}\right).
$$
Doing something similar in 3 dimensions is impossible with real numbers: we can't give
three orthogonal vectors in $\{+1,-1\}^3$.  However, using \emph{complex} numbers allows us to define
the Fourier transform for any $N$.
Let $\omega_N=e^{2\pi i/N}$ be an $N$-th primitive root of unity (``root of unity'' means that $\omega_N^k=1$ for some integer $k$, in this case $k=N$).
The rows of the matrix will be indexed by $j\in\{0,\ldots,N-1\}$ and the columns by $k\in\{0,\ldots,N-1\}$.
Define the $(j,k)$-entry of the matrix $F_N$ by $\frac{1}{\sqrt{N}}\omega_N^{jk}$, where the exponent $jk$ is the usual product of two integers:
$$
F_N=\frac{1}{\sqrt{N}}\left(\begin{array}{ccc}  & \vdots & \\ \cdots & \omega_N^{jk} &\cdots\\ & \vdots & \end{array}\right)
$$
This $F_N$ is a unitary matrix, because each column has norm~1 and any two distinct columns (say those indexed by $k$ and $k'$) are orthogonal:
$$
\sum_{j=0}^{N-1}\frac{1}{\sqrt{N}}(\omega_N^{jk})^*\frac{1}{\sqrt{N}}\omega_N^{jk'}=\frac{1}{N}\sum_{j=0}^{N-1}\omega_N^{j(k'-k)}
=\left\{\begin{array}{ll}1 & \mbox{ if } k=k'\\0 & \mbox{otherwise}\end{array}\right.
$$
using the formula for geometric sums from Appendix~\ref{appsec:eqs}.

Since $F_N$ is unitary and symmetric, 
the inverse $F_N^{-1}=F_N^*$ only differs from $F_N$ by having minus signs in the exponent of the entries.
For a vector $v\in\mathbb{R}^N$, the vector $\widehat{v}=F_Nv$ is called the Fourier transform of $v$.%
\footnote{One often sees the Fourier transform of a \emph{function} rather than of a vector, 
but on finite domains that's just a notational variant: 
a vector $v\in\mathbb{R}^N$ can also be viewed as a function $v:\{0,\ldots,N-1\}\rightarrow\mathbb{R}$ defined by $v(i)=v_i$.
Also, sometimes the term ``Fourier transform'' is used for for what we call the inverse Fourier transform.}
Its entries are given by $\widehat{v}_j = \frac{1}{\sqrt{N}}\sum_{k=0}^{N-1}\omega_N^{jk}v_k$.

\section{The Fast Fourier Transform}

The naive way of computing the Fourier transform $\widehat{v}=F_Nv$ of $v\in\mathbb{R}^N$ just does the 
matrix-vector multiplication to compute all the entries of $\widehat{v}$.
This would take $O(N)$ steps (additions and multiplications) per entry, and $O(N^2)$ steps to compute the whole vector $\widehat{v}$.
However, there is a more efficient way of computing $\widehat{v}$.
This algorithm is called the \emph{Fast Fourier Transform} (FFT, due to Cooley and Tukey in 1965~\cite{cooley&tukey:fft}), and takes only $O(N\log N)$ steps.
This difference between the quadratic $N^2$ steps and the near-linear $N\log N$ is tremendously important in practice when $N$ is large,
and is the main reason that Fourier transforms are widely used in practice.

We will assume $N=2^n$, which is usually fine because we can add zeroes to our vector to make its dimension a power of~2
(but similar FFTs can be given also directly for most $N$ that aren't a power of~2).
The key to the FFT is to rewrite the entries of $\widehat{v}$ as follows:
\begin{eqnarray}
\widehat{v}_j & = & \frac{1}{\sqrt{N}}\sum_{k=0}^{N-1}\omega_N^{jk}v_k
              = \frac{1}{\sqrt{N}}\left(\sum_{{\rm even\; }k}\omega_N^{jk}v_k + \omega_N^j\sum_{{\rm odd\; }k}\omega_N^{j(k-1)}v_k \right)\\
              & = & \frac{1}{\sqrt{2}}\left(\frac{1}{\sqrt{N/2}}\sum_{{\rm even\; }k}\omega_{N/2}^{jk/2}v_k + \omega_N^j\frac{1}{\sqrt{N/2}}\sum_{{\rm odd\; }k}\omega_{N/2}^{j(k-1)/2}v_k \right)\label{eq:fftformula}
\end{eqnarray}
Note that we've rewritten the entries of the $N$-dimensional Fourier transform $\widehat{v}$ 
in terms of two $N/2$-dimensional Fourier transforms,
one of the even-numbered entries of $v$, and one of the odd-numbered entries of $v$.

This suggests a recursive procedure for computing $\widehat{v}$:
first separately compute the Fourier transform $\widehat{v_{\rm even}}$ of the $N/2$-dimensional vector 
of even-numbered entries of $v$, and the Fourier transform $\widehat{v_{\rm odd}}$ of the $N/2$-dimensional 
vector of odd-numbered entries, then compute the $N$ entries 
$$
\widehat{v}_j=\frac{1}{\sqrt{2}}(\widehat{v_{\rm even}}_j+\omega_N^j\widehat{v_{\rm odd}}_j).
$$
Strictly speaking this is not well-defined, because $\widehat{v_{\rm even}}$ and $\widehat{v_{\rm odd}}$ are just $N/2$-dimensional vectors while $\widehat{v}$ is $N$-dimensional.
However, note that $\omega_{N/2}^{jk/2}$ (for even $k$) and $\omega_{N/2}^{j(k-1)/2}$ (for odd $k$) don't change if we add $N/2$ to $j$. Hence we can just put two identical copies $\widehat{v_{\rm even}}$ on top of each other to get an $N$-dimensional vector (i.e., defining $\widehat{v_{\rm even}}_{j+N/2}=\widehat{v_{\rm even}}_{j}$), and similarly for $\widehat{v_{\rm odd}}$,  then it all works out in the sense of reproducing the $N$ entries given by Eq.~\eqref{eq:fftformula}.

The time $T(N)$ it takes to implement $F_N$ this way can be written recursively as $T(N)=2T(N/2)+O(N)$,
because we need to compute two $N/2$-dimensional Fourier transforms and do $O(N)$ additional operations to compute $\widehat{v}$.  
This recursion works out to time $T(N)=O(N\log N)$, as promised.
Similarly, we have an equally efficient algorithm for the \emph{inverse} Fourier transform $F_N^{-1}=F_N^*$,
whose entries are $\frac{1}{\sqrt{N}}\omega_N^{-jk}$.

\section{Application: multiplying two polynomials}

Suppose we are given two real-valued polynomials $p$ and $q$, each of degree at most $d$:
$$
p(x)=\sum_{j=0}^d a_jx^j\mbox{ and }
q(x)=\sum_{k=0}^d b_kx^k
$$
We would like to compute the product of these two polynomials, which is
$$
(p\cdot q)(x)=\left(\sum_{j=0}^d a_jx^j\right)\left(\sum_{k=0}^d b_kx^k\right)=\sum_{\ell=0}^{2d}(\underbrace{\sum_{j=0}^{2d}a_jb_{\ell-j}}_{c_\ell})x^\ell,
$$
where implicitly we set $a_j=b_j=0$ for $j>d$ and $b_{\ell-j}=0$ if $j>\ell$.
Clearly, each coefficient $c_\ell$ by itself takes $O(d)$ steps (additions and multiplications)
to compute, which suggests an algorithm for computing the coefficients of $p\cdot q$ that takes $O(d^2)$ steps.  However, using the fast Fourier transform we can do this in $O(d\log d)$ steps, as follows.

The \emph{convolution} of two vectors $a,b\in\mathbb{R}^N$ is a vector $a*b\in\mathbb{R}^N$ whose $\ell$-th entry is defined by 
$(a*b)_\ell=\frac{1}{\sqrt{N}}\sum_{j=0}^{N-1} a_j b_{\ell-j {\rm mod } N}$.
Let us set $N$ to $2d+1$ (the number of nonzero coefficients of $p\cdot q$), rounded up to the nearest power of~2.
Make the above $(d+1)$-dimensional vectors of coefficients $a$ and $b$ $N$-dimensional by adding zeroes. 
Then the coefficients of the polynomial $p\cdot q$ are proportional to the entries of the convolution: $c_\ell=\sqrt{N}(a*b)_\ell$.
It is easy to show (Exercise~\ref{ex:convolutionproduct}) that the Fourier coefficients of the convolution of $a$ and $b$ are the products of the Fourier coefficients of $a$ and $b$: 
for every $\ell\in\{0,\ldots,N-1\}$ we have $\left(\widehat{a * b}\right)_\ell=\widehat{a}_\ell\cdot\widehat{b}_\ell$.
This immediately suggests an algorithm for computing the vector of coefficients~$c_\ell$: apply the FFT to $a$ and $b$ to get $\widehat{a}$ and $\widehat{b}$, multiply those two vectors entrywise to get $\widehat{a*b}$, apply the inverse FFT to get $a*b$, and finally multiply $a*b$ with $\sqrt{N}$ to get the vector $c$ of the coefficients of $p\cdot q$.  Since the FFTs and their inverse take $O(N\log N)$ steps, and pointwise multiplication of two $N$-dimensional vectors takes $O(N)$ steps, this algorithm takes $O(N\log N)=O(d\log d)$ steps.

Note that if two numbers $a_d\cdots a_1a_0$ and $b_d\cdots b_1b_0$ are given in decimal notation, then we can interpret their digits
as coefficients of single-variate degree-$d$ polynomials $p$ and $q$, respectively: $p(x)=\sum_{j=0}^d a_jx^j$ and  $q(x)=\sum_{k=0}^d b_kx^k$. 
The two numbers will now be $p(10)$ and $q(10)$.
Their product is the evaluation of the product-polynomial $p\cdot q$ at the point $x=10$.
This suggests that we can use the above procedure (for fast multiplication of polynomials) to multiply two numbers in $O(d\log d)$ steps, 
which would be a lot faster than the standard $O(d^2)$ algorithm for multiplication that one learns in primary school.
However, in this case we have to be careful since the steps of the above algorithm are themselves multiplications between numbers, 
which we cannot count at unit cost anymore if our goal is to implement a multiplication between numbers!  
Still, it turns out that implementing this idea carefully allows one to multiply two $d$-digit numbers 
in $O(d\log d\log\log d)$ elementary operations.  This is known as the Sch\"onhage-Strassen algorithm~\cite{schonhage&strassen} (slightly improved further by F\"urer~\cite{furer:fastermultj} and Harvey and van der Hoeven~\cite{harvey&hoeven:fastermultj}, who made it $O(d\log d)$ elementary operations), and is one of the ingredients of Shor's algorithm (Chapter~\ref{chap:factoring}). We'll skip the details.

\section{The quantum Fourier transform}

Since $F_N$ is an $N\times N$ unitary matrix, we can interpret it as a quantum operation,
mapping an $N$-dimensional vector of amplitudes to another $N$-dimensional vector of amplitudes.
This is called the quantum Fourier transform (QFT).
In case $N=2^n$ (which is the only case we will care about), this will be an $n$-qubit unitary.
Notice carefully that this quantum operation does something different from the classical Fourier transform:
in the classical case we are given a vector $v$, written on a piece of paper so to say, and we compute
the vector $\widehat{v}=F_Nv$, and also write the result on a piece of paper.
In the quantum case, we are working on \emph{quantum states}; these are vectors of amplitudes, 
but we don't have those written down anywhere---they only exist as the amplitudes in a superposition.
We will see below that the QFT can be implemented by a quantum circuit using $O(n^2)$ elementary gates.
This is exponentially faster than even the FFT (which takes $O(N\log N)=O(2^n n)$ steps),
but it achieves something different: computing the QFT won't give us the entries of the Fourier
transform written down on a piece of paper, but only as the amplitudes of the resulting state.

\section{An efficient quantum circuit}\label{sec:qftcircuit}

Here we will describe the efficient circuit for the $n$-qubit QFT.
The elementary gates we will allow ourselves are Hadamards and controlled-$R_s$ gates, where 
$$
R_s=\left(\begin{array}{cc}1 & 0\\ 0 & e^{2\pi i/2^s}\end{array}\right). 
$$
Note that $R_1=Z=\left(\begin{array}{rr}1 & 0\\ 0 & -1\end{array}\right)$, $R_2=\left(\begin{array}{cc}1 & 0\\ 0 & i\end{array}\right)$.
For large~$s$, $e^{2\pi i/2^s}$ is close to~1 and hence the $R_s$-gate is close to the identity-gate~$I$. For simplicity we will just treat each $R_s$ as an elementary gate, rather than building them up out of some finite gate set.

Since the QFT is linear, it suffices if our circuit implements it correctly on all $n$-qubit basis states $\ket{k}$, i.e., it should map
$$
\ket{k}\mapsto F_N\ket{k}=\frac{1}{\sqrt{N}}\sum_{j=0}^{N-1}\omega_N^{jk}\ket{j}.
$$
The key to doing this efficiently is to rewrite $F_N\ket{k}$, which turns out to be a \emph{product state}
(so $F_N$ does not introduce entanglement when applied to a basis state~$\ket{k}$), as follows.
Let $\ket{k}=\ket{k_1\ldots k_n}$, $k_1$ being the most significant bit.
Note that for integer $j=j_1\ldots j_n$, we can write $j/2^n=\sum_{\ell=1}^n j_\ell 2^{-\ell}$.
For example, binary $0.101$ is $1\cdot 2^{-1}+0\cdot 2^{-2}+1\cdot 2^{-3}=5/8$.
We have the following sequence of equalities (which is probably most easily verified by working backwards from the last formula):
\begin{eqnarray*}
F_N\ket{k} & = & \frac{1}{\sqrt{N}}\sum_{j=0}^{N-1}e^{ 2\pi i j k/2^n}\ket{j}\\
           & = & \frac{1}{\sqrt{2^n}}\sum_{j\in\01^n}e^{ 2\pi i (\sum_{\ell=1}^n j_\ell 2^{-\ell})k}\ket{j_1\ldots j_n}\\
           & = & \frac{1}{\sqrt{2^n}}\sum_{j\in\01^n}\prod_{\ell=1}^n e^{2\pi i j_\ell k/2^\ell}\ket{j_1\ldots j_n}\\
           & = & \bigotimes_{\ell=1}^{n}\frac{1}{\sqrt{2}}\left(\ket{0}+e^{2\pi i k/2^\ell}\ket{1}\right).
\end{eqnarray*}
Note that $e^{2\pi i k/2^\ell}=e^{2\pi i\, k_1\ldots k_{n-\ell}.k_{n-\ell+1}\ldots k_{n}}=e^{2\pi i\, 0.k_{n-\ell+1}\ldots k_{n}}$: the $n-\ell$ most significant bits of $k$ don't matter for this value, because $e^{2\pi i m}=1$ if $m$ is an integer.

\begin{figure*}[htb]
\centering
\includegraphics[height=3.8cm]{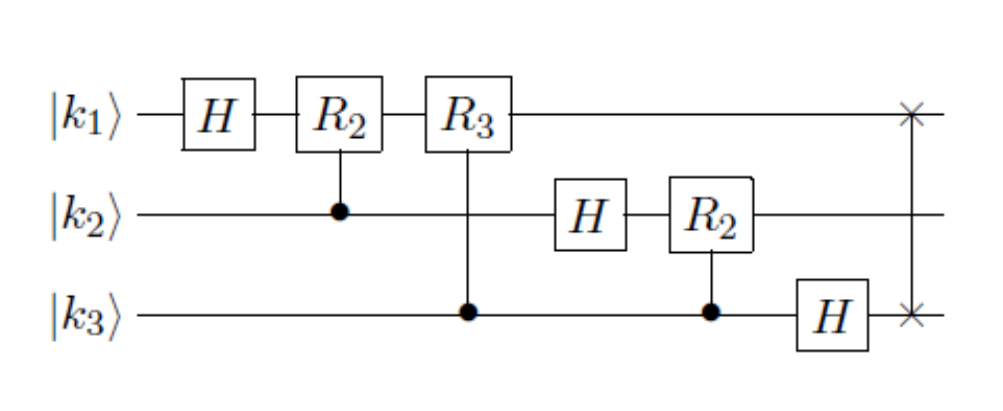}
\caption{The circuit for the 3-qubit QFT}\label{fig3qft}
\end{figure*}

As an example, for $n=2$ we have the 2-qubit product state
$$
F_4\ket{k_1k_2}=\frac{1}{\sqrt{2}}(\ket{0}+e^{2\pi i 0.k_2}\ket{1})\otimes\frac{1}{\sqrt{2}}(\ket{0}+e^{2\pi i 0.k_1k_2}\ket{1}).
$$
And for $n=3$ we have the 3-qubit product state
$$
F_8\ket{k_1k_2k_3}=\frac{1}{\sqrt{2}}(\ket{0}+e^{2\pi i 0.k_3}\ket{1})\otimes\frac{1}{\sqrt{2}}(\ket{0}+e^{2\pi i 0.k_2k_3}\ket{1})\otimes\frac{1}{\sqrt{2}}(\ket{0}+e^{2\pi i 0.k_1k_2k_3}\ket{1}).
$$
These examples suggest what the circuit should be.
To prepare the first qubit of the desired state $F_8\ket{k_1k_2k_3}$, we can just apply a Hadamard to $\ket{k_3}$, 
giving state $\frac{1}{\sqrt{2}}(\ket{0}+(-1)^{k_3}\ket{1})$ and observe that $(-1)^{k_3}=e^{2\pi i 0.k_3}$.
To prepare the second qubit of the desired state, apply a Hadamard to $\ket{k_2}$, 
giving $\frac{1}{\sqrt{2}}(\ket{0}+e^{2\pi i 0.k_2}\ket{1})$,
and then conditioned on $k_3$ (\emph{before} we apply the Hadamard to $\ket{k_3}$) apply $R_2$. 
This multiplies $\ket{1}$ with a phase $e^{2\pi i 0.0k_3}$, producing
the correct qubit $\frac{1}{\sqrt{2}}(\ket{0}+e^{2\pi i 0.k_2k_3}\ket{1})$.
Finally, to prepare the third qubit of the desired state, we apply a Hadamard to $\ket{k_1}$, 
apply $R_2$ conditioned on $k_2$, and $R_3$ conditioned on~$k_3$. 
This produces the correct qubit $\frac{1}{\sqrt{2}}(\ket{0}+e^{2\pi i 0.k_1k_2k_3}\ket{1})$.
We have now produced all three qubits of the desired state $F_8\ket{k_1k_2k_3}$, 
\emph{but in the wrong order}: the first qubit should be the third and vice versa. So the final step is just to swap qubits~1 and~3.
Figure~\ref{fig3qft} illustrates the circuit in the case $n=3$.  
Here the black circles indicate the control-qubits for each of the controlled-$R_s$ operations,
and the operation at the end of the circuit swaps qubits~1 and~3.
The general case works analogously: starting with $\ell=1$,
we apply a Hadamard to $\ket{k_\ell}$ and then ``rotate in'' the additional phases required, 
conditioned on the values of the later bits $k_{\ell+1}\ldots k_n$.
Some swap-gates at the end then put the qubits in the right order (recall that we can implement a swap with 3 CNOTs by  Exercise~\ref{chap:qircuitdj}.\ref{ex:swapcnot}).

Since the circuit involves $n$ qubits, and at most $n$ gates are applied to each qubit, the overall circuit uses at most $n^2$ gates.
In fact, many of those gates are phase gates $R_s$ with $s\gg\log n$, which are very close to the identity and hence
don't do much anyway. As observed by Coppersmith~\cite{coppersmith:fourier}, we can actually omit those from the circuit, keeping only $O(\log n)$ gates per qubit and $O(n\log n)$ gates overall.
Intuitively, the overall error caused by these omissions will be small (Exercise~\ref{ex:approxqft} asks you to make this precise).
Finally, note that by inverting the circuit (i.e., reversing the order of the gates and taking the adjoint $U^*$ of each gate $U$)
we obtain an equally efficient circuit for the inverse Fourier transform $F_N^{-1}=F_N^*$.

\section{Application: phase estimation}\label{ssecphaseestimation}

An important application of the QFT is in \emph{phase estimation}. This was originally due to Kitaev~\cite{kitaev:stabilizer}, it was put in a broader context by Cleve et al.~\cite{cemm:revisited}, and is now a very common subroutine in many quantum algorithms.

\begin{figure}[hbt]
\centering
\includegraphics[width=15cm]{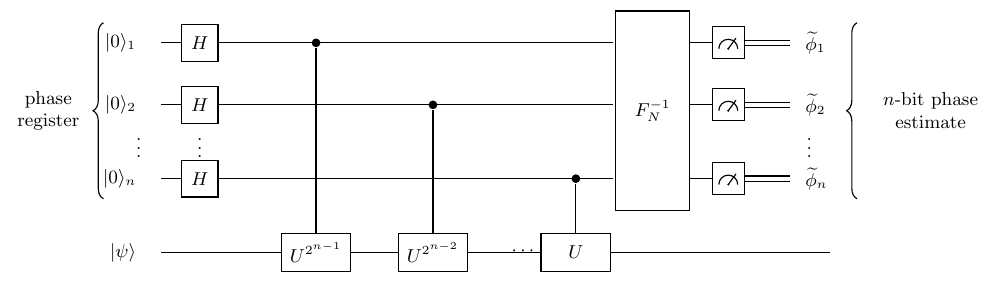}
\caption{Estimating the phase of the eigenvalue of $U$'s eigenstate $\ket{\psi}$ with $n$ bits of precision}\label{fig:QPEcircuit}
\end{figure}

Suppose we can apply a unitary $U$ and we are given an eigenvector $\ket{\psi}$ of $U$ with corresponding unknown  eigenvalue $\lambda$ (i.e., $U\ket{\psi}=\lambda\ket{\psi}$),
and we would like to compute or at least approximate
this~$\lambda$.  Since $U$ is unitary, $\lambda$ must have magnitude~1,
so we can write it as $\lambda=e^{2\pi i\phi}$ for some real number $\phi\in[0,1)$.
Here is the algorithm for phase estimation, illustrated in Figure~\ref{fig:QPEcircuit}:
\begin{enumerate}
\item Start with $\ket{0^n}\ket{\psi}$.
\item For $N=2^n$, apply $F_N$ to the first $n$ qubits to get $\frac{1}{\sqrt{2^n}}\sum_{j=0}^{N-1}\ket{j}\ket{\psi}$;\\
in fact, $H^{\otimes n}\otimes I$ would have the same effect.
\item Apply the map $\ket{j}\ket{\psi}\mapsto\ket{j}U^j\ket{\psi}=e^{2\pi i \phi j}\ket{j}\ket{\psi}$.
In other words, apply~$U$ to the second register for a number of times given by the first register. We can split this into a number of controlled applications of powers of 2 of $U$: controlled on the most significant bit of $j$, apply $U^{2^{n-1}}$; conditioned on the next most significant bit of $j$, apply $U^{2^{n-2}}$, etc.
\item Apply the inverse Fourier transform $F_N^{-1}$ to the first $n$ qubits and measure the result.
\end{enumerate}

There are $1+2+\cdots+2^{n-1}=2^n-1$ controlled applications of $U$ in total.
The actual cost of this depends on the application.
Note that after step~3, the first $n$ qubits are in state $\frac{1}{\sqrt{N}}\sum_{j=0}^{N-1}e^{2\pi i \phi j}\ket{j}=F_N\ket{2^n\phi}$. Hence, in the special case that $\phi$ can be written exactly with $n$ bits, the inverse Fourier transform is going to give us $\ket{2^n\phi}=\ket{\phi_1\ldots\phi_n}$ with probability~1!

In the general case that $\phi$ can\emph{not} be written exactly with $n$ bits of precision, then one can
show that this procedure still (with high probability) spits out a good $n$-bit approximation to $\phi$,  with additive error $O(2^{-n})$; we'll omit the reasonably straightforward calculation that proves this. Accordingly, if we want to compute an estimate of $\phi$ within additive error $\pm \eps$, then we could run phase estimation with $n=\ceil{\log(1/\eps)}$ bits of precision.
The main cost of this would be the number of applications of $U$ in the phase-estimation circuit, which is $\leq 2^n=O(1/\eps)$.

\section*{Exercises}

\begin{enumerate}

\item
For $\omega=e^{2\pi i/3}$ and $F_3=\frac{1}{\sqrt{3}}\left(\begin{array}{lll}1 & 1 & 1\\ 1 & \omega & \omega^2\\ 1 & \omega^2 & \omega\end{array}\right)$,
calculate $F_3\left(\begin{array}{l}0\\ 1\\ 0\end{array}\right)$ and
$F_3\left(\begin{array}{l}1\\ \omega^2\\ \omega\end{array}\right)$

\item\label{ex:convolutionproduct}
Prove that the Fourier coefficients of the convolution of vectors $a$ and $b$ are the product of the Fourier coefficients of $a$ and $b$. In other words, prove that for every $a,b\in\mathbb{R}^N$ and every $\ell\in\{0,\ldots,N-1\}$ we have $\left(\widehat{a * b}\right)_\ell=\widehat{a}_\ell\cdot\widehat{b}_\ell$. Here the Fourier transform $\widehat{a}$ is defined as the vector $F_N a$, and the $j$-th entry of the convolution-vector $a*b$ is
$(a*b)_j=\frac{1}{\sqrt{N}}\sum_{k=0}^{N-1} a_k b_{(j-k) {\rm mod } N}$.

\item\label{ex:tvddistance} (H)
  The \emph{total variation distance} between two probability distributions $P$ and $Q$ on the same set, is defined as $d_{TVD}(P,Q)=\frac{1}{2}\sum_i\left|P(i)-Q(i)\right|$. An equivalent alternative way to define this:
  $d_{TVD}(P,Q)$ is the maximum, over all events~$E$, of $|P(E)-Q(E)|$. Hence $d_{TVD}(P,Q)$ is small iff all events have roughly the same probability under~$P$ and under~$Q$.
  
  The \emph{Euclidean distance} between two states $\ket{\phi}=\sum_i\alpha_i\ket{i}$
and $\ket{\psi}=\sum_i\beta_i\ket{i}$ is defined as $\norm{\ket{\phi}-\ket{\psi}}=\sqrt{\sum_i|\alpha_i-\beta_i|^2}$.
Assume the two states are unit vectors with (for simplicity) real amplitudes.
Suppose the Euclidean distance is small: $\norm{\ket{\phi}-\ket{\psi}}=\epsilon$.
If we measure $\ket{\phi}$ in the computational basis then the probability distribution over the outcomes is given by the $|\alpha_i|^2$, and if we measure $\ket{\psi}$ then the probabilities are $|\beta_i|^2$. Show that these distributions are close: the total variation distance $\frac{1}{2}\sum_i\left|\alpha_i^2-\beta_i^2\right|$ is $\leq \epsilon$.

\item\label{ex:approxqft} (H)
The \emph{operator norm} of a matrix~$A$ is defined as $\displaystyle\norm{A}=\max_{v:\norm{v}=1}\norm{Av}$.\\ 
An equivalent definition is that $\norm{A}$ is the largest singular value of~$A$ (see Appendix~\ref{secsvd}).\\[0.3em]
The \emph{distance} between two matrices $A$ and $B$ is defined as $\displaystyle\norm{A-B}$.
\begin{enumerate}
\item  What is the distance between the $2\times 2$ identity matrix and the phase-gate $\left(\begin{array}{rr}1 & 0\\ 0 & e^{i\phi}\end{array}\right)$?
 \item  What is the distance between the $4\times 4$ identity matrix and the \emph{controlled} version of the phase gate of (a)?
\item
What is the distance between the $2^n\times 2^n$ identity matrix $I_{2^n}$ and the controlled phase gate of (b) tensored with $I_{2^{n-2}}$?
\item Suppose we have a product of $n$-qubit unitaries $U=U_TU_{T-1}\cdots U_1$ (for instance, each $U_i$ could be an elementary gate on a few qubits, tensored with identity on the other qubits). Suppose we drop the $j$-th gate from this sequence:
$U'=U_TU_{T-1}\cdots U_{j+1}U_{j-1}\cdots U_1$.  Show that $\norm{U'-U}=\norm{I-U_j}$.
\item Now we also drop the $k$-th unitary: $U''=U_TU_{T-1}\cdots U_{j+1}U_{j-1}\cdots\cdots U_{k+1}U_{k-1}\cdots U_1$.
Show that $\norm{U''-U}\leq\norm{I-U_j}+\norm{I-U_k}$.
\item Give a quantum circuit with $O(n\log n)$ elementary gates that has distance less than $1/n$ from the Fourier transform $F_{2^n}$.
\end{enumerate}
{\footnotesize Comment: The above exercise shows the important fact that if we have a quantum circuit $C$ that has various subparts (``subroutines''), then a circuit $\tilde{C}$ where those subroutines  are implemented with small operator-norm error, rather than perfectly, still works well: if $\norm{C-\tilde{C}}$ is small then (by definition of operator norm) for all initial states $\ket{\phi}$ the states $C\ket{\phi}$ and $\tilde{C}\ket{\phi}$ are close in Euclidean distance. By Exercise~\ref{ex:tvddistance} then also the final output distributions are close (in total variation distance).}

\item Suppose $a\in\mathbb{R}^N$ is a vector (indexed by $\ell=0,\ldots,N-1$) which is $r$-periodic in the following sense:
there exists an integer $r$ such that $a_\ell=1$ whenever $\ell$ is an integer multiple of $r$, and $a_\ell=0$ otherwise.
Compute the Fourier transform $F_N\, a$ of this vector, i.e., write down a formula for the entries of the vector $F_N a$.  
Assuming $r$ divides $N$, write down a simple closed form for the formula for the entries. Which are the nonzero entries in the vector $F_N\, a$, and what is their magnitude?

\item 
\begin{enumerate}
    \item The squared Fourier transform, $F_N^2$, turns out to map computational basis states to computational basis states. Describe this map, i.e., determine to which basis state a basis state $\ket{k}$ gets mapped for each $k\in\01^n$. 
    \item Show that $F_N^4=I$. What can you conclude about the eigenvalues of $F_N$?
\end{enumerate}

\end{enumerate}

\chapter{Shor's Factoring Algorithm}\label{chap:factoring}

\section{Factoring}
Probably the most important quantum algorithm so far is Shor's factoring
algorithm~\cite{shor:factoring}.\index{Shor, P.}\index{factoring}
It can find a factor of a composite number $N$ in roughly $(\log N)^2$ 
steps, which is polynomial in the length $\log N$ of the input.
On the other hand, there is no known classical (deterministic
or randomized) algorithm that can factor $N$ in polynomial time.
The best known classical randomized algorithms run in time roughly
$$
2^{(\log N)^\alpha},
$$
where $\alpha=1/3$ for a heuristic upper bound~\cite{lenstras:sieve}\index{Lenstra, A.}\index{Lenstra, H.}
and $\alpha=1/2$ for a less-heuristic but still not fully proven upper bound~\cite{lenstra&pomerance:rigorous}.\index{Lenstra, H.}\index{Pomerance, C.}
In fact, much of modern cryptography\index{cryptography} is based on the conjecture 
that no fast classical factoring algorithm exists~\cite{rivest:crypto}.\index{Rivest, R.}
All this cryptography (for example RSA\index{RSA}) would be broken if Shor's algorithm 
could be physically realized.
In terms of complexity classes: factoring (rather, the decision problem
equivalent to it) is provably in $\BQP$\index{BQP} but is not known to be in $\BPP$.\index{BPP}
If indeed factoring is not in $\BPP$, 
then the quantum computer would be the first counterexample 
to the ``strong'' Church-Turing thesis,\index{Church-Turing thesis!strong} 
which states that all ``reasonable'' models of computation are polynomially equivalent
(see~\cite{emdeboas:mm}\index{Emde Boas, P.~van} and~\cite[p.31,36]{papadimitriou:cc}).\index{Papadimitriou, C.}

\section{Reduction from factoring to period-finding}\label{ssecperiodfinding}

The crucial observation of Shor was that there is an efficient
quantum algorithm for the problem of \emph{period-finding}\index{period-finding} 
and that factoring can be reduced to this, in the sense that an efficient algorithm for period-finding implies an efficient algorithm for factoring.

We first explain the reduction.
Suppose we want to find factors of the composite number $N>1$.
We may assume $N$ is odd and not a prime power, since those cases can easily be filtered out by a classical algorithm.
Now choose some integer $x\in\{2,\ldots,N-1\}$ uniformly at random, and compute its greatest common divisor with~$N$.%
\footnote{The \emph{greatest common divisor} (gcd) of two integers $a$ and $b$ is the largest positive integer $c$ that divides both $a$ and $b$. If $gcd(a,b)=1$, then $a$ and $b$ are called \emph{coprime}. The gcd can be computed in time roughly quadratic in the number of bits of $a$ and $b$ on a classical computer by the over-2000-year-old \emph{Euclid's algorithm}, and in nearly linear time by a slightly more recent improved algorithm~\cite{SZ:binarygcd}.
More generally, the \emph{extended Euclidean algorithm} allows to efficiently find integers $x$ and $y$ such that $ax+by=gcd(a,b)$.}\index{greatest common divisor} If $x$ is not coprime to~$N$, then $gcd(x,N)$ is a nontrivial factor of~$N$, so then we are already done.
From now on consider $x$ and $N$ are coprime, so $x$ is an element of the multiplicative group $\mathbb{Z}_N^*$. 
Consider the sequence 
$$
1=x^0\pmod N,~~~x^1\pmod N,~~~x^2\pmod N,\ldots
$$
Because $x$ generates a cyclic subgroup of $\mathbb{Z}_N^*$, this sequence will cycle after a while: 
there is a least $0<r\leq N$ such that $x^r=1\pmod N$. 
This $r$ is called the \emph{period} of the sequence (a.k.a.\ the \emph{order} of the element $x$ in the group $\mathbb{Z}_N^*$).
Assuming $N$ is odd and not a prime power (those cases are easy to factor anyway), it can be shown that with probability $\geq 1/2$, the period~$r$ is even and $x^{r/2}+1$ and $x^{r/2}-1$ are not multiples of $N$~\cite[Theorem~A4.13]{nielsen&chuang:qc}.
In that case we have:
\begin{eqnarray*}
x^r                    & \equiv & 1 \mod N \hspace{5mm} \Longleftrightarrow \\
(x^{r/2})^2            & \equiv & 1 \mod N \hspace{5mm} \Longleftrightarrow \\
(x^{r/2}+1)(x^{r/2}-1) & \equiv & 0 \mod N \hspace{5mm} \Longleftrightarrow \\
(x^{r/2}+1)(x^{r/2}-1) & = & kN    \mbox{ for some $k$}.
\end{eqnarray*}
Note that $k>0$ because both $x^{r/2}+1>0$ and $x^{r/2}-1>0$ ($x>1$).
Hence $x^{r/2}+1$ or $x^{r/2}-1$ will share a factor with $N$.
Because $x^{r/2}+1$ and $x^{r/2}-1$ are not multiples of $N$
this factor will be $<N$, and in fact \emph{both} these numbers will share 
a non-trivial factor with $N$.       
Accordingly, if we have $r$ then we can compute the \emph{greatest common divisors} $gcd(x^{r/2}+1,N)$ and $gcd(x^{r/2}-1,N)$, 
and both of these two numbers will be non-trivial factors of~$N$. 
If we are unlucky we might have chosen an $x$ that does not give a factor (which we can detect efficiently), but trying a few different random $x$ 
gives a high probability of finding a factor.

Thus the problem of factoring reduces to finding the period $r$
of the function given by modular exponentiation $f(a)=x^a\mod N$.
In general, the period-finding problem can be stated as follows:

\bigskip

\noindent
{\bf The period-finding problem}:\index{period-finding}\\
We are given some function $f:\mathbb{N}\rightarrow\{0,\ldots,N-1\}$ 
with the property that there is some unknown $r\in\{1,\ldots,N\}$ 
such that $f(a)=f(b)$ iff $a=b\mod r$.
The goal is to find $r$.

\bigskip

\noindent
One might think that if $f$ itself is efficiently computable, then period-finding is an easy problem to solve even on a classical computer:
just compute $f(0),f(1),f(2),\ldots$ until we encounter the value $f(0)$ for the second time.
The input at which this happens is the period~$r$ that we're trying to find.
The problem with this approach is that $r$ could be huge, for instance $N^{1/2}$ or $N^{1/100}$, which is exponentially large in the number of inputs bits. To be efficient, we would like a runtime that is polynomial in $\log N$, since that is the bitsize of the inputs to~$f$. It is generally believed that classical computers cannot solve period-finding efficiently.

We will show below how we \emph{can} solve this problem efficiently on a quantum computer,
using only $O(\log\log N)$ evaluations of $f$ and $O(\log\log N)$ 
quantum Fourier transforms.\index{quantum Fourier transform} 
An evaluation of $f$ can be viewed as analogous to the
application of a query in the algorithms of the previous chapters.
Even a somewhat more general kind of period-finding can be solved
by Shor's algorithm with very few $f$-evaluations, whereas any
classical bounded-error algorithm would need to evaluate the function 
$\Omega(N^{1/3}/\sqrt{\log N})$ times in order to find 
the period~\cite{cleve:orderfinding}.\index{Cleve, R.}

How many steps (elementary gates) does Shor's algorithm take?
For $a=N^{O(1)}$, we can compute $f(a)=x^a\mod N$
in $O((\log N)^2\log\log N\log\log\log N)$ steps by the ``square-and-multiply'' method, using known algorithms for fast integer multiplication mod~$N$, see Exercise~\ref{ex:modexp}. 

Moreover, as explained in the previous chapter, the quantum Fourier transform can be implemented using $O((\log N)^2)$ steps.
Accordingly, Shor's algorithm finds a factor of $N$ using an
expected number of $O((\log N)^2(\log\log N)^2\log\log\log N)$ gates, 
which is only slightly worse than quadratic in the input-length.

\section{Shor's period-finding algorithm}

Now we will show how Shor's algorithm finds the period $r$ of the function $f$,
given a ``black-box'' that maps $\ket{a}\ket{0^n}\mapsto\ket{a}\ket{f(a)}$.
We can always efficiently pick some $q=2^\ell$ such that 
$N^2<q\leq 2N^2$. Then we can implement the Fourier transform $F_q$ 
using $O((\log N)^2)$ gates.
Let $O_f$ denote the unitary that maps $\ket{a}\ket{0^n}\mapsto\ket{a}\ket{f(a)}$,
where the first register consists of $\ell$ qubits, and the second of $n=\ceil{\log N}$ qubits.

\begin{figure}[hbt]
\centering
\setlength{\unitlength}{0.35mm}
\begin{picture}(280,145)
\put(0,130){\makebox(20,20){$\ket{0}$}}
\put(10,107){\makebox(20,20){$\vdots$}}
\put(0,80){\makebox(20,20){$\ket{0}$}}
\put(0,50){\makebox(20,20){$\ket{0}$}}
\put(0,0){\makebox(20,20){$\ket{0}$}}
\put(260,107){\makebox(20,20){measure}}
\put(200,20){\makebox(20,20){measure}}
\put(230,107){\makebox(20,20){$\vdots$}}
\put(20,140){\line(1,0){25}}
\put(20,90){\line(1,0){25}}
\put(20,60){\line(1,0){90}}
\put(20,10){\line(1,0){90}}
\put(85,140){\line(1,0){25}}
\put(85,90){\line(1,0){25}}
\put(45,80){\framebox(40,70){}}
\put(65,115){\makebox(0,0){$F_q$}}
\put(65,28){\makebox(20,20){$\vdots$}}
\put(165,28){\makebox(20,20){$\vdots$}}
\put(215,140){\line(1,0){25}}
\put(215,90){\line(1,0){25}}
\put(110,0){\framebox(40,150){}}
\put(130,70){\makebox(0,0){$O_f$}}
\put(150,140){\line(1,0){25}}
\put(150,90){\line(1,0){25}}
\put(150,60){\line(1,0){25}}
\put(150,10){\line(1,0){25}}
\put(175,80){\framebox(40,70){}}
\put(195,115){\makebox(0,0){$F_q$}}
\end{picture}
\caption{Shor's period-finding algorithm}\label{figshor}
\end{figure}

Shor's period-finding algorithm is illustrated in Figure~\ref{figshor}.%
\footnote{Notice the resemblance of the basic structure (Fourier, $f$-evaluation, Fourier) with the basic 
structure of Simon's\index{Simon, D.} algorithm (Hadamard, query, Hadamard). This is not a coincidence, because Shor was inspired by reading Simon's paper. The number of qubits used is roughly $3\log N$: $\log q\approx 2\log N$ qubits for the first register, and $\ceil{\log N}$ for the second register. This number can be reduced to slightly more than $2\log N$ qubits~\cite{beauregard:shor,hallgren:phd}. Accordingly, to factor for instance a 2048-bit integer~$N$, slightly more than 4096 (perfect) qubits suffice.}
Start with $\ket{0^\ell}\ket{0^n}$. Apply the QFT (or just $\ell$ Hadamard gates) to the first register to build 
the uniform superposition
$$
\frac{1}{\sqrt{q}}\sum_{a=0}^{q-1}\ket{a}\ket{0^n}.
$$
The second register still consists of zeroes.
Now use the ``black-box'' to compute $f(a)$ in quantum parallel:
$$
\frac{1}{\sqrt{q}}\sum_{a=0}^{q-1}\ket{a}\ket{f(a)}.
$$
Observing the second register gives some value $f(s)$, with $s<r$.
Let $m$ be the number of elements of $\{0,\ldots,q-1\}$ that map to the observed value $f(s)$. 
Because $f(a)=f(s)$ iff $a=s\mod r$, the $a$ of the form $a=jr+s$ 
($0\leq j<m$) are exactly the $a$ for which $f(a)=f(s)$.
Thus the first register collapses to a superposition of 
$\ket{s},\ket{r+s},\ket{2r+s},\ket{3r+s},\ldots$; this superposition runs until the last number of the form $jr+s$ that is $<q$, let's define $m$ to be the number of elements in this superposition, i.e., the number of integers $j$ such that $jr+s\in\{0,\ldots,q-1\}$ (depending on $s$, this~$m$ will be $\ceil{q/r}$ or $\floor{q/r}$). 
The second register collapses to the classical state $\ket{f(s)}$.
We can now ignore the second register, and have in the first:
$$
\frac{1}{\sqrt{m}}\sum_{j=0}^{m-1}\ket{jr+s}.
$$
Applying the QFT again gives
$$
\frac{1}{\sqrt{m}}\sum_{j=0}^{m-1}\frac{1}{\sqrt{q}}\sum_{b=0}^{q-1}e^{2\pi i\frac{(jr+s)b}{q}}\ket{b}=
\frac{1}{\sqrt{mq}}\sum_{b=0}^{q-1}e^{2\pi i\frac{sb}{q}}\left(\sum_{j=0}^{m-1}e^{2\pi i\frac{jrb}{q}}\right)\ket{b}.
$$
We want to see which $\ket{b}$ have amplitudes with large squared absolute value---those are the $b$ we are likely to see if we now measure.
Using that $\sum_{j=0}^{m-1} z^j=(1-z^m)/(1-z)$ for $z\neq 1$ (see Appendix~\ref{appusefulmath}), we compute:
\begin{equation}\label{eqshoramplitudes}
\sum_{j=0}^{m-1}e^{2\pi i\frac{jrb}{q}}=
\sum_{j=0}^{m-1}\left(e^{2\pi i\frac{rb}{q}}\right)^j=
\left\{
\begin{array}{ll}
m                                       & \mbox{if } e^{2\pi i\frac{rb}{q}}=1\\
\frac{1-e^{2\pi i\frac{mrb}{q}}}{1-e^{2\pi i\frac{rb}{q}}} & \mbox{if } e^{2\pi i\frac{rb}{q}}\neq 1
\end{array}\right.
\end{equation}

\paragraph{Easy case: $r$ divides $q$.}
Let us do an easy case first. Suppose $r$ divides $q$, so the whole period ``fits'' an
integer number of times in the domain $\{0,\ldots,q-1\}$ of $f$, and $m=q/r$. 
For the first case of Eq.~(\ref{eqshoramplitudes}), 
note that $e^{2\pi i rb/q}=1$ iff $rb/q$ is an integer iff $b$ is a multiple of $q/r$.
Such $b$ will have squared amplitude equal to $(m/\sqrt{mq})^2=m/q=1/r$.
Since there are exactly $r$ such basis states $b$, together they have all the amplitude: the sum of squares of those amplitudes is~1, so the amplitudes of $b$ that are \emph{not} integer multiples of $q/r$ must all be~0.
Thus we are left with a superposition where only the $b$ that are integer multiples 
of $q/r$ have nonzero amplitude. Observing this final superposition gives 
some random multiple $b=cq/r$, with $c$ a uniformly random number in $\{0,\ldots,r-1\}$.
Thus we get a $b$ such that 
$$
\frac{b}{q}=\frac{c}{r},
$$
where $b$ and $q$ are known to the algorithm, and $c$ and $r$ are not.
There are $\phi(r)\in\Omega(r/\log\log r)$ numbers smaller 
than $r$ that are coprime to $r$~\cite[Theorem~328]{hardy&wright}, 
so $c$ will be coprime to $r$ with probability $\Omega(1/\log\log r)\geq \Omega(1/\log\log N)$. Accordingly, an expected number of $O(\log\log N)$ repetitions of the 
procedure of this section suffices to obtain a $b=cq/r$ with $c$ 
coprime to $r$.\footnote{The number of required $f$-evaluations for period-finding can
actually be reduced from $O(\log\log N)$ to $O(1)$.}
Once we have such a~$b$, we can obtain $r$ as the denominator by writing $b/q$ in lowest terms.
Of course, our algorithm doesn't actually know whether $c$ and~$r$ are coprime in some particular run of the algorithm, but it can efficiently check if the purported factors $gcd(x^{r/2}+1,N)$ and $gcd(x^{r/2}-1,N)$ are actual factors of~$N$ by division (which, like multiplication, can be done classically with a near-linear number of gates).

\paragraph{Hard case: $r$ does not divide $q$.}
Because our $q$ is a power of~2, it is actually quite likely that $r$ does not divide $q$.
However, the same algorithm will still yield with high probability a $b$ which is \emph{close} to a multiple of $q/r$.
Note that $q/r$ is no longer an integer, and $m=\floor{q/r}$, possibly $+1$.
All calculations up to and including Eq.~(\ref{eqshoramplitudes}) are still valid.
Using $|1-e^{i\theta}|=2|\sin(\theta/2)|$, we can rewrite the absolute value of the second case of Eq.~(\ref{eqshoramplitudes}) to
$$
\frac{|1-e^{2\pi i \frac{mrb}{q}}|}{|1-e^{2\pi i\frac{rb}{q}}|}=\frac{|\sin(\pi mrb/q)|}{|\sin(\pi rb/q)|}.
$$
The right-hand side is the ratio of two sine-functions of $b$, 
where the numerator oscillates much faster than the denominator because of the additional factor of $m$.
Note that the denominator is close to~0 (making the ratio large) iff $b$ is close to an integer multiple of $q/r$.
For most of those~$b$, the numerator won't be close to~0.
Hence, roughly speaking, the ratio will be small if $b$ is far from an integer multiple of $q/r$, 
and large for most $b$ that are close to a multiple of $q/r$.
Doing the calculation precisely, one can show that with probability at least $4/\pi^2$ (see~\cite{shor:factoring,nielsen&chuang:qc} for details) the measurement yields a $b$ such that
$$
\left|\frac{b}{q}-\frac{c}{r}\right|\leq\frac{1}{2q},
$$
for a random $c\in \{0,\ldots,r-1\}$. 
Equivalently, $|b-cq/r|\leq 1/2$, so the measurement outcome $b$ will be an integer multiple of $q/r$ rounded up or down to an integer.
As in the easy case, $b$ and $q$ are known to us while $c$ and $r$ are unknown.

Because the known ratio $b/q$ is now not exactly equal to the unknown ratio $c/r$, we cannot just try to find $r$ by writing $b/q$ in lowest terms like we did in the easy case.
However, two distinct fractions, each with denominator $\leq N$, must be
at least $1/N^2>1/q$ apart.\footnote{Consider two fractions 
  $c/r$ and $c'/r'$ with integer $c,c',r,r'$, and $r,r'\leq N$.
  If $c/r\neq c'/r'$ then 
$cr'-c'r$ is a nonzero integer, and hence $|c/r-c'/r'|=|(cr'-c'r)/rr'|\geq 1/|rr'|\geq 1/N^2$.}
Therefore $c/r$ is the \emph{only} fraction with 
denominator $\leq N$ at distance $\leq 1/2q$ from the known ratio~$b/q$.
Applying a classical method called ``continued-fraction expansion''
to $b/q$ efficiently gives us the fraction with denominator $\leq N$ 
that is closest to $b/q$ (see next section).  
This fraction must be $c/r$.
Again, $c$ and $r$ will be coprime with probability $\Omega(1/\log\log r)$,
in which case writing $c/r$ in lowest terms gives~$r$.

\section{Continued fractions}

Let $[a_0,a_1,a_2,\ldots]$ (finite or infinite) denote the real number
$$
a_0+\frac{1}{
a_1+\frac{1}{
a_2+\frac{1}{\ldots}}}
$$
This is called a \emph{continued fraction} (CF). The $a_i$ are the
\emph{partial quotients}. We assume these to be positive natural
numbers (\cite[p.131]{hardy&wright} calls such CF ``simple'').
$[a_0,\ldots,a_n]$ is the $n$-th \emph{convergent} of the fraction.
\cite[Theorem~149 \&\ 157]{hardy&wright} gives a simple way to compute
numerator and denominator of the $n$-th convergent from the partial quotients:
\begin{quote}
If
$$
\begin{array}{lll}
p_0=a_0, & p_1=a_1a_0+1, & p_n=a_np_{n-1}+p_{n-2}\\
q_0=1,   &  q_1=a_1,     & q_n=a_nq_{n-1}+q_{n-2}
\end{array}
$$
then $[a_0,\ldots,a_n]=\displaystyle\frac{p_n}{q_n}$.
Moreover, this fraction is in lowest terms.
\end{quote}
Note that $q_n$ increases at least exponentially with
$n$ (because $q_{n}\geq 2q_{n-2}$).
Given a real number $x$, the following ``algorithm'' gives a continued
fraction expansion of $x$~\cite[p.135]{hardy&wright}:
$$
\begin{array}{ll}
a_0:=\floor{x},   & x_1:=1/(x-a_0)\\
a_1:=\floor{x_1}, & x_2:=1/(x_1-a_1)\\
a_2:=\floor{x_2}, & x_3:=1/(x_2-a_2)\\
\ldots &
\end{array}
$$
Informally, we just take the integer part of the number as the partial quotient
and continue with the inverse of the decimal part of the number.
The convergents of the CF approximate $x$ as follows~\cite[Theorem~164 \&\ 171]{hardy&wright}:
\begin{quote}
If $x=[a_0,a_1,\ldots]$ then
$\displaystyle\left|x-\frac{p_n}{q_n}\right|<\frac{1}{q_n^2}$.
\end{quote}
Recall that $q_n$ increases exponentially with $n$, so this convergence is quite fast. Moreover, $p_n/q_n$ provides the \emph{best} approximation of $x$ among
all fractions with denominator $\leq q_n$~\cite[Theorem~181]{hardy&wright}:
\begin{quote}
If $n>1$, $q\leq q_n$, $p/q\neq p_n/q_n$, then
$\displaystyle\left|x-\frac{p_n}{q_n}\right|<\left|x-\frac{p}{q}\right|$.
\end{quote}

\section*{Exercises}

\begin{enumerate}

\item\label{ex:modexp} 
This exercise is about efficient classical implementation of modular exponentiation. 
\begin{enumerate}
\item (H) Given $n$-bit numbers $x$ and $N$, compute the whole sequence\\
$x^0$ mod $N$, $x^1$ mod $N$, $x^2$ mod $N$, $x^4$ mod $N$, $x^8$ mod $N$, $x^{16}$ mod $N,\ldots,x^{2^{n-1}}$ mod $N$,\\
using $O(n^2\log(n)\log\log(n))$ steps.
\item Suppose $n$-bit number $a$ can be written as $a=a_{n-1}\ldots a_1a_0$ in binary.  Express $x^a$ mod $N$ as a product of the numbers computed in part (a).
\item Show that you can compute $f(a)=x^a$ mod $N$ in $O(n^2\log(n)\log\log(n))$ steps.
\end{enumerate}

\item Consider the function $f(a)=7^a \mbox{ mod }10$.
\begin{enumerate}
\item What is the period $r$ of $f$?
\item Show how Shor's algorithm finds the period of~$f$, using a Fourier transform over $q=128$ elements.
Write down all intermediate superpositions of the algorithm for this case (don't just copy the general expressions from the notes, but instantiate them with actual numbers as much as possible, incl.\ with the value of the period found in~(a)).
You may assume you're lucky, meaning the first run of the algorithm already gives a measurement outcome $b=cq/r$ with $c$ coprime to~$r$.
\end{enumerate}

\item\label{ex:rsa}
(H)
This exercise explains basic RSA encryption.  Suppose Alice wants to allow other people to send encrypted messages to her, such that she is the only one who can decrypt them. She believes that factoring an $n$-bit number can't be done efficiently (efficient $=$ in time polynomial in $n$).  So in particular, she doesn't believe in quantum computing.

Alice chooses two large random prime numbers, $p$ and $q$, and computes their product $N=p\cdot q$ (a typical size is to have $N$ a number of $n=2048$ bits, which corresponds to both $p$ and $q$ being numbers of roughly 1024 bits).
She computes the so-called Euler $\phi$-function: $\phi(N)=(p-1)(q-1)$; she also chooses an \emph{encryption exponent} $e$, which doesn't share any nontrivial factor with $\phi(N)$ (i.e., $e$ and $\phi(N)$ are coprime).
Group theory guarantees there is an efficiently computable \emph{decryption exponent} $d$ such that $de=1$ mod $\phi(N)$. The \emph{public key} consists of $e$ and~$N$ (Alice puts this on her homepage), while the \emph{secret key} consists of $d$ and the factorization of~$N$.
Any number $m\in\{1,\ldots,N-1\}$ that is coprime to $N$, can be used as a message. There are $\phi(N)$ such $m$, and these numbers form a group under the operation of multiplication mod $N$.
The number of bits $n=\lceil\log N\rceil$ of $N$ is the maximal length (in bits) of a message $m$ and also the length (in bits) of the encryption.
The encryption function is defined as $C(m)=m^e$ mod $N$, and the decryption function is $D(c)=c^d$ mod $N$.
\begin{enumerate}
\item Give a randomized algorithm by which Alice can efficiently generate the secret and public key.
\item Show that Bob can efficiently compute the encoding $C(m)$ of the message $m$ that he wants to send to Alice, knowing the public key but not the private key.
\item Show that $D(C(m))=m$ for all possible messages.
\item Show that Alice can efficiently decrypt the encryption $C(m)$ she receives from Bob.
\item Show that if Charlie could factor $N$, then he could efficiently decrypt Bob's message.
\end{enumerate}

\end{enumerate}

\chapter{Hidden Subgroup Problem}\label{chap:hsp}

\section{Hidden Subgroup Problem}

\subsection{Group theory reminder}\label{sec:grouptheory}

A group $G$ consists of a set of elements (which is usually denoted by $G$ as well) and an operation $\circ:G\times G\to G$ (often written as addition or multiplication), such that 
\begin{enumerate}
\item the operation is associative: $g\circ(h\circ k)=(g\circ h)\circ k$ for all $g,h,k\in G$;
\item there is an \emph{identity element} $e\in G$ satisfying $e\circ g=g\circ e=g$ for every $g\in G$;
\item and every $g\in G$ has an \emph{inverse} $g^{-1}\in G$, such that $g\circ g^{-1}=g^{-1}\circ g=e$ (if the group operation is written as addition, then $g^{-1}$ is written as $-g$). 
\end{enumerate}
We often abbreviate $g\circ h$ to $gh$. The group is \emph{Abelian} (or \emph{commutative}) if $gh=hg$ for all $g,h\in G$. Simple examples of finite additive Abelian groups are $G=\01^n$ with bitwise addition mod~2 as the group operation, and $G=\mathbb{Z}_N$, the ``cyclic group'' of integers mod~$N$. The set $G=\mathbb{Z}_N^*$ is the multiplicative group consisting of all integers in $\{1,\ldots,N-1\}$ that are coprime to $N$, with multiplication mod~$N$ as the group operation.\footnote{Euler's $\phi$-function counts the number of elements of $\{1,\ldots,N-1\}$ that are coprime to $N$, so $|\mathbb{Z}_N^*|=\phi(N)$. Note that for prime~$p$, the multiplicative Abelian group $\mathbb{Z}_p^*$ is isomorphic to the additive group $\mathbb{Z}_{p-1}$. However, for general~$N$, $\mathbb{Z}_N^*$ need not be isomorphic to $\mathbb{Z}_{\phi(N)}$.} 
An important example of a non-Abelian group is the ``symmetric group'' $S_n$, which is the group of $n!$ permutations of $n$ elements, using composition as the group operation. 

A \emph{subgroup} $H$ of $G$, denoted $H\leq G$, is a subset of $G$ that is itself a group, i.e., it contains $e$ and is closed under taking products and inverses. A (left) \emph{coset} of $H$ is a set $gH=\{gh\mid h\in H\}$, i.e., a translation of $H$ by the element~$g$. All cosets of $H$ have size $|H|$, and it is easy to show that two cosets $gH$ and $g'H$ are either equal or disjoint, so the set of cosets partitions~$G$ into equal-sized parts.\footnote{This also proves Lagrange's theorem for finite groups: if $H\leq G$ then $|H|$ divides $|G|$.}
Note that $g$ and $g'$ are in the same coset of $H$ iff $g^{-1}g'\in H$.

If $T\subseteq G$, then we use $\langle T\rangle$ to denote the set of elements of $G$ that we can write as products of elements from $T$ and their inverses. This $H=\langle T\rangle$ is a subgroup of $G$, and $T$ is called a \emph{generating set} of $H$. Note that adding one more element $t\not\in\langle T\rangle$ to $T$ at least doubles the size of the generated subgroup, because $H$ and $tH$ are disjoint and $H\cup tH\subseteq\langle T\cup\{t\}\rangle$. This implies that every $H\leq G$ has a generating set of size $\leq\log|H|\leq\log|G|$.
We abbreviate $\langle\{\gamma\}\rangle$ to~$\langle \gamma\rangle$, which is the cyclic group generated by $\gamma$; every cyclic group of size $N$ is isomorphic to~$\mathbb{Z}_N$.

\subsection{Definition and some instances of the HSP}

The \emph{Hidden Subgroup Problem} is the following:
\begin{quote}
Given a known group $G$ and a function $f:G\to S$ where $S$ is some finite set.\\ 
Suppose $f$ has the property that there exists a subgroup $H\leq G$ such that $f$ is constant within each coset, and distinct on different cosets:
$f(g)=f(g') \text{ iff } gH=g'H.$\\
Goal: find $H$.
\end{quote}
We assume $f$ can be computed efficiently, meaning in time polynomial in $\log|G|$ (the latter is the number of bits needed to describe an input $g\in G$ for $f$). Since $H$ may be large, ``finding~$H$'' typically means finding a generating set for $H$.

This looks like a rather abstract algebraic problem, but many important problems can be written as an instance of the HSP. 
We will start with some examples where $G$ is Abelian.

\paragraph{Simon's problem.}
This is a very natural instance of HSP. Here $G$ is the additive group $\mathbb{Z}_2^n=\01^n$ of size $2^n$, $H=\{0,s\}$ for a ``hidden'' $s\in\01^n$, and $f$ satisfies $f(x)=f(y)$ iff $x-y\in H$. Clearly, finding the generator of $H$ (i.e., finding~$s$) solves Simon's problem.

\paragraph{Period-finding.}
As we saw in Chapter~\ref{chap:factoring}, we can factor a large number~$N$ if we can solve the following: given an $x$ that is coprime to $N$ and associated function $f:\mathbb{Z}\to\mathbb{Z}^*_N$ by $f(a)=x^a$~mod~$N$, find the period~$r$ of $f$.\footnote{This $r$ is also known as the \emph{order} of the element $x$ in the group $\mathbb{Z}_N^*$, so this problem is also known as \emph{order-finding}.}
Since $\langle x\rangle$ is a size-$r$ subgroup of the group $\mathbb{Z}_N^*$, the period $r$ divides $|\mathbb{Z}_N^*|=\phi(N)$. Hence we can restrict the domain of~$f$ to $\mathbb{Z}_{\phi(N)}$.

Period-finding is an instance of the HSP as follows. Let $G=\mathbb{Z}_{\phi(N)}$ and consider its subgroup $H=\langle r\rangle$ of all multiples of $r$ up to $\phi(N)$ (i.e., $H=r\mathbb{Z}_{\phi(N)}=\{0,r,2r,\ldots,\phi(N)-r\}$). Note that because of its periodicity, $f$ is constant on each coset $s+H$ of $H$, and distinct on different cosets. Also, $f$ is efficiently computable by repeated squaring. Since the hidden subgroup~$H$ is generated by $r$, finding the generator of~$H$ solves the period-finding problem.

\paragraph{Discrete logarithm.}
Another problem often used in classical public-key cryptography is the discrete logarithm problem: given a generator $\gamma$ of a cyclic multiplicative group~$C$ of size~$N$ (so $C=\{\gamma^a\mid a\in\{0,\ldots,N-1\}\}$), and $A\in C$, can we find the unique $a\in\{0,1,\ldots,N-1\}$ such that $\gamma^a=A$? This $a$ is called the discrete logarithm of $A$ (w.r.t.\ generator $\gamma$). It is generally believed that classical computers need time roughly exponential in $\log N$ to compute $a$ from $A$ (and one can actually \emph{prove} this in a model where we can only implement group operations via some ``black-box''~\cite{shoup:lowerbounddiscrlog}). This assumption underlies for instance the security of Diffie-Hellman key exchange (where $C=\mathbb{Z}_p^*$ for some large prime~$p$, see Exercise~\ref{ex:diffie}), as well as elliptic-curve cryptography.

Discrete log is an instance of the HSP as follows. 
We take $G=\mathbb{Z}_N\times\mathbb{Z}_N$ and define function $f:G\to C$ by
$f(x,y)=\gamma^x A^{-y}$, which is efficiently computable by repeated squaring. For group elements $g_1=(x_1,y_1),g_2=(x_2,y_2)\in G$ we have
$$
f(g_1)=f(g_2)\Longleftrightarrow \gamma^{x_1-ay_1}=\gamma^{x_2-ay_2} \Longleftrightarrow (x_1-x_2)=a(y_1-y_2)\mbox{ mod }N\Longleftrightarrow g_1-g_2\in\langle(a,1)\rangle.
$$
Let $H$ be the subgroup of $G$ generated by the element $(a,1)$, then we have an instance of the HSP. Finding the generator of the hidden subgroup~$H$ gives us~$a$, solving the discrete log problem.

\section{An efficient quantum algorithm if $G$ is Abelian}

In this section we show that HSPs where $G$ (and hence $H$) is Abelian, and where $f$ is efficiently computable, can be solved efficiently by a quantum algorithm. This generalizes Shor's factoring algorithm, and will also give an efficient quantum algorithm for computing discrete logarithms.

\subsection{Representation theory and the quantum Fourier transform}

We start by quickly explaining the basics of representation theory.
The idea here is to replace group elements by matrices, so that linear algebra can be used as a tool in group theory.
A $d$-dimensional \emph{representation} of a multiplicative group $G$ is a map $\rho:g\mapsto\rho(g)$ from $G$ to the set of $d\times d$ invertible complex matrices, satisfying $\rho(gh)=\rho(g)\rho(h)$ for all $g,h\in G$. The latter property makes the map~$\rho$ a \emph{homomorphism}. It need not be an \emph{iso}morphism (i.e., bijective), for example the constant-1 function is a trivial representation of any group.
A representation of~$G$ is \emph{irreducible} if it cannot be decomposed further into the direct sum of lower-dimensional representations of~$G$.
A 1-dimensional representation of~$G$ is called a \emph{character} of~$G$ (sometimes \emph{linear character}). Note that a character~$\chi$ is irreducible, and the complex values $\chi(g)$ must have modulus~1  in finite groups, because $|\chi(g^k)|=|\chi(g)|^k$ for all integers~$k$.
For example, the multiplicative group $\mathbb{Z}_2=\{1,-1\}$ (which would isomorphically be $\mathbb{Z}_2=\{0,1\}$ when written additively) has two characters: the $\chi$ that maps both elements to~1, and the $\chi$ that maps 1 to~1 and $-1$ to $-1$. 

In the remainder of this section we will restrict attention to the case where $G$ is Abelian (and usually finite).
In the Abelian case the characters are exactly the irreducible representations (irreps): there are no irreps of dimension $>1$. The ``Basis Theorem'' of group theory says that every finite Abelian group $G$ is isomorphic to a direct product $\mathbb{Z}_{N_1}\times\cdots\times\mathbb{Z}_{N_\ell}$ of cyclic groups. First consider just one cyclic group $\mathbb{Z}_N$, written additively. Consider the discrete Fourier transform (Chapter~\ref{chap:ft}), which is an $N\times N$ matrix. Ignoring the normalizing factor of $1/\sqrt{N}$, its $k$-th column may be viewed as a map $\chi_k:\mathbb{Z}_N\to\mathbb{C}$ defined by $\chi_k(j)=\omega_N^{jk}$, where $\omega_N=e^{2\pi i/N}$. Note that $\chi_k(j+j')=\chi_k(j)\chi_k(j')$, so $\chi_k$ is actually a 1-dimensional representation (i.e., a character) of~$\mathbb{Z}_N$. In fact, the $N$ characters corresponding to the $N$ columns of the Fourier matrix are \emph{all} the characters of $\mathbb{Z}_N$. For Abelian groups $G$ that are (isomorphic to) a product $\mathbb{Z}_{N_1}\times\cdots\times\mathbb{Z}_{N_\ell}$ of cyclic groups, the $|G|=N_1\cdots N_\ell$ characters are just the products of the characters of the individual cyclic groups $\mathbb{Z}_{N_j}$. Note that the characters are pairwise orthogonal.

The set of all characters of $G$ forms a group $\widehat{G}$ with the operation of pointwise multiplication. This is called the \emph{dual group} of~$G$. If $H\leq G$, then the following is a subgroup of $\widehat{G}$ of size $|G|/|H|$:
$$
H^\perp=\{\chi_k\mid \chi_k(h)=1\mbox{ for all }h\in H\}.
$$ 
Let us interpret the quantum Fourier transform in terms of the characters. For $k\in \mathbb{Z}_N$, define the state whose amplitudes are the (normalized) values of $\chi_k$:
$$
\ket{\chi_k}=\frac{1}{\sqrt{N}}\sum_{j=0}^{N-1}\chi_k(j)\ket{j}
=\frac{1}{\sqrt{N}}\sum_{j=0}^{N-1}\omega_N^{jk}\ket{j}.
$$
With this notation, the QFT just maps the standard (computational) basis of $\mathbb{C}^N$ to the orthonormal basis corresponding to the characters:
$$
F_N:\ket{k}\mapsto\ket{\chi_k}.
$$
As we saw in Chapter~\ref{chap:ft}, this map can be implemented by an efficient quantum circuit if $N$ is a power of~2. The QFT corresponding to a group $G$ that is isomorphic to $\mathbb{Z}_{N_1}\times\cdots\times\mathbb{Z}_{N_\ell}$ is just the tensor product of the QFTs for the individual cyclic groups.
For example, the QFT corresponding to $\mathbb{Z}_2$ is the Hadamard gate~$H$, so the QFT corresponding to $\mathbb{Z}_2^n$ is $H^{\otimes n}$ (which is of course very different from the QFT corresponding to $\mathbb{Z}_{2^n}$).

\subsection{A general algorithm for Abelian HSP}\label{ssec:genalgoAbHSP}

The following is an efficient quantum algorithm for solving the HSP for some Abelian group $G$ (written additively) and function $f:G\to S$. This algorithm, sometimes called the ``standard algorithm'' for HSP, was first observed by Kitaev~\cite{kitaev:stabilizer} (inspired by Shor's algorithm) and worked out further by many, for instance Mosca and Ekert~\cite{mosca&ekert:hidden}.
\begin{enumerate}
\item Start with $\ket{0}\ket{0}$, where the two registers have dimension $|G|$ and $|S|$, respectively.
\item Create a uniform superposition over $G$ in the first register: $\displaystyle\frac{1}{\sqrt{|G|}}\sum_{g\in G}\ket{g}\ket{0}$.\\[-1em]
\item Compute $f$ in superposition: $\displaystyle\frac{1}{\sqrt{|G|}}\sum_{g\in G}\ket{g}\ket{f(g)}$.
\item Measure the second register. This yields some value $f(s)$ for unknown $s\in G$. The first register collapses to a superposition over the $g$ with the same $f$-value as $s$ (i.e., the coset $s+H$):
$\displaystyle\frac{1}{\sqrt{|H|}}\sum_{h\in H}\ket{s+h}$.
\item Apply the QFT corresponding to $G$ to this state, giving
$\displaystyle\frac{1}{\sqrt{|H|}}\sum_{h\in H}\ket{\chi_{s+h}}$.
\item Measure and output the resulting $g$.
\end{enumerate}
The key to understanding this algorithm is to observe that step~5 maps the uniform superposition over the coset $s+H$ to a uniform superposition over the labels of $H^\perp$:
\begin{align*}
\frac{1}{\sqrt{|H|}}\sum_{h\in H}\ket{\chi_{s+h}} & =\frac{1}{\sqrt{|H||G|}}\sum_{h\in H}\sum_{g\in G}\chi_{s+h}(g)\ket{g}\\
 & =\frac{1}{\sqrt{|H||G|}}\sum_{g\in G}\chi_s(g)\sum_{h\in H}\chi_{h}(g)\ket{g}=\sqrt{\frac{|H|}{|G|}}\sum_{g:\chi_g\in H^\perp}\chi_s(g)\ket{g},
\end{align*}
where the last equality follows from the orthogonality of characters of the group $H$ (note that $\chi_g$ restricted to $H$ is a character of $H$, and it's the constant-1 character iff $\chi_g\in H^\perp$): 
$$
\sum_{h\in H}\chi_{h}(g)=\sum_{h\in H}\chi_{g}(h)=\left\{
\begin{array}{cl}
|H| & \mbox{if }\chi_g\in H^\perp\\
0 & \mbox{if }\chi_g\not\in H^\perp
\end{array}
\right.
$$
The phases $\chi_s(g)$ do not affect the probabilities of the final measurement, since $|\chi_s(g)|^2=1$.
The above algorithm thus samples uniformly from the (labels of) elements of $H^\perp$. Each such element $\chi_g\in H^\perp$ gives us a constraint on $H$ because $\chi_g(h)=1$ for all $h\in H$.\footnote{This is a \emph{linear} constraint mod~$N$. For example, say $G=\mathbb{Z}_{N_1}\times \mathbb{Z}_{N_2}$, and $g=(g_1,g_2)$ is the label of an element of $H^\perp$. Then $1=\chi_g(h)=\omega_{N_1}^{g_1h_1}\omega_{N_2}^{g_2h_2}$ for all $h=(h_1,h_2)\in H$, equivalently $g_1h_1N_2+g_2h_2N_1=0$ mod~$N_1N_2$.} 
Generating a small number of such elements will give sufficient information to find the generators of $H$ itself. Consider our earlier examples of Abelian HSP:

\paragraph{Simon's problem.}
Recall that $G=\mathbb{Z}_2^n=\01^n$ and $H=\{0,s\}$ for the HSP corresponding to Simon's problem. Setting up the uniform superposition over $G$ can be done by applying $H^{\otimes n}$ to the initial state $\ket{0^n}$ of the first register. The QFT corresponding to $G$ is just $H^{\otimes n}$. The $2^n$ characters are $\chi_g(x)=(-1)^{x\cdot g}$. The algorithm will uniformly sample from labels of elements of
$$
H^\perp=\{\chi_g\mid \chi_g(h)=1\mbox{ for all }h\in H\}=\{\chi_g\mid g\cdot s=0\}.
$$
Accordingly, the algorithm samples uniformly from the $g\in\01^n$ such that $g\cdot s=0$ (mod~2). Doing this an expected $O(n)$ times gives $n-1$ linearly independent equations about $s$, from which we can find $s$ using Gaussian elimination.

\paragraph{Period-finding.}
For the HSP corresponding to period-finding, $G=\mathbb{Z}_{\phi(N)}$ and $H=\langle r\rangle$, and 
$$
H^\perp=\{\chi_b \mid e^{2\pi i b h/\phi(N)}=1\mbox{ for all }h\in H\}=\{\chi_b \mid br/\phi(N)\in\{0,\ldots,r-1\}\}.
$$ 
Accordingly, the output of the algorithm is an integer multiple $b=c\phi(N)/r$ of $\phi(N)/r$, for uniformly random $c\in\{0,\ldots,r-1\}$. 

Notice that the algorithm doesn't actually know $\phi(N)$, which creates two problems. First, of the 4 numbers $b,c,\phi(N),r$ involved in the equation $b=c\phi(N)/r$ we only know the measurement outcome~$b$, which is not enough to compute~$r$. Second, step~5 of the algorithm wants to do a QFT corresponding to the group $\mathbb{Z}_{\phi(N)}$ but it doesn't know $\phi(N)$; and even if we knew $\phi(N)$, we've only seen how to efficiently implement a QFT over $\mathbb{Z}_q$ when $q$ is a power of~2. Fortunately, if we actually use the QFT over $\mathbb{Z}_q$ for $q$ a power of~2 that is roughly $N^2$ (and in step~1 set up a uniform superposition over $\mathbb{Z}_q$ instead of over~$G$), then one can show that the above algorithm still works: the measurement yields an integer~$b$ that (with constant probability) is close to an integer multiple of $q/r$.\footnote{There is something that needs to be proved here, but we will skip the details. In fact one can even use a Fourier transform for $q=O(N)$ instead of $O(N^2)$~\cite{hales&hallgren:improvedfourier}. Note that this also reduces the number of qubits used by Shor's algorithm from roughly $3\log N$ to roughly $2\log N$.}
This is basically just Shor's algorithm as described in Chapter~\ref{chap:factoring}.

\paragraph{Discrete logarithm.}
For the HSP corresponding to the discrete log problem, where $G=\mathbb{Z}_N\times\mathbb{Z}_N$ and $H=\langle(a,1)\rangle$, a small calculation shows that $H^\perp=\{\chi_{(c,-ac)}\mid c\in\mathbb{Z}_N\}$ (see Exercise~\ref{ex:HperpDL}). Hence sampling from $H^\perp$ yields some label $(c,-ac)\in G$ of an element of $H^\perp$, from which we can compute the discrete logarithm~$a$ if $gcd(c,N)=1$ (repeating a small number of times will give such a $c$ with high probability). The QFT corresponding to $G$ is $F_N\otimes F_N$, which we don't know how to implement efficiently for arbitrary~$N$, but which we can replace by $F_q\otimes F_q$ for some power-of-2 $q$ chosen to be somewhat larger than~$N$.

\bigskip

\noindent
In the above algorithm we assumed $G$ is a \emph{finite} Abelian group. These techniques have been much extended to the case of infinite groups such as $G=\mathbb{Z}$ and even $\mathbb{R}^d$, to obtain efficient quantum algorithms for problems like Pell's equation~\cite{hallgren:pellj}, and computing properties in number fields~\cite{biasse&song:princideal}.

\section{General non-Abelian HSP}

\subsection{The symmetric group and the graph isomorphism problem}

The Abelian HSP covers a number of interesting computational problems, including period-finding and discrete log. However, there are also some interesting computational problems that can be cast as an instance of HSP with a \emph{non-Abelian}~$G$. Unfortunately we do not have an efficient algorithm for most non-Abelian HSPs.

A good example is the \emph{graph isomorphism} (GI) problem: given two undirected $n$-vertex graphs ${\cal G}_1$ and ${\cal G}_2$, decide whether there exists a bijection taking the vertices of ${\cal G}_1$ to those of ${\cal G}_2$ that makes the two graphs equal. 
No efficient classical algorithm is known for GI, so it would be great if we could solve this efficiently on a quantum computer.\footnote{For a long time, the best algorithm for GI took time roughly $2^{\sqrt{n}}$~\cite{babai&luks:gi}, but in a recent breakthrough Babai gave a ``quasi-polynomial'' algorithm, which is $2^{(\log n)^{O(1)}}$ time~\cite{babai:quasipolygi}. That's not yet polynomial, but a lot faster than before.}

How can we try to solve this via the HSP?
Let ${\cal G}$ be the $2n$-vertex graph that is the disjoint union of the two graphs ${\cal G}_1$ and~${\cal G}_2$. Let $G=S_{2n}$. Let $f$ map $\pi\in S_{2n}$ to $\pi({\cal G})$, which means that edge $(i,j)$ becomes edge $(\pi(i),\pi(j))$. Let $H$ be the automorphism group $\Aut({\cal G})$ of ${\cal G}$, which is the set of all $\pi\in S_{2n}$ that map ${\cal G}$ to itself. This gives an instance of the HSP, and solving it would give us a generating set of $H=\Aut({\cal G})$.

Assume for simplicity that each of ${\cal G}_1$ and ${\cal G}_2$ is connected.
If ${\cal G}_1$ and ${\cal G}_2$ are not isomorphic, then the only automorphisms of ${\cal G}$ are the ones that permute vertices inside ${\cal G}_1$ and inside ${\cal G}_2$: $\Aut({\cal G})=\Aut({\cal G}_1)\times\Aut({\cal G}_2)$. However, if the two graphs are isomorphic, then $\Aut({\cal G})$ will also contain a permutation that swaps the first $n$ with the second $n$ vertices.
Accordingly, if we were able to find a generating set of the hidden subgroup $H=\Aut({\cal G})$, then we can just check whether all generators are in $\Aut({\cal G}_1)\times \Aut({\cal G}_2)$ and decide graph isomorphism. 

\subsection{Non-Abelian QFT on coset states}

One can try to design a quantum algorithm for general, non-Abelian instances of the HSP along the lines of the earlier standard algorithm: set up a uniform superposition over a random coset of~$H$, apply the QFT corresponding to~$G$, measure the final state, and hope that the result gives useful information about~$H$. QFTs corresponding to non-Abelian~$G$ are much more complicated than in the Abelian case, because the irreducible representations $\rho$ can now have dimension $d>1$. For completeness, let's write down the QFT anyway. Let $\widehat{G}$ denote the set of irreducible representations of $G$, and $\dim(\rho)$ be the dimension of a particular $\rho\in\widehat{G}$. We can assume without loss of generality that the $\dim(\rho)\times \dim(\rho)$ matrices~$\rho(g)$ are unitary. The QFT corresponding to $G$ is defined as follows:
$$
\ket{g}\longmapsto \sum_{\rho\in\widehat{G}}\sqrt{\frac{\dim(\rho)}{|G|}}\ket{\rho}\sum_{i,j=1}^{\dim(\rho)}\rho(g)_{ij}\ket{i,j},
$$
where $\ket{\rho}$ denotes a name or label of~$\rho$.
It can be shown that this map is unitary. In particular, $|G|=\sum_{\rho\in\widehat{G}}\dim(\rho)^2$, which implies that the dimensions on the left and the right are the same, and that the right-hand state has norm~1.
In many cases this QFT can still be implemented with an efficient quantum circuit, including for the symmetric group $G=S_{2n}$ that is relevant for graph isomorphism~\cite{beals:fourier,mrr:genericQFT}. However, that is not enough for an efficient algorithm: the standard algorithm does not always yield much information about the hidden $H\leq S_{2n}$~\cite{grigni:qhsp,mrs:defiesfourier,hallgren:limcoset}.

There are some special cases of non-Abelian HSP that can be computed efficiently, for instance for normal subgroups~\cite{hrt:normal}, solvable groups~\cite{watrous:solvable,ims:nonAb}, and nil-2 groups~\cite{iss:nil2groups}.

\subsection{Query-efficient algorithm}

While we do not have a general efficient quantum algorithm for the non-Abelian HSP, there does exist an algorithm that needs to compute~$f$ only a few times, i.e., a \emph{query-efficient} algorithm, due to Ettinger et al.~\cite{ehk:hsppoly}.
We will sketch this now.

Consider steps 1--3 of the standard algorithm for the Abelian case. Even in the general non-Abelian case, this produces a coset state, i.e., a two-register superposition where the second register ranges over the values of~$f$, and the first register will be a uniform superposition over the coset of $H$ that corresponds to that value of~$f$.
Suppose we do this $m$ times, producing a state $\ket{\psi_H}$ which is the tensor product of $m$ coset states for the same unknown~$H$ (for simplicity, below we'll ignore the fact that this state also depends on the particular values $f$ takes on the cosets of~$H$).
One can show that the coset states corresponding to different possible~$H$ are pairwise almost orthogonal: $|\inp{\psi_H}{\psi_{H'}}|$ is exponentially small in~$m$. 
How large should we take $m$ to ensure that these states are ``sufficiently orthogonal'' to enable us to learn $H$ from $\ket{\psi_H}$?
The hidden subgroup~$H$ is generated by a set of $\leq \log|G|$ elements. Hence the total number of possible $H$ that we want to distinguish is at most $\binom{|G|}{\log|G|}\leq 2^{(\log |G|)^2}$. This upper bound on the number of possible~$H$ allows us to define a POVM measurement $\{E_H\}$ (see Section~\ref{ssecmeasurement} for the definition of POVM), with one element for each possible hidden subgroup~$H$, such that if we measure $\ket{\psi_H}$ with this POVM, then we are likely to get the correct outcome~$H$. Choosing $m=O((\log|G|)^2)$ makes the states ``sufficiently orthogonal'' for this idea to work (see Exercise~\ref{ex:povm}). This POVM need not be efficiently implementable: circuits to implement it (using only a computational-basis measurement at the end) may require a number of elementary gates that's polynomial in~$|G|$. But at least the number of times we need to \emph{query} the function~$f$ is only polylogarithmic in $|G|$.

\medskip

\noindent
For those interested in more HSP results,
a good source is Childs's lecture notes~\cite[Chapter~4--14]{childs:qalgonotes}.

\section*{Exercises}

\begin{enumerate}

\item Show that the Deutsch-Jozsa problem for $n=1$ (i.e., where $f:\01\to\01$) is an instance of the HSP.
Explicitly say what $G$, $f$, $H$, and $H^\perp$ are, and how sampling from $H^\perp$ allows you to solve the problem.

\item\label{ex:HperpDL}
Show that for the HSP corresponding to discrete log, we indeed have  $H^\perp=\{\chi_{(c,-ac)}\mid c\in\mathbb{Z}_N\}$ as claimed near the end of Section~\ref{ssec:genalgoAbHSP}.

\item\label{ex:diffie} 
This exercise explains Diffie-Hellman key exchange, which is secure under the assumption that the adversary cannot efficiently compute discrete logarithms.  
Alice and Bob choose a public key consisting of a large prime $p$ (say, of 1000 or 2000 bits) and generator $\gamma$ of the group~$\mathbb{Z}_p^*$, which has size $\phi(p)=p-1$. To agree on a shared secret key~$K$, Alice chooses a uniformly random $a\in\{0,\ldots,p-2\}$ and sends Bob the group element $A=\gamma^a$; Bob chooses a uniformly random $b\in\{0,\ldots,p-2\}$ and sends Alice $B=\gamma^b$. Alice and Bob use $K=\gamma^{ab}$ as their secret key, which they can use for instance to encrypt messages using a one-time pad. 
\begin{enumerate}
\item Show that both Alice and Bob can efficiently compute $K$ given the communication.
\item Show that an adversary who can efficiently compute discrete logarithms, can compute $K$ from the public key and the communication tapped from the channel (i.e., $A$, $B$, $p$ and $\gamma$, but not $a$ and $b$).
\end{enumerate}

\item\label{ex:povm} Suppose we are given an unknown state $\ket{\psi_i}$ from a known set of $K$ states $\{\ket{\psi_j}\mid j\in[K]\}$.
\begin{enumerate}
\item Suppose the states are pairwise orthogonal: $\inp{\psi_j}{\psi_k}=\delta_{jk}$. 
Give a projective measurement that determines $i$ with probability~1.
\item (H) Suppose the states are pairwise \emph{almost} orthogonal: $|\inp{\psi_j}{\psi_k}|\leq 1/100K^2$ for all distinct $j,k\in[K]$. 
Define $E_i=\frac{2}{3}\ketbra{\psi_i}{\psi_i}$. Show that $I-\sum_{i=1}^K E_i$ is positive semidefinite.
\item Under the same assumption as (b), 
give a POVM that determines $i$ with success probability at least 2/3.
\end{enumerate}

\item\label{ex:GIswap} (H)
Suppose we have an efficient algorithm to produce, from a given undirected $n$-vertex graph $\cal G$, the following $n^2$-qubit state:
$$
a_{\cal G}\sum_{\pi\in S_n}\ket{\pi({\cal G})},
$$
where the basis states correspond to $n\times n$ adjacency matrices.
Here $a_{\cal G}$ is a scalar that makes the norm equal to~1.
Use this procedure to efficiently decide (with high success probability) whether two given graphs ${\cal G}_1$ and ${\cal G}_2$ are isomorphic or not.  

\end{enumerate}

\chapter{Grover's Search Algorithm}\label{chap:grover}\index{Grover, L.}

The second-most important quantum algorithm after Shor's is Grover's search algorithm~\cite{grover:search}.
It doesn't provide an exponential speed-up, only a quadratic speed-up, but it is much more widely applicable than Shor.

\section{The problem}

{\bf The search problem}:\\ 
For $N=2^n$, we are given an arbitrary $x\in\01^N$.
The goal is to find an $i$ such that $x_i=1$\index{search} 
(and to output `no solutions' if there are no such $i$). We denote the number of solutions in $x$ by~$t$ (i.e., $t$ is the Hamming weight of~$x$).

\bigskip

This problem may be viewed as a simplification of the problem 
of searching an $N$-slot unordered database or search space, modeled by an $N$-bit string. 
Classically, a randomized algorithm
would need $\Theta(N)$ queries to solve the search problem.
Grover's algorithm solves it in $O(\sqrt{N})$ queries, and $O(\sqrt{N}\log N)$ other gates (the number of gates can be reduced a bit further, see Exercise~\ref{ex:grovergates}).

\section{Grover's algorithm}\label{sec:grover}

Let $O_{x,\pm}\ket{i}=(-1)^{x_i}\ket{i}$ denote the $\pm$-type oracle for 
the input $x$ (i.e., a phase-query), and $R_0$ be the unitary transformation that puts a 
$-1$ in front of all basis states $\ket{i}$ where $i\neq 0^n$, 
and that does nothing to the basis state $\ket{0^n}$.\footnote{This unitary $R_0$ is independent of~$x$, and can be implemented using $O(n)$ elementary gates (Exercise~\ref{chap:qircuitdj}.\ref{ex:negate0circuit}.a).}
The \emph{Grover iterate} is 
\begin{equation}\label{eq:groveriteration}
{\cal G}=H^{\otimes n}R_0H^{\otimes n}O_{x,\pm}.
\end{equation}
Note that 1 Grover iterate makes 1 query, and uses $O(\log N)$ other gates.

Grover's algorithm starts in the $n$-bit state $\ket{0^n}$,
applies a Hadamard transformation to each qubit to get the uniform 
superposition $\ket{U}=\frac{1}{\sqrt{N}}\sum_i\ket{i}$ of all $N$ indices,
applies $\cal G$ to this state $k$ times (for some $k$ to 
be chosen later), and then measures the final state.
Intuitively, what happens is that in each iteration some amplitude is
moved from the indices of the 0-bits to the indices of the 1-bits.
The algorithm stops when almost all of the amplitude is on the 1-bits,
in which case a measurement of the final state 
will probably give the index of a 1-bit.
Figure~\ref{figgrover} illustrates this.

\begin{figure}[hbt]
\centering
\setlength{\unitlength}{0.3mm}
\begin{picture}(390,150)
\put(30,75){\makebox(10,30){$\left\{\makebox(10,50){ }\right.$}}
\put(5,75){\makebox(20,30){$n$}}
\put(40,120){\makebox(20,20){$\ket{0}$}}
\put(40,80){\makebox(20,20){$\ket{0}$}}
\put(40,40){\makebox(20,20){$\ket{0}$}}
\put(350,75){\makebox(10,30){$\left.\makebox(10,50){ }\right\}$}}
\put(390,75){\makebox(20,30){\footnotesize measure}}
\put(60,130){\line(1,0){25}}
\put(60,90){\line(1,0){25}}
\put(60,50){\line(1,0){25}}
\put(105,130){\line(1,0){25}}
\put(105,90){\line(1,0){25}}
\put(105,50){\line(1,0){25}}
\put(85,120){\framebox(20,20){$H$}}
\put(85,80){\framebox(20,20){$H$}}
\put(85,40){\framebox(20,20){$H$}}
\put(130,40){\framebox(40,100){}}
\put(150,90){\makebox(0,0){$\cal G$}}
\put(170,130){\line(1,0){25}}
\put(170,90){\line(1,0){25}}
\put(170,50){\line(1,0){25}}
\put(210,90){\makebox(0,0){$\cal G$}}
\put(195,40){\framebox(40,100){}}
\put(260,130){\makebox(0,0){$\dots$}}
\put(260,90){\makebox(0,0){$\dots$}}
\put(260,50){\makebox(0,0){$\dots$}}
\put(300,90){\makebox(0,0){$\cal G$}}
\put(285,40){\framebox(40,100){}}
\put(325,130){\line(1,0){25}}
\put(325,90){\line(1,0){25}}
\put(325,50){\line(1,0){25}}
\put(220,5){\makebox(10,30){$\underbrace{\makebox(200,5){ }}_k$}}
\end{picture}
\caption{Grover's algorithm, with $k$ Grover iterates}\label{figgrover}
\end{figure}

In order to analyze this, define the following ``good'' and ``bad'' states, corresponding to the solutions and non-solutions, respectively:
$$
\ket{G}=\frac{1}{\sqrt{t}}\sum_{i:x_i=1}\ket{i}
\mbox{ and }
\ket{B}=\frac{1}{\sqrt{N-t}}\sum_{i:x_i=0}\ket{i}.
$$
Then the uniform state over all indices can be written as
$$
\ket{U}=\frac{1}{\sqrt{N}}\sum_{i=0}^{N-1}\ket{i}=\sin(\theta)\ket{G}+\cos(\theta)\ket{B}, \ \ \mbox{ for }\theta=\arcsin\sqrt{t/N}.
$$
The Grover iterate $\cal G$ is actually the product of two \emph{reflections}. A reflection through a subspace $V$ is
a unitary $A$ such that $Av=v$ for all vectors $v\in V$, and $Aw=-w$ for all $w$ that are orthogonal to~$V$. Note that we can write $A=2P_V-I$, where $P_V$ is the projector onto subspace~$V$. If $V$ is 1-dimensional, spanned by vector $u$, we also call this a reflection through~$u$.

Firstly, $O_{x,\pm}$ is a reflection through the subspace~$V$ spanned by the basis states that are not solutions; restricted to the 2-dimensional space spanned by $\ket{G}$ and $\ket{B}$ 
this is in fact just a reflection through the state~$\ket{B}$.\footnote{Note that $O_{x,\pm}$ is not (and is not claimed to be, and does not need to be for our purposes) a reflection through $\ket{B}$ in the full $N$-dimensional space. For example, if 3 and 4 are non-solutions, then the state $\frac{1}{\sqrt{2}}(\ket{3}-\ket{4})$ is orthogonal to $\ket{B}$, yet $O_{x,\pm}$ maps that state to itself rather than to minus itself (the latter is what a reflection through $\ket{B}$ in the full $N$-dimensional space would do).} 
Secondly,
$$
H^{\otimes n}R_0H^{\otimes n}=H^{\otimes n}(2\ketbra{0^n}{0^n}-I)H^{\otimes n}=2H^{\otimes n}\ketbra{0^n}{0^n}H^{\otimes n} - H^{\otimes n}I H^{\otimes n}=2\ketbra{U}{U}-I
$$
is a reflection through $\ket{U}$.

Here is Grover's algorithm restated, assuming we know the fraction of solutions is $\eps=t/N$:
\begin{enumerate}
\item Set up the starting state $\ket{U}=H^{\otimes n}\ket{0}$
\item Repeat the following $k=O(1/\sqrt{\eps})$ times:
\begin{enumerate}
\item Reflect through $\ket{B}$ (i.e., apply $O_{x,\pm}$)
\item Reflect through $\ket{U}$ (i.e., apply $H^{\otimes n}R_0H^{\otimes n}$)
\end{enumerate}
\item Measure the first register and check that the resulting $i$ is a solution
\end{enumerate}

\pagebreak

\begin{figure}[hbt]
\centering
\includegraphics{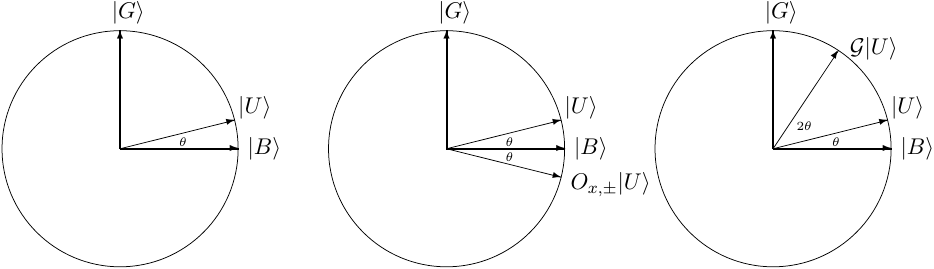}
\caption{The first iteration of Grover: (picture on the left) start with $\ket{U}$; (middle) reflect through $\ket{B}$ to get $O_{x,\pm}\ket{U}$; (right) reflect through $\ket{U}$ to get ${\cal G}\ket{U}$}\label{fig:groveriterate}
\end{figure}

\paragraph{Geometric argument:}
There is a fairly simple geometric argument why the algorithm works.
The analysis is in the 2-dimensional real plane spanned by $\ket{B}$ and $\ket{G}$.
We start with 
$$
\ket{U}=\sin(\theta)\ket{G}+\cos(\theta)\ket{B}.
$$
The two reflections (a) and (b) increase the angle from $\theta$ to $3\theta$, moving us towards the good state
as illustrated in Figure~\ref{fig:groveriterate}. 

The next two reflections (a) and (b) increase the angle with another $2\theta$, etc.
More generally, after $k$ applications of (a) and (b) our state has become 
$$
\sin((2k+1)\theta)\ket{G}+\cos((2k+1)\theta)\ket{B}.
$$
If we now measure, then the probability of seeing a solution is $P_k=\sin((2k+1)\theta)^2$.
We want $P_k$ to be as close to~1 as possible.
Note that if we can choose $\tilde{k}=\frac{\pi}{4\theta} - 1/2$,
then $(2\tilde{k}+1)\theta=\pi/2$ and hence
$P_{\tilde{k}}=\sin(\pi/2)^2=1$.
An example where this works is if $t=N/4$,
for then $\theta=\pi/6$ and $\tilde{k}=1$.

Unfortunately $\tilde{k}=\frac{\pi}{4\theta} - 1/2$ will usually not be an integer, and we can only do an integer number of Grover iterations.
However, if we choose $k$ to be the integer closest to
$\tilde{k}$, then our final state will still be close to~$\ket{G}$ and the failure probability will still be small (assuming $t\ll N$): 
\begin{eqnarray*}
1-P_k & = & \cos((2k+1)\theta)^2=\cos((2\tilde{k}+1)\theta+2(k-\tilde{k})\theta)^2\\
      & = & \cos(\pi/2+2(k-\tilde{k})\theta)^2=\sin(2(k-\tilde{k})\theta)^2\leq \sin(\theta)^2=\frac{t}{N},
\end{eqnarray*}
where we used $|k-\tilde{k}|\leq 1/2$.
Since $\theta=\arcsin\sqrt{t/N}\geq\sqrt{t/N}$, the number of queries
is $k\leq\frac{\pi}{4\theta}\leq\frac{\pi}{4}\sqrt{\frac{N}{t}}$.

\paragraph{Algebraic argument:}
For those who don't like geometry, here's an alternative (but equivalent) algebraic argument.
Let $a_k$ denote the amplitude of the indices of the $t$ 1-bits after 
$k$ Grover iterates, and $b_k$ the amplitude of the indices of the 0-bits.
Initially, for the uniform superposition $\ket{U}$ we have $a_0=b_0=1/\sqrt{N}$.
Using that $H^{\otimes n}R_0H^{\otimes n}=\frac{2}{N}J-I$, 
where $J$ is the $N\times N$ all-1s matrix, we find the following recursion:
\begin{eqnarray*}
a_{k+1} & = & \frac{N-2t}{N}a_k + \frac{2(N-t)}{N}b_k\\
b_{k+1} & = & \frac{-2t}{N}a_k + \frac{N-2t}{N}b_k
\end{eqnarray*}
The following formulas, due to Boyer et~al.~\cite{bbht:bounds},\index{Brassard, G.}\index{Hoyer, P.@H\protect{{\o}yer,} P.}\index{Boyer, M.}\index{Tapp, A.}
provide a closed form for $a_k$ and $b_k$ (which may be
verified by substituting them into the recursion).
With $\theta=\arcsin\sqrt{t/N}$ as before, define
\begin{eqnarray*}
a_k & = & \frac{1}{\sqrt{t}}\sin((2k+1)\theta)\\
b_k & = & \frac{1}{\sqrt{N-t}}\cos((2k+1)\theta)
\end{eqnarray*}
Accordingly, after $k$ iterations the success probability (the sum of squares of the amplitudes of the locations of the $t$ 1-bits) 
is the same as in the geometric analysis
$$
P_k=t\cdot a_k^2=(\sin((2k+1)\theta))^2.
$$
Thus we have a bounded-error quantum search algorithm with
$O(\sqrt{N/t})$ queries, \emph{assuming we know~$t$}. We now list (without full proofs) a number of useful variants of Grover:
\begin{itemize}
\item If we know $t$ exactly, then the algorithm can be tweaked to end up in \emph{exactly} the good state. Roughly speaking, you can make the angle $\theta$ slightly smaller, such that $\tilde{k}=\pi/4\theta - 1/2$ becomes an integer
(see Exercise~\ref{ex:exactgrover}).
\item If we do \emph{not} know $t$, then there is a problem: we do not know which~$k$ to use, so we do not know when to stop doing the Grover iterates. 
Note that if $k$ gets too big, the success probability
$P_k=(\sin((2k+1)\theta))^2$ goes down again! 
However, a slightly more complicated algorithm due to 
\cite{bbht:bounds} (basically running the above algorithm with exponentially increasing guesses for $k$) shows that an \emph{expected} number of $O(\sqrt{N/t})$ queries still suffices to find a solution if there are $t$ solutions. If there is no solution ($t=0$), then we can easily detect that by checking $x_i$ for the $i$'s that the algorithm outputs.
\item If we know a lower bound $\tau$ on the actual (possibly unknown) number of solutions~$t$, then the above algorithm  uses an \emph{expected} number of $O(\sqrt{N/\tau})$ queries. If we run this algorithm for up to three times its expected number of queries, then (by Markov's inequality) with probability at least $2/3$ it will have found a solution. This way we can turn an \emph{expected} runtime into a \emph{worst-case} runtime.
\item If we do not know $t$ but would like to reduce the probability of not finding a solution to some small $\eps>0$, then we can do this using $O(\sqrt{N\log(1/\eps)})$ queries (see Exercise~\ref{ex:log1overeps}).\\ 
NB: The important part here is that the $\log(1/\eps)$ is inside the square-root; usual error-reduction by $O(\log(1/\eps))$ repetitions of basic Grover would give the worse upper bound of $O(\sqrt{N}\log(1/\eps))$ queries.
\end{itemize}

\section{Amplitude amplification}\label{sec:amplamplif}

The analysis that worked for Grover's algorithm is actually much more generally applicable
(we will also see it again in the next chapter).
In this section we describe a very similar procedure that allows us to amplify the ``good'' part of the outcome of an algorithm.
Quite abstractly, suppose we have a quantum circuit $A$ (without measurements) that acts on $m$ qubits, such that
\[
A\ket{0^m}=\sqrt{p}\ket{\psi_1}+\sqrt{1-p}\ket{\psi_0},
\]
where $\ket{\psi_1}$ and 
$\ket{\psi_0}$ are normalized $m$-qubit states that are orthogonal to each other; it could for instance be that the last qubit of $\ket{\psi_1}$ is $\ket{1}$ and the last qubit of $\ket{\psi_0}$ is $\ket{0}$. For some reason we like the state $\ket{\psi_1}$ and we want to increase its ``weight'' $\sqrt{p}$ in the superposition.
The following procedure achieves this.

In analogy with the analysis of Grover, think of $\ket{\psi_1}$ as the ``good state'' and $\ket{\psi_0}$ as the ``bad state,'' and view these two states as the vertical and horizontal axes in a 2-dimensional picture. Our starting state will be $\ket{U}=A\ket{0^m}$, which plays the role that the uniform state played in Grover, and which is of course easy to obtain by applying $A$ once to basis state $\ket{0^m}$. The angle between $\ket{U}$ and the horizontal axis is $\theta=\arcsin\sqrt{p}$.
We would like to rotate this initial state towards the good state, i.e., towards the vertical axis. As for Grover, we could implement the desired rotation as a product of two reflections: a reflection through the bad state and a reflection through~$\ket{U}$.

For the first reflection, suppose we have a circuit $R_G$ that can somehow distinguish the good state from the bad state by putting a ``$-$'' in front of $\ket{\psi_1}$ and leaving $\ket{\psi_0}$ alone. For example, if the last qubit of $\ket{\psi_1}$ is $\ket{1}$ and the last qubit of $\ket{\psi_0}$ is $\ket{0}$, then $R_G$ would be extremely easy: it would just apply a $Z$-gate to the last qubit. 
The second reflection can be implemented as 
\[
AR_0A^{-1}.
\]
It is easy to check that this maps the state $\ket{U}=A\ket{0^m}$ to itself, while every state orthogonal to $\ket{U}$ gets a ``$-$'' in front of it, so indeed this reflects through~$\ket{U}$.
Like before, the product of these two reflections corresponds to a rotation by an angle of~$2\theta$ in the 2-dimensional picture.
Since we need $A^{-1}$ for this reflection, it is important that $A$ is just a unitary, and does not involve any measurements.

The following \emph{amplitude amplification} procedure  from~\cite{bhmt:countingj} increases the amplitude of the good state to be close to~1:
\begin{enumerate}
\item Setup the starting state $\ket{U}=A\ket{0^m}$
\item Repeat the following $O(1/\sqrt{p})$ times:
\begin{enumerate}
\item Reflect through the bad state $\ket{\psi_0}$ (i.e., apply $R_G$)
\item Reflect through $\ket{U}$ (i.e., apply $AR_0A^{-1}$)
\end{enumerate}
\end{enumerate}
The analysis is the same as for Grover: the initial angle between our algorithm's state and the horizontal axis is $\theta$ and every iteration increases this angle by $2\theta$, so after $k$ iterations our state is
\[
\sin((2k+1)\theta)\ket{\psi_1}+\cos((2k+1)\theta)\ket{\psi_0}.
\]
We would like to end up with angle $(2k+1)\theta\approx \pi/2$, because then the amplitude of the good state $\ket{\psi_1}$ would be close to $\sin(\pi/2)=1$. Hence, like before, we choose $k$ to be $\frac{\pi}{4\theta} - 1/2$ rounded to the nearest integer. This is $O(1/\sqrt{p})$. 
If we do not know in advance what $p$ is, then we can try out exponentially decreasing guesses for its value, similar to how we handle the case of Grover with unknown number of solutions.

Note that the Hadamard transform $H^{\otimes n}$ can be viewed as an algorithm with success probability $p=t/N$ for a search problem of size $N$ with $t$ solutions, because $H^{\otimes n}\ket{0^n}$ is the uniform superposition over all $N$ locations. Hence Grover's algorithm is a special case of amplitude amplification, where $m=n$, $A=A^{-1}=H^{\otimes n}$, and $R_G$ corresponds to a phase-query to~$x$.

Amplitude amplification allows to speed up a large class of classical algorithms: any unitary algorithm~$A$ that has some non-trivial probability~$p$ of finding a solution to whatever problem we're trying to solve, and where we can efficiently recognize a solution, can be amplified to success probability nearly~1 by $O(1/\sqrt{p})$ runs of $A$ and $A^{-1}$. In contrast, classically we would need to repeat $A$ $O(1/p)$ times before we have success probability close to~1.

\section{Application: faster solution for satisfiability}

Grover's algorithm has many applications: basically any classical algorithm that has some search-component
can be improved using Grover's algorithm as a subroutine.  This includes many basic computer applications
such as finding shortest paths and minimum spanning trees, various other graph algorithms, etc.

We can also use it to speed up the computation of \NP-complete problems (see Chapter~\ref{chap:qcomplexitytheory} for the complexity class \NP), albeit only quadratically, not exponentially.
As an example, consider the satisfiability problem: we are given a Boolean formula $\phi(i_1,\ldots,i_n)$
and want to know if it has a satisfying assignment, i.e., a setting of the bits $i_1,\ldots,i_n$ that
makes $\phi(i_1,\ldots,i_n)=1$. 
A classical brute-force search along all $2^n$ possible assignments takes time roughly $2^n$.

To find a satisfying assignment faster, define the $N=2^n$-bit input to Grover's 
algorithm by $x_i=\phi(i)$, where $i\in\01^n$. For a given assignment $i=i_1\ldots i_n$ it is easy to compute $\phi(i)$ classically
in polynomial time. We can write that computation as a reversible circuit (using only Toffoli gates), corresponding to a
unitary $U_\phi$ that maps $\ket{i,0,0}\mapsto\ket{i,\phi(i),w_i}$, where the third register holds some classical workspace
the computation may have needed.  To apply Grover we need an oracle that puts the answer in the phase and doesn't
leave workspace around (because that could mess up the intended interference effects, see Exercise~\ref{chap:qircuitdj}.\ref{ex:workspacesinterference} for an example). Define $O_x$ as the unitary that
first applies $U_\phi$, then applies a $Z$-gate to the second register, and then applies $U_\phi^{-1}$ to ``clean up''
the workspace again.  This has the form we need for Grover: $O_{x,\pm}\ket{i}=(-1)^{x_i}\ket{i}$; here we did not explicitly write the workspace qubits, which start and end in $\ket{0}$.
Now we can run Grover and find a satisfying assignment with high probability if there is one,
using a number of elementary operations that is $\sqrt{2^n}$ times some polynomial factor.

If brute-force search is basically the best thing we can do classically to solve some particular \NP-hard problem, then that computation can be sped up quadratically on a quantum computer using Grover search like above. However, there are also \NP-hard problems where we know algorithms that still run in exponential time, but that are much faster than brute-force search. For example, consider the famous Traveling Salesman Problem (TSP): given an $n$-vertex graph with weights (distances) on the edges, find the shortest tour in this graph that visits every node exactly once. Since there are $(n-1)!$ many different tours, classical brute-force search would take time $(n-1)!$, times some polynomial in~$n$. Grover's algorithm could speed this up quadratically. However, there are much more clever classical algorithms for TSP. In particular, the Bellman-Held-Karp dynamic programming algorithm solves TSP in time $2^n$, times a polynomial in $n$. This algorithm is \emph{much} faster than the Grover-based  $O(\sqrt{n!})$ ($n!\approx(n/e)^{n}$ by Stirling's approximation), and is not amenable to a straightforward speed-up using Grover. Nevertheless, it turns out quantum computers can still solve TSP polynomially faster than the best known classical algorithms, albeit in a much more complicated way than by just applying Grover~\cite{ambainis2019quantum}.

\section*{Exercises}

\begin{enumerate}


\item
\begin{enumerate}
\item
Suppose $n=2$, and $x=x_{00}x_{01}x_{10}x_{11}=0001$.
Give the specific initial state, three intermediate states, and final state in Grover's algorithm, for $k=1$ iterations (using the decomposition of one Grover iterate into a product of four unitaries from Eq.~\eqref{eq:groveriteration}).
What is the success probability?
\item
Give the final state  after $k=2$ iterations.
What is now the success probability?
\end{enumerate}

\item 
\begin{enumerate}
    \item 
Suppose you have a quantum algorithm for some computational problem that takes $\sqrt{N}$ operations on inputs of size $N$, each operation of constant cost $C$. And the best-possible classical algorithm for the same computational problem takes $N$ operations, each of constant cost $c$. Suppose $C$ is much larger than $c$ (which is certainly the case in the current state of quantum technology: doing one elementary quantum gate is much more expensive than doing one classical logic gate). How large does the input-size $N$ have to be before the quantum algorithm has lower cost than the best-possible classical algorithm?
\item Suppose you have a quantum algorithm of cost $C\sqrt{2^n}$ for satisfiability of $n$-variable Boolean formulas, where the best-possible classical algorithm has cost $c 2^n$, and again $C$ is much larger than $c$. 
How large does $n$ have to be before the quantum algorithm has lower cost than the best-possible classical algorithm?
\end{enumerate}

\item
Show that if the number of solutions is $t=N/4$, then Grover's algorithm
always finds a solution with certainty after just one query. How many queries would
a classical algorithm need to find a solution with certainty if $t=N/4$?
And if we allow the classical algorithm error probability 1/10?

\item\label{ex:findallsols} Suppose we have a string of $N=2^n$ bits,
containing $t$ ones (solutions) and $N-t$ zeroes.  You may assume you know the number~$t$.
\begin{enumerate}
\item Show that we can use Grover's algorithm
to find the positions of \emph{all} $t$ ones, using an expected number
of $O(t\sqrt{N})$ queries.
You can argue on a high level, no need to draw actual quantum circuits.
\item (H) Show that this can be improved to an expected number
of $O(\sqrt{tN})$ queries.
\end{enumerate}

\item\label{ex:exactgrover}
At the end of Section~\ref{sec:grover} we claimed without proof that Grover's algorithm can be tweaked to work \emph{with probability~1} if we know the number of solutions exactly.  For $N=2^n$, this question asks you to provide such an exact algorithm for an $x\in\01^N$ with a unique solution (so we are promised that there is exactly one $i\in\01^n$ with $x_i=1$, and our goal is to find this $i$).
\begin{enumerate}
\item Give the success probability of the basic version of Grover's algorithm after $k$ iterations.
\item Suppose the optimal number of iterations $\tilde{k}=\frac{\pi}{4\arcsin(1/\sqrt{N})}-\frac{1}{2}$ is not an integer.  Show that if we round $\tilde{k}$ up to the nearest integer, doing $\ceil{\tilde{k}}$ iterations, then the algorithm will have success probability strictly less than~1.
\item Define a new $2N$-bit string $y\in\01^{2N}$, indexed by $(n+1)$-bit strings $j=j_1\ldots j_n j_{n+1}$, by setting
$$
y_j=\left\{
\begin{array}{ll}
1 & \mbox{if }x_{j_1\ldots j_n}=1 \mbox{ and } j_{n+1}=0,\\
0 & \mbox{otherwise.}
\end{array}
\right.
$$
Show how you can implement the following $(n+1)$-qubit unitary
$$
S_y:\ket{j}\mapsto (-1)^{y_j}\ket{j},
$$
using one query to $x$ (of the usual form $O_x:\ket{i,b}\mapsto\ket{i,b\oplus x_i}$) and a few elementary gates.
\item Let $\gamma\in[0,2\pi)$ and let $U_\gamma=\left(
\begin{array}{cc}
\cos\gamma & -\sin\gamma\\
\sin\gamma & \cos\gamma
\end{array}
\right)$ be the corresponding rotation matrix. Let ${\cal A}=H^{\otimes n}\otimes U_\gamma$ be an $(n+1)$-qubit unitary. What is the probability (as a function of $\gamma$) that measuring the state ${\cal A}\ket{0^{n+1}}$ in the computational basis gives a solution $j\in\01^{n+1}$ for~$y$ (i.e., such that $y_j=1$)?
\item (H) Give a quantum algorithm that finds the unique solution in string~$x$ with probability~1 using $O(\sqrt{N})$ queries to $x$.
\end{enumerate}

\item\label{ex:log1overeps} 
Given query access to $x\in\01^N$, with unknown Hamming weight $t=|x|$, we want to find a solution, i.e., an index $i\in\{0,\ldots,N-1\}$ such that $x_i=1$. If $x=0^N$ then our search algorithm should output ``no solution.''
\begin{enumerate}
\item (H) Suppose we know an integer~$s$ such that $t\in\{1,\ldots,s\}$. Give a quantum algorithm that finds a solution with probability~1, using $O(\sqrt{sN})$ queries to~$x$.
\item Suppose we know that $t\in\{s+1,\ldots,N\}$.  Give a quantum algorithm that finds a solution with probability at least $1-2^{-s}$, using $O(\sqrt{sN})$ queries to~$x$.
\item For given $\eps>2^{-N}$, give a quantum algorithm that solves the search problem with probability $\geq 1-\eps$ using $O(\sqrt{N\log(1/\eps)})$ queries, without assuming anything about $t$.
\end{enumerate}

\item\label{ex:approxcount} (H) Here we will \emph{approximately count} the number of 1s in a string $x\in\01^N$. Let $t=|x|$ denote that (unknown) number.
\begin{enumerate}
\item Given an integer $m\in\{1,\ldots,N\}$, describe a quantum algorithm that makes $O(\sqrt{N/m})$ queries to~$x$ and decides between the cases $t\leq m/2$ and $t\in[m,2m]$ with probability at least $2/3$. That is, the algorithm has to output 0 with probability $\geq 2/3$ whenever $t\leq m/2$, has to output 1 with probability $\geq 2/3$ whenever $t\in[m,2m]$, and can output whatever it wants for other values of $t$.
\item Give a quantum algorithm that uses $O(\sqrt{N}\log\log N)$ queries to~$x$ and that outputs an integer $m$ such that, with probability $\geq 2/3$, the unknown $t$ lies between $m/2$ and $2m$.
\end{enumerate}

\item\label{ex:amplitudeestimation} 
Suppose we have a quantum circuit $A$ acting on $m$ qubits, such that
$$
A\ket{0^m}=\sqrt{a}\ket{\phi_1}\ket{1}+\sqrt{1-a}\ket{\phi_0}\ket{0},
$$
where $\ket{\phi_1}$ and 
$\ket{\phi_0}$ are arbitrary normalized $(m-1)$-qubit states, and $a\in[0,1/4]$ is an unknown number. Our goal is to estimate~$a$ (this is known as \emph{amplitude estimation}).
Let $\cal S$ be the 2-dimensional subspace spanned by $\ket{\phi_1}\ket{1}$ and 
$\ket{\phi_0}\ket{0}$.
\begin{enumerate}
    \item Show that in $\cal S$, the unitary $I\otimes Z$ (where $I$ is the identity on $m-1$ qubits and $Z$ is the phase-flip gate) is a reflection through $\ket{\phi_0}\ket{0}$.
    \item Let $R_0=2\ketbra{0^m}{0^m}-I$ be a reflection through $\ket{0^m}$. Show that in $\cal S$, the unitary $AR_0A^{-1}$ is a reflection through $A\ket{0^m}$.
    \item Show that 
 in $\cal S$, the unitary $U=AR_0A^{-1}\cdot (I\otimes Z)$ is a rotation over angle $2\theta$, where $\theta=\arcsin\sqrt{a}$.
 \item (H) Given some $\eps\in(0,1/2)$, show how you can use phase estimation (Section~\ref{ssecphaseestimation}) with $O(1/\eps)$ applications of $U$ to find an approximation $\tilde{a}$ of $a$ such that $|\sqrt{\tilde{a}}-\sqrt{a}|\leq\eps$.
 \item (H) Suppose we have query-access to a string $x\in\01^N$ of unknown Hamming weight $t=|x|$. Use (d) to compute an integer $\tilde{t}$ such that $|\tilde{t}-t|\leq\sqrt{N}$ with success probability $\geq 2/3$, using $O(\sqrt{N})$ queries to~$x$.
 \item Suppose $x\in\01^N$ has $|x|\in\{0,1\}$. Use (d) to compute $|x|$ with success probability $\geq 2/3$, using $O(\sqrt{N})$ queries to~$x$ (you may \emph{not} invoke Grover here).
 \end{enumerate}

\item\label{ex:grovergates} Suppose you are given query access to $x\in\01^N$, where $|x|=1$ and $N=2^n$. You want to find the unique index $i=i_{n-1}\ldots i_0\in\01^n$ such that $x_i=1$.
\begin{enumerate}
\item Let $k\in\{1,\ldots,n-1\}$.
Fix the first $n-k$ bits of the $n$-bit index $i_{n-1}\ldots i_0$ to specific values $i^*_{n-1}\ldots i^*_k$.
Give a quantum algorithm to find a solution with success probability~1 (if one exists) among the $2^k$ indices $i\in\01^n$ that start with $i^*_{n-1}\ldots i^*_k$, using $O(\sqrt{2^k})$ queries and $O(\sqrt{2^k}\, k)$ other gates.
\item Give a quantum algorithm that solves the search problem on $x$ using $O(\sqrt{N})$ queries and $O(\sqrt{N}\log\log N)$ other gates.\\
{\footnotesize Comment: The $O(\sqrt{N}\log\log N)$ gates achieved here is better than standard Grover, which uses $O(\sqrt{N}\log N)$ gates. The $\log\log N$ can be reduced a bit further, to nearly-constant~\cite{arunachalam&wolf:grovergate}.}
\end{enumerate}

\item\label{ex:minfind}
(H)
Let $x=x_0\ldots x_{N-1}$ be a sequence of distinct real numbers, where $N=2^n$, and each $x_i$ can be written exactly using $b$ bits.
We can query these in the usual way, i.e., we can apply $(n+b)$-qubit unitary $O_x:\ket{i,0^b}\mapsto\ket{i,x_i}$, as well as its inverse.
The \emph{minimum} of $x$ is defined as $\min\{x_i\mid i\in\{0,\ldots,N-1\}\}$.
Give a quantum algorithm that finds (with probability $\geq 2/3$) an index achieving the minimum,
using at most $O(\sqrt{N}\log N)$ queries to the input, and prove that this algorithm works.\\[1mm]
Bonus: give a quantum algorithm that uses $O(\sqrt{N})$ queries.

\item\label{ex:collisionfind}
Let $x=x_0\ldots x_{N-1}$, where $N=2^n$ and $x_i\in\01^n$, be an input that
we can query in the usual way.
We are promised that this input is 2-to-1: for each $i$ there is exactly one other $j$ such that $x_i=x_j$.%
\footnote{The 2-to-1 inputs for Simon's algorithm are a very special case
of this, where the collisions are determined by a secret string $s\in\01^n$: $x_i$ equals $x_j$ if $i=j\oplus s$.}
Such an $(i,j)$-pair is called a \emph{collision}.
\begin{enumerate}
\item Suppose $S$ is a uniformly  randomly chosen set of $s\leq N/2$ elements of $\{0,\ldots,N-1\}$.
What is the probability that there exists a collision in $S$?
\item (H) Give a classical randomized algorithm that finds a collision (with probability $\geq 2/3$) using $O(\sqrt{N})$ queries to $x$.
\item (H) Give a quantum algorithm that finds a collision (with probability $\geq 2/3$) using $O(N^{1/3})$ queries.
\end{enumerate}

\item 
Consider an undirected graph $G=(V,E)$, with vertex set $V=\{1,\ldots,n\}$ and edge-set $E$.  We say $G$ is \emph{connected} if, for every pair of vertices $i,j \in V$, there is a path between $i$ and $j$ in the graph. The \emph{adjacency matrix} of $G$ is the $n\times n$ Boolean matrix $M$ where $M_{ij}=1$ iff $(i,j)\in E$ (note that $M$ is a symmetric matrix because $G$ is undirected).
Suppose we are given input graph $G$ in the form of a unitary that allows us to query whether an edge $(i,j)$ is present in~$G$ or not:
$$
O_M:\ket{i,j,b}\mapsto\ket{i,j,b\oplus M_{ij}}.
$$
\begin{enumerate}
\item Assume $G$ is connected. Suppose we have a set $A$ of edges which we already know to be in the graph (so $A\subseteq E$; you can think of $A$ as given classically, you don't have to query it).  Let $G_A=(V,A)$ be the subgraph induced by only these edges, and suppose $G_A$ is not connected, so it consists of $c>1$ connected components. Call an edge $(i,j)\in E$ ``good'' if it connects two of these components.  Give a quantum algorithm that finds a good edge with an \emph{expected} number of $O(n/\sqrt{c-1})$ queries to~$M$. 
\item Give a quantum algorithm that uses at most $O(n^{3/2})$ queries to~$M$ and decides (with success probability at least $2/3$) whether $G$ is connected or not.
\item Show that classical algorithms for deciding (with success probability at least $2/3$) whether $G$ is connected, need to make $\Omega(n^2)$ queries to~$M$.
\end{enumerate}

\item\label{ex:exactsimon} This exercise is about reducing the error probability in Simon's algorithm (Chapter~\ref{chap:simon}) to~0.
\begin{enumerate}
\item (H) Suppose we already have $k\leq n-2$ linearly independent strings $j_1,\ldots,j_k\in\01^n$ that are orthogonal to the hidden string~$s$ (i.e., $j_\ell\cdot s=0$ mod~2, for all $\ell\in\{1,\ldots,k\}$). Let $V$ be the subspace of $2^k$ strings that they span (with bitwise addition mod~2).
In the setting of Simon's problem, give a quantum algorithm that uses $O(1)$ queries and \emph{with probability~1} outputs some $j\not\in V$ such that $j\cdot s=0$ mod~2. You don't need to write out the circuit fully, but be precise about what you're applying, what the different components are etc.
\item Give a quantum algorithm that finds $s$ with probability~1, using $O(n)$ queries.
\end{enumerate}

\end{enumerate}

\chapter{Quantum Walk Algorithms}\label{chap:qwalks}

\section{Classical random walks}

Consider an undirected graph $G$ with $N$ vertices. 
Suppose at least an $\eps$-fraction of the vertices are ``marked,''
and we would like to find a marked vertex. One way to do this is with a \emph{random walk}: 
\begin{quote}
Start at some specific vertex $y$ of the graph.\\
Repeat the following a number of times:
Check if $y$ is marked, and if not then choose one of its neighbors at random and set $y$ to be that neighbor. 
\end{quote}
This may seem like a stupid algorithm, but it has certain advantages.
For instance, it only needs space $O(\log N)$, because you only need to keep track of the current vertex $y$, and maybe a counter that keeps track of how many steps you've already taken.\footnote{Here we're assuming the neighbors of a given vertex are efficiently computable, so you don't actually need to keep the whole graph in memory. This will be true for all graphs we consider here.}
Such an algorithm can for example decide whether there is a path from a specific vertex $y$ to a specific vertex $x$ using $O(\log N)$ space. We'd start the walk at $y$ and only $x$ would be marked; one can show that if there exists a path from $y$ to $x$ in $G$, then we will reach $x$ in poly$(N)$ steps.

Let us restrict attention to $d$-regular graphs without self-loops, so each vertex has exactly $d$ neighbors. 
A random walk on such a graph~$G$ corresponds to an $N\times N$ symmetric matrix $P$, where $P_{x,y}=1/d$ if $(x,y)$ is an edge in $G$, and $P_{x,y}=0$ otherwise. This $P$ is the normalized adjacency matrix of~$G$. If $v\in\mathbb{R}^N$ is a vector with a~1 at position~$y$ and 0s elsewhere, then $Pv$ is a vector whose $x$-th entry is $(Pv)_x=1/d$ if $(x,y)$ is an edge, and $(Pv)_x=0$ otherwise. In other words, $Pv$ is the uniform probability distribution over the neighbors of $y$, which is what you get by taking one step of the random walk starting at~$y$. More generally, if $v$ is a probability distribution on the vertices, then $Pv$ is the new probability distribution on vertices after taking one step of the random walk, and $P^k v$ is the probability distribution after taking $k$ steps.

Suppose we start with some probability-distribution vector $v$ (which may or may not be concentrated at one vertex $y$). We will assume $G$ is connected and not bipartite. Then $P^k v$ will converge to the uniform distribution over all vertices, and the speed of convergence is determined by the ``gap'' between the first eigenvalue of $P$ and all other eigenvalues. This can be seen as follows.
Let $\lambda_1\geq\lambda_2\geq\cdots\geq\lambda_N$ be the eigenvalues of~$P$, ordered by size, and $v_1,\ldots,v_N$ be corresponding orthonormal eigenvectors.\footnote{Analyzing graphs by looking at the eigenvalues and eigenvectors of their adjacency matrix is called ``algebraic graph theory'' or ``spectral graph theory,'' see for instance~\cite{brouwer&haemers:spectra}.}
The largest eigenvalue is~$\lambda_1=1$, and corresponds to the eigenvector $v_1$, which is $\sqrt{N}$ times $u=(1/N)$, the uniform distribution over all vertices. One can show that our assumption that $G$ is connected implies $\lambda_2<1$, and our assumption that $G$ is not bipartite implies $\lambda_N>-1$.
Hence all eigenvalues $\lambda_i$ for $i\in\{2,\ldots,N\}$ will be in $(-1,1)$; the corresponding eigenvector $v_i$ will be orthogonal to the uniform vector~$v_1$, so the sum of its entries is~0. 
Let $\delta>0$ be the difference between $\lambda_1=1$ and $\max_{i\geq 2}|\lambda_i|$ (hence $|\lambda_i|\leq 1-\delta$ for all $i\geq 2$). This $\delta$ is called the ``spectral gap'' of the graph.

Now decompose the starting distribution $v$ as $v=\sum_{i=1}^N\alpha_iv_i$.  Since the sum of $v$'s entries is~1, 
and each eigenvector $v_i$ with $i\geq 2$ has entries summing to~0, it follows that $\alpha_1v_1=u$.
Now let us see what happens if we apply the random walk for $k$ steps, starting from $v$:
$$
P^k v=P^k\left(\sum_i\alpha_i v_i\right)=\sum_i\alpha_i\lambda_i^k v_i=u+\sum_{i\geq 2}\alpha_i\lambda_i^k v_i.
$$
Consider the squared norm of the difference between $P^k v$ and $u$:
$$
\norm{P^k v - u}^2=\norm{\sum_{i\geq 2}\alpha_i\lambda_i^k v_i}^2=\sum_{i\geq 2}|\alpha_i|^2|\lambda_i|^{2k}\leq \norm{v}^2(1-\delta)^{2k}.
$$
Since $v$ is a probability distribution, we have $\norm{v}^2\leq 1$. By choosing $k=\ln(1/\eta)/\delta$, we get
$\norm{P^k v - u}\leq\eta$.  In particular, if $\delta$ is not too small, then we get quick convergence of the random walk 
to the uniform distribution $u$, no matter which distribution $v$ we started with.\footnote{Convergence in total 
variation distance can be derived from this by Cauchy-Schwarz, choosing $\eta\ll 1/\sqrt{N}$.}
Once we are close to the uniform distribution, we have probability roughly $\eps$ of hitting a marked vertex.
Of course, the same happens if we just pick a vertex uniformly at random, but that may not always be an option if the graph is given implicitly.

Suppose it costs $\Sc$ to set up an initial state $v$; it costs $\Uc$ to update the current vertex, i.e., to perform one step of the random walk;
and it costs $\Cc$ to check whether a given vertex is marked. 
``Cost'' is left undefined for now, but typically
it will count number of queries to some input, or number of elementary operations.
Consider a classical search algorithm that starts at $v$, and then repeats the following until it finds a marked vertex:
check if the current vertex is marked, and if not run a random walk for roughly $1/\delta$ steps to get close to the uniform distribution.
Ignoring constant factors, the expected cost before this procedure finds a marked item, is on the order of
\begin{equation}\label{eqcrwcost}
\Sc+\frac{1}{\eps}\left(\Cc + \frac{1}{\delta}\Uc\right).
\end{equation}

\section{Quantum walks}
We will now modify the classical random walk algorithm preceding Eq.~(\ref{eqcrwcost}) to a quantum algorithm, 
where the distribution-preserving matrix $P$ is changed to a norm-preserving matrix $W(P)$ (i.e., a unitary). This is due to Magniez et al.~\cite{mnrs:searchwalk}, inspired by Szegedy~\cite{szegedy:qmarkov}; our presentation is mostly based on Santha's survey paper~\cite{santha:qrwsurvey},
to which we refer for more details and references.

While the basis state of a classical random walk is the current vertex we are at,
a basis state of a quantum walk has \emph{two} registers, the first corresponding 
to the \emph{current} vertex and the second corresponding to the \emph{previous} vertex.
Equivalently, a basis state of a quantum walk corresponds to an \emph{edge} of the graph.

Our resulting quantum walk algorithm for search will actually be quite analogous to Grover's algorithm.
We'll call a basis state $\ket{x}\ket{y}$ ``good'' if $x$ is a marked vertex, and ``bad'' otherwise.
Define $\ket{p_x}=\sum_y \sqrt{P_{xy}}\ket{y}$ to be the uniform superposition over the neighbors of $x$.
As for Grover, define ``good'' and ``bad'' states as the superpositions over good and bad basis states:
$$
\ket{G}=\frac{1}{\sqrt{|M|}}\sum_{x\in M}\ket{x}\ket{p_x}
\mbox{ and }
\ket{B}=\frac{1}{\sqrt{N-|M|}}\sum_{x\not\in M}\ket{x}\ket{p_x},
$$
where $M$ denotes the set of marked vertices. 
Note that $\ket{G}$ is just the uniform superposition over all edges $(x,y)$ where the first coordinate is marked, and $\ket{B}$  
is just the uniform superposition over all edges $(x,y)$ where the first coordinate is not marked.

If $\eps=|M|/N$ and $\theta:=\arcsin\sqrt{\eps}$, then the uniform state over all edges can be written as
$$
\ket{U}=\frac{1}{\sqrt{N}}\sum_x\ket{x}\ket{p_x}=\sin(\theta)\ket{G}+\cos(\theta)\ket{B}.
$$
Here is the algorithm for searching a marked vertex if an $\eps$-fraction is marked\footnote{As in Grover, if we don't know $\eps$ 
then we just run the algorithm repeatedly with exponentially decreasing guesses for $\eps$ 
(1/2, 1/4, 1/8, \ldots). If at the end we still haven't found a marked item, we'll conclude that probably none exists.}:
\begin{enumerate}
\item Setup the starting state $\ket{U}$
\item Repeat the following $O(1/\sqrt{\eps})$ times:
\begin{enumerate}
\item Reflect through $\ket{B}$
\item Reflect through $\ket{U}$
\end{enumerate}
\item Measure the first register and check that the resulting vertex $x$ is marked.
\end{enumerate}
The description and analysis of this algorithm takes places in the 2-dimensional space spanned by~$\ket{G}$ and~$\ket{B}$.
We'll explain in a moment how to implement (a) and (b).
Assuming we know how to do that, the proof that this algorithm finds a marked vertex is the same as for Grover and for amplitude amplification (Chapter~\ref{chap:grover}).
We start with $\ket{U}=\sin(\theta)\ket{G}+\cos(\theta)\ket{B}$.
The two reflections (a) and (b) increase the angle from $\theta$ to $3\theta$, moving us towards the good state 
(similarly to the analysis for Grover, you can draw a 2-dimensional picture with axes $\ket{B}$ and $\ket{G}$ to see this).
More generally, after $k$ applications of (a) and (b) our state has become 
$$
\sin((2k+1)\theta)\ket{G}+\cos((2k+1)\theta)\ket{B}.
$$
Choosing $k\approx\frac{\pi}{4\theta}=O(1/\sqrt{\eps})$, we will have $\sin((2k+1)\theta)\approx 1$,
at which point measuring the first register will probably yield a vertex~$x$ that is marked. We now look more closely how to implement the two kinds of reflections.

\smallskip

\paragraph{(a) Reflect through $\ket{B}$.}
Reflecting through $\ket{B}$ is relatively straightforward:
we just have to ``recognize'' whether the first register contains a marked $x$, and put a $-1$ if so
(note that this is really a reflection through the subspace spanned by the bad basis states, but restricted to the 2-dimensional subspace spanned by $\ket{G}$ and $\ket{B}$ that's the same as a reflection through~$\ket{B}$).

\smallskip

\paragraph{(b) Reflect through $\ket{U}$.}
This is where the quantum walk comes in.
Let $\cal A$ be the subspace span$\{\ket{x}\ket{p_x}\}$ and $\cal B$ be span$\{\ket{p_y}\ket{y}\}$.
Let ${\rm ref}({\cal A})$ and ${\rm ref}({\cal B})$ denote reflections through $\cal A$ and~$\cal B$, respectively.
Define $W(P)={\rm ref}({\cal B}){\rm ref}({\cal A})$ to be the product of these two reflections.
This may be viewed as an indirect unitary analogue of the classical random-walk operator~$P$.
Suppose we are able to implement the following two operations (in a controlled manner):
\begin{quote}
(1) $\ket{x}\ket{0}\mapsto\ket{x}\ket{p_x}$\\
(2) $\ket{0}\ket{y}\mapsto\ket{p_y}\ket{y}$
\end{quote}
Since (1) and (2) prepare a uniform superposition over the neighbors of $x$ and $y$, respectively, one can think of them as taking one classical walk step ``in superposition.''
Note that ${\rm ref}({\cal A})$ can be implemented by applying the inverse of (1), putting a minus if the second register is not $\ket{0}$, and applying~(1).
We can similarly implement ${\rm ref}({\cal B})$ using~(2) and its inverse.
Hence we can think of $W(P)={\rm ref}({\cal B}){\rm ref}({\cal A})$ as corresponding to \emph{four} steps of the classical walk in superposition.

Let us sketch the main idea for how to implement the reflection through $\ket{U}$ up to small error.
Consider the eigenvalues and eigenvectors of $W(P)$.
The eigenvalues of $W(P)$ can be related to the eigenvalues $\lambda_1,\lambda_2,\ldots$ of~$P$ as follows.
Let $\theta_j\in[0,\pi/2]$ be such that $|\lambda_j|=\cos(\theta_j)$.
We won't prove it here, but it turns out that the eigenvalues of $W(P)$ are of the form $e^{\pm 2 i \theta_j}$. The state $\ket{U}$ is an eigenvalue-1 eigenvector of~$W(P)$, corresponding to $\theta_1=0$.
The spectral gap of $P$ is $\delta$, so all eigenvectors of $W(P)$ that do not have eigenvalue~1, have eigenvalue $e^{\pm 2 i \theta_j}$ where $\theta_j\geq\sqrt{2\delta}$,
because $1-\delta\geq|\lambda_j|=\cos(\theta_j)\geq 1-\theta_j^2/2$.
We now want to implement a reflection~$R(P)$ through the subspace spanned by the eigenvalue-1 eigenvectors of $W(P)$; restricted to the 2-dimensional subspace spanned by $\ket{G}$ and $\ket{B}$ this will be the desired reflection through~$\ket{U}$.

We will implement $R(P)$ by using \emph{phase estimation} (see Section~\ref{ssecphaseestimation}) with precision $\sqrt{\delta}/2$ on $W(P)$ to distinguish the eigenvalue-1 eigenvectors from the other eigenvectors.
This precision requires $O(1/\sqrt{\delta})$ applications of~$W(P)$, and $O(\log(1/\delta))$ auxiliary qubits that start in~$\ket{0}$ and where the estimate will be stored. 
Assume for simplicity that phase estimation always gives an estimate $\tilde{\theta}_j$ of $\theta_j$ that is within precision $\sqrt{\delta}/2$.\footnote{Phase estimation will actually give a superposition over estimates $\tilde{\theta}_j$, with small but nonzero amplitudes on bad estimates, but we'll skip the technical details that are needed to deal with this.}
Because the nonzero $\theta_j$ are at least $\sqrt{2\delta}$, approximating them within additive error $\sqrt{\delta}/2$
is good enough to determine whether the actual value $\theta_j$ itself is~0 or not. We then multiply the state with a $-1$ if the estimate is sufficiently far from~0, and finally reverse the phase estimation to put the auxiliary qubits back to~0. Applied to some eigenvector $\ket{w}$ of $W(P)$ with corresponding eigenvalue $e^{\pm 2 i \theta_j}$, $R(P)$ maps
$$
R(P): \ket{w}\ket{0}\stackrel{\mbox{~PE~~}}{\mapsto}\ket{w}\ket{\tilde{\theta}_j}\mapsto(-1)^{[\theta_j\neq 0]}\ket{w}\ket{\tilde{\theta}_j}\stackrel{\mbox{PE}^{-1}}{\mapsto}(-1)^{[\theta_j\neq 0]}\ket{w}\ket{0}.
$$
This has the desired effect: ignoring the auxiliary qubits (which start and end in~0), $R(P)$ maps eigenvalue-1 eigenvectors of~$W(P)$ to themselves, and puts a $-1$ in front of the other eigenvectors.

\bigskip

Now that we know how to implement the algorithm, let us look at its complexity.
Consider the following setup, update, and checking costs:
\begin{itemize}
\item Setup cost $\Sc$: the cost of constructing $\ket{U}$
\item Checking cost $\Cc$: the cost of the unitary map $\ket{x}\ket{y}\mapsto m_x\ket{x}\ket{y}$, where $m_x=-1$ if $x$ is marked, and $m_x=1$ otherwise
\item Update cost $\Uc$: the cost of one step of the quantum walk, i.e., of $W(P)$
\end{itemize}
The cost of part (a) of the algorithm is $\Cc$.
Since $R(P)$ uses $O(1/\sqrt{\delta})$ applications of $W(P)$, and a few other gates, the cost of part (b) of the algorithm is essentially $O(\Uc/\sqrt{\delta})$.
Ignoring constant factors, the total cost of the algorithm is then
\begin{equation}\label{eqqrwcost}
\Sc+\frac{1}{\sqrt{\eps}}\left(\Cc + \frac{1}{\sqrt{\delta}}\Uc\right).
\end{equation}
Compare this with the classical cost of Eq.~(\ref{eqcrwcost}): quantum search square-roots both $\eps$ and $\delta$.

\section{Applications}

There are a number of interesting quantum walk algorithms that beat the best classical algorithms.
We'll give three examples here.  More can be found in~\cite{santha:qrwsurvey}.

\subsection{Grover search}

Let us first derive a quantum algorithm for search.
Suppose we have an $N$-bit string $x$ of weight $t$, and we know $t/N\geq\eps$. 
Consider the complete graph $G$ on $N$ vertices. Then the matrix $P$ for the random walk on $G$ 
has 0s on its diagonal, and its off-diagonal entries are all equal to $1/(N-1)$.
This can be written as $P=\frac{1}{N-1}J-\frac{1}{N-1}I$, where $J$ is the all-1 matrix and $I$ is the identity.  The only nonzero eigenvalue of $J$ is $N$, and adding a multiple of $I$ just shifts the eigenvalues of a matrix, hence the largest eigenvalue of $P$ is $\lambda_1=N/(N-1)-1/(N-1)=1$ (corresponding to the uniform vector) and all its other eigenvalues are $-1/(N-1)$.
Note that the spectral gap~$\delta$ is very large here: $\delta=1-1/(N-1)\approx 1$.  We'll mark a vertex $i$ iff $x_i=1$.
Then, measuring cost by number of queries, a quantum walk on $G$ will have $\Sc=\Uc=0$ and $\Cc=1$.
Plugging this into Eq.~(\ref{eqqrwcost}), the quantum walk will find a marked vertex (with high probability) using $O(1/\sqrt{\eps})$ queries.  
The worst case is $\eps=1/N$, in which case we'll use $O(\sqrt{N})$ queries.
Not surprisingly, we've essentially rederived Grover's algorithm.

\subsection{Finding collisions}\label{sseccollision}

Consider the following problem: 
\begin{quote}
Input: $x=x_0,\ldots,x_{n-1}$, where each $x_i$ is an integer.\footnote{Say, all $x_i\leq n^2$ to avoid having to use too much space to store these numbers.}\\
Goal: find distinct $i$ and $j$ such that $x_i=x_j$ if these exist, otherwise output ``all elements are distinct.''
\end{quote}
The problem of finding collisions is important in cryptography, in particular if the map $i\mapsto x_i$ is a so-called ``hash function.'' The decision version of this problem (deciding if there exists at least one collision) is also known as \emph{element distinctness}.

Consider the graph whose vertices correspond to the sets $R\subseteq\{0,\ldots,n-1\}$ of $r$ elements. 
The total number of vertices is $N=\binom{n}{r}$.
We'll put an edge between the vertices for $R$ and $R'$ iff these two sets differ in exactly two elements;
in other words, you can get from $R$ to $R'$ by removing one element $i$ from $R$ and replacing it by a new element $j$.
The resulting graph $J(n,r)$ is known as the \emph{Johnson graph}.
It is $r(n-r)$-regular, since every $R$ has $r(n-r)$ different neighbors $R'$.
Its spectral gap is known to be $\delta=\frac{n}{r(n-r)}$~\cite[Sec.~12.3.2]{brouwer&haemers:spectra}; we won't prove that here, just note that if $r\ll n$, then $\delta\approx 1/r$.
For each set $R$ we also keep track of the corresponding sequence of $x$-values, $x_R=(x_i)_{i\in R}$.
Hence the full ``name'' of a vertex is the pair $(R,x_R)$.

We'll call a vertex in $J(n,r)$ \emph{marked} if it contains a collision, i.e., the corresponding set $R$ contains distinct $i,j$ such that $x_i=x_j$. In the worst case there is exactly one colliding pair $i,j$ (more collisions only make the problem easier).
The probability that $i$ and $j$ are both in a random $r$-set~$R$, is $\eps=\frac{r}{n}\frac{r-1}{n-1}$.
Hence the fraction of marked vertices is at least $\eps\approx(r/n)^2$.

We will now determine the setup, checking, and update costs.
The setup cost (measured in terms of queries) is $\Sc=r+1$: we have to create a uniform superposition $\ket{U}$ over all edges $R,R'$, 
and for each such basis state query all $r+1$ elements of~$R\cup R'$ to add the information $x_R$ and $x_{R'}$.
Checking whether a given vertex $R,x_R$ contains a collision doesn't take any queries because we already have $x_R$, hence $\Cc=0$.
To determine the update cost, note that mapping the second register of $\ket{R,x_R}\ket{0}$ to a superposition of all neighbors $R',x_{R'}$ requires querying (in superposition for all neighbors $R'$) the value $x_j$ of the element $j$ that was added to get $R'$, as well as ``unquerying'' the value $x_i$ of the element $i$ that was removed from $R$ (since $x_i$'s value shouldn't be present anymore in the information we write down for $R'$). Hence $\Uc=2$.
Plugging this into Eq.~(\ref{eqqrwcost}), the cost of a quantum walk algorithm for collision-finding is
$$
\Sc+\frac{1}{\sqrt{\eps}}\left(\Cc + \frac{1}{\sqrt{\delta}}\Uc\right)=O(r+n/\sqrt{r}).
$$
This cost is $O(n^{2/3})$ if we choose to set $r=n^{2/3}$ (rounded to an integer). What the quantum walk produces at the end is a superposition where, if we measure the first register, with high probability we'll see a marked vertex. That way we obtain a set $R$ that contains a collision; and because $R$ is small, we can now cheaply find the colliding indices $i,j\in R$.

This query complexity $O(n^{2/3})$ turns out to be the optimal quantum query complexity for finding collisions~\cite{aaronson&shi:collision}.
By some more work involving efficient data structures, using a quantum-accessible classical RAM, the \emph{time} complexity (= total number of elementary quantum gates plus total number of queries) can be brought down to $n^{2/3}(\log n)^{O(1)}$~\cite{ambainis:edj}.

\subsection{Finding a triangle in a graph}

Consider the following triangle-finding problem: 
\begin{quote}
Input: the adjacency matrix of a graph $H$ on $n$ vertices.\\
Goal: find vertices $u,v,w$ that form a triangle (i.e., $(u,v),(v,w),(w,u)$ are all edges in the graph), if they exist.
\end{quote}
We'll assume we have query access to the entries of the adjacency matrix of $H$, which tells us whether $(u,v)$ is an edge or not.
There are $\binom{n}{2}$ bits in this oracle, one for each potential edge of $H$.
It is not hard to see that a classical algorithm needs $\Omega(n^2)$ queries before it can decide with good probability 
whether a graph contains a triangle or not.  For example, take a bipartite graph consisting of 2 sets of $n/2$ vertices each, 
such that any pair of vertices from different sets is connected by an edge.  Such a graph is triangle-free, but adding 
any one edge will create a triangle.  A classical algorithm would have to query all those edges separately.

Let us try a quantum walk approach.  Again consider the Johnson graph $J(n,r)$.
Each vertex will correspond to a set $R\subseteq\{0,\ldots,n-1\}$ of $r$ vertices, annotated with the result of querying all possible $\binom{r}{2}$ edges having both endpoints in $R$. We will call the vertex for set $R$ \emph{marked} if it contains \emph{one edge} of a triangle.
If there is at least one triangle in the graph, then the fraction of marked vertices is at least $\eps\approx (r/n)^2$.

The setup cost will be $\Sc=\binom{r+1}{2}$: for an edge $(R,R')$ of the Johnson graph we query the $\binom{r+1}{2}$ possible edges induced by the $r+1$ $H$-vertices of $R\cup R'$. The update cost will be $\Uc=2r-2$, because if we remove one vertex $i$ from $R$ then we have to remove information about $r-1$ edges in $H$, and if we add a new~$j$ to $R$ we have to query $r-1$ new edges in $H$.

Getting a good upper bound for the checking cost $\Cc$ requires some more work---namely Grover search plus another quantum walk!
Suppose we are given a set $R$ of $r$ vertices.  How do we decide whether $R$ contains an edge of a triangle?
If we can efficiently decide, for a given $u$ and $R$, whether $R$ contains vertices $v,w$ such that $u,v,w$ form a triangle in $H$,  
then we could combine this with a Grover search over all $n$ possible vertices $u$ of $H$.
Given $u$ and $R$, let us design a subroutine based on another quantum walk, this time on the Johnson graph $J(r,r^{2/3})$. 
Each vertex of this Johnson graph corresponds to a subset $R'\subseteq R$ of $r'=r^{2/3}$ vertices.  
Its spectral gap is $\delta'=r/r'(r-r')\approx 1/r^{2/3}$.
We'll mark $R'$ if it contains vertices $v,w$ such that $u,v,w$ form a triangle. 
If there is at least one triangle involving $u$ and some $v,w\in R$, then the fraction of marked vertices 
$R'$ in $J(r,r^{2/3})$ is at least $\eps'\approx (r'/r)^2=1/r^{2/3}$.
For this subroutine, the setup cost is $O(r^{2/3})$ (for each $v\in R'$, query whether $(u,v)$ is an edge in $H$); the update cost is $O(1)$ (if we replace $v$ in $R$ by $w$, then we need to ``unquery'' edge $(u,v)$ and query edge $(u,w)$); and the checking cost is~0.
Plugging this into Eq.~(\ref{eqqrwcost}), we can decide for a fixed $u$ whether there exist two vertices in $R'$ that complete $u$ to a triangle, using $O(r^{2/3})$ queries.
Let's ignore the small error probability of the latter subroutine (it can be dealt with, but that's technical).
We combine this with Grover search over all $n$ vertices $u$ to get $\Cc=O(\sqrt{n}r^{2/3})$.

Plugging these $\Sc$, $\Uc$, and $\Cc$ into Eq.~(\ref{eqqrwcost}), the overall cost of a quantum walk algorithm for triangle-finding is
$$
\Sc+\frac{1}{\sqrt{\eps}}\left(\Cc + \frac{1}{\sqrt{\delta}}\Uc\right)=O\left(r^2+\frac{n}{r}(\sqrt{n}r^{2/3} + r^{3/2})\right).
$$
This is $O(n^{13/10})$ if we set $r=n^{3/5}$~\cite{mss:triangle}. The quantum walk algorithm ends with a superposition where most of the amplitude sits on sets~$R$ containing one edge of a triangle (i.e., two vertices of $H$ that are part of a triangle). Now a measurement of that final state gives us such a set~$R$ with high probability, and then it's relatively cheap to find the third vertex of the triangle by another Grover search over the $n-r$ vertices of $H$ that are not in~$R$.

The exponent $13/10$ can be slightly improved further~\cite{belovs:learninggraphs,lms:improvedtrianglej,jkm:nested}, and the best known exponent is $5/4$~\cite{legall:triangle}. The optimal quantum query complexity for triangle-finding is an open question;
the best lower bound is only $\Omega(n)$.  The optimal quantum \emph{time} complexity of this problem is still open.

\section*{Exercises}

\begin{enumerate}

\item Let $d<n$, and $P$ be the projector on a $d$-dimensional subspace $V\subseteq\mathbb{R}^n$ that is spanned by orthonormal vectors $v_1,\ldots,v_d$.
This means that $Pv=v$ for all $v\in V$, and $Pw=0$ for all $w$ that are orthogonal to $V$.
\begin{enumerate}
\item Show that $P$ can be written in Dirac notation as $P=\sum_{i=1}^d \ketbra{v_i}{v_i}$.
\item Show that $R=2P-I$ is a reflection through the subspace corresponding to $P$, i.e., $Rv=v$ for all $v$ in that subspace, and $Rw=-w$ for all $w$ orthogonal to that subspace.
\end{enumerate}

\item Let $G$ be a $d$-regular graph that is bipartite, so its vertex set $V=[N]$ can be partitioned into disjoint sets $A$ and $B$, and all its edges are in $A\times B$. Give an eigenvector with eigenvalue~1 of the associated $N\times N$ normalized adjacency matrix $P$, and another eigenvector with eigenvalue~$-1$.

\item This exercise is about obtaining a quantum algorithm for finding collisions with a slightly different quantum walk. Consider the problem of Section~\ref{sseccollision}: we can query elements of the sequence of integers $x_0,\ldots,x_{n-1}$, and want to find distinct $i$ and $j$ such that $x_i=x_j$ (or report that there are no collisions). Again consider the Johnson graph $J(n,r)$, for some $r$ to be optimized over later. Deviating from Section~\ref{sseccollision}, now call a vertex $R$ \emph{marked} if there exist $i\in R$ and $j\in\{0,\ldots,n-1\}\setminus R$ such that $x_i=x_j$. Show that we can find a marked vertex in this graph with high probability using $O(n^{2/3})$ queries to~$x$. You may ignore small error probabilities, for example when using Grover's algorithm. Be explicit about what data you store about~$x$ at each vertex~$R$.

\item\label{ex:matrixtest} (H)
Let $A$, $B$, and $C$ be $n\times n$ matrices with real entries.
We'd like to decide whether or not $AB=C$.  Of course, you could multiply $A$ and $B$ and compare the result with $C$,
but matrix multiplication is expensive (the current best algorithm takes time roughly $O(n^{2.38})$).
\begin{enumerate}
\item Give a classical randomized algorithm that verifies whether $AB=C$ (with success probability at least $2/3$) using $O(n^2)$ steps,
using the fact that matrix-vector multiplication can be done in $O(n^2)$ steps.
\item Show that if we have query-access to the entries of the matrices (i.e., oracles that map $i,j,0\mapsto i,j,A_{i,j}$ and similarly for $B$ and $C$), then every classical algorithm needs at least $\Omega(n^2)$ queries to detect a difference between $AB$ and $C$ with error probability $\leq 1/3$.
\item Give a quantum walk algorithm that verifies whether $AB=C$ (with success probability at least $2/3$) using $O(n^{5/3})$ queries to matrix-entries.
\end{enumerate}

\item\label{ex:3sat} 
A 3-SAT instance $\phi$ over $n$ Boolean variables $x_1,\ldots,x_n$ is a formula which is the AND of a number of clauses, each of which is an OR of 3 variables or their negations. For example, $\phi(x_1,\ldots,x_4)=(x_1\vee x_2\vee \overline{x_3})\wedge(x_2\vee x_3\vee \overline{x_4})$ is a 3-SAT formula with 2 clauses.  A satisfying assignment is a setting of the $n$ variables such that $\phi(x_1,\ldots,x_n)=1$ (i.e, TRUE). You may assume the number of clauses is at most some polynomial in~$n$.  In general it is NP-hard to find a satisfying assignment to such a formula.  Brute force would try out all $2^n$ possible truth-assignments, but something much better is possible: consider the following simple algorithm of Sch\"oning~\cite{schoning:ksat}, 
which is a classical random walk on the set of all $N=2^n$ truth assignments:
\begin{quote}
Start with a uniformly random $x\in\01^n$.\\
Repeat the following at most $3n$ times: if $\phi(x)=1$ then STOP, else find the leftmost clause that is false, randomly choose one of its 3 variables and flip its value.
\end{quote}
One can show that this algorithm has probability at least $(3/4)^n/\sqrt{5n}$ of finding a satisfying assignment (if $\phi$ is satisfiable).  You may assume this without proof.
\begin{enumerate}
\item Use the above to give a classical algorithm that finds a satisfying assignment with high probability in time $(4/3)^n\cdot p(n)$, where
$p(n)$ is some polynomial factor.
\item (H) Give a quantum algorithm that finds a satisfying assignment (with high probability) in time $\sqrt{(4/3)^n}\cdot p(n)$. 
\end{enumerate}

\end{enumerate}

\chapter{Hamiltonian Simulation}\label{chap:hamsim}

\section{Hamiltonians}\label{sec:hamiltonians}

Thus far, we have viewed the dynamics of quantum systems from the perspective of \emph{unitary transformations}: apart from measurement, the only way a quantum state (i.e., a vector of amplitudes) can change is by multiplication with a unitary matrix, for instance a 2-qubit gate tensored with identities on the other qubits. But which unitary will actually occur in a given physical system? This is determined by the \emph{Hamiltonian} of the system, which is the observable~$H$ corresponding to the \emph{total energy} in the system.
The expectation value $\bra{\psi}H\ket{\psi}$ is called the energy of state $\ket{\psi}$.
Typically, this total energy is the sum of several different terms, corresponding to kinetic energy, potential energy, etc. Also typically, it is the sum of many \emph{local} terms that each act on only a few of the particles (qubits) of the system, for example if all interactions are between pairs of particles. 

One can think of the Hamiltonian~$H$ as describing the physical characteristics of the system. These do not determine the initial state $\ket{\psi(0)}$ of the system, but they do determine the \emph{evolution} of the state in time, i.e., the state $\ket{\psi(t)}$ as a function of the time-parameter~$t$, given initial state~$\ket{\psi(0)}$.
 This is governed by the most important equation in quantum mechanics: the \emph{Schr\"odinger equation}. It is a linear differential equation that relates the time-derivative of the current state to that state itself and to the Hamiltonian:
$$
i\hbar\frac{d\ket{\psi(t)}}{dt} = H\ket{\psi(t)}.
$$
Here $\hbar$ is a very small (at least in standard units) yet very important physical constant: Planck's constant divided by $2\pi$. We can set it to~1 by choosing appropriate units, and hence will ignore it from now on.
In general $H$ may itself change with~$t$, but for simplicity we will only consider here the case where $H$ is time-independent. Then, if we start in some state $\ket{\psi(0)}$, the solution to this differential equation is the following unitary evolution of the state:\footnote{Applying a function, for instance $f(x)=e^{-ix}$, to a normal matrix means applying $f$ to its eigenvalues: if $A$ has diagonalization $VDV^{-1}$ then $f(A)=Vf(D)V^{-1}$, where $f(D)$ is the diagonal matrix obtained by applying $f$ to the diagonal entries of~$D$. For example, if $A=\sum_j\lambda_j a_ja_j^T$ and $f(x)=e^{-ix}$, then $f(A)=\sum_je^{-i\lambda_j}a_ja_j^*$. Note that if $A$ is Hermitian, then $e^{-iA}$ is unitary: it maps $a_j$ to $e^{-i\lambda_j}a_j$.} 
$$
\ket{\psi(t)}=U\ket{\psi(0)},~~~\mbox{ where }U=e^{-iHt}.
$$
So evolution for time~$t$ induced by Hamiltonian~$H$, corresponds to applying the unitary matrix $e^{-iH}$ $t$~times.  Note, however, that $t$ need not be integer here: this evolution is continuous in time, in contrast to the discrete picture one gets from the circuit model with elementary quantum gates.

In areas like quantum chemistry (i.e., the study of properties of molecules and their interaction) and material sciences, it is often important to figure out how a quantum system will evolve from some given initial state, for instance starting from a basis state.\footnote{It is also very important in chemistry to be able to find out \emph{global} properties of a given Hamiltonian like its lowest energy, a.k.a.~\emph{ground state energy}. Unfortunately this problem seems to be hard to solve (in fact it is so-called \QMA-hard, see Chapter~\ref{ch:QMAlocham}) even for a quantum computer, even for the special case of 2-local Hamiltonians~\cite{kitaev:qnp,KKR:localham}.}
This is typically hard to do on classical computers, since the number of parameters (amplitudes) is exponential in the number of particles.
However, a \emph{quantum} computer is like a universal quantum system, and should be able to efficiently simulate every efficient quantum process, in the same way that a classical universal Turing machine can efficiently simulate other (classical) physical processes.\footnote{In Chapter~\ref{chap:qcomplexitytheory} we will see that it is actually possible to classically simulate quantum computers (and hence quantum systems more generally) with a polynomial amount of \emph{space}, but our best methods still use an exponential amount of \emph{time}. If factoring a large integer is a hard problem for classical computers (which is widely believed), then Shor's efficient quantum factoring algorithm (Chapter~\ref{chap:factoring}) implies that it is \emph{impossible} to simulate a quantum computer in polynomial time on a classical computer.}
In fact, this was the main reason why Feynman invented quantum computers: as a controllable quantum system that can be used to simulate other quantum systems.

In order to realize that idea, we need methods to efficiently implement the unitary evolution that is induced by a given Hamiltonian. In other words, we need methods to implement $U=e^{-iHt}$ as a quantum circuit of gates (say, up to some small error~$\eps$
\footnote{If $n$-qubit unitary $\widetilde{U}$ (e.g., a quantum circuit with not too many gates) is meant to approximate $n$-qubit unitary~$U$, then we can measure the error by the operator norm of their difference  $\norm{U-\widetilde{U}}=\max_\psi\norm{U\ket{\psi}-\widetilde{U}\ket{\psi}}$. 
However, we will also see simulation methods that allow some auxiliary qubits, say $a$ of them, which start in state $\ket{0^a}$ and should end in something close to state $\ket{0^a}$. In this case $\widetilde{U}$ acts on more qubits than $U$, so we cannot use the operator norm of their difference; instead we can measure the error on the subspace of $(n+a)$-qubit states where the last $a$ qubits are $\ket{0}$: $\max_{\psi}\norm{(U\ket{\psi})\ket{0^a}-\widetilde{U}(\ket{\psi}\ket{0^a})}$.
This way of measuring error still allows you to analyze a sequence of approximate unitaries using triangle inequality in a way that the errors add up at most linearly (as in Ex~\ref{chap:ft}.\ref{ex:approxqft}).}), and to apply this to a given initial state~$\ket{\psi}$. 
For simplicity we'll  ignore the minus sign in Hamiltonian simulation, implementing $U=e^{iHt}$ rather than $e^{-iHt}$.
This is known as the problem of ``Hamiltonian simulation'': 
\begin{quote}
    given $H$, $t$, and $\eps$, find a small circuit $\widetilde{U}$ that is $\eps$-close to $U=e^{iHt}$.
    \end{quote}
Note that we're asking for a unitary $\widetilde{U}$ that works for every possible $\ket{\psi}$, since the operator-norm error is defined by a worst-case $\ket{\psi}$.
In some cases it is possible to improve the solution by making use of specific properties of the $\ket{\psi}$ that we start from, but in this chapter we'll just focus on finding a $\widetilde{U}$ that works for all $\ket{\psi}$.

In this chapter we will cover several methods for Hamiltonian simulation.
We will also assume that our quantum system consists of $n$ qubits. Some physical systems, for instance electron spins, naturally correspond to qubits. More complicated Hilbert spaces, for instance with basis states labeled by the positions ($x,y,z$ coordinates) of all particles involved, can be encoded (approximately) in binary to reduce them to the case of qubits. This encoding can be done in many ways; much of the art in quantum chemistry is in how best to do this for specific systems, but we won't study that here (see for instance~\cite{caoetal:qchemistry}).

\section{Method 1: Lie-Suzuki-Trotter methods}\label{sec:trotter}

Note that an $n$-qubit Hamiltonian is a $2^n\times 2^n$ matrix, which is huge even for moderate~$n$. Typically in Hamiltonian simulation we are dealing with very structured Hamiltonians that have a much shorter classical description.
Suppose our Hamiltonian is of the form $H=\sum_{j=1}^m H_j$, where $m$ is not too big (say, polynomial in $n$) and each $H_j$ acts only on a few of the $n$ qubits. For concreteness assume each $H_j$ acts non-trivially on only \emph{two} of the qubits.\footnote{This means $H$ can be described efficiently by $m$ $4\times 4$ matrices, rather than by a $2^n\times 2^n$ matrix. A different assumption that is often made on Hamiltonians and that we will see later, is that $H$ is \emph{$s$-sparse}, meaning each of the $2^n$ columns has at most $s$ nonzero entries, and we have some efficient ``sparse access'' to these nonzero entries.
Note that if $H=\sum_j H_j$ and each $H_j$ acts on only 2 qubits (so it's effectively a $4\times 4$ matrix)), then $H$ is $4m$-sparse.
Thus, roughly speaking, the locality assumption implies the sparsity assumption.}
Such a Hamiltonian is called \emph{2-local}.
Note that, for fixed~$t$, the unitary $e^{iH_j t}$ is really just a 2-qubit gate, acting like identity on the other $n-2$ qubits; this 2-qubit gate could in turn be constructed from CNOTs and single-qubit gates. 

Our goal is to implement $U=e^{iHt}=e^{i\sum_j H_j t}$. It is now tempting to view this exponential of a sum of matrices as a product $\prod_{j=1}^m e^{i H_j t}$, which is just a product of $m$ 2-qubit gates. If all terms~$H_j$ are diagonal, or if there is some basis in which all terms are diagonal (equivalently, if all $H_j$ commute), then this indeed works out.
However, in general matrix exponentials do not work that way: $e^{A+B}$ need not equal $e^A e^B$ if $A$ and $B$ do not commute (see Exercise~\ref{ex:noncommute}).
The Lie-Suzuki-Trotter decomposition gives us a way to handle this. It uses the fact that if $A$ and $B$ have small operator norm, then $e^{A+B}$ and $e^{A}e^{B}$ are \emph{approximately} equal: $e^{A+B}=e^{A}e^{B} + E$, where the error-term $E$ is a matrix whose operator norm is $O(\norm{A}\cdot\norm{B})$.\footnote{A non-rigorous but reasonably convincing way to see this is to approximate the term $e^M$ by its first-order Taylor series $I+M$, which is a good approximation if $M$ has small norm (the error of the approximation will be quadratic in that norm). Then $e^Ae^B-e^{A+B}\approx (I+A)(I+B)-(I+A+B)=AB$. In case you ever need it: the so-called Baker-Campbell-Hausdorff formula gives a much more precise expression.}

How can we use this to approximate $U$ by a circuit $\widetilde{U}$ of 2-qubit gates? 
Assume each of the terms $H_j$ has operator norm $\leq 1$ (see Exercise~\ref{ex:moreenergy} for why such normalization matters).
First consider the simple case $m=2$, so $H=H_1+H_2$. 
We can now implement $U=e^{iHt}$ by doing a little bit of $H_1$, a little bit of $H_2$, a little bit of $H_1$, etc. More precisely, for every integer $r\geq 1$ of our choice, we have
\begin{equation}
U=e^{iHt}=(e^{iHt/r})^r
=(e^{iH_1t/r + iH_2t/r})^r
=(e^{iH_1t/r}e^{iH_2t/r}+E)^r.
\end{equation}
Here the error-term $E$ has norm $\norm{E}=O(\norm{iH_1t/r}\cdot \norm{iH_2t/r})=O(\norm{H_1}\cdot\norm{H_2}t^2/r^2)$. 
Our approximating circuit will be $\widetilde{U}=(e^{iH_1t/r}e^{iH_2t/r})^r$,
which uses $2r=O(t^2/\eps)$ 2-qubit gates.
Since errors in a product of unitaries add at most linearly (see Exercise~\ref{chap:ft}.\ref{ex:approxqft}), we have approximation error  
\[
\norm{U-\widetilde{U}}\leq r\norm{E}=O(\norm{H_1}\cdot\norm{H_2}t^2/r)=O(t^2/r).
\]
Choosing $r=O(t^2/\eps)$, we can make this error $\leq\eps$.  

The same idea works for the general case where we have $m>2$ Hamiltonian terms:
\begin{equation}\label{eq:mtermTrotter}
U=e^{iHt}=(e^{iHt/r})^r
=(e^{iH_1 t/r +\cdots + iH_m t/r})^r
=(e^{iH_1 t/r}\cdots e^{iH_m t/r}+E)^r,
\end{equation}
where $\norm{E}=
O(m^2t^2/r^2)$  (Exercise~\ref{ex:TrotterE}). Choosing $r=O(m^2t^2/\eps)$, we have an approximating circuit $\widetilde{U}=(e^{iH_1t/r}\cdots e^{iH_mt/r})^r$ with $mr=O(m^3t^2/\eps)$ 2-qubit gates, and error $\norm{U-\widetilde{U}}\leq r\norm{E}\leq\eps$.

This is the \emph{first-order} Lie-Suzuki-Trotter approach to Hamiltonian simulation, due to Lloyd~\cite{lloyd:sim}.  
The number of gates of the circuit $\widetilde{U}$ depends quadratically on the time~$t$ for which we want to simulate the evolution, which is not optimal.
One can do fancier higher-order Lie-Suzuki-Trotter decompositions that make the dependence on $t$ nearly linear, but we won't explain those here (see Exercise~\ref{ex:improvertrotter}). Instead we will describe two methods below with linear $t$-dependence. The number of gates of~$\widetilde{U}$ depends polynomially on~$\eps$; this can be very much improved as well, as we will see.\footnote{While the upper bounds on the number of gates for Trotter methods are theoretically worse (in their dependence on $t$ and $\eps$) than the other two methods explained in this chapter, in practice Trotter is quite competitive~\cite{childsetal:trotter}. Trotter also has the advantages of being relatively simple and of not requiring any auxiliary qubits.}

\section{Method 2: Linear combination of unitaries (LCU)}\label{sec:LCU}

Here we will describe a method for Hamiltonian simulation whose complexity depends linearly on the time~$t$ for which we want to evolve the state, and only logarithmically on the desired error~$\eps$.

Let's start with a more general problem.
Suppose we have a $2^n\times 2^n$ matrix $M$ and an $n$-qubit state $\ket{\psi}$, and we would like to prepare the state $M\ket{\psi}/\norm{M\ket{\psi}}$. Here $M$ need not be unitary, but suppose we can write $M$ as a linear combination of $m$ unitaries $V_0,\ldots,V_{m-1}$:%
\footnote{In fact \emph{every} $M$ can be written in such a way, because the $4^n$ $n$-qubit Pauli matrices (each of which is unitary) form a basis for the linear space of all $2^n\times 2^n$ matrices. See Appendix~\ref{app:pauli}.} 
$$
M=\sum_{j=0}^{m-1} \alpha_j V_j,
$$
with the $\alpha_j$ being nonnegative reals (we can always absorb complex phases into the $V_j$). Let $\norm{\alpha}_1=\sum_j\alpha_j$, and let $W$ be a unitary acting on $\ceil{\log m}$ qubits that maps 
\[
W:\ket{0}\mapsto\frac{1}{\sqrt{\norm{\alpha}_1}}\sum_{j=0}^{m-1}\sqrt{\alpha_j}\ket{j}.
\]
We don't care what $W$ does on other states than the all-0 state (so we are not fully specifying $W$; every unitary $W$ that maps $\ket{0}$ as above will work for LCU).
Suppose each $V_j$ is an ``easy'' unitary, for instance a 2-qubit gate tensored with identity on the other $n-2$ qubits, or a small circuit. Also suppose we can implement these unitaries in a controlled way: we have access to a 2-register unitary $$
V=\sum_{j=0}^{m-1} \ketbra{j}{j}\otimes V_j. 
$$
This maps $\ket{j}\ket{\phi}\mapsto\ket{j}V_j\ket{\phi}$, and we can think of the first register as ``selecting'' which unitary $V_j$ to apply to the second register.\footnote{In the literature, this $V$ is often called ``select-$V$.'' One might expect the cost of $V$ to be not much higher than the costliest~$V_j$, just like the cost of a classical ``if $A$ then $B$, else $C$'' statement is not much bigger than the largest of the costs of $B$ and~$C$. However, if we measure circuit size, then the cost of $V$ could be roughly the \emph{sum} of the costs of the $V_j$s because circuits for each $V_j$ should be ``included'' in the circuit for~$V$.} 

We want to use $V$ and $W$ to implement $M$ on a given state $\ket{\psi}$.
Consider the following algorithm:
\begin{enumerate}\label{eq:lcu4step}
\item Start with two-register state $\ket{0}\ket{\psi}$, where the first register has $\ceil{\log m}$ qubits.
\item Apply $W$ to the first register.
\item Apply $V$ to the whole state.
\item Apply $W^{-1}$ to the first register.
\end{enumerate}
A small calculation (see Exercise~\ref{ex:justifyMpsi}) shows that the resulting state can be written as \begin{equation}\label{eq:Mpsi}
\frac{1}{\norm{\alpha}_1}\ket{0}M\ket{\psi}+\sqrt{1-\frac{\norm{M\ket{\psi}}^2}{\norm{\alpha}_1^2}}\ket{\phi}, 
\end{equation}
where $\ket{\phi}$ is some other normalized state that we don't care about, but that has no support on basis states where the first register is~$\ket{0}$ (i.e., all such basis states have amplitude~0 in $\ket{\phi}$).
Note that the state of \eqref{eq:Mpsi}
has norm~1, because the squared norm of the first term is $\norm{M\ket{\psi}}^2/\norm{\alpha}_1^2$.

If we were to measure the first register, the probability of outcome~0 is $p=\norm{M\ket{\psi}}^2/\norm{\alpha}_1^2$.
In case of that measurement outcome, the second register would collapse to the normalized version of $M\ket{\psi}$, as desired. 
The success probability $p$ may be small, but we could use $O(1/\sqrt{p})=O(\norm{\alpha}_1/\norm{M\ket{\psi}})$ rounds of amplitude amplification to amplify the part of the state that starts with $\ket{0}$.\footnote{If we do not know the value of $p$ in advance, then we can try out exponentially decreasing guesses for~$p$, like we do for Grover's algorithm in Chapter~\ref{chap:grover} when we don't know the number of solutions.}
Thus we would prepare (the normalized version of) $M\ket{\psi}$ in the second register. Unfortunately this usage of amplitude amplification assumes the ability to implement a unitary (as well as its inverse) to prepare $\ket{\psi}$ from a known initial state, say~$\ket{0}$. Regular amplitude amplification won't work if instead of a unitary that prepares $\ket{\psi}$ we just have one copy of the state~$\ket{\psi}$ available, which is the typical situation in Hamiltonian simulation. However, Exercise~\ref{ex:oblivamplampl} gives us a variant called \emph{oblivious} amplitude amplification, which circumvents this problem: it works even with just one copy of~$\ket{\psi}$, as long as $M$ is proportional to a unitary (or close to that). Fortunately, this is the situation when we use LCU for Hamiltonian simulation, where $M\approx e^{iHt}$.

\subsection{Hamiltonian simulation via LCU}\label{ssec:hamsimLCU}

Recall that our goal is to efficiently implement the unitary $e^{iHt}$ that is induced by a given Hamiltonian~$H$, normalized so that $\norm{H}\leq 1$.
The following approach is due to Berry et al.~\cite{Berryetal:expimpr,BerryChilds:hamsim,BCK:nearlyoptimal}. Suppose, somewhat paradoxically, that we can write out the \emph{Hermitian} matrix $H$ as a linear combination of \emph{unitaries}: $H=\sum_j\alpha_j V_j$. For example, if $H$ is the sum of $m$ 2-local terms like before, then every 2-local term can be written as the sum of at most 16 $n$-qubit Pauli matrices (each of which is unitary and acts non-trivially on only two qubits). Thus we would decompose $H$ as a sum of at most $16m$ unitaries, each acting non-trivially on only two of the $n$ qubits. The sum of coefficients $\norm{\alpha}_1$ will be $O(m)$.

Using the Taylor series $e^x=\sum_{k=0}^\infty x^k/k!$, we write the unitary we want to implement as 
\begin{equation}\label{eq:LCUeiHt}
e^{iHt}=\sum_{k=0}^\infty \frac{(iHt)^k}{k!}
=\sum_{k=0}^\infty \frac{(it)^k}{k!}\left(\sum_{j\in[m]} \alpha_j V_j\right)^k
=\sum_{k=0}^\infty \frac{(it)^k}{k!}\sum_{j_1,\ldots,j_k\in[m]}\alpha_{j_1}\cdots\alpha_{j_k}V_{j_1}\cdots V_{j_k}.
\end{equation}
Note that if each $V_j$ is easy to implement and $k$ is not too big, then the product $V_{j_1}\cdots V_{j_k}$ is also not too hard to implement. For example, if each $V_j$ is one elementary gate (tensored with identity on the non-affected qubits), then that product is a circuit of only $k$ elementary gates.

Exercise~\ref{ex:errorTaylor} shows that if we truncate the Taylor series at $k=O(t+\log(1/\eps))$, dropping the terms of higher order, then the induced error (i.e., the dropped part) has operator norm at most~$\eps$. Accordingly, we can take the part of the right-hand side of Eq.~(\ref{eq:LCUeiHt}) for $k=O(t+\log(1/\eps))$ and then use the linear combination of unitaries approach to approximately implement $e^{iHt}$.
The unitaries in this decomposition are of the form $V_{j_1,\ldots,j_k}=i^k V_{j_1}\cdots V_{j_k}$; let 
\[
{\cal V}=\sum_{j_1,\ldots,j_k}\ketbra{j_1,\ldots,j_k}{j_1,\ldots,j_k}\otimes V_{j_1,\ldots,j_k}
\]
denote the controlled operation of the $V_{j_1,\ldots,j_k}$ unitaries, each of which involves $k$ $V_j$'s; we can also think of ${\cal V}$ as $k$ applications of the basic controlled unitary $V=\sum_{j}\ketbra{j}{j}\otimes iV_{j}$.
The corresponding nonnegative coefficients in the decomposition of Eq.~\eqref{eq:LCUeiHt} are
$$
\beta_{j_1,\ldots,j_k}=\frac{t^k}{k!}\alpha_{j_1}\cdots\alpha_{j_k},~~~~\mbox{for }k\leq O(t+\log(1/\eps)).
$$
These $\beta$-coefficients add up to 
$$
\norm{\beta}_1=\sum_{k=0}^{O(t+\log(1/\eps))}\frac{t^k}{k!}\sum_{j_1,\ldots,j_k}\alpha_{j_1}\cdots\alpha_{j_k}\leq\sum_{k=0}^\infty\frac{t^k}{k!}\sum_{j_1,\ldots,j_k}\alpha_{j_1}\cdots\alpha_{j_k}=\sum_{k=0}^\infty\frac{(t\norm{\alpha}_1)^k}{k!}=e^{t\norm{\alpha}_1},
$$
so straightforward application of the LCU method with oblivious amplitude amplification uses  $O(\norm{\beta}_1)=O(e^{t\norm{\alpha}_1})$ applications of~${\cal V}$ and ${\cal V}^{-1}$, each of which involves as $k=O(t+\log(1/\eps))$ applications of the basic controlled unitary $V$ or its inverse.

The logarithmic $\eps$-dependence of the complexity of the above method is excellent. The exponential dependence on~$t\norm{\alpha}_1$ is quite terrible for large~$t$, but not too bad for very small~$t$. So what we'll do if we want to do a simulation for large~$t$, is to divide that~$t$ into $b=t\norm{\alpha}_1$ blocks of time  $\tau=1/\norm{\alpha}_1$ each, run the above algorithm for time~$\tau$ with error $\eps'=\eps/b$, and then glue $b$ time-$\tau$ simulations together. This will simulate $(e^{iH\tau})^b=e^{iHt}$, with error $\leq b\eps'=\eps$. The cost of each time-$\tau$ simulation is $O(e^{\tau\norm{\alpha}_1})=O(1)$ applications of $\cal V$ and ${\cal V}^{-1}$, each of which involves $O(\tau+\log(1/\eps'))=O(\log(t\norm{\alpha}_1/\eps))$ applications of the $V$ and $V^{-1}$.  The overall cost will be $b$ times that, since we'll run $b$ subsequent time-$\tau$ simulations in order to implement a time-$t$ simulation.

To give a more concrete example, consider again the special case where $H=\sum_i H_i$ consists of 2-local terms, so the unitaries $V_j$ in the induced linear combination of unitaries $H=\sum_{j=1}^{m}\alpha_j V_j$ only act nontrivially  on 2 qubits each.
Then we approximate the time-$\tau$ unitary  $e^{iH\tau}$ by a linear combination of unitaries 
\begin{equation}\label{eq:LCUforHamsim}
M=\sum_{k=0}^{O(\tau+\log(1/\eps'))}\sum_{j_1,\ldots,j_k\in[m]}\beta_{j_1,\ldots,j_k}V_{j_1,\ldots,j_k},
\end{equation}
where each $V_{j_1,\ldots,j_k}$ is a product of $k=O(\tau+\log(1/\eps'))=O(\log(t\norm{\alpha}_1/\eps))$ 2-qubit gates.
We can implement this using the linear combination of unitaries approach, and repeat this $b=t\norm{\alpha}_1$ times.
The cost of the unitary $W$ is typically relatively small (see Exercise~\ref{ex:statepreparation}), so we can $\eps$-approximate the unitary $e^{iHt}$ using a circuit of roughly $O(t\norm{\alpha}_1\log(t\norm{\alpha}_1/\eps))=O(mt\log(mt/\eps))$ applications of~$V$ and $V^{-1}$, and slightly more other 2-qubit gates.
Note that the cost depends linearly on the evolution-time~$t$, and logarithmically on the error~$\eps$, both of which are much better than Lie-Suzuki-Trotter methods.

\section{Method 3: Transforming block-encoded matrices}\label{sec:blockenc}

In this section we'll describe a recent approach that is very general and flexible.  Suppose $A$ is an $n$-qubit matrix with operator norm $\norm{A}\leq 1$, and we can implement an $(n+1)$-qubit unitary
\begin{equation}\label{eq:UblockA}
U=\left(\begin{array}{cc} A & \cdot\\
\cdot & \cdot
\end{array}\right).
\end{equation}
The `$\cdot$'s are unspecified $2^n\times 2^n$-dimensional matrices, the only constraint on which is that $U$ is unitary. Such a $U$ is called a unitary \emph{block-encoding} of~$A$. 
Note that $$
U:\ket{0}\ket{\psi}\mapsto\ket{0}A\ket{\psi}+\ket{1}\ket{\phi},
$$ 
where we can't say much about the (subnormalized) state $\ket{\phi}$.
Written more technically, the defining property of such a block-encoding is 
$(\bra{0}\otimes I)U(\ket{0}\otimes I)=A$, where the first register is one qubit.
More generally we can define an \emph{$a$-qubit} block-encoding of $A$, which is an ($a+n)$-qubit unitary $U$ with the property that $(\bra{0^a}\otimes I)U(\ket{0^a}\otimes I)=A$.

\paragraph{Example 1: LCU does block-encoding.}
From Eq.~(\ref{eq:Mpsi}) we can see that LCU (without the final amplitude amplification) implements a $\ceil{\log m}$-qubit  block-encoding of the matrix $A=M/\norm{\alpha}_1$.

\paragraph{Example 2: Block-encoding a sparse Hermitian matrix.}
Let $A$ be a $2^n\times 2^n$ Hermitian matrix of operator norm $\norm{A}\leq 1$ that is $s$-sparse, so each row and column of $A$ has at most $s$ nonzero entries (for simplicity assume \emph{exactly} $s$ nonzero entries). Since this matrix~$A$ is still an exponentially large object, we have to be careful how we can access such sparse matrices. First, we assume we can query the entries of $A$ in the usual way: we have an oracle
$$
O_A:\ket{i,j}\ket{0}\mapsto \ket{i,j}\ket{A_{ij}},
$$
where we assume the last register has sufficiently many qubits to write down the complex entry~$A_{ij}$ either exactly or with sufficiently many bits of precision.
Of course, since $A$ is sparse, $A_{ij}$ will actually be~0 for most $(i,j)$.
Let $\nu(j,\ell)\in\{0,\ldots,N-1\}$ denote the location of the $\ell$-th nonzero entry of the $j$-th column of $A$;
so the $s$~nonzero entries in the $j$-th column are at positions $\nu(j,0),\ldots,\nu(j,s-1)$.
We also assume we have another oracle that allows us to find these locations:
$$
O_{A,loc}:\ket{j,\ell}\mapsto\ket{j,\nu(j,\ell)}.
$$
Here both registers have $n$ qubits. We think of $\ell\in\{0,\ldots,s-1\}$ as an $n$-bit integer whose most significant bits are 0; of course, the unitary $O_{A,loc}$ will also do something when $\ell\geq s$, but we don't care about that (this is an incomplete specification of the unitary $O_{A,loc}$, similar to how in LCU we only specified what the unitary $W$ does on the all-0 state, not on other states).
We also assume we can run $O_A^{-1}$ and $O_{A,loc}^{-1}$. Together these assumptions are called having ``sparse access'' to~$A$.

We will now show how to implement a block-encoding of the matrix $A/s$.
Exercise~\ref{ex:sparseblock} shows how  we can implement two $(2n+1)$-qubit unitaries that create superpositions over the locations of the nonzero entries in the $j$-th column and $i$-th row of $A$, respectively:
$$
W_1:\ket{0}\ket{0^n}\ket{j}\mapsto\frac{1}{\sqrt{s}}\ket{0}\sum_{k:A_{kj}\neq 0}\ket{k,j},\mbox{~~~~~~~}W_3:\ket{0}\ket{0^n}\ket{i}\mapsto\frac{1}{\sqrt{s}}\sum_{\ell:A_{i\ell}\neq 0}\ket{0}\ket{i,\ell},
$$
using one $O_{A,loc}$-query and a few other $A$-independent gates.
We can also implement the following unitary using one query to each of $O_A$ and $O_A^{-1}$, and a few other $A$-independent gates (and some auxiliary qubits that start and end in $\ket{0}$):
$$
W_2:\ket{0}\ket{k,j}\mapsto
A_{kj}\ket{0}\ket{k,j} + \sqrt{1-|A_{kj}|^2}\ket{1}\ket{k,j}.
$$
By following the action on initial state $\ket{0^{n+1}j}$ step-by-step (Exercise~\ref{ex:sparseblock}), one can show that the $(0^{n+1}i,0^{n+1}j)$-entry of $U=W_3^{-1} W_2W_1$ equals $A_{ij}/s$.
In other words, $U$ is an $(n+a)$-qubit block-encoding of matrix $A/s$ for some~$a$ (depending on how many auxiliary qubits are actually used).

How can we use a given  block-encoding $U$ of $A$?
Suppose that for some function $f:\R\to\R$ we want to implement a unitary~$V$ that looks like
$$
V=\left(\begin{array}{cc} f(A) & \cdot\\
\cdot & \cdot
\end{array}\right),
$$
using a small number of applications of the block-encoding of~$A$.
Here we don't care what submatrices sit at the `$\cdot$' entries of $U$ or $V$, as long as the upper-left block of~$V$ is $f(A)$ and $V$ as a whole is unitary.
For example, in Hamiltonian simulation $A$ would be the Hamiltonian~$H$ and $f(x)$ would be $e^{ixt}$, so that we are effectively implementing $f(H)=e^{iHt}$. In the HHL algorithm in the next chapter, $f(x)$ will be $1/x$, so that we effectively implement~$A^{-1}$. 

It turns out that we can implement a good approximation of $V$ efficiently if we have a \emph{low-degree polynomial}~$P$ approximating~$f$. 
The idea is that we can let $P$ act on the eigenvalues of~$A$, thus transforming a block-encoding of~$A$ into one of~$P(A)$. 
We state without proof the following theorem by Gily\'en et al.~\cite[follows by applying  their Theorem~56 to the real and to the imaginary part of the polynomial]{gilyenea:svtrans}, which extends work of Low et al.~\cite{LowChuang:Qubitization,LowChuang:OptHamSim,LowChuang:Uniform,LowYoderChuang:SignalProc}.

\begin{theorem}\label{thm:SVT}
Let $P:[-1,1]\to\{c\in\mathbb{C} \mid |c|\leq 1/4\}$ 
be a degree-$d$ polynomial, and let $U$ be a unitary $a$-qubit block-encoding of Hermitian matrix~$A$.
We can implement a unitary $O(a)$-qubit block-encoding $V$ of $P(A)$ using $d$ applications of $U$ and $U^{-1}$, one controlled application of $U$, and $O(ad)$ other 2-qubit gates.
\end{theorem}

This theorem can be generalized to a powerful technique called ``quantum singular-value transformation'' (QSVT)~\cite{gilyenea:svtrans}, where $A$ can be an arbitrary matrix, non-Hermitian and even non-square.

\subsection{Hamiltonian simulation via transforming block-encoded matrices}

Let's see how we can use Theorem~\ref{thm:SVT} for Hamiltonian simulation for a given sparse Hamiltonian~$H$.
Like in Section~\ref{ssec:hamsimLCU}, we approximate the function $f(x)=e^{ixt}$ using a polynomial~$P$ of degree $d=O(t+\log(1/\eps))$, which is the first $d$ terms of the Taylor series of~$f$ (see Exercise~\ref{ex:errorTaylor}), divided by~$4(1+\eps)$ to ensure that its range satisfies the condition of Theorem~\ref{thm:SVT}.
If $H$ is $s$-sparse and we have sparse access to it, then Example~2 of Section~\ref{sec:blockenc} shows how to efficiently implement a block-encoding~$U$ of the scaled Hamiltonian~$H/s$, using $O(1)$ queries to~$H$ and $O(n)$ other gates. Note that evolving Hamiltonian~$H$ for time $t$ is the same as evolving $H/s$ for time~$st$.
Theorem~\ref{thm:SVT} now gives us a block-encoding $V$ of $P(H)\approx \frac{1}{4}e^{iHt}$. This $V$ invokes $U$ and $U^{-1}$ $O(st+\log(1/\eps))$ times, and maps:
$$
V:\ket{0}\ket{\psi}\mapsto\ket{0}P(H)\ket{\psi}+\ket{\phi},
$$
where $\ket{\phi}$ has no support on basis states starting with $\ket{0}$.
Since $P(H)\approx \frac{1}{4}e^{iHt}$ is essentially proportional to a unitary, we can now apply $O(1)$ rounds of oblivious amplitude amplification to boost the factor $\frac{1}{4}$ to essentially~1, using only one copy of $\ket{\psi}$.

This implements the desired unitary $e^{iHt}$ on one copy of $\ket{\psi}$, up to small error.
The complexity of $\eps$-precise Hamiltonian simulation of an $s$-sparse Hamiltonian $H$ of operator norm $\leq 1$ then becomes $O(st+\log(1/\eps))$ queries to~$H$ and $O(n(st+\log(1/\eps)))$ 2-qubit gates.

\section*{Exercises}

\begin{enumerate}
  
\item\label{ex:noncommute}
Compute the following four $2\times 2$ unitaries: 
$e^{i\frac{\pi}{2} Z}$, $e^{i\frac{\pi}{2} X}$, $e^{i \frac{\pi}{2} X}e^{i\frac{\pi}{2} Z}$, 
and $e^{i\frac{\pi}{2}(X+Z)}$.\\ 
Here $X$ and $Z$ are the usual Pauli matrices.\\ 
{\footnotesize Comment: 
Note that $e^{i \frac{\pi}{2} X}e^{i\frac{\pi}{2}Z}\neq e^{i\frac{\pi}{2}(X+Z)}$, illustrating the issue that Trotter methods deal with.}

\item\label{ex:moreenergy} Suppose we want to implement a certain unitary $U$, and we can do that by switching on a Hamiltonian $H$ for some time $t$: $U=e^{-iHt}$. Now suppose $H'$ is another Hamiltonian, with 100 times as much energy as $H$: $H'=100H$.
Show that using $H'$ we can implement $U$ a 100 times faster than with~$H$.\\
{\footnotesize Comment: This exercise is about time in the physical sense of the word, not about ``time complexity'' in the sense of circuit size. It shows why some kind of normalization of $H$ is needed if we want to talk about the time it takes to implement something. We can always ``speed up'' a computation by a factor $k$ if we can multiply our Hamiltonian with a factor~$k$ (which is like injecting $k$ times as much energy).}

\item\label{ex:TrotterE}
(H) This exercise justifies the error bound of Eq.~\eqref{eq:mtermTrotter}.
Let $H_1,\ldots,H_m$ satisfy $\norm{H_j}\leq 1$ and let $\delta>0$ be small (in Eq.~\eqref{eq:mtermTrotter} we'd set $\delta=t/r$). Show that there is a constant $c$, independent of $\delta$ and $H_1,\ldots,H_m$, such that 
\[
e^{iH_1 \delta +\cdots + iH_m \delta}
=e^{iH_1 \delta}\cdots e^{iH_m \delta}+E,
\]
for some $E$ of norm $\norm{E}\leq c\delta^2\sum_{j=1}^{m-1}j$.

\item\label{ex:improvertrotter} (H) This exercise gives a version of Trotter methods with a better dependence of the circuit-size on the evolution-time $t$ and the approximation error $\eps$ than the $O(t^2/\eps)$ 2-qubit gates of basic Trotter (Section~\ref{sec:trotter}). Let $A,B$ be Hermitian matrices of the same dimension and operator norm $\norm{A},\norm{B}\leq 1$. You may assume the following fact without proof\footnote{This fact can be proved by taking 2nd-order Taylor expansions of the four exponentials involved in Eq.~\eqref{eq:bettertrotter}, and showing that the left- and right-hand sides match up to $O(\eta^3)$ terms. But you don't need to provide that proof. In the formula used for basic Trotter, the right-hand side was $e^{i\eta A}e^{i\eta B}$ with a bigger error term, of norm $O(\eta^2)$.}: for every $\eta\in[0,1]$ there exists a matrix~$E$ of operator norm $\norm{E}=O(\eta^3)$ such that
\begin{equation}\label{eq:bettertrotter}
e^{i\eta(A+B)}=e^{i\eta A/2}e^{i\eta B}e^{i\eta A/2}+E.
\end{equation}
Suppose $H=H_1+H_2$ is an $n$-qubit Hamiltonian where $H_1,H_2$ are 2-local terms of operator norm $\leq 1$. For every $t\geq 0$ and $\eps \in(0,1/2)$, give a circuit $\widetilde{U}$ of $O(t^{1.5}/\sqrt{\eps})$ 2-qubit gates that approximates the unitary $U=e^{iHt}$ up to error~$\eps$ (i.e., $\norm{U-\widetilde{U}}\leq\eps$). Every 2-qubit unitary counts as a 2-qubit gate here (don't worry about writing out in CNOTs and 1-qubit gates).

\item Consider the simple case of the linear-combination-of-unitaries trick where $m=2$ and $M=V_0+V_1$. Describe the unitaries $V$ and $W$, and track how the initial state $\ket{0}\ket{\psi}$ changes through the 4-step algorithm in Section~\ref{sec:LCU} (you don't have enough information to fully describe the final state because you only know what $W$ does on basis state $\ket{0}$, but you should end with something of the form Eq.~\eqref{eq:Mpsi}).

\item
\begin{enumerate}
\item Write the gate $\displaystyle G=\frac{1}{\sqrt{2}}\left(
\begin{array}{rr}
1 & -i\\ 
i & -1
\end{array}
\right)$ as a linear combination of the Pauli matrices (see Appendix~\ref{app:pauli}).
\item Suppose you want to implement $G$ via LCU, using the linear combination of (a). What are $W\ket{0}$ and $V$?
\item Consider the final state after the 4-step algorithm of Section~\ref{sec:LCU}. Calculate the part of that state that starts with $\ket{0}$ (without writing out that final state fully!)
\end{enumerate}

\item\label{ex:justifyMpsi}
(H) Give a calculation to justify that the 4-step algorithm in Section~\ref{sec:LCU} indeed always produces a state of the form of Eq.~(\ref{eq:Mpsi}).

\item\label{ex:statepreparation}
Let $v\in[-1,1]^N$ be a vector with real entries, of dimension $N=2^n$, indexed by $i\in\01^n$. Suppose we can query the entries of this vector by a unitary that maps
$$
                O_v:\ket{i}\ket{0^p}\mapsto\ket{i}\ket{v_i},
$$
so where the binary representation of the $i$-th entry of $v$ is written into the second register. We assume this second register has $p$ qubits, and the numbers $v_i$ can all be written exactly with $p$ bits of precision (it doesn't matter how, but for concreteness say that the first bit indicates the sign of the number, followed by the $p-1$ most significant bits after the decimal dot).
Our goal is to prepare the $n$-qubit quantum state
$\displaystyle
                \ket{\psi}=\frac{1}{\norm{v}}\sum_{i\in\01^n}v_i\ket{i}.$
\begin{enumerate}
\item Show how you can implement the following 3-register map (where the third register is one qubit) using one application of $O_v$ and one of $O_v^{-1}$, and some $v$-independent unitaries (you don't need to draw detailed circuits for these unitaries, nor worry about how to write those in terms of elementary gates).
$$
\ket{i}\ket{0^p}\ket{0}\mapsto
\ket{i}\ket{0^p}(v_i\ket{0}+\sqrt{1-v_i^2}\ket{1}).
$$
\item Suppose you apply the map of (a) to a uniform superposition over all $i\in\01^n$. Write the resulting state, and calculate the probability that measuring the last qubit in the computational basis gives outcome~0.
\item What is the resulting 3-register state if the previous measurement gave outcome 0?
\item Assume you know $\norm{v}$ exactly. Give an algorithm that prepares $\ket{\psi}$ exactly, using $\displaystyle O\left(\frac{\sqrt{N}}{\norm{v}}\right)$ applications of $O_v$ and $O_v^{-1}$, and some $v$-independent unitaries.
\end{enumerate}

\item\label{ex:oblivamplampl}
  (H) This exercise explains \emph{oblivious} amplitude amplification.\\
Let $M$ be an $n$-qubit unitary. We start from $\ket{\Psi}=\ket{0^a}\ket{\psi}$ for unknown $n$-qubit state $\ket{\psi}$. Our goal is to prepare the state $\ket{\Phi}=\ket{0^a}M\ket{\psi}$ (this $\ket{\Phi}$ is the analogue of the ``good state'' in amplitude amplification). Let $U$ be an $(a+n)$-qubit unitary, independent of $\ket{\psi}$, such that
$$
U\ket{\Psi}=\sin(\theta)\ket{\Phi}+\cos(\theta)\ket{\Phi^\perp},
$$
where $\theta$ is some angle that's independent of $\ket{\psi}$, while $\ket{\Phi^\perp}$ is some normalized state that depends on $\ket{\psi}$ and has no support on basis states starting with $0^a$ (this $\ket{\Phi^\perp}$ is the analogue of the ``bad state'').
If $\theta$ is close to $\pi/2$, then we can just apply $U$ to our starting state $\ket{\Psi}$ and measure the first register; we'll see $0^a$ with probability $\sin(\theta)^2\approx 1$ and in that case end up with the desired state $\ket{\Phi}$. But suppose $\theta$ is quite small. Here we will see how we can amplify the angle $\theta$ to roughly $\pi/2$, without assuming a unitary to prepare $\ket{\Psi}$.
\begin{enumerate}
 \item 
 Let $\cal S$ be the 2-dimensional space  spanned by $\ket{\Phi}$ and $\ket{\Phi^\perp}$.
 Let $R=(I-2\ketbra{0^a}{0^a})\otimes I$ be a unitary that puts a `$-$' in front of every basis state that starts with~$0^a$.
    Show that $R$, restricted to $\cal S$, is a reflection through $\ket{\Phi^\perp}$. 
    \item Define
$\ket{\Psi^\perp}=U^{-1}\left(\cos(\theta)\ket{\Phi}-\sin(\theta)\ket{\Phi^\perp}\right)$. Show $U\ket{\Psi}$ and $U\ket{\Psi^\perp}$ are orthogonal.\\[1.1mm]
    One can also show with a bit more work~\cite[Lemma~3.7]{Berryetal:expimpr} the stronger statement that $\ket{\Psi^\perp}$ has no support on basis states starting with $0^a$. You may assume this fact without proof in the remainder of this exercise.
       \item Show that $-URU^{-1}$, restricted to $\cal S$, is a reflection through $U\ket{\Psi}$ (note the minus sign!)
    \item Show that 
    $(-URU^{-1} R)^k U\ket{0^a}\ket{\psi}=\sin((2k+1)\theta)\ket{\Phi}+\cos((2k+1)\theta)\ket{\Phi^\perp}$.
\item How large should we take $k$ in order to end up with (approximately) the state $\ket{\Phi}$?\\
{\footnotesize NB: If you know $\theta$ exactly, then you can even exactly prepare $\ket{\Phi}$ (along the lines of Exercise~\ref{chap:grover}.\ref{ex:exactgrover}) but you don't need to show that.}
\end{enumerate}

\item\label{ex:errorTaylor}
(H) Let $\eps\in(0,1/2]$. Show that you can choose a sufficiently large constant $c$ (independent of $t$ and $\eps$) such that for all Hermitian $H$ with operator norm $\norm{H}\leq 1$, we have
$$
\norm{e^{iHt}-\sum_{k=0}^{c(t+\log(1/\eps))-1}\frac{(iHt)^k}{k!}}=\norm{\sum_{k=c(t+\log(1/\eps))}^\infty \frac{(iHt)^k}{k!}}\leq\eps.
$$

\item\label{ex:sparseblock}This exercise looks at the details of block-encoding an $s$-sparse matrix $A$ with $\norm{A}\leq 1$ from Section~\ref{sec:blockenc}. Assume for simplicity that the entries of $A$ are real and can each be written down exactly with $p$ bits of precision, that each row and column contains \emph{exactly} $s$ non-zero entries (indexed by $0,\ldots,s-1$), and that $s$ is a power of~2.
\begin{enumerate}
\item Show how to implement $W_1$ using an $O_{A,loc}$-query and a few other $A$-independent gates. For simplicity you may assume $s$ is a power of~2 here, and you can use arbitrary single-qubit gates, possibly controlled by another qubit.\\
(Note that the same method allows to implement $W_3$.)
\item (H) Show how to implement $W_2$ using an $O_A$-query, an  $O_A^{-1}$-query, and a few other $A$-independent gates (you may use auxiliary qubits as long as those start and end in $\ket{0}$). 
\item Show that the $(0^{n+1}i,0^{n+1}j)$-entry of $W_3^{-1} W_1$ is $1/s$ if $A_{ij}\neq 0$, and is~0 if $A_{ij}=0$.
\item Show that the $(0^{n+1}i,0^{n+1}j)$-entry of $W_3^{-1} W_2W_1$ is exactly $A_{ij}/s$.
\end{enumerate}

\item\label{ex:simPaulimatrix}
\begin{enumerate}
\item Give a quantum circuit on $n+1$ qubits that uses $O(n)$ gates, no auxiliary qubits, and computes the parity of the first $n$ bits, in the sense that it maps
$$
\ket{x}\ket{0}\mapsto\ket{x}\ket{\sum_{i=1}^nx_i\mbox{ mod 2}}\text{ for all }x\in\01^n.
$$
\item Let $P = Z \otimes Z\otimes \cdots\otimes Z$ be an $n$-qubit Hamiltonian, where $Z$ is the usual phase-flip Pauli matrix. What is the result of applying matrix $P$ to an $n$-qubit basis state $\ket{x}$?
\item Fix some positive real number $t$.
Let $U=e^{iPt}$ be the $n$-qubit unitary induced by applying the above Hamiltonian for time $t$ (via the Schr\"odinger equation, dropping the minus sign in the exponent like in Section~\ref{sec:hamiltonians}).  
What is the result of applying $U$ to an $n$-qubit basis state $\ket{x}$?
\item (H) Give a quantum circuit with $O(n)$ gates that implements $U$ exactly using one auxiliary qubit that starts and ends in $\ket{0}$.
You may use arbitrary single-qubit gates, which may be controlled by another qubit.
\item (H) Now suppose $P$ is a product of $n$ arbitrary Pauli matrices, not necessarily all-$Z$. Show how to implement $U=e^{iPt}$.
\end{enumerate}

\item Suppose you have a classical description of an  $n$-qubit Hamiltonian $H$ that is the sum of $m=n^2$ 2-local terms. Assume the eigenvalues of the Hermitian matrix $H$ lie in $[0,1)$, and can all be written exactly with $2\log n$ bits of precision. You would like to exactly determine the smallest eigenvalue $\lambda_{\min}$ of $H$, corresponding to unknown $n$-qubit eigenstate $\ket{\psi_{\min}}$. You're given (as a quantum state) an $n$-qubit state $\ket{\psi}$ that has a significant overlap with $\ket{\psi_{\min}}$: $|\inp{\psi}{\psi_{\min}}|^2\geq 0.7$.
Give a polynomial-size quantum circuit that, with probability $\geq 2/3$, outputs $\lambda_{\min}$ exactly.\\
\emph{\footnotesize NB: You don't need to write down the circuit to the last detail; a clear description of the different parts of the circuit (possibly with some reference to details in the lecture notes) suffices.}

\end{enumerate}

\chapter{The HHL Algorithm}\label{chap:hhl}

\section{The linear-system problem}\label{sec:LSP}

In this chapter we present the Harrow-Hassidim-Lloyd (HHL~\cite{hhl:lineq}) algorithm for solving large systems of linear equations. Such a system is given by an $N\times N$ matrix $A$ with real or complex entries, and an $N$-dimensional nonzero vector $b$. Assume for simplicity that $N=2^n$. The linear-system problem is
\begin{quote}
{\bf LSP}: find an $N$-dimensional vector $x$ such that $Ax=b$.
\end{quote}
Solving large systems of linear equations is extremely important in many computational problems in industry, in science, in optimization, in machine learning, etc.
In many applications it suffices to find a vector $\tilde{x}$ that is close to the actual solution~$x$.

We will assume $A$ is invertible (equivalently, has rank~$N$) in order to guarantee the existence of a unique solution vector~$x$, which is then just $A^{-1}b$.
This assumption is just for simplicity: if $A$ does not have full rank, then the methods below would still allow to invert it on its support, replacing $A^{-1}$ by the ``Moore-Penrose pseudoinverse.''

The HHL algorithm can solve ``well-behaved'' large linear systems very fast (under certain assumptions), but in a rather weak sense: instead of outputting the $N$-dimensional solution vector $x$ itself, its goal is to output the $n$-qubit state
$$
\ket{x}:=\frac{1}{\norm{x}}\sum_{i=0}^{N-1} x_i\ket{i},
$$
or some other $n$-qubit state close to~$\ket{x}$. This state $\ket{x}$ has the solution vector \emph{as its vector of amplitudes}, up to normalization and global phase.
This is called the \emph{quantum} linear-system problem:
\begin{quote}
{\bf QLSP}: find an $n$-qubit state $\ket{\tilde{x}}$ for which there is a $\theta\in[0,2\pi)$ such that\\ $\norm{\ket{x}-e^{i\theta}\ket{\tilde{x}}}\leq\eps$ and $Ax=b$.
\end{quote}
Note that the QLSP is an inherently quantum problem, since the goal is to produce an $n$-qubit state whose amplitude-vector (up to normalization and up to $\eps$-error) is a solution to the linear system. In general this is not as useful as just having the $N$-dimensional vector $x$ written out on a piece of paper, but in some cases where we only want some partial information about $x$, it may suffice to just (approximately)  construct~$\ket{x}$.

We will assume without loss of generality that $A$ is Hermitian (see Exercise~\ref{ex:wloghermitian}). 
Let us state the more restrictive assumptions that will make the linear system ``well-behaved'' and suitable for the HHL algorithm:
\begin{enumerate}
\item We have a unitary that can prepare the vector $b$ as an $n$-qubit quantum state $\ket{b}=\frac{1}{\norm{b}}\sum_i b_i\ket{i}$ using a circuit of $B$ 2-qubit gates.
We also assume for simplicity that $\norm{b}=1$.
\item The matrix $A$ is \emph{$s$-sparse} and we have sparse access to it, like in Section~\ref{sec:blockenc}. Such sparsity is not essential to the algorithm, and could be replaced by other properties that enable an efficient block-encoding of~$A$.
\item The matrix $A$ is \emph{well-conditioned}: the ratio between its largest and smallest singular value is at most some $\kappa$.\footnote{This ratio is called the \emph{condition number}. Note that the assumption that $A$ is invertible is equivalent to having a finite condition number. The stronger assumption that the condition number is small, intuitively says that $A$ is invertible in a stable or robust way, so that small errors (say due to noise or to finite-precision rounding) don't lead to massive errors in the solution vector~$x$.}
For simplicity, assume the smallest singular value is $\geq 1/\kappa$ while the largest is $\leq 1$.
In other words, all eigenvalues of~$A$ lie in the interval $[-1,-1/\kappa]\cup[1/\kappa,1]$.
The smaller the ``condition number'' $\kappa$ is, the better it will be for the algorithm.
Let's assume our algorithm knows~$\kappa$, or at least knows a reasonable upper bound on~$\kappa$. 
\end{enumerate}

\section{The basic HHL algorithm for linear systems}\label{ssec:HHLalgos}

Let us start with some intuition. The solution vector $x$ that we are looking for is $A^{-1}b$, so we would like to apply $A^{-1}$ to $b$. Because $A$ is assumed to be Hermitian, it has spectral decomposition $A=\sum_{j=0}^{N-1}\lambda_j a_j a_j^*$, where the vectors $a_j$ are an orthonormal basis of eigenvectors for the whole $N$-dimensional space, and $\lambda_j\in\mathbb{R}$ are the corresponding eigenvalues.
Then the map $A^{-1}$ is the same as the map $a_j\mapsto\frac{1}{\lambda_j}a_j$: we just want to multiply the eigenvector $a_j$ with the scalar $1/\lambda_j$. The vector $b$ can also be written as a linear combination of the eigenvectors $a_j$: $b=\sum_j\beta_j a_j$ (we don't need to know the coefficients $\beta_j$ for what follows). We want to apply $A^{-1}$ to $b$ to obtain $A^{-1}b=\sum_j\beta_j \frac{1}{\lambda_j}a_j$, normalized, as an $n$-qubit quantum state.

Unfortunately the maps $A$ and $A^{-1}$ are not unitary (unless $|\lambda_j|=1$ for all $j$), so we cannot just apply $A^{-1}$ as a quantum operation to state $\ket{b}$ to get state~$\ket{x}$. Fortunately  $U=e^{iA}=\sum_j e^{i\lambda_j} a_j a_j^*$ \emph{is} unitary, and has the same eigenvectors as $A$ and $A^{-1}$. We can implement $U$ and powers of~$U$ by Hamiltonian simulation, and then use phase estimation (Section~\ref{ssecphaseestimation}) to estimate the $\lambda_j$ associated with eigenvector $\ket{a_j}$ with some small approximation error (for the present sketch, assume for simplicity that the eigenvalues are in $[0,1)$ and that the approximation error is~0).
Conditioned on our estimate of~$\lambda_j$ we can then rotate an auxiliary $\ket{0}$-qubit to $\frac{1}{\kappa\lambda_j}\ket{0}+\sqrt{1-\frac{1}{(\kappa\lambda_j)^2}}\ket{1}$ (this is a valid state because $|\kappa\lambda_j|\geq 1$). Next we undo the phase estimation to set the register that contained the estimate back to $\ket{0}$.
Suppressing the auxiliary qubits containing the temporary results of the phase estimation (these qubits start and end in state $\ket{0}$), we have now unitarily mapped
$$
\ket{a_j}\ket{0}\mapsto \ket{a_j}\left(\frac{1}{\kappa\lambda_j}\ket{0}+\sqrt{1-\frac{1}{(\kappa\lambda_j)^2}}\ket{1}\right).
$$
If we prepare a copy of $\ket{b}\ket{0}=\sum_j\beta_j\ket{a_j}\ket{0}$ and apply the above unitary map to it, then we obtain
\begin{equation}\label{eq:hhlstatebeforeAA}
\sum_j\beta_j\ket{a_j}\left(\frac{1}{\kappa\lambda_j}\ket{0}+\sqrt{1-\frac{1}{(\kappa\lambda_j)^2}}\ket{1}\right)=
\frac{1}{\kappa}\underbrace{\sum_j\beta_j\frac{1}{\lambda_j}\ket{a_j}}_{\propto\ket{x}}\ket{0} + \ket{\phi}\ket{1},
\end{equation}
where we don't care about the (subnormalized) state $\ket{\phi}$.
Note that because $\sum_j|\beta_j/\lambda_j|^2\geq\sum_j|\beta_j|^2=1$, the norm of the part of the state ending in qubit $\ket{0}$ is at least $1/\kappa$.
Accordingly, we can now apply $O(\kappa)$ rounds of amplitude amplification to amplify this part of the state to have amplitude essentially~1. This prepares state $\ket{x}$ to good approximation, as intended.

This rough sketch (which Exercise~\ref{ex:hhldetails} asks you to make more precise) is the basic idea of HHL. It leads to an algorithm that produces a state $\ket{\tilde{x}}$ that is $\eps$-close to $\ket{x}$, using
roughly $\kappa^2 s/\eps$ queries to~$A$ and roughly $\kappa(s\kappa n/\eps + B)$ other 2-qubit gates.

\section{Improving the efficiency of the HHL algorithm}

The complexity of the above basic HHL algorithm can be improved further. 
Gily\'{e}n et al.~\cite{gilyenea:svtrans} used the singular-value transformation technique of Section~\ref{sec:blockenc}
to implement $A^{-1}$, improving on an LCU construction due to Childs et al.~\cite{CKS:optimalHHL}.
We would like to apply the function $f(x)=1/x$ to a block-encoding of $A$ in order to get a block-encoding of $A^{-1}$ (up to normalization) that we can then apply to $\ket{b}\ket{0}$. 

\begin{figure}[hbt]
\centering
\includegraphics[width=12cm]{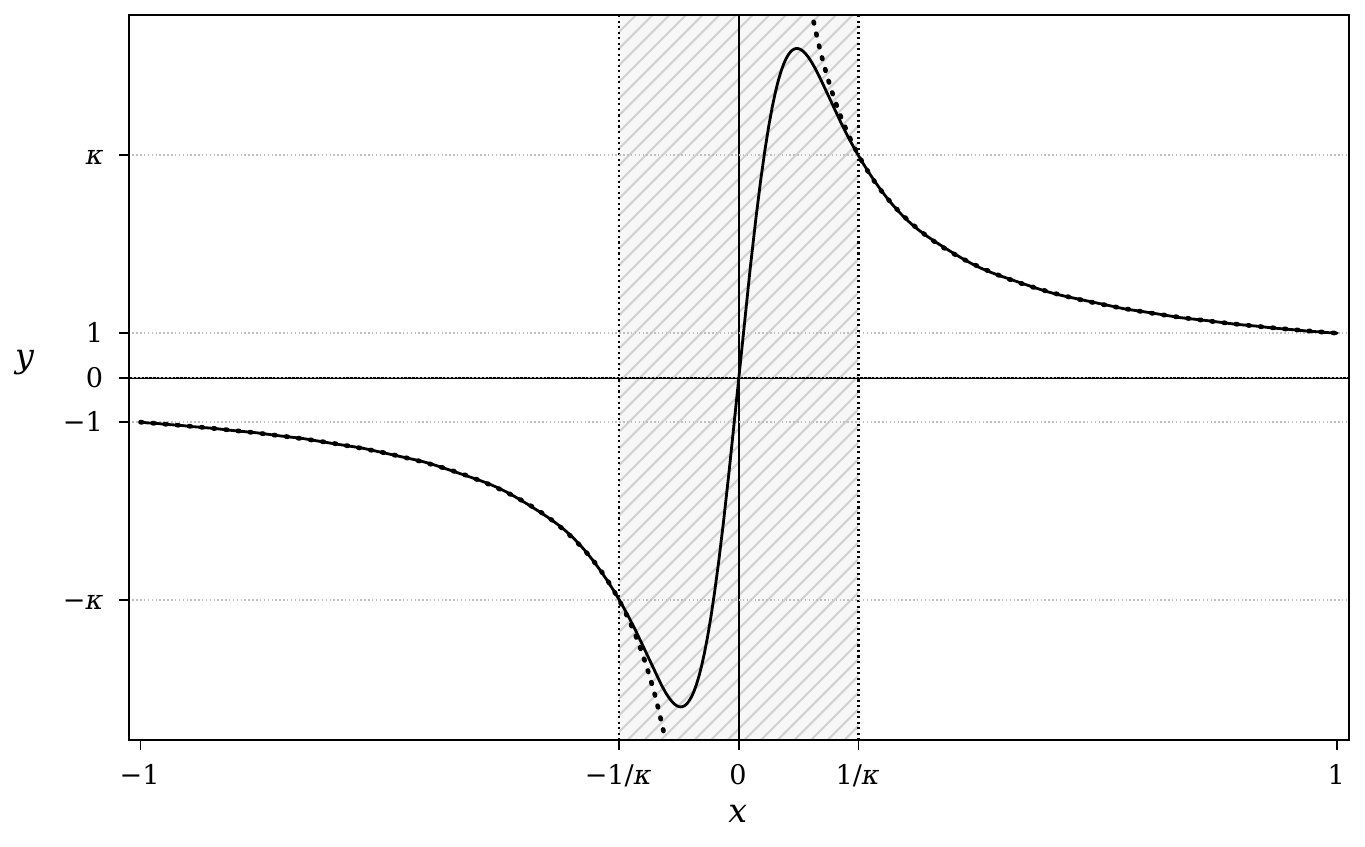}
\caption{Low-degree polynomial approximation of $1/x$ on the domain $E_\kappa=[-1,1]\backslash[-1/\kappa,1/\kappa]$}\label{fig:polyapproxinverse}
\end{figure}

The function $f(x)=1/x$ is not itself a polynomial, so we need to approximate it by a low-degree polynomial in order to be able to apply Theorem~\ref{thm:SVT} of Chapter~\ref{chap:hamsim}.
Childs et al.~\cite[Lemmas~17-19]{CKS:optimalHHL} constructed such a polynomial, as we sketch now. They start from the following function with $b=O(\kappa^2\log(\kappa/\eps))$: 
$$
\frac{1-(1-x^2)^b}{x}.
$$
This is in fact a polynomial of degree $D=2b-1$, because the numerator is a polynomial of degree~$2b$ where all terms have degree $\geq 1$, so we can divide out the~$x$ of the denominator.
Since $(1-x^2)^b$ is close to~0 for $x\in[-1,1]$ (unless $|x|\approx 0$), this polynomial is indeed close to $1/x$ for $x\in[-1,1]$ (we won't care about the case $|x|\approx 0$ because we'll apply this polynomial to a matrix whose eigenvalues aren't very close to~0). 
More precisely, this polynomial is $\eps/2$-close to $1/x$ whenever $x$ lies in the interval $E_\kappa=[-1,-1/\kappa]\cup[1/\kappa,1]$. Its range on this domain is $[-\kappa,-1]\cup[1,\kappa]$ (ignoring the small~$\eps$ for simplicity). 
See Figure~\ref{fig:polyapproxinverse} for illustration.

Like every degree-$D$ polynomial, it can be written exactly as a sum of the first $D+1$ Chebyshev polynomials of the first kind.\footnote{These univariate polynomials are defined recursively as follows: $T_0(x)=1$, $T_1(x)=x$, and $T_{d+1}=2xT_d(x)-T_{d-1}(x)$. Note that $T_d$ has degree $d$, and maps $[-1,1]$ to $[-1,1]$. The polynomials $T_0,\ldots,T_D$ are linearly independent (even orthonormal in a certain way) and hence span the set of all univariate polynomials of degree $\leq D$.}
Childs et al.\ show that the coefficients in this sum decrease quickly for larger degree, and that dropping the Chebyshev polynomials of degree higher than $d=O(\kappa\log(\kappa/\eps))$ incurs only small error~$\eps/2$. The resulting degree-$d$ polynomial $p$ $\eps$-approximates $1/x$ on the interval~$E_\kappa$, and its largest value (in absolute value) on this domain is~$\kappa$.
Now define the polynomial $P=p/(4\kappa)$. This has the same degree $d$ as the polynomial~$p$, but a range~$[-1/4,1/4]$ that fits the assumption of Theorem~\ref{thm:SVT} of Chapter~\ref{chap:hamsim} (there's a trick to ensure the values of $P$ are within that range even for $x\approx 0$, which we'll skip here).

As we saw in Section~\ref{sec:blockenc}, we can implement a block-encoding of the $s$-sparse matrix~$A/s$ using $O(1)$ sparse-access queries to~$A$ and $O(n)$ other gates. Using a factor $O(s)$ more work, there is way to turn this into a block-encoding of~$A$ itself; alternatively, we could directly invert the matrix $A/s$, whose singular values are $\geq 1/(\kappa s)$. 
We now apply Theorem~\ref{thm:SVT} with this block-encoding of $A$, and the polynomial~$P=p/(4(1+\eps)\kappa)$, of degree $d=O(\kappa\log(\kappa/\eps))$. Note that all eigenvalues of $A$ lie in the interval $E_\kappa$, where $p(x)\approx 1/x$, hence $p(A)\approx A^{-1}$ and $P(A)\approx\frac{1}{4\kappa}A^{-1}$. Theorem~\ref{thm:SVT} then gives us a block-encoding of $P(A)$, at the expense of running the block-encoding of $A$ $O(d)$ times. Using $O(\kappa)$ rounds of amplitude amplification on top of this, we can get rid of the $1/(4\kappa)$ factor and end up with essentially the state $A^{-1}\ket{b}$, normalized.\footnote{Note that we need to assume a unitary to prepare $\ket{b}$ here, having just one copy of~$\ket{b}$ is not enough. We cannot use \emph{oblivious} amplitude amplification here because that assumes we have a block-encoding of a matrix that is proportional to a unitary (or close to that), which $A^{-1}$ is not.} This gives a quantum algorithm that solves the QLSP using $O(d\kappa s)=O(\kappa^2 s \log(\kappa/\eps))$ queries to~$A$, and $O(\kappa s(\kappa n \log(\kappa/\eps)+ B))$ 2-qubit gates. Note that compared to basic HHL, the dependence on $1/\eps$ has been improved from linear to logarithmic. The dependence on $\kappa$ can be improved from quadratic to linear, using a technique called ``variable-time amplitude amplification''~\cite{ambainis:variabletime,CKS:optimalHHL,CGJ:blockenc,linlin&tong:qls} that we won't explain here.

The HHL algorithm can in some cases solve the QLSP exponentially faster than classical algorithms can solve the LSP.
In particular, if the sparsity~$s$, the condition number~$\kappa$, and the cost~$B$ of preparing~$\ket{b}$ are all $\leq\polylog(N)$, and the allowed error is $\eps\geq 2^{-\polylog(N)}$, then this improved version of the HHL algorithm uses $\polylog(N)$ queries and gates to solve (in a quantum way) an $N$-dimensional linear system.
It can also be used for other tasks, for instance approximately solving differential equations and other applications in scientific computing, see the lecture notes of Lin Lin~\cite{linlin:qscomp} and references therein.

\section*{Exercises}

\begin{enumerate}
  
\item\label{ex:wloghermitian} Suppose we are given an arbitrary invertible $N\times N$ matrix $A$ and an $N$-dimensional vector~$b$.
\begin{enumerate}
\item Give a \emph{Hermitian} $2N\times 2N$ matrix $A'$ (depending on $A$ but not on $b$) and $2N$-dimensional vector~$b'$ (depending on $b$ but not on $A$), such that a solution $x$ to the linear system $Ax=b$ can be read off from a solution to the system $A'x'=b'$.
  \item How does the condition number of your $A'$ relate to that of~$A$?
\end{enumerate}

\item\label{ex:hhldetails}
This exercise asks you to add more details to the sketch of the basic HHL algorithm given at the start of Section~\ref{ssec:HHLalgos}. 
For simplicity we will only count queries, not gates.
\begin{enumerate}
\item Use Hamiltonian simulation and phase estimation to implement the following unitary map:
$$
\ket{a_j}\ket{0}\mapsto\ket{a_j}\ket{\widetilde{\lambda_j}},
$$
where $\ket{\widetilde{\lambda_j}}$ is a superposition over estimates of $\lambda_j$, which (if measured) gives with probability $\geq 0.99$ an estimator~$\ell\in[-1,1]$ such that $|\lambda_j-\ell|\leq\eps/\kappa$.
Your implementation is allowed to use $O(\kappa s/\eps + \log(\kappa/\eps))$ queries to the sparse matrix~$A$.
You may invoke the best Hamiltonian simulator for sparse matrices from Section~\ref{sec:blockenc}.
\item Show that the basic HHL algorithm can be implemented using $O(\kappa^2 s/\eps + \kappa\log(\kappa/\eps))$ sparse-access queries to~$A$.
To make your life easier, you may assume that $\ket{\widetilde{\lambda_j}}$ is just one basis state, so one estimator which is close to $\lambda_j$ rather than a superposition over estimators (and hence the success probability~0.99 is actually~1). You may also assume for simplicity that the amplitude amplification at the end works perfectly.
\end{enumerate}

\item Suppose $A$ and $B$ are sparse, well-conditioned, Hermitian $N\times N$ matrices, and we can efficiently generate unit vector $b\in\R^N$ as a quantum state~$\ket{b}=\sum_{i=0}^{N-1} b_i\ket{i}$. Here we will see how we can efficiently find the solution to the linear system $ABx=b$ as a quantum state~$\ket{x}$.

More precisely, assume $N=2^n$. Assume we have a unitary circuit to produce the $n$-qubit state $\ket{b}$ with a number of elementary gates that's polynomial in $n$. Let $s_A$ and $s_B$ be the sparsities of the matrices $A$ and $B$, respectively, and $\kappa_A$ and $\kappa_B$ be their condition numbers (ratio of largest over smallest singular value). 
You may assume (like in the 3rd item at the end of Section~\ref{sec:LSP}) that the singular values of $A$ are in the interval $[1/\kappa_A,1]$, and the singular values of $B$ are in the interval $[1/\kappa_B,1]$.
Let $x\in\mathbb{C}^N$ be $B^{-1}A^{-1}b$, which is the unique solution to the linear system $ABx=b$. Show how you can produce an $n$-qubit state $\ket{\tilde{x}}$ that is $\eps$-close (in the usual Euclidean distance) to the $n$-qubit state $\ket{x}=\frac{1}{\norm{x}}\sum_{i \in \01^n} x_i \ket{i}$, using a number of queries (to matrix entries) and elementary gates that is polynomial in $s_A$, $s_B$, $\kappa_A$, $\kappa_B$, $1/\eps$, and $n$.\\ 
{\footnotesize Comment: Again, the point here is to avoid the polynomial dependence on the dimension $N$ that classical linear solvers would have, and replace that by a polynomial dependence on $n=\log N$.}
\end{enumerate}

\chapter{Quantum Query Lower Bounds}\label{chap:lowerb}

\section{Introduction}

Most of the algorithms we have seen so far worked in the \emph{query} model.
Here the goal usually is to compute some function $f:\01^N\rightarrow\01$ on a given input $x=x_0\ldots x_{N-1}\in \01^N$.
The distinguishing feature of the query model is the way $x$ is accessed:
$x$ is not given explicitly, but is stored in a random access memory, and we're being charged unit
cost for each \emph{query} that we make to this memory.
Informally, a query asks for and receives the $i$-th element $x_i$ of the input.
Formally, we model a query unitarily as the following 2-register quantum operation $O_x$,
where the first register is $N$-dimensional and the second is $2$-dimensional\footnote{If the input $x$ consists of
non-binary items $x_i$ (as is the case for instance with the input for Simon's algorithm) then those can be simulated by querying individual bits of each $x_i$.}:
$$
O_x:\ket{i,b}\mapsto\ket{i,b\oplus x_i}.
$$
In particular, $\ket{i,0}\mapsto\ket{i,x_i}$.
This only states what $O_x$ does on basis states, but by linearity this determines the full unitary.
Note that a quantum algorithm can apply $O_x$ to a superposition of basis states,
gaining some sort of access to several input bits $x_i$ at the same time.

A $T$-query quantum algorithm starts in a fixed state, say the all-0 state $\ket{0\ldots 0}$, and then interleaves
fixed unitary transformations $U_0,U_1,\ldots,U_T$ with queries.
The algorithm's fixed unitaries may act on a workspace-register, in addition to the two registers on which $O_x$ acts.
In this case we implicitly extend $O_x$ by tensoring it with the identity operation on this extra register, so it maps
$$
O_x:\ket{i,b,w}\mapsto\ket{i,b\oplus x_i,w}.
$$
Hence the final state of the algorithm can be written as the following matrix-vector product:
$$
U_TO_xU_{T-1}O_x\cdots O_xU_1O_xU_0\ket{0\ldots 0}.
$$
This state depends on the input $x$ only via the $T$ queries.
The output of the algorithm is obtained by a measurement of the final state.
For instance, if the output is Boolean, the algorithm could just measure the final state
in the computational basis and output the first bit of the result.

The query complexity of some function $f$ is now the minimal number of queries needed
for an algorithm that outputs the correct value $f(x)$ for every $x$ in the domain of $f$
(with error probability at most $1/3$, say).
Note that we just count queries to measure the complexity of the algorithm\footnote{Clearly, $N$ queries always suffice since we can just query each of the $N$ input bits separately, thus learning $x$ completely, and then look up and output whatever the correct value is for that input.}, 
while the intermediate fixed unitaries are treated as costless.

In many cases, the overall computation time of quantum query algorithms
(as measured by the total number of elementary gates, say) is not much bigger than the query complexity.
This justifies analyzing the latter as a proxy for the former.
This is the model in which essentially all the quantum algorithms we've seen work:
Deutsch-Jozsa, Simon, Grover, the various random walk algorithms. 
Even the period-finding algorithm that is the quantum core of Shor's algorithm works because
it needs only few queries to the periodic function.

\section{The polynomial method}\label{sec:polymethod}

\paragraph{From quantum query algorithms to polynomials.}
An \emph{$N$-variate multilinear polynomial} $p$ is a function $p:\mathbb{C}^N\rightarrow\mathbb{C}$
that can be written as
$$
p(x_0,\ldots,x_{N-1})=\sum_{S\subseteq\{0,\ldots,N-1\}} a_S\prod_{i\in S}x_i,
$$
for some complex numbers $a_S$.
The \emph{degree} of $p$ is $deg(p)=\max\{|S| : a_S\neq 0\}$.
It is easy to show that every function $f:\01^N\rightarrow\mathbb{C}$
has a unique representation as such a polynomial; $deg(f)$ is defined as the degree of that polynomial (see Exercise~\ref{ex:uniquepoly}).
For example, the 2-bit AND function is $p(x_0,x_1)=x_0x_1$, and the 2-bit Parity function is $p(x_0,x_1)=x_0+x_1-2x_0x_1$.
Both polynomials have degree~2.
Sometimes a lower degree suffices for a polynomial to \emph{approximate} the function. For example, $p(x_0,x_1)=\frac{1}{3}(x_0+x_1)$ approximates the 2-bit AND function up to error 1/3 for all inputs, using degree~1.

A very useful property of $T$-query algorithms is that the
amplitudes of their final state are degree-$T$ $N$-variate polynomials of $x$~\cite{fortnow&rogers:limitations,bbcmw:polynomialsj}.
More precisely:
consider a $T$-query algorithm with input $x\in\01^N$ acting on an $m$-qubit space.
Then its final state can be written
$$
\sum_{z\in\01^m}\alpha_z(x)\ket{z},
$$
where each $\alpha_z$ is a multilinear complex-valued polynomial in $x$ of degree at most~$T$.

\smallskip

\begin{proof}
The proof is by induction on $T$. The base case ($T=0$) trivially holds: the algorithm's state $
U_0\ket{0\ldots 0}$ is independent of $x$, so its amplitudes are constants
(i.e., polynomials of degree~0).

For the induction step, suppose we have already done $T$ queries.
Then by the induction hypothesis the state after $U_T$ can be written as
$$
\sum_{z\in\01^m}\alpha_z(x)\ket{z},
$$
where each $\alpha_z$ is a multilinear polynomial in $x$ of degree at most $T$.
Each basis state $\ket{z}=\ket{i,b,w}$ consists of 3 registers: the two registers $\ket{i,b}$ of the query,
and a workspace register containing basis state $\ket{w}$.
The algorithm now makes another query $O_x$ followed by a unitary $U_{T+1}$. 
The query swaps basis states $\ket{i,0,w}$ and $\ket{i,1,w}$ if $x_i=1$, and
doesn't do anything to these basis states if $x_i=0$.
This changes amplitudes as follows:
\begin{align*}
\alpha_{i,0,w}(x)\ket{i,0,w}+\alpha_{i,1,w}(x)\ket{i,1,w}\mapsto\\
((1-x_i)\alpha_{i,0,w}(x)+x_i\alpha_{i,1,w}(x))\ket{i,0,w}+(x_i\alpha_{i,0,w}(x)+(1-x_i)\alpha_{i,1,w}(x))\ket{i,1,w}.
\end{align*}
Now the new amplitudes are of the form $(1-x_i)\alpha_{i,0,w}(x)+x_i\alpha_{i,1,w}(x)$ or 
$x_i\alpha_{i,0,w}(x)+(1-x_i)\alpha_{i,1,w}(x)$. The new amplitudes are still polynomials in $x_0,\ldots,x_{N-1}$.
Their degree is at most 1 more than the degree of the old amplitudes, so at most $T+1$.
Finally, since $U_{T+1}$ is a linear map that is independent of $x$, it does not increase the degree of the amplitudes further
(each amplitude after $U_{T+1}$ is a linear combination of the amplitudes before $U_{T+1}$).
This concludes the induction step.

Note that this construction could introduce degrees higher than~1, e.g., terms of the form $x_i^2$.
However, our inputs $x_i$ are 0/1-valued, so we have $x_i^k=x_i$ for all integers $k\geq 1$.
Accordingly, we can reduce higher degrees to~1, making the polynomials multilinear without increasing degree.
\end{proof}

Suppose our algorithm acts on an $m$-qubit state.
If we measure the first qubit of the final state and output the resulting bit, then the probability of output~1 is given by 
$$
p(x)=\sum_{z\in \{1\}\times\01^{m-1}}|\alpha_z(x)|^2.
$$ 
This is a real-valued polynomial of $x$ of degree at most $2T$, because $|\alpha_z(x)|^2$ is the sum of the squares of the real and imaginary parts of the amplitude $\alpha_z(x)$, and each of those two parts is a polynomial of degree $\leq T$.
Note that if the algorithm computes $f$ with error $\leq 1/3$, then $p$ is an approximating polynomial for $f$:
if $f(x)=0$ then $p(x)\in[0,1/3]$ and if $f(x)=1$ then $p(x)\in[2/3,1]$.
This gives a method to lower bound the minimal number of queries needed 
to compute $f$: if one can show that every polynomial that approximates $f$ has degree at least $d$,
then every quantum algorithm computing $f$ with error $\leq 1/3$ must use at least $d/2$ queries.

\paragraph{Applications of the polynomial method.}
For our examples we will restrict attention to \emph{symmetric} functions.%
\footnote{One can also use the polynomial method for non-symmetric functions, for instance 
to prove a tight lower bound of $\Omega(N^{2/3})$ queries for the general problem of collision-finding; 
this matches the quantum walk algorithm of Section~\ref{sseccollision}.
However, that lower bound proof is substantially more complicated and we won't give it here (see~\cite{aaronson&shi:collision}).}
Those are the ones where the function value $f(x)$ only depends on the Hamming weight (number of 1s) in the input $x$.  
Examples are $N$-bit OR, AND, Parity, Majority, etc.

Suppose we have a polynomial $p(x_0,\ldots,x_{N-1})$ that approximates $f$ with error $\leq 1/3$. Then it is easy to see
that a polynomial that averages over all permutations~$\pi$ of the $N$ input bits $x_0,\ldots,x_{N-1}$:
$$
q(x)=\frac{1}{N!}\sum_{\pi\in S_N}p(\pi(x)),
$$
still approximates $f$.
As it turns out, we can define a single-variate polynomial $r(z)$ of the same degree as $q$,
such that $q(x)=r(|x|)$.\footnote{To see why this is the case, note that for every degree $i$, 
all degree-$i$ monomials in the symmetrized polynomial $q$ have the same coefficient $a_i$.  
Moreover, on input $x\in\01^N$ of Hamming weight $z$, exactly $\binom{z}{i}$ of the degree-$i$ monomials are~1, while the others are~0.
Hence $q(x)=\sum_{i=0}^d a_i\binom{|x|}{i}$.  Since $\binom{z}{d}=z(z-1)\cdots (z-d+1)/d!$ is a single-variate 
polynomial in $z$ of degree $d$, we can define $r(z)=\sum_{i=0}^d a_i\binom{z}{i}$.
For example, if $q(x_0,x_1)=x_0x_1$ then $r$ would be the unique univariate polynomial such that $r(0)=0$, $r(1)=0$ and $r(2)=1$, i.e., $r(z)=z(z-1)/2=\binom{z}{2}$.
The degree of $r$ will be equal to that of $q$, while the degree of $q$ is less than or equal to that of $p$; an example where the degree actually goes down is $p(x_0,x_1)=x_0-x_1$, which symmetrizes to the 0 polynomial (so the degree goes down from 1 to~0).}
This $r$ is defined on all real numbers, and we know something about its behavior on integer points $\{0,\ldots,N\}$. 
Thus it suffices to lower bound the degree of single-variate polynomials with the appropriate behavior.

For an important example, consider the $N$-bit OR function.  
Grover's algorithm can find an $i$ such that $x_i=1$ (if such an $i$ exists)
and hence can compute the OR function with error probability $\leq 1/3$ using $O(\sqrt{N})$ queries.
By the above reasoning, any $T$-query quantum algorithm that computes the OR with error $\leq 1/3$
induces a single-variate polynomial $r$ satisfying
\begin{quote}
$r(0)\in[0,1/3]$, and $r(t)\in[2/3,1]$ for all integers $t\in\{1,\ldots,N\}$.
\end{quote}
This polynomial $r(x)$ ``jumps'' between $x=0$ and $x=1$ (i.e., it has a derivative $r'(x)\geq 1/3$ for some $x\in[0,1]$), while it remains fairly constant on the domain $\{1,\ldots,N\}$.
By a classical theorem from approximation theory (proved independently around the same time by Ehlich and Zeller~\cite{ehlich&zeller:schwankung}, and by Rivlin and Cheney~\cite{rivlin&cheney:approx}), 
such polynomials must have degree $d\geq\Omega(\sqrt{N})$. 
Hence $T\geq\Omega(\sqrt{N})$ as well.
Accordingly, Grover's algorithm is optimal (up to a constant factor)
in terms of number of queries.

What about \emph{exact} algorithms for OR?  Could we tweak Grover's algorithm so that it always finds a
solution with probability~1 (if one exists), using $O(\sqrt{N})$ queries?
This turns out not to be the case: a $T$-query exact algorithm for OR induces a polynomial $r$ of degree $\leq 2T$ that satisfies 
\begin{quote}
$r(0)=0$, and $r(t)=1$ for all integers $t\in\{1,\ldots,N\}$.
\end{quote}
It is not hard to see that such a polynomial needs degree at least $N$: 
observe that $r(x)-1$ is a non-constant polynomial with at least $N$ roots.\footnote{A ``root'' is an $x$ 
such that $r(x)=0$. It is a well-known fact from algebra that every univariate non-constant polynomial 
of degree $d$ has at most $d$ roots (over any field). Note that this is not true for \emph{multi}variate polynomials; for example the polynomial $x_0\cdots x_{N-1}$ has $2^N-1$ roots in $\01^N$ but its degree is only~$N$.} 
Hence $T\geq N/2$ (this can be improved to $T\geq N$, see Exercise~\ref{ex:ANDinN}).  Accordingly, Grover cannot be made exact without losing the square-root speed-up!

Using the polynomial method, one can in fact show for \emph{every symmetric function} $f$ that is defined on all $2^N$ inputs, 
that quantum algorithms cannot provide a more-than-quadratic speed-up over classical algorithms.
More generally,  for \emph{every} function $f$ (symmetric or non-symmetric)  that is defined on all inputs\footnote{Note that this doesn't include functions where the input has to satisfy a certain promise, such as Deutsch-Jozsa and Simon's problem.}, 
quantum algorithms cannot provide more than a power-6 speed-up over classical algorithms (see Exercise~\ref{ex:6thpower}).
The polynomial method has recently been strengthened by Arunachalam et al.~\cite{ABP:cbforms} to an \emph{optimal} lower bound method, by imposing more constraints on the polynomial (which can increase the required degree, while still giving a lower bound on quantum query complexity).

\section{The quantum adversary method}\label{sec:adversarybound} 

The polynomial method has a strength which is also a weakness: it applies even to a stronger (and less physically meaningful) model of computation where we allow \emph{any linear transformation} on the state space, not just unitary ones.  As a result, it does not always provide the strongest possible lower bound for quantum query algorithms.

Ambainis~\cite{ambainis:lowerboundsj,ambainis:degreevsqueryj} provided an alternative method for quantum lower bounds, the \emph{quantum adversary}. This exploits unitarity in a crucial way and in certain cases yields a provably better bound than the polynomial method~\cite{ambainis:degreevsqueryj}.
We will present a very simple version of the adversary method here, a much stronger (in fact optimal!) version is given in Chapter~\ref{chap:adversary}.

Recall that a quantum query algorithm is a sequence 
$$
U_TO_xU_{T-1}O_x\cdots O_xU_1O_xU_0,
$$
applied to the fixed starting state $\ket{0\ldots 0}$, where the basic ``query transformation'' $O_x$ depends on the input~$x$, and $U_0,U_1,\ldots,U_T$ are arbitrary unitaries that don't depend on~$x$.  Consider the evolution of our quantum state under all possible choices of $x$. Let $\ket{\psi^t_x}$ denote the state after applying $U_t$ when the input is~$x$.  In particular, $\ket{\psi^0_x} = U_0\ket{0\ldots 0}$ for all $x$ (and hence $\inp{\psi^0_x}{\psi^0_y} = 1$ for each $x, y$), and $\ket{\psi_x^T}$ is the final state of the algorithm on input~$x$ before the final measurement.  Now if the algorithm computes the Boolean function $f$ with success probability $2/3$ on every input, then the final measurement must accept (i.e., output~1) every $x \in f^{-1}(0)$ with probability $\leq 1/3$, and must accept every $y \in f^{-1}(1)$ with probability $\geq 2/3$. This means the two states $\ket{\psi^T_x}$ and $\ket{\psi^T_y}$ cannot be too close together, or equivalently their inner product cannot be too close to~1. Specifically, we must have $|\inp{\psi^T_x}{\psi^T_y}| \leq \frac{17}{18}$.\footnote{Remember Exercise~\ref{ex:tvddistance} from Chapter~\ref{chap:ft} for states $\ket{\phi}$ and $\ket{\psi}$: 
if $\norm{\phi-\psi}=\eps$, then the total variation distance between the probability distributions 
you get from measuring $\ket{\phi}$ and $\ket{\psi}$, respectively, is at most $\eps$.
Hence, if we know there is a two-outcome measurement that accepts $\ket{\phi}$ with probability $\leq 1/3$ and accepts
$\ket{\psi}$ with probability $\geq 2/3$, then we must have total variation distance at least $1/3$ and hence $\eps\geq 1/3$. Assume for simplicity that the inner product $\inp{\phi}{\psi}$ is real.
Via the equation $\eps^2=\norm{\phi-\psi}^2=2-2\inp{\phi}{\psi}$, 
this translates into an upper bound $|\inp{\phi}{\psi}|\leq 1-\eps^2/2\leq 17/18$ (this upper bound can be improved to $2\sqrt{2}/3$ with more careful analysis).} 
This suggests that we find a set $R \subseteq f^{-1}(0) \times f^{-1}(1)$ of hard-to-distinguish $(x,y)$-pairs, 
and consider the following \emph{progress measure}
$$
S_t = \sum_{(x, y) \in R}|\inp{\psi^t_x}{\psi^t_y}|
$$
as a function of $t$.  By our observations, initially we have $S_0 = |R|$, and in the end we must have $S_T \leq \frac{17}{18}|R|$.  Also, crucially, the progress measure is \emph{unaffected} by each application of a unitary $U_t$, since each $U_t$ is independent of the input and unitary transformations preserve inner products.

If we can determine an upper bound $\Delta$ on the change $|S_{t + 1} - S_t|$ in the progress measure at each step, we can conclude that the number $T$ of queries is at least $\frac{|R|}{18\Delta}$.  Ambainis proved the following. Suppose that
\begin{itemize}
\item[(i)] each $x \in f^{-1}(0)$ appearing in $R$, appears at least $m_0$ times in pairs $(x,y)$ in $R$;
\item[(ii)] each $y \in f^{-1}(1)$ appearing in $R$, appears at least $m_1$ times in pairs $(x,y)$ in $R$;
\item[(iii)] for each $x \in f^{-1}(0)$ and $i \in\{0,\ldots,N-1\}$, there are at most $\ell_0$ inputs $y \in f^{-1}(1)$ such that $(x, y) \in R$ and $x_i\neq y_i$;
\item[(iv)] for each $y \in f^{-1}(1)$ and $i \in\{0,\ldots,N-1\}$, there are at most $\ell_1$ inputs $x \in f^{-1}(0)$ such that $(x, y) \in R$ and $x_i\neq y_i$.
\end{itemize}
Then for all $t \geq 0$, $|S_{t + 1} - S_t| = O \left(\sqrt{\frac{\ell_0}{m_0} \cdot \frac{\ell_1}{m_1}} \cdot|R|\right)=:\Delta$.
We will not prove this inequality here, though it is a reasonably straightforward generalization of the answer to Exercise~\ref{ex:hybridmethod}, and we will see a stronger result in the next chapter.
This upper bound $\Delta$ on the progress-per-query immediately implies a lower bound on the number of queries:
\begin{equation}\label{eqadversary}
T = \Omega \left(\sqrt{\frac{m_0}{\ell_0}\cdot \frac{m_1}{\ell_1}}\right).
\end{equation}
Intuitively, conditions (i)-(iv) imply that $|S_{t + 1} - S_t|$ is small relative to $|R|$ by bounding the ``distinguishing ability'' of any query.
The art in applying this technique lies in choosing the relation $R$ carefully to maximize this quantity, i.e.,
make $m_0$ and/or $m_1$ large, while keeping $\ell_0$ and $\ell_1$ small. 

Note that for the $N$-bit OR function this method easily gives the optimal $\Omega(\sqrt{N})$ lower bound, as follows.
Choose $R=\{(x,y): x=0^N, y\mbox{ has Hamming weight }1\}$. Then $m_0=N$ while $m_1=\ell_0=\ell_1=1$. 
Plugging this into Eq.~(\ref{eqadversary}) gives the right $\Omega(\sqrt{N})$ bound.

Let us give another application, a lower bound that is much harder to prove using the polynomial method.
Suppose $f:\01^N\rightarrow\01$ is a 2-level AND-OR tree, with $N=k^2$ input bits:
$f$ is the AND of $k$ ORs, each of which has its own set of $k$ inputs bits.
By carefully doing 2~levels of Grover search (search for a subtree which is $0^k$), 
one can construct a quantum algorithm that computes $f$ with
small error probability and $O(\sqrt{k}\cdot\sqrt{k})=O(\sqrt{N})$ queries.
It was long an open problem to give a matching lower bound on the approximate degree, and this was proved only in~2013~\cite{sherstov:andor,bun&thaler:approxdeg}. In contrast, the adversary method gives the optimal lower bound on the quantum query complexity quite easily: choose the relation~$R$ as follows
\begin{quote}
$R$ consists of those pairs $(x,y)$ where\\ 
$x$ has one subtree with input $0^k$ and the other $k-1$ subtrees have an arbitrary $k$-bit input of Hamming weight 1 (note $f(x)=0$)\\
$y$ is obtained from $x$ by changing one of the bits of the $0^k$-subtree to~1 (note $f(y)=1$).
\end{quote}
Then $m_0=m_1=k$ and $\ell_0=\ell_1=1$, and we get
a lower bound of $\Omega\left(\sqrt{\frac{m_0m_1}{\ell_0\ell_1}}\right)=\Omega(k)=\Omega(\sqrt{N})$.
Another lower bound one can prove fairly easily using a strengthened version of the adversary method is for inverting a permutation, see Exercise~\ref{ex:LBinvpermutation}.

\section*{Exercises}

\begin{enumerate}

\item\label{ex:uniquepoly}
Consider a function $f:\01^N\to\mathbb{R}$. Show that this function can be represented by an $N$-variate multilinear polynomial of degree $\leq N$, and that this representation is unique.
  
\item\label{ex:parityin2}
Consider a 2-bit input $x=x_0x_1$ with phase-oracle $O_{x,\pm}:\ket{i}\mapsto(-1)^{x_i}\ket{i}$.
Write out the final state of the following 1-query quantum algorithm: $HO_{x,\pm}H\ket{0}$.
Give a degree-2 polynomial $p(x_0,x_1)$ that equals the probability that this algorithm outputs~1 on input $x$.
What function does this algorithm compute?

\item Consider polynomial $p(x_0,x_1)=0.3+0.4x_0+0.5x_1$, which approximates the 2-bit OR function.
Write down the symmetrized polynomial $q(x_0,x_1)=\frac{1}{2}(p(x_0,x_1)+p(x_1,x_0))$.
Give a single-variate polynomial $r$ such that $q(x)=r(|x|)$ for all $x\in\01^2$.

\item\label{ex:paritylower} (H)
Let $f$ be the $N$-bit Parity function, which is 1 if its input $x\in\01^N$ has odd Hamming weight, and 0 if the input has even Hamming weight (assume $N$ is an even number).
\begin{enumerate}
\item Give a quantum algorithm that computes Parity with success probability~1 on every input $x$,
using $N/2$ queries.
\item Show that this is optimal, even for quantum algorithms that have error probability $\leq 1/3$ on every input
\end{enumerate}

\item\label{ex:ANDinN} 
Suppose we have a $T$-query quantum algorithm that computes the $N$-bit AND function with success probability~1 on all inputs $x\in\01^N$.
In Section~\ref{sec:polymethod} we showed that such an algorithm has $T\geq N/2$ (we showed it for OR, but the same argument works for AND).
Improve this lower bound to $T\geq N$.

\item\label{ex:allequal} 
Consider the following 3-bit function $f:\01^3\rightarrow\01$:\\
$f(x_0,x_1,x_2)=1$ if $x_0=x_1=x_2$, and $f(x_0,x_1,x_2)=0$ otherwise
\begin{enumerate}
\item How many queries does a classical deterministic algorithm need to compute $f$? Explain your answer.
\item Give a quantum algorithm that computes $f$ with success probability~1 using~2 queries.
\item (H) Show that 2 queries is optimal: there is no quantum algorithm that computes $f$ with success probability~1 using only 1~query.
\end{enumerate}

\item\label{ex:maj}
Let $f$ be the $N$-bit Majority function, which is 1 if its input $x\in\01^N$ has Hamming weight $>N/2$, and 0 if the input has Hamming weight $\leq N/2$ (assume $N$ is even).
\begin{enumerate}
\item Prove that $deg(f)\geq N/2$. What does this imply for the query complexity of exact quantum algorithms that compute majority?
\item (H) Use the adversary method to show that every bounded-error quantum algorithm for computing Majority, needs $\Omega(N)$ queries. Be explicit about what relation $R$ you're using, and about the values of the parameters $m_0,m_1,\ell_0,\ell_1$.
\end{enumerate}

\item 
Let $k$ be an odd natural number, $N=k^2$, and define the Boolean function $f:\01^N\to\01$ as the $k$-bit majority of $k$ separate $k$-bit OR functions. In other words, the $N$-bit input is $x=x^{(1)}\ldots x^{(k)}$ with $x^{(i)}\in\01^k$ for each $i\in[k]$, and $f(x)$ is the majority value of the $k$ bits $\OR(x^{(1)}),\ldots,\OR(x^{(k)})$. Use the adversary method to prove that computing this~$f$ with error probability $\leq 1/3$ requires $\Omega(N^{3/4})$ quantum queries. Be explicit about what relation $R$ you're using, and about the values of the parameters $m_0,m_1,\ell_0,\ell_1$.

\item\label{ex:LBinvpermutation}
This question is about the quantum complexity of inverting a permutation, which is an important problem in cryptography.
Let $N$ be a power of~2 and $S=\{0,\ldots,N-1\}$.
Let $x\in S^N$ correspond to a permutation on $S$, meaning that each $j\in S$ occurs exactly once as an entry of $x$ (so the map $i\mapsto x_i$ is a permutation). Suppose we can query $x$, i.e., we have a unitary $O_x$ that maps $\ket{i,j}\to\ket{i,x_i+j\mbox{ mod }N}$ for all $i,j\in S$, and we can also apply $O_x^{-1}$.
\begin{enumerate}
\item Show how we can find the unique index $i\in S$ for which $x_i=0$, with success probability $\geq 2/3$, using $O(\sqrt{N})$ queries to $O_x$ and $O_x^{-1}$.
\item (H) The adversary lower bound of Section~\ref{sec:adversarybound} still works with the following modifications:\\ 
(1) the $x$'s and $y$'s are not binary strings, but strings over a larger alphabet, such as~$S$, and (2) let $\ell_{x,i}$ be the number of $y\in Y$ such that $(x,y)\in R$ and $x_i\neq y_i$; $\ell_{y,i}$ be the number of $x\in X$ such that $(x,y)\in R$ and $x_i\neq y_i$; and $\ell_{\max}=\max\{\ell_{x,i}\cdot\ell_{y,i}: (x,y)\in R, i\in \{0,\ldots,N-1\}, x_i\neq y_i\}$.\\
In this case the quantum query lower bound is $\Omega(\sqrt{m_0m_1/\ell_{\max}})$. You may assume this without proof.

Use this strengthened adversary bound to show a lower bound of $\Omega(\sqrt{N})$ quantum queries for computing the task of part (a).
\end{enumerate}

\item\label{ex:sortLB} (H) 
Consider the \emph{sorting} problem: there are $N$ numbers $a_1,\ldots,a_N$
and we want to sort these. We can only access the numbers by making
\emph{comparisons}. A comparison is similar to a black-box query:
it takes 2 indices $i,j$ as input and outputs whether $a_i<a_j$ or not.
The output of a sorting algorithm should be the list of $N$ indices,
sorted in increasing order.
It is known that for classical computers,
$N\log(N) + O(N)$ comparisons are necessary
and sufficient for sorting. Prove that a quantum algorithm needs
at least $\Omega(N)$ comparisons for sorting, even if it is allowed an error probability~$\leq 1/3$.

\item\label{ex:6thpower} Consider a total Boolean function $f:\01^N\rightarrow\01$. Given an input $x\in\01^N$ and subset $B\subseteq\{0,\ldots,N-1\}$ of indices of variables, let $x^B$ denote the $N$-bit input obtained from~$x$ by flipping all bits~$x_i$ whose index~$i$ is in~$B$.
The \emph{block sensitivity} $bs(f,x)$ of $f$ at input $x$, is the maximal integer~$k$ such that there exist disjoint sets $B_1,\ldots,B_k$ satisfying $f(x)\neq f(x^{B_i})$ for all $i\in[k]$.
The \emph{block sensitivity} $bs(f)$ of $f$ is $\max_{x\in\01^N} bs(f,x)$.
\begin{enumerate}
\item (H) Show that the bounded-error quantum query complexity of $f$ is $\Omega(\sqrt{bs(f)})$.
\item It is known that for every total Boolean function~$f$, there is a classical deterministic algorithm that computes it using $O(bs(f)^3)$ many queries. What can you conclude from this and part~(a) about the relation between deterministic and quantum query complexity for total functions?
\end{enumerate}

\item\label{ex:hybridmethod} (H)
In this exercise we will derive the quantum lower bound for the search problem in a self-contained way, without using the polynomial or adversary method (this exercise uses what is called the ``hybrid method'').

    Let $N=2^n$. Consider an input $x\in\01^N$ that we can query. Assume $x$ has Hamming weight 0 or 1, and suppose we would like to find the unique solution to the search problem (if a solution exists). Let $\cal A$ be any $T$-query quantum algorithm for this. Suppose for simplicity that the algorithm acts on only $n$ qubits (so there are no auxiliary qubits), and ${\cal A}=U_TO_{x,\pm}U_{T-1}O_{x,\pm}\cdots U_1O_{x,\pm}U_0$, so $\cal A$ interleaves phase-queries to~$x$ and unitaries that are independent of~$x$. The initial state is $\ket{0^n}$. Let $\ket{\psi^t_x}$ denote the $n$-qubit state right after applying $U_t$, when we run $\cal A$ on input $x$, so the final state is $\ket{\psi_x^T}$. Let $e_i\in\01^N$ be the input that has a~1 only at position~$i$. Assume the algorithm $\cal A$ is successful in finding the right solution~$i$ after $T$ queries in the following sense: $\norm{\ket{\psi_{e_i}^T}-\ket{i}}\leq 1/4$ and $\norm{\ket{\psi_{0^N}^T}-\ket{i}}\geq 3/4$ for all $i\in\{0,\ldots,N-1\}$ (note that the basic Grover algorithm is an example of such an $\cal A$).
    \begin{enumerate}
    \item Consider the run of algorithm $\cal A$ on input $x=0^N$, and for $t\in\{0,\ldots,T-1\}$ let the amplitudes $\alpha_{t,i}$ be such that $\ket{\psi_{0^N}^t}=\sum_{i=0}^{N-1}\alpha_{t,i}\ket{i}$.\\[0.5mm]
      Prove that $\norm{\ket{\psi^1_{0^N}} - \ket{\psi^1_{e_i}}}\leq 2|\alpha_{0,i}|$, for all $i\in\{0,\ldots,N-1\}$.
    \item  Prove that $\norm{\ket{\psi^T_{0^N}} - \ket{\psi^T_{e_i}}}\leq 2\sum_{t=0}^{T-1}|\alpha_{t,i}|$, for all $i\in\{0,\ldots,N-1\}$.
    \item Prove that $1/2\leq \norm{\ket{\psi^T_{0^N}} - \ket{\psi^T_{e_i}}}$, for all $i\in\{0,\ldots,N-1\}$.
    \item Prove that $T\geq \sqrt{N}/4$.
      \end{enumerate}

\item\label{ex:kwiseindep} 
Consider a standard quantum query algorithm: it makes $T$ queries to a string~$x\in\01^N$, with arbitrary unitaries $U_0,U_1,\ldots,U_T$ (that are independent of $x$) around the queries, and then measures a two-outcome POVM $\{M,I-M\}$ on the final $m$-qubit state $\ket{\psi_x}$.
\begin{enumerate}
  \item Show that the probability $P(x)$ of getting the first measurement outcome (on input $x$) is $\bra{\psi_x}M\ket{\psi_x}$, and that this can be written as an $N$-variate multilinear polynomial in the bits of $x$ of degree $\leq 2T$.
  \item (H) A \emph{$k$-wise independent distribution} $D$ is a probability distribution over $\01^N$, such that for each set $S\subseteq[N]$ of at most $k$ coordinates, the distribution on the $k$-bit substring $x_S=(x_i)_{i\in S}$ is uniformly random (i.e, for each $z\in\01^{|S|}$, the probability under distribution $D$ of the event that $x_S=z$, is $1/2^{|S|}$).
  
Show that a $T$-query quantum algorithm cannot distinguish the uniform distribution $U$ on its input $x$ from a $2T$-wise independent distribution $D$ on $x$, in the sense that no matter what binary measurement the algorithm does at the end, the probability of output~1 is the same under $U$ and under $D$.
\end{enumerate}

\end{enumerate}

\chapter{Quantum Algorithms from the Generalized Adversary Bound}\label{chap:adversary}

\section{The generalized adversary bound}\label{sec:ADVlower}

In the previous chapter we saw two different lower bound methods for the quantum query complexity of a given function $f$: the polynomial method and the adversary method.
Neither is optimal for every possible~$f$. For example, the polynomial method doesn't give optimal lower bounds for iterations of some small functions \cite{ambainis:degreevsqueryj} (see Exercise~\ref{ex:polynottight}), while the adversary bound of 
Section~\ref{sec:adversarybound} cannot prove optimal lower bounds for instance for distinguishing 2-to-1 from 1-to-1 inputs.

In this chapter we will look at a stronger version of the adversary bound, which turns out to give \emph{optimal} quantum query complexity lower bounds for all Boolean functions.
The beauty of an optimal lower bound method is that it can also produce algorithms: if the best-possible lower bound on the query complexity of~$f$ is $T$, then there must actually exist a $T$-query algorithm for $f$!

Suppose $f:{\cal D}\to\01$, with ${\cal D}\subseteq\01^N$, is a Boolean function whose quantum query complexity we'd like to determine.\footnote{If the domain is ${\cal D}=\01^N$ then $f$ is called a \emph{total} function, otherwise it's a \emph{partial} function.}
Consider a $T$-query quantum algorithm ${\cal A}=U_TO_xU_{T-1}\cdots U_1O_xU_0$, with initial state $\ket{0^m}$, that computes $f$ with error probability $\leq \eps<1/2$ for each $x\in{\cal D}$. Let $\ket{\psi_x^t}$ denote the algorithm's state after $U_t$ has been applied, given input~$x$. Note that $\ket{\psi_x^0}=U_0\ket{0^m}$ is independent of~$x$. 
The crucial property, already used in the earlier version of the adversary bound, is that $\inp{\psi_x^t}{\psi_y^t}$ is~1 at the start ($t=0$), but has to be small at the end ($t=T$) for every $(x,y)$-pair with different function values $f(x)\neq f(y)$.

The generalized adversary matrix puts weights $\alpha_x\in\mathbb{C}$ on the inputs $x\in{\cal D}$, with the constraint $\sum_{x\in{\cal D}}|\alpha_x|^2=1$ for normalization. It also puts real (but possibly negative!)\ weights $\Gamma_{xy}$ on $(x,y)$-pairs with different function values. We impose the constraint that $\Gamma_{xy}=\Gamma_{yx}$, and $\Gamma_{xy}=0$ whenever $f(x)=f(y)$. A $|{\cal D}|\times|{\cal D}|$ matrix $\Gamma$ with these properties is called an \emph{adversary matrix}.

Let us use these weights to define a progress measure:\footnote{We could have just absorbed the $\alpha_x^*\alpha_y$ into the $\Gamma_{xy}$ and dispensed with the $\alpha_x$'s altogether, but it will be cleaner to have separate weights on the $x$'s and separate weights $\Gamma_{xy}$ on the pairs.}
\[
S_t=\sum_{x,y\in{\cal D}}\Gamma_{xy}\alpha^*_x\alpha_y\inp{\psi_x^t}{\psi_y^t}.
\]
In the same spirit as Section~\ref{sec:adversarybound}, we will show that $|S_0|$ is large, that $|S_T|$ is much smaller, and that $S_{t+1}$ can't be much smaller than~$S_t$ (i.e., $S_t$ can't decrease very fast if we spend one more query). This will give the lower bound on~$T$.

At the start of the algorithm ($t=0$), before any queries have been made, we have $\inp{\psi_x^0}{\psi_y^0}=1$ for all $x,y$ and hence $S_0=\sum_{x,y}\Gamma_{xy}\alpha^*_x\alpha_y=\alpha^*\Gamma\alpha$. Since $\alpha$ is restricted to a unit vector, the biggest we can make $|S_0|$ is 
\[
|S_0|=\norm{\Gamma},
\]
the operator norm (largest singular value) of $\Gamma$, by choosing $\alpha$ to be an eigenvector of $\Gamma$ corresponding to the largest eigenvalue in absolute value.

At the end of the algorithm ($t=T$), the final states $\ket{\psi_x^T}$ and $\ket{\psi_y^T}$ must be distinguishable with success probability $\geq 1-\eps$
whenever $f(x)\neq f(y)$. The following claim, proved in Exercise~\ref{ex:advST}, shows that this forces $|S_T|$ to be significantly smaller than $|S_0|$.

\begin{claim}\label{cl:advST}   
$|S_T|\leq 2\sqrt{\eps(1-\eps)}\,\norm{\Gamma}$.
\end{claim}

\noindent
For example, if our algorithm has error probability $\eps=1/3$, then 
$|S_T|< 0.95\,\norm{\Gamma}$.
Accordingly, the progress measure has to change significantly in the course of the $T$-query algorithm. How much can one more query change $S_t$? This is upper bounded by the following claim, proved in Exercise~\ref{ex:advDelta}.

\begin{claim}\label{cl:advDelta}   
Let $\Gamma_i$ denote the $|{\cal D}|\times|{\cal D}|$ matrix obtained from $\Gamma$ by setting $\Gamma_{xy}$ to 0 if $x_i=y_i$. For all $t\in\{0,\ldots,T-1\}$ we have $|S_t-S_{t+1}|\leq 2\max_{i\in[N]} \norm{\Gamma_i}$.
\end{claim}

\noindent
These two claims, together with our value for the initial $|S_0|$, imply
\[
\left(1-2\sqrt{\eps(1-\eps)}\right)\norm{\Gamma}\leq |S_0|-|S_T|\leq |S_0-S_T|=|\sum_{t=0}^{T-1} S_t - S_{t+1}|\leq \sum_{t=0}^{T-1} |S_t - S_{t+1}|\leq 2T\max_{i\in[N]} \norm{\Gamma_i}.
\]
We get the following lower bound on quantum query complexity, due to H{\o}yer, Lee, and \v{S}palek~\cite{hls:madv}:

\begin{theorem}[Generalized adversary bound]\label{th:genadv}
Let $f:{\cal D}\to\01$, with ${\cal D}\subseteq\01^N$, and $\Gamma$ be an adversary matrix for $f$. Every quantum algorithm that computes $f$ with worst-case error probability $\leq \eps$, needs at least $\displaystyle
\left(\frac{1}{2}-\sqrt{\eps(1-\eps)}\right)\frac{\norm{\Gamma}}{\max_{i\in [N]}\norm{\Gamma_i}}$
queries.
\end{theorem}

As an example, let us (again) prove the $\Omega(\sqrt{N})$ lower bound for search. 
Consider the domain ${\cal D}=\{0^N,e_1,\ldots,e_N\}$ of inputs of weight 0 or~1 ($e_j$ is the $N$-bit string that has a 1 only at position~$j$). Define $(N+1)\times(N+1)$ adversary matrix 
\[
\Gamma=\left(\begin{array}{cccc}
0 & 1 &\cdots & 1\\
1 & 0 &\cdots & 0\\
\vdots & \vdots & \ddots & \vdots\\
1 & 0 & \cdots & 0 
\end{array}\right).
\]
Then $\norm{\Gamma}\geq\norm{\Gamma (1,0,\ldots,0)^T}=\sqrt{N}$. Each $\Gamma_i$ is the $2\times 2$ $X$-matrix padded with extra rows and columns of 0s, so $\norm{\Gamma_i}=1$.
Hence we obtain the familiar lower bound of $\Omega(\sqrt{N})$ queries for search.

More generally, we can recover the lower bound of Section~\ref{sec:adversarybound} by constructing an appropriate adversary matrix~$\Gamma$ based on the relation~$R$.
The rows and columns of $\Gamma$ are indexed by ${\cal D}=f^{-1}(0)\cup f^{-1}(1)$.
Define $d_x=|\{y:(x,y)\in R\}|$ and similarly define $d_y$. Define $\Gamma_{xy}=1/\sqrt{d_xd_y}$ if $(x,y)\in R$ or $(y,x)\in R$, and $\Gamma_{xy}=0$ otherwise.
Let $v$ be the vector with entries $\sqrt{d_x}$, and note that $\norm{v}^2=2|R|$.
We have $\norm{\Gamma}\geq \frac{1}{\norm{v}^2}v^T\Gamma v=\frac{1}{2|R|}\sum_{(x,y)\in R}2=1$.
We can also show
$\norm{\Gamma_i}\leq\sqrt{
\frac{\ell_0\ell_1}{m_0m_1}}$ for all $i$.\footnote{Define two $|{\cal D}|\times|{\cal D}|$ matrices $A,B$ by $A_{xy}=1/\sqrt{d_x}$ if $x_i\neq y_i$ and ($(x,y)\in R$ or $(y,x)\in R$), and $A_{xy}=0$ otherwise;
and $B_{xy}=1/\sqrt{d_y}$ if $x_i\neq y_i$ and ($(x,y)\in R$ or $(y,x)\in R$), and $B_{xy}=0$ otherwise. Note that $\Gamma_i=A\circ B$, where `$\circ$' denotes the entrywise product. For matrices with nonnegative entries, it is known~\cite[Appendix~A]{spalek&szegedy:adversaryj} that $\norm{A\circ B}$ is at most  the largest norm among the rows of $A$ (which is $\leq\sqrt{\ell_0/m_0}$, because each row of $A$ has at most $\ell_0$ nonzero entries, each of which is at most $1/\sqrt{m_0}$) times the largest norm among the columns of $B$ (which is $\leq\sqrt{\ell_1/m_1}$).
Hence $\norm{\Gamma_i}\leq\sqrt{
\frac{\ell_0\ell_1}{m_0m_1}}$.}
Now Theorem~\ref{th:genadv} gives the $\Omega\left(\sqrt{\frac{m_0m_1}{\ell_0\ell_1}}\right)$  bound of Eq.~(\ref{eqadversary}).

\section{The dual of the generalized adversary bound}

Consider the best-possible lower bound that we can obtain by optimizing over adversary matrix $\Gamma$:
\[
\begin{array}{lll}
\max        & \norm{\Gamma}&\\
\mbox{s.t.} & \norm{\Gamma_i}\leq 1 & \forall i\in[N]\\
            & \Gamma\mbox{ is symmetric}& \\
            & \Gamma_{xy}=0 &\forall x,y\in{\cal D}\mbox{ with } f(x)=f(y)
            \end{array}
\]
We will call the optimal value the \emph{generalized adversary bound} for $f$ (a.k.a.\ the \emph{negative-weights adversary bound}), and denote it by $\ADV(f)$.
Because the statement $\norm{\Gamma_i}\leq 1$ is equivalent to the matrix inequality $-I\preceq \Gamma_i\preceq I$, the above maximization problem can be written in the form of a so-called \emph{semidefinite program}: an optimization problem over real-valued variables, typically arranged in one or more matrices, with an objective function that's linear in the variables, and psd constraints that are linear in the variables as well. Every maximization-SDP has an associated minimization-SDP which (under mild assumptions that hold in our case) has the \emph{same optimal value}. The first SDP is called the \emph{primal} SDP, the second is called the \emph{dual} SDP.\footnote{Strictly speaking the maximum in the primal should be a supremum and the minimum in the dual should be an infimum, because the optimal value need not always be attained; it could be that there's only an infinite sequence of feasible solutions whose objective values converge to the optimum without ever reaching it.} The equality of the optimal values of these two SDPs is called \emph{strong duality} (see Exercise~\ref{ex:weakduality}
for a proof of the easy half of this equality).
This generalizes the better-known strong duality of linear programs, which correspond to SDPs with diagonal matrices.

With some effort that we will skip here, one can show that the dual of the above maximization-SDP can be written as the following minimization-SDP:
\[
\begin{array}{lll}

\min & \displaystyle\max_{x\in{\cal D}}\max\left(\sum_{j\in[N]} \norm{u_{xj}}^2, \sum_{j\in[N]} \norm{v_{xj}}^2\right)&\\[3mm]
\mbox{s.t.} & \displaystyle\sum_{j:x_j\neq y_j} \inp{u_{xj}}{v_{yj}}=[f(x)\neq f(y)] & \forall x,y\in{\cal D}\\
\end{array}
\]
The truth-value $[f(x)\neq f(y)]$ is~1 if $f(x)\neq f(y)$, and 0 if $f(x)=f(y)$.
This SDP associates with every $x\in{\cal D}$ and every $j\in[N]$ two vectors
$u_{xj}$ and $v_{xj}$ (of some dimension~$d$ that will implicitly be optimized over). We can write this more explicitly as an optimization problem over psd matrices by defining, for each $j\in[N]$, a $2|{\cal D}|\times 2|{\cal D}|$ psd matrix $Z_j$ whose entries are given by the pairwise inner products of the $2|{\cal D}|$ vectors $u_{xj},v_{xj}$ (these vectors are the ``Gram vectors'' of $Z_j$). Then the optimization is over psd matrices $Z_1,\ldots,Z_N$, the objective function is the largest diagonal entry of the matrix $\sum_{j\in[N]}Z_j$, and the constraints are linear functions of the entries of the $Z_j$'s.

By strong duality, the optimal value of this minimization-SDP is $\ADV(f)$ as well. A feasible solution $\Gamma$ to the primal gives a lower bound on $\ADV(f)$, while a feasible solution $\{u_{xj},v_{xj}\}_{x\in{\cal D},j\in[N]}$ for the dual gives an upper bound on $\ADV(f)$. The central result of this chapter is that $\ADV(f)$ is not only a lower bound on the quantum query complexity of $f$ (which follows from Theorem~\ref{th:genadv}) but also an \emph{upper} bound, as we will see in the next section.
This means that \emph{a feasible solution to the dual SDP actually gives us an algorithm for $f$}!

\section{$\ADV$ is an upper bound on quantum query complexity}

In this section we will construct a bounded-error quantum algorithm for computing~$f$, derived from a feasible solution 
$\{u_{xj},v_{xj}\}_{x\in{\cal D},j\in[N]}$
of the dual SDP for $\ADV(f)$.\footnote{Note that we only focus on query complexity here, not total number of gates. Our presentation follows the approach of~\cite{lmrss:stateconv}, modified and simplified for our special case of computing Boolean functions rather than their more general case of state-transformation. For other generalizations of the adversary bound to different scenarios, see~\cite{belovs:variations}. Reichardt's first proof of the optimality of the generalized adversary bound~\cite{reichardt:tight,reichardt:equiv} went through so-called ``span programs,'' but we won't need those here.}
Let's say the objective value of this feasible solution is $A$; the query complexity of our algorithm will turn out to be $O(A)$.
Below $x,y$ always range over~$\cal D$, and $j$ always ranges over $[N]$.

Our algorithm will act on 3 registers. The first register is spanned by $\ket{j}, j\in[N]$, and $\ket{0}$, the second is 1 qubit, and the third contains the states $\ket{v_{xj}}\in\mbox{span}\{\ket{1},\ldots,\ket{d}\}$ and the special state $\ket{0}$ (so the third register has $\ceil{\log (d+1)}$ qubits).
For each $x$, define the following two 3-register states:
\[
\ket{t_x^+}=\frac{1}{\sqrt{2}}(\ket{0}\ket{0}\ket{0}+\ket{1}\ket{f(x)}\ket{0})
\mbox{~~~and~~~}
\ket{t_x^-}=\frac{1}{\sqrt{2}}(\ket{0}\ket{0}\ket{0}-\ket{1}\ket{f(x)}\ket{0}).
\]
The algorithm starts with the all-0 state
\[
\ket{0}\ket{0}\ket{0}=\frac{1}{\sqrt{2}}(\ket{t_x^+}+\ket{t_x^-}).
\]
The goal of the algorithm is to (approximately) multiply $\ket{t_x^+}$ with $+1$ and $\ket{t_x^-}$ with $-1$, which rather magically gives a final state that tells us $f(x)$: 
\[
\frac{1}{\sqrt{2}}(\ket{t_x^+}-\ket{t_x^-})=\ket{1}\ket{f(x)}\ket{0}.
\]
The key will be to use phase estimation (Section~\ref{ssecphaseestimation}) with a well-chosen unitary~$U_x$ that depends on~$x$, to distinguish $\ket{t_x^+}$ and $\ket{t_x^-}$.
Define $U_x=(2\Pi_x-I)(2\Lambda-I)$ as the product of two reflections:
\begin{itemize}
\item Consider (unnormalized) states
$\displaystyle
\ket{\psi_y}=\frac{0.01}{\sqrt{A}}\ket{t_y^-}-\sum_j\ket{j}\ket{\overline{y_j}}\ket{v_{yj}}$, where $\overline{y_j}=1-y_j$.\\
Note that $\norm{\ket{\psi_y}}\leq \frac{0.01}{\sqrt{A}} + \sqrt{A}$, using triangle inequality and the fact that $\sum_j\norm{v_{yj}}^2\leq A$.
Let $\Lambda$ be the projector on the subspace that is orthogonal to the span of these $\ket{\psi_y}$'s, and $2\Lambda-I$ be the reflection through this subspace. In other words, the unitary $2\Lambda-I$ puts a $-$ in front of all $\ket{\psi_y}$'s and leaves states alone if they are orthogonal to all $\ket{\psi_y}$'s. This reflection costs no queries to implement, since it doesn't depend on the actual input~$x$.
\item Let $\Pi_x$ be the projector on the subspace spanned by states that have $\ket{j}\ket{x_j}$ in their first two registers (with arbitrary states in the third register) and by states having $\ket{0}$ in their third register. Then the reflection $2\Pi_x-I$ through this subspace puts a $-$ in front of states $\ket{j}\ket{\overline{x_j}}\ket{v}$ if $\inp{v}{0}=0$, and leaves the states alone that are in the subspace of~$\Pi_x$.
This reflection can be implemented with 1 query to~$x$  (Exercise~\ref{ex:2PiminI}).
\end{itemize}
We now relate $\ket{t_x^+}$ and $\ket{t_x^-}$ to the eigenstates of this unitary $U_x$. In the next two claims, the informal ``close'' should be read as ``within small constant Euclidean distance.''

\begin{claim}\label{cl:ADV1} 
$\ket{t_x^+}$ is close to an eigenstate $\ket{\phi}$ of $U_x$ that has eigenvalue~1 (i.e., phase~0).
\end{claim}

\begin{proof}
Define
\[
\ket{\phi}=\ket{t_x^+}+
\frac{0.01}{\sqrt{A}}\sum_j\ket{j}\ket{x_j}\ket{u_{xj}}.
\]
The second term on the right-hand side has norm $\leq 0.01$ because $\sum_j\norm{u_{xj}}^2\leq A$, so $\ket{t_x^+}$ is indeed close to $\ket{\phi}$ ($\ket{\phi}$'s norm is close to but not equal to~1, but this doesn't matter).

Note that $\inp{\phi}{\psi_y}=0$ for all $y$, because $\inp{t_x^+}{t_y^-}=\frac{1}{2}[f(x)\neq f(y)]$ and 
\[
\left(
\sum_j\ket{j}\ket{x_j}\ket{u_{xj}}
\right)^*
\sum_j\ket{j}\ket{\overline{y_j}}\ket{v_{yj}}
=\sum_{j}\inp{x_j}{\overline{y_j}}\cdot \inp{u_{xj}}{v_{yj}}
=\sum_{j:x_j\neq y_j}\inp{u_{xj}}{v_{yj}}=[f(x)\neq f(y)],
\]
and because  $\ket{t_x^+},\ket{t_x^-}$ have $\ket{0}$ in the third register and so are orthogonal to all $\ket{j}\ket{b}\ket{u_{xj}},  \ket{j}\ket{b}\ket{v_{xj}}$, $b\in\01$.
This shows that $\ket{\phi}$ lies in the subspace of $\Lambda$, so it is an eigenvalue-1 eigenvector of~$2\Lambda-I$.
Also, $\ket{\phi}$ is a linear combination of $\ket{t_x^+}$ (which has $\ket{0}$ in its third register) and states that have $\ket{j}\ket{x_j}$ in their first two registers, so $\ket{\phi}$ lies in the subspace of~$\Pi_x$ and hence is also an eigenvalue-1 eigenvector of $2\Pi_x-I$. Hence $\ket{\phi}$ is an eigenvalue-1 eigenvector of $U_x=(2\Pi_x-I)(2\Lambda-I)$.
\end{proof}

\begin{claim}\label{cl:ADV2}
$\ket{t_x^-}$ is close to a superposition of eigenstates of $U_x$ with eigenvalues of the form $e^{i\theta}$ with $\theta\in(-\pi,\pi]$ and $|\theta|> \Theta=1/(1000A)$. 
\end{claim}

\begin{proof}
Let $\{\ket{\beta}\}$ be a complete orthonormal set of eigenvectors of $U_x$, with respective eigenvalues $e^{i\theta_\beta}$, $\theta_\beta\in(-\pi,\pi]$. 
Let $P_\Theta=\sum_{\beta:|\theta_\beta|\leq\Theta}\ketbra{\beta}{\beta}$ be the projector on the eigenvectors with small eigenphase.
Define vectors $w=100\sqrt{A}\ket{\psi_x}$ and $v=P_\Theta\Pi_x w=P_\Theta\ket{t_x^-}$. Our goal is to show that $v$ has small norm.

To that end, define $v'=(2\Lambda-I)v$ and $v''=(2\Pi_x-I)v'=U_x v$, and note that
\[
\norm{v-v''}^2
=\norm{\sum_{\beta:|\theta_\beta|\leq\Theta}(1-e^{i\theta_\beta})\inp{\beta}{v}\ket{\beta}}^2
=\sum_{\beta:|\theta_\beta|\leq\Theta}|1-e^{i\theta_\beta}|^2\cdot|\inp{\beta}{v}|^2
\leq 2(1-\cos\Theta)\norm{v}^2\leq \Theta^2\norm{v}^2.
\]
Because $v+v'=2\Lambda v$, the vector $v+v'$ lies in the subspace corresponding to $\Lambda$ and hence is orthogonal to $\ket{\psi_x}$ and to $w=100\sqrt{A}\ket{\psi_x}$. 
We then have
\[
0=\inp{v+v'}{w}=\bra{v+v'}\Pi_x \ket{w}
+\bra{v+v'}(I-\Pi_x)\ket{w}
=\bra{v+v''}\Pi_x\ket{w}+
\bra{v-v''}(I-\Pi_x)\ket{w},
\]
where the last equality used that $v'+v''=2\Pi_x v'$ (hence $\Pi_x(v'+v'')=2\Pi_x v'$ and so $\Pi_x v'=\Pi_x v''$) and $v'-v''=2(I-\Pi_x)v'$ (hence $(I-\Pi_x) v'=-(I-\Pi_x) v''$). 
We can now upper bound $\norm{v}$ by
\begin{align*}
\norm{v}^2 & = |\bra{v}P_\Theta\Pi_x\ket{w}|
 = |\bra{v}\Pi_x\ket{w}|\\
& = \frac{1}{2}|\bra{v-v''}\Pi_x\ket{w}+\bra{v+v''}\Pi_x\ket{w}|
 = \frac{1}{2}|\bra{v-v''}(2\Pi_x-I)\ket{w}|\\
& \leq \frac{1}{2}\norm{v-v''}\cdot\norm{w}
 \leq \frac{1}{2}\Theta\norm{v}\cdot 100 \sqrt{A}\norm{\psi_x}\\
& \leq \frac{1}{2}\frac{1}{1000A}\norm{v}\cdot 100 \sqrt{A}\left(\frac{0.01}{\sqrt{A}} + \sqrt{A}\right)=\left(\frac{1}{20}+\frac{1}{2000A}\right)\norm{v},
\end{align*}
where the first inequality is Cauchy-Schwarz.
This implies $\norm{v}\leq 1/20+1/(2000A)$, which is small (we may assume $A\geq 1$).
\end{proof}

Phase estimation with precision $\Theta/2$ can distinguish between these two cases (eigenphase 0 vs $\geq \Theta$). This allows us to put a $+$ in front of $\ket{t_x^+}$ and a $-$ in front  of $\ket{t_x^-}$: run phase estimation, multiply with $-1$ whenever the absolute value of the phase estimate is $>\Theta/2$,
and then invert the phase estimation.
Phase estimation uses $O(1/\Theta)=O(A)$ applications of $U_x$, and hence $O(A)$ queries to~$x$, as promised.
There are small errors in this process due to the fact that Claims~\ref{cl:ADV1} and~\ref{cl:ADV2} say ``close to'' rather than ``equal,'' and due to the small approximation errors of phase estimation. Accordingly, our final state will be close to  $\ket{1}\ket{f(x)}\ket{0}$ but not necessarily equal to it, and we end up with an $O(A)$-query quantum algorithm for~$f$ that has a small error probability.

It should be noted that the upper bound is on the algorithm's query complexity, not on its gate complexity. The number of gates of the algorithm is $O(A)$ times the number of gates needed to implement the reflection $2\Lambda-I$ (the reflection $2\Pi_x-I$ is relatively easy to implement, see Exercise~\ref{ex:2PiminI}). In general this number of gates could be very large, though in some cases it can be made quite small, for instance~\cite{abrw:qjunta}.

\section{Applications}

Let us see how we can derive an $O(\sqrt{N})$-query quantum algorithm for the $N$-bit OR function from the dual adversary, for the special case where the $N$-bit input $x$ is promised to have at most one 1-bit. Consider the set ${\cal D}=\{0^N,e_1,\ldots,e_N\}$ of possible inputs. For the dual adversary bound we need to choose vectors $u_{xj},v_{xj}$ for each $x\in{\cal D}$ and $j\in[N]$. Here vectors of dimension~1 (i.e., numbers) already suffice: we define $u_{0^Nj}=v_{0^Nj}=1/N^{1/4}$ for all~$j$,
and $u_{e_kj}=v_{e_kj}=N^{1/4}$ if $j=k$ and $u_{e_kj}=v_{e_kj}=0$ otherwise.
For the objective function, note that for each $x$, $\sum_j u_{e_kj}^2=\sum_j v_{e_kj}^2=\sqrt{N}$; for $x=0^N$ this is because each of the $N$ $j$'s contributes $(1/N^{1/4})^2=1/\sqrt{N}$ to the sum, while for $x=e_k$ there is only one nonzero contribution, namely $(N^{1/4})^2=\sqrt{N}$ for $j=k$. It is also easy to verify that $\sum_{j:x_j\neq y_j} u_{xj}v_{yj}=[f(x)\neq f(y)]$ for all $x,y\in{\cal D}$.

A number of new quantum algorithms have been derived from the dual SDP for $\ADV$, for instance for finding $k$-collisions~\cite{belovs:kdistinctness}, learning and testing ``juntas'' (functions that only depend on few coordinates)~\cite{belovs:learningjuntas,abrw:qjunta}, $st$-connectivity in graphs~\cite{belovs&reichardt:conn}, and formula evaluation~\cite{reichardt&spalek:formula,acrsz:andorj}.
In general it is often quite hard and non-intuitive to come up with a feasible solution $\{u_{xj},v_{xj}\}$ for the dual SDP with a small objective value, but Belovs's learning graphs~\cite{belovs:learninggraphs} can sometimes help with more intuitive constructions of feasible solutions.

\section{Perfect composition and AND-OR trees}

If we have two Boolean functions $f:\01^n\to\01$ and
$g:\01^m\to\01$, then we can define the function $F=f\circ g^n$  (on $N=nm$ bits) which is their composition, as
\[
F(x^1,\ldots,x^n)=f(g(x^1),\ldots,g(x^n))
\]
where each $x^i\in\01^m$.
One beautiful property of the adversary bound is \emph{perfect composition}: 
\[
\ADV(F)=\ADV(f)\ADV(g).
\]
There are no hidden constant factors here!
The upper bound $\ADV(F)\leq\ADV(f)\ADV(g)$ of this composition property  can be proved by combining feasible solutions for the dual SDPs for $f$ and $g$ to a feasible solution for the dual SDP for $F$. Similarly, the lower bound $\ADV(F)\geq\ADV(f)\ADV(g)$ can be proved by combining feasible solutions for the primal SDPs for $f$ and~$g$. We will skip the rather technical details, see~\cite{BL20} for proof of a stronger and more general result.

Because of the optimality of the generalized adversary bound, it follows from the composition property that the quantum query complexity of $F$ equals the product of the query complexities of $f$ and $g$, up to a constant factor. 
For example, for the 2-level AND-OR tree on $N=k^2$ bits mentioned at the end of Chapter~\ref{chap:lowerb}, we immediately get an optimal $O(\sqrt{N})$-query quantum algorithm from the fact that the $k$-bit AND and OR functions each have quantum query complexity (and hence $\ADV$) equal to $\Theta(\sqrt{k})$.
Note that we are not directly composing bounded-error algorithms here (e.g., trying to put Grover on top of another Grover to compute the AND-OR tree):
reasoning about the composed function at the level of the adversary bound and then only translating to quantum query algorithms at the end, cleanly circumvents the problem of how the error probabilities of composed bounded-error algorithms for $f$ and $g$ affect the error probability of the resulting algorithm for~$F$.

This composition result, used in a more subtle way, can also give a quantum speed-up for evaluating \emph{game trees}. Imagine a two-player game, such as chess. First white chooses one of several possible moves, then black chooses one of several moves, etc. We can picture this as a tree where the root is the initial position and the leaves are the final positions (which are win, lose, or draw). If we assign a binary value to each leaf indicating whether white wins, then the evaluation of the game as a whole is a large, multilevel, unbalanced, AND-OR tree. If it's white's turn and at least one subtree evaluates to~1, then the current position is a~1 as well: there is a winning move for white (this corresponds to an OR function). 
If it's black's turn and one of the subtrees from the current position is labeled~0, then the current position is also labeled 0 because black has a non-losing move (this is an AND function). The value at the root of the tree indicates whether white has a sequence of moves guaranteed to win or not. Using the adversary bound to do a more subtle AND-OR composition, there is a quantum algorithm that evaluates this tree using roughly $\sqrt{N}$ queries to the binary values at the leaves~\cite{acrsz:andorj,reichardt:gametrees}. In contrast, a classical algorithm has to evaluate nearly all $N$ leaves in the worst case unless the fan-out of the tree is very small~\cite{saks&wigderson:trees}.

\section*{Exercises}

\begin{enumerate}

\item\label{ex:polynottight} Consider the 4-bit Boolean function $g(x_0,x_1,x_2,x_3)$ which is 1 iff the 4-bit input~$x$ is increasing or decreasing, i.e., if $x\in\{0000,0001,0011,0111,1111,1110,1100,1000\}$.
Let $f$ be the function on $N=4^d$ input bits obtained by composing $g$ with itself $d$ times, in a tree of depth $d$, where the value of each internal node is obtained by applying $g$ to the values of its 4 children, and with the input bits at the $N$ leaves.
\begin{enumerate}
    \item Show that $deg(g)\leq 2$ and $deg(f)\leq 2^d$.
    \item Show that the polynomial method cannot prove a lower bound better than $O(\sqrt{N})$ on the bounded-error quantum query complexity of $f$.
    \item It is known that $\ADV(g)>2.51\ldots$ (see~\cite[
Example 3.35]{belovs:phd} for a proof, or you could use an SDP-solver). Use this to show that $\ADV(f)\geq N^{c}$ for some $c>1/2$.
\end{enumerate}
{\footnotesize Comment: This exercise shows that the generalized adversary bound can sometimes prove substantially stronger lower bounds than the polynomial method. Incidentally, ~\cite[Example 3.3]{belovs:phd} also shows that if we restrict the adversary matrix to have nonnegative entries, then the best adversary lower bound we can prove for $g$ is 2.5, and hence the best lower bound we can prove for the quantum query complexity $f$ with such restricted $\Gamma$, is at most $\Omega(2.5^d)$. So this example also shows that the ``negative weights'' $\ADV$ can give substantially better bounds than the ``nonnegative weights'' version of the adversary bound.}

\item\label{ex:advST} (H) This exercise justifies Claim~\ref{cl:advST}.
Below, the $x,y$ always range over $\cal D$.
\begin{enumerate}
\item Let $P_1,P_0$ denote the projectors on the subspaces corresponding to outputs 1 and 0, respectively. Suppose inputs $x,y$ have $f(x)\neq f(y)$.\\ 
Show that 
$\inp{\psi_x^t}{\psi_y^t}=
\bra{\psi_x^t}P_{f(x)} \cdot P_{1-f(y)}\ket{\psi_y^t}+
\bra{\psi_x^t}P_{1-f(x)} \cdot P_{f(y)}\ket{\psi_y^t}$ for all $t,x,y$.
\item Define unnormalized states 
$\ket{\phi}=\sum_x\alpha_x P_{f(x)}\ket{\psi_x^T}\ket{x}$ and  $\ket{\phi^\perp}=\sum_x\alpha_x P_{1-f(x)}\ket{\psi_x^T}\ket{x}$.
Show that $\norm{\ket{\phi}}^2+\norm{\ket{\phi^\perp}}^2=1$, $\norm{\ket{\phi^\perp}}^2\leq\eps$, and $\norm{\ket{\phi}}\cdot\norm{\ket{\phi^\perp}}\leq \sqrt{\eps(1-\eps)}$.
\item Show that $|S_T|\leq 2|\bra{\phi}(I\otimes \Gamma)\ket{\phi^\perp}|$.
\item Show that $|S_T|\leq 2\sqrt{\eps(1-\eps)}\,\norm{\Gamma}$.
\end{enumerate}

\item\label{ex:advDelta} (H) 
This exercise justifies Claim~\ref{cl:advDelta}. It will be convenient for the proof to assume a phase-oracle, so the $T$-query algorithm will be of the form $U_TO_{x,\pm}U_{T-1}\cdots U_1O_{x,\pm}U_0$, applied to initial state $\ket{0^m}$ and followed by a measurement of the first qubit to produce the output bit. Below, the $x,y$ always range over $\cal D$ and the $i$ ranges over $[N]$. 
\begin{enumerate}
    \item Show that $S_t-S_{t+1}=\sum_{x,y}\Gamma_{xy}\alpha_x^*\alpha_y\bra{\psi_x^t}(I-O_{x,\pm}O_{y,\pm})\ket{\psi_y^t}$. 
\item Let $P_i=\ketbra{i}{i}\otimes I$ be the projector on the subspace where the query register is~$\ket{i}$. Define $\ket{\phi_i^t}=\sum_x\alpha_x P_i\ket{\psi_x^t}\ket{x}$.
Show that $\sum_i \norm{\ket{\phi_i^t}}^2=1$.
\item Show that $S_t-S_{t+1}=2\sum_i\bra{\phi_i^t}(I\otimes \Gamma_i)\ket{\phi_i^t}$.
\item Show that $|S_t-S_{t+1}|\leq 2\max_i\norm{\Gamma_i}$.
\end{enumerate}

\item\label{ex:weakduality}
(H) The following is a primal-dual pair of SDPs in so-called standard form:
\[
\begin{array}{lllll}
\min        & \Tr(CX) &                        \hspace*{2em} & \max & b^T y\\
\mbox{s.t.} & \Tr(A_iX)=b_i~~\forall i\in[m] &  \hspace*{2em} & \mbox{s.t.}& \sum_{i=1}^m y_i A_i \preceq C\\
            & X\succeq 0 & & &\\
\end{array}
\]
The input here consists of Hermitian $n\times n$ matrices $C,A_1,\ldots,A_m$ and vector $b\in\mathbb{R}^m$.
The $n\times n$ matrix $X$ is the variable of the primal, and the vector $y\in\mathbb{R}^m$ is the variable of the dual.
Prove that ``weak duality'' always holds: for every feasible solution $X$ for the primal and every feasible solution $y$ for the dual, we have $\Tr(CX) \geq b^T y$.


\item\label{ex:2PiminI}
Show how the reflection $2\Pi_x-I$ can be implemented with 1 phase-query to~$x$, a $Z$-gate, and a circuit that decides if the third register is $\ket{0}$. 

\end{enumerate}

\chapter{Quantum Complexity Theory}\label{chap:qcomplexitytheory}

\section{Most functions need exponentially many gates}

As we have seen, quantum computers seem to provide enormous speed-ups for problems like factoring,
and square-root speed-ups for various search-related problems. Could they be used to significantly speed up all or almost all problems?
Here we will show that this is not the case: quantum computers are not significantly better
than classical computers for most problems.

Consider the problem of computing a Boolean function $f:\01^n\rightarrow\01$ by means of a quantum circuit.
Ideally, most such functions would be computable by efficient quantum circuits  (i.e., using at most poly$(n)$ elementary gates).
Instead, we will show by means of a simple counting argument that almost all such functions $f$ have
circuit complexity nearly $2^n$. This is a variant of a well-known counting argument for classical
Boolean circuits due to Riordan and Shannon~\cite{Riordan&Shannon:circuit}.

Let us fix some finite set of elementary gates, for instance the Shor basis $\{H,T,\CNOT\}$ 
or $\{H,\mbox{Toffoli}\}$.
Suppose this set has $k$ types of gates, each acting on at most 3~qubits.
Let us try to count the number of distinct circuits that have at most $C$ elementary gates. 
For simplicity we include the initial qubits (the $n$ input bits as well as workspace qubits, which are initially $\ket{0}$) 
as a $(k+1)$st type among those $C$ gates.  
First we need to choose which type of elementary gate each of the $C$ gates is; this can be done in $(k+1)^C$ ways.
Now every gate has at most 3 ingoing and 3 outgoing wires.
For each of its 3 outgoing wires we can choose an ingoing wire into one of the gates in the following level;
this can be done in at most $(3C)^3$ ways.
Hence the total number of circuits with up to $C$ elementary gates is at most $(k+1)^C (3C)^{3C}=C^{O(C)}$.  
We are clearly overcounting here, but that's OK because we want an upper bound on the number of circuits. 

We'll say that a specific circuit \emph{computes} a Boolean function $f:\01^n\rightarrow\01$ if for every input $x\in\01^n$,
a measurement of the first qubit of the final state (obtained by applying the circuit to
initial state $\ket{x,0}$) gives value $f(x)$ with probability at least $2/3$.
Each of our $C^{O(C)}$ circuits can compute at most one $f$  
(in fact some of those circuits don't compute any Boolean function at all).
Accordingly, with $C$ gates we can compute at most $C^{O(C)}$ distinct Boolean functions $f:\01^n\rightarrow\01$.
Hence even if we just want to be able to compute 1\%\ of all $2^{2^n}$ Boolean functions, then we already need 
$$
C^{O(C)}\geq \frac{1}{100}2^{2^n}\mbox{, which implies }C\geq\Omega(2^n/n).
$$
Accordingly, very few computational problems will be efficiently solvable on a quantum computer.
Below we will try to classify those that \emph{are} efficiently solvable, using the tools of complexity theory.

\section{Classical and quantum complexity classes}

A computational decision problem on binary strings corresponds to what is often called a ``language'' in complexity theory: a language $L\subseteq\01^*$ is a set of binary strings of arbitrary lengths, and the corresponding decision problem is to determine whether a given string is an element of~$L$ or not.
For example, $L$ could be the set of prime numbers encoded in binary (corresponding to the problem of deciding whether a given number is prime or not) or the set of satisfiable Boolean formulas. It is often convenient to think of such a decision problem as corresponding to a sequence of Boolean functions $f_n:\01^n\to\01$, one for each input-length~$n$, where $f_n$ takes value~1 exactly on the $n$-bit strings that are in~$L$.

A ``complexity class'' is a set of decision problems (i.e., languages) that all have similar complexity in some sense,
for instance the ones that can be solved with polynomial time or polynomial space.
Let us first mention four of the most important \emph{classical} complexity classes:
\begin{itemize}
\item \Pp. The class of problems that can be solved by classical deterministic computers using polynomial time.
\item \BPP. The class of problems that can be solved by classical randomized computers using polynomial time (and with error probability $\leq 1/3$ on every input).
\item \NP. The class of problems where the `yes'-instances can be verified in polynomial time if some prover gives us a polynomial-length ``witness.''
More precisely, a language $L$ is in \NP\ iff there exists a deterministic polynomial-time algorithm $A$, with two inputs $x,y$ (where $y$ is at most polynomially longer than $x$), such that $x\in L$ iff there is a $y$ such that $A(x,y)$ outputs~1. 

Some problems $L$ in this class are \emph{\NP-complete}, meaning that any other problem $L'\in\NP$ can be reduced to $L$ in polynomial time:
there exists a polynomial-time computable function $f$ such that $x\in L'$ iff $f(x)\in L$.
Hence the \NP-complete problems are the hardest problems in \NP.
An example is the problem of satisfiability: we can verify that a given Boolean formula is satisfiable if 
a prover gives us a satisfying assignment~$y$, so the satisfiability-problem is in \NP, but one can even show that it is \NP-complete.
Other examples of \NP-complete problems are  integer linear programming, travelling salesman, graph-colorability, etc.
\item \PSPACE. The class of problems that can be solved by classical deterministic computers using polynomial space.
\end{itemize}
We can consider quantum analogues of all such classes, an enterprise that was started by Bernstein and Vazirani~\cite{bernstein&vazirani:qcomplexity}:
\begin{itemize}
\item \EQP. The class of problems that can be solved exactly by quantum computers using polynomial time.  This class depends on the set of elementary gates one allows, and therefore is not so interesting.
\item \BQP. The class of problems that can be solved by quantum computers using polynomial time (and with error probability $\leq 1/3$ on every input). This class is the accepted formalization of ``efficiently solvable by quantum computers.''
\item ``quantum \NP''. In analogy with the above definition of \NP, one could define quantum \NP\ as the class of problems where the `yes'-instances can be verified efficiently if some prover gives us a ``quantum witness'' of a polynomial number of qubits. For every `yes'-instance there should be a quantum witness that passes the verification with probability~1, while for `no'-instances every quantum witness should be rejected with probability~1. This class is again dependent on the elementary gates one allows, and not so interesting.
  
  Allowing error probability $\leq 1/3$ on every input, we get a class called \QMA\ (``quantum Merlin-Arthur'').  This is a more robust and more interesting quantum version of \NP. In particular, like \NP, \QMA\ has \emph{complete} problems: problems in \QMA\ to which every other \QMA-problem can be efficiently reduced. The most famous example of such a problem is deciding whether the \emph{ground state energy} (i.e., lowest eigenvalue) of a given \emph{$k$-local Hamiltonian} (see Chapter~\ref{chap:hamsim}) is at most some given number $a$ or at least $a+1/\poly(n)$. Determining the ground state energy of a given physical system is extremely important in physics and chemistry.
  It is not hard to see that the problem is \emph{in} \QMA: we can just let the quantum witness be the ground state (i.e., an eigenstate for the lowest eigenvalue) and measure its energy using the Hamiltonian, which is the observable corresponding to total energy. The problem turns out to be \QMA-complete already for $k=2$~\cite{kitaev:qnp,KKR:localham}, so every other problem in \QMA\ can be efficiently reduced to the 2-local Hamiltonian problem.  We will devote Chapter~\ref{ch:QMAlocham} to this.
\item \QPSPACE. The class of problems that can be solved by quantum computers using polynomial space, with error probability $\leq 1/3$ on every input. Despite allowing quantum computing and small error probabilities on top of the classical complexity class \PSPACE, it turns out we don't get a larger class: $\QPSPACE=\PSPACE$.
\end{itemize}
As explained in Appendix~\ref{appusefulCS}, in all the above cases the error probability $1/3$ can be reduced efficiently to much smaller constant~$\eps>0$:
just run the computation $O(\log(1/\eps))$ times and take the majority of the answers given by these runs.

We should be a bit careful about what we mean by a ``polynomial-time [or space] quantum algorithm.''
Our model for computation has been quantum circuits, and we need a separate quantum circuit for each new input-length.
So a quantum algorithm of time $p(n)$ would correspond to a \emph{family} of quantum circuits $\{C_n\}$,
where $C_n$ is the circuit that is used for inputs of length $n$; it should have at most $p(n)$ elementary gates.%
\footnote{To avoid smuggling loads of hard-to-compute information into this definition (e.g., $C_n$ could contain information about whether the $n$-th Turing machine halts or not), we will require this family to be efficiently describable: there should be a classical
Turing machine which, on input $n$ and $j$, outputs (in time polynomial in $n$) the $j$-th elementary gate
of $C_n$, with information about where its incoming and outgoing wires go.}

We have $\BPP\subseteq\BQP$, because a \BPP-machine on a fixed input-length $n$
can be written as a polynomial-size reversible circuit (i.e., consisting of Toffoli gates) that starts from a state that involves some coin flips.
Quantum computers can generate those coin flips using Hadamard transforms, 
then run the reversible circuit, and measure the final answer bit.
It is believed that \BQP\ contains problems that aren't in \BPP, for example factoring large integers:
this problem (or rather the decision-version thereof) is in \BQP\ because of Shor's algorithm, 
and is generally believed not to be in \BPP.
In the next section we will prove that $\BQP\subseteq\PSPACE$.
Thus we get the following sequence of inclusions:
$$
\Pp\subseteq \BPP\subseteq \BQP\subseteq \PSPACE.
$$
It is generally believed that $\Pp=\BPP$~\cite{impagliazzo&wigderson:PisBPP}, while the other inclusions are believed to be strict.
Note that a \emph{proof} that \BQP\ is strictly greater than \BPP\ 
(for instance, a proof that factoring cannot be solved efficiently by classical randomized computers) 
would imply that $\Pp\neq \PSPACE$, solving what has been one of the main open problems
in computer science since the 1960s. Hence such a proof---if it exists at all---will probably be very hard.

What about the relation between \BQP\ and \NP?
It's generally believed that \NP-complete problems are probably not in \BQP.
The main evidence for this is the lower bound for Grover search:
a quantum brute-force search on all $2^n$ possible assignments
to an $n$-variable formula gives a square-root speed-up, but not more.
This is of course not a proof, since there might be some more clever, non-brute-force 
methods that exploit the structure of the problem to solve satisfiability. However, neither in the classical nor in the quantum case do we know
clever methods that solve the general satisfiability problem much faster than brute-force search.

Finally, there could also be problems in \BQP\ that are not in \NP, so it may well be that \BQP\ and \NP\ are incomparable.
Much more can be said about quantum complexity classes; see for instance Watrous's survey~\cite{watrous:qcomplexitysurvey}.

\section{Classically simulating quantum computers in polynomial space}\label{sec:bqpinpspace}

When Richard Feynman first came up with quantum computers~\cite{feynman:simulating}, he motivated them by
\begin{quote}
``the full description of quantum mechanics
for a large system with R particles is given by a function $q(x_1,x_2,\ldots,x_R,t)$
which we call the amplitude to find the particles $x_1,\ldots,x_R$ [RdW: think of $x_i$ as one qubit], 
and therefore, because it has too many variables, it cannot be simulated with a normal
computer with a number of elements proportional to R or proportional to N.''
[\ldots]\\ 
``Can a quantum system be probabilistically simulated by
a classical (probabilistic, I'd assume) universal computer? In other words, a
computer which will give the same probabilities as the quantum system
does. If you take the computer to be the classical kind I've described so far
(not the quantum kind described in the last section) and there are no changes
in any laws, and there's no hocus-pocus, the answer is certainly, No!''
\end{quote}
The suggestion to devise a \emph{quantum} computer to simulate quantum physics is of course a brilliant one,
but the main motivation is not quite accurate.  As it turns out, it is not necessary to keep track of all (exponentially many)
amplitudes in the state to classically simulate a quantum system.  Here will prove the result of Bernstein and Vazirani~\cite{bernstein&vazirani:qcomplexity} that quantum computers can actually be simulated
efficiently in terms of \emph{space} (though not necessarily in terms of \emph{time}).

Consider a circuit with $T=\poly(n)$ gates that acts on $S$ qubits, where the first $n$ of those $S$ qubits give the classical input string.
Assume for simplicity that all gates are either the 1-qubit Hadamard or the 3-qubit Toffoli gate
(as mentioned in Section~\ref{sec:universalgatesets}, these two gates together suffice for universal quantum computation),
and that the classical output (0 or~1) of the algorithm is determined by a measurement of the first qubit of the final state.
Without loss of generality $S\leq 3T$, because $T$ Toffoli gates won't affect more than $3T$ qubits. 
Let $U_j$ be the unitary that applies the $j$-th gate to its (1 or~3) qubits, and applies identity to all other qubits.
The entries of this matrix are of a simple form ($0$, $1/\sqrt{2}$, or $-1/\sqrt{2}$ for Hadamard; $0$ or $1$ for Toffoli)
and easy to compute.  Let $\ket{i_0}=\ket{x}\ket{0^{S-n}}$ be the starting state, where $x\in\01^n$ is the classical input, 
and the second register contains the workspace qubits the algorithm uses.
The final state will be
$$
\ket{\psi_x}=U_TU_{T-1}\cdots U_2U_1\ket{i_0}.
$$
The amplitude of basis state $\ket{i_T}$ in this final state is
$$
\inp{i_T}{\psi_x}=\bra{i_T}U_TU_{T-1}U_{T-2}\cdots U_2U_1\ket{i_0}.
$$
Inserting an identity matrix $I=\sum_{i\in\01^S}\ketbra{i}{i}$ between the gates, we can rewrite this as\footnote{For the physicists: this is very similar to a path integral.}
\begin{align*}
\inp{i_T}{\psi_x} & = \bra{i_T}U_T\left(\sum_{i_{T-1}\in\01^S}\hspace*{-0.7em}\ketbra{i_{T-1}}{i_{T-1}}\right)U_{T-1}\left(\sum_{i_{T-2}}\ketbra{i_{T-2}}{i_{T-2}}\right)U_{T-2}\cdots U_2\left(\sum_{i_1}\ketbra{i_1}{i_1}\right)U_1\ket{x,0}\\
 & = \sum_{i_{T-1}\in\01^S,\ldots,i_1\in\01^S}\,\prod_{j=1}^T\bra{i_j}U_j\ket{i_{j-1}}.
\end{align*}
The term $\bra{i_j}U_j\ket{i_{j-1}}$ is just one entry of the matrix $U_j$ and hence easy to calculate because $U_j$ acts non-trivially on only 1 or 3 qubits (see Exercise~\ref{ex:pickingoutentryofU}).
Then $\prod_{j=1}^T\bra{i_j}U_j\ket{i_{j-1}}$ is also easy to compute, in polynomial space (and even in polynomial \emph{time}). 
If $\ell$ of the $T$ gates are Hadamards, then each such term is either 0 or $\pm 1/\sqrt{2^\ell}$.

Adding up $\prod_{j=1}^T\bra{i_j}U_j\ket{i_{j-1}}$ for all $i_{T-1},\ldots,i_1$
is also easy to do in polynomial space if we reuse space for each new $i_{T-1},\ldots,i_1$. 
Hence the amplitude $\inp{i_T}{\psi_x}$ can be computed exactly using polynomial space.%
\footnote{Of course, the calculation will take exponential \emph{time}, because there are $2^{S(T-1)}$ different sequences 
$i_{T-1},\ldots,i_1$ that we need to go over sequentially.}
We assume that the \BQP\ machine's answer is obtained by measuring the first qubit of the final state.
Then its acceptance probability is the sum of squares of all amplitudes of basis states starting with a~1:
$\sum_{i_T:(i_T)_1=1}|\inp{i_T}{\psi_x}|^2$.  Since we can compute each amplitude $\inp{i_T}{\psi_x}$ in polynomial space,
and we can loop over all $i_T\in\01^S$ whose first bit is~1 to sum their squared amplitudes, the acceptance probability of a \BQP-circuit on classical input $x$ can also be computed in polynomial space.
This proves the inclusion $\BQP\subseteq \PSPACE$.

\section*{Exercises}

\begin{enumerate}

\item\label{ex:factoringdec} (H)
The following problem is a decision version of the factoring problem:
\begin{quote}
Given positive integers $N$ and $k$, decide if $N$ has a prime factor $p\in\{k,\ldots,N-1\}$.
\end{quote}
Show that if you can solve this decision problem efficiently (i.e., in time polynomial in the input-length $n=\ceil{\log N}$),
then you can also find the prime factors of $N$ efficiently.

\item\label{ex:pickingoutentryofU}
\begin{enumerate}
\item Let $U$ be an $S$-qubit unitary which applies a Hadamard gate to the $k$-th qubit, and identity gates to the other $S-1$ qubits.
Let $i,j\in\01^S$. Show an efficient way (i.e., using time polynomial in $S$) to classically calculate the matrix-entry $U_{i,j}=\bra{i}U\ket{j}$ (note: even though $U$ is a tensor product
of $2\times 2$ matrices, it's still a $2^S\times 2^S$ matrix, so calculating $U$ completely isn't efficient).
\item Let $U$ be an $S$-qubit unitary which applies a CNOT gate to the $k$-th and $\ell$-th qubits, and identity gates to the other $S-2$ qubits.
Let $i,j\in\01^S$. Show an efficient way to classically calculate the matrix-entry $U_{i,j}=\bra{i}U\ket{j}$.
Here $k$ and $\ell$ need not be adjacent, but to simplify your notation you may assume that they are adjacent.
\end{enumerate}

\item\label{ex:bqptobqp}
This exercise shows how to use \BQP-algorithms as subroutines in other \BQP-algorithms.
\begin{enumerate}
\item (H) Suppose $L$ is a language in \BQP. Let $f$ be the corresponding Boolean function, so $f(x)=1$ iff $x\in L$. Show that there is a $w\leq\poly(n)$ and a polynomial-size quantum circuit $U$ that implements the following map for all $x\in\01^n$:
\[
\ket{x,0^{w+1}} \mapsto \sqrt{p}\ket{x,f(x)}\ket{\phi(x)} + \sqrt{1-p}\ket{x,1-f(x)}\ket{\psi(x)},
\]
where $p\geq 1-\exp(-n)$, and $\ket{\phi(x)}$ and $\ket{\psi(x)}$ are states of the $w$-qubit workspace. 
\item Show that there is a polynomial-size quantum circuit $V$ that (when restricted to the subspace of input states  where the workspace qubits are $\ket{0}$) is $\exp(-n)$-close in operator norm to the following unitary:
\[
O_f: \ket{x,b,0^w} \mapsto\ket{x,b\oplus f(x),0^w},
\]
for all $x\in\01^n$ and $b\in\01$.
\item (H) Suppose $L$ is a language in \BQP, and you have a polynomial-size quantum circuit for another language $L'$ that uses queries to the language $L$ (i.e., applications of the unitary $O_f$). Show that the language $L'$ is also in \BQP: there is a polynomial-size quantum circuit for $L'$ that doesn't need queries to $L$. 
\end{enumerate}

\item\label{eq:BQPinPP} (H)
Consider a circuit $C$ with $T=\poly(n)$ elementary gates (only Hadamards and Toffolis) acting on $S=\poly(n)$ qubits.
Suppose this circuit computes $f:\01^n\rightarrow\01$ with bounded error probability: for every $x\in\01^n$,
when we start with basis state $\ket{x,0^{S-n}}$, run the circuit and measure the first qubit, then the result equals $f(x)$ with probability at least $2/3$.
\begin{enumerate}
\item Consider the following quantum algorithm:
start with basis state $\ket{x,0^{S-n}}$, run the above circuit $C$ without the final measurement, apply a $Z$ gate to the first qubit, and reverse the circuit $C$.
Denote the resulting final state by $\ket{\psi_x}$.
Show that if $f(x)=0$ then the amplitude of basis state $\ket{x,0^{S-n}}$ in $\ket{\psi_x}$ is in the interval $[1/3,1]$, while if $f(x)=1$ then the amplitude
of $\ket{x,0^{S-n}}$ in $\ket{\psi_x}$ is in $[-1,-1/3]$.
\item $\PP$ is the class of computational decision problems that can be solved by classical randomized polynomial-time computers with success probability $>1/2$
(however, the success probability could be exponentially close to 1/2, i.e., $\PP$ is $\BPP$ without the `B' for bounded-error).
Show that $\BQP\subseteq\PP$.
\end{enumerate}

\item\label{ex:entdetection} (H)
Consider the following computational problem, called the ``entanglement-detection problem'': given a classical description of a polynomial-sized quantum circuit $C$ (say with $H$, $T$ and CNOT gates), decide whether the $(n+1)$-qubit state $C\ket{0^{n+1}}$ is entangled, where you think of the first $n$ qubits as ``Alice'' and the last qubit as ``Bob''. Show that if you can efficiently (i.e., in time polynomial in the length of the input, which is the number of bits used in the description of $C$) solve this decision problem on a classical computer, then you can efficiently solve every problem in $\NP$ on a classical computer.

\end{enumerate}

\chapter{\QMA\ and the Local Hamiltonian Problem}\label{ch:QMAlocham}

\section{Quantum Merlin-Arthur (\QMA)}\label{sec:qmadef}

One can think of the complexity class \NP\ as formalizing the standard notion of an efficiently-verifiable proof.
For example, to prove that a given formula $\phi$ is satisfiable we give a satisfying assignment: this is easy to verify but hard to find (unless \Pp=\NP).
The theory of \NP-completeness shows that many very different computational problems (satisfiability, TSP, integer linear programming, etc.) are essentially the same computational problem, in the sense that instances from one can easily be translated to instances of another in deterministic polynomial time. 

We can relax the notion of proof slightly by allowing the prover and verifier the use of randomness, and allowing them some error probability, say 1/3. For every $x\in L$ the prover should be able to provide a polynomial-size proof or ``witness'' that convinces the randomized verifier (with probability $\geq 2/3$) that $x\in L$; while for $x\not\in L$, no matter what purported ``witness'' the prover sends, the verifier should only accept with probability $\leq 1/3$.
For historical reasons~\cite{babai&moran:am}, the complexity class corresponding to such~$L$ is called  \emph{Merlin-Arthur} (\MA), with Merlin referring to the omniscient prover and Arthur referring to the mere mortal (i.e., randomized and polynomial-time) king who is supposed to verify Merlin's proofs. 

Quantum Merlin-Arthur (\QMA) is the proper quantum analogue of \NP.
We already mentioned it in the previous chapter, but let us formally define \QMA\ here.
In contrast to \NP, which consists of languages~$L$ where every string $x$ is either in or out of $L$, \QMA\ is a class of \emph{promise problems}. A promise problem $L$ partitions the set $\01^*$ of all binary strings into $L_1$, $L_0$, and $L_*$. An algorithm is ``promised'' that it never receives inputs from $L_*$; if it gets an input from $L_b$ for $b\in\01$, then it has to determine~$b$. The usual languages are promise problems where $L_*=\emptyset$.

\begin{definition}\label{def:QMA}
A promise problem $L=(L_1,L_0,L_*)$ is in the class \QMA, if there exists a uniform family $\{C_n\}$ of polynomial-size quantum circuits with two input registers and one output qubit, and a polynomial $w$ (for ``witness length''), such that for all $x\in\01^*$:
\begin{itemize}
    \item {\bf Completeness:} If $x\in L_1\cap\01^n$, then there exists a $w(n)$-qubit state $\ket{\psi}$ (a ``proof'' or ``witness'' state) such that the circuit $C_n$ outputs 1 with probability $\geq 2/3$ when run on $x,\ket{\psi}$.
    \item {\bf Soundness:} If $x\in L_0\cap\01^n$, then for every $w(n)$-qubit state $\ket{\psi}$, the circuit $C_n$ outputs~1 with probability $\leq 1/3$ when run on $x,\ket{\psi}$.
\end{itemize}
\end{definition}
If we force $\ket{\psi}$ to be classical, then we get a class called \QCMA. If we additionally force the verifier to be classical, then we get \MA.
And if we additionally replace success probability $2/3$ by~1, then we get \NP.%
\footnote{As mentioned in the previous chapter, requiring success probability~1 instead of 2/3 in \QMA\ itself (like in \NP) leads to an unsatisfactory complexity class because it depends on what set of elementary gates one uses.}
Hence $\NP\subseteq\MA\subseteq\QCMA\subseteq\QMA$ follows immediately from the definitions.
It is believed that \MA=\NP~\cite{klivans&melkebeek:GIsubexp}, for similar reasons as why we believe \BPP=\Pp~\cite{impagliazzo&wigderson:PisBPP}. 
However, \QMA\ is strongly believed to be a larger class than \NP, meaning that ``quantum proofs'' can prove more than classical proofs can. Moreover, as we will see later, we can identify important \QMA-complete problems, none of which is in \NP\ (unless \NP=\QMA). 

We have fixed the error probability here to 1/3 rather arbitrarily. We can easily reduce this to a much smaller $\delta$ like we would for a randomized algorithm: repeat the protocol $O(\log(1/\delta))$ times and output the majority output bit of those runs (see Appendix~\ref{appusefulCS}). 
If done naively, we'd need a new witness state $\ket{\psi}$ in each run, because the measurement that produces the verifier's output bit in each run can collapse the state. One has to be careful about soundness here, since the prover can send a large entangled state instead of a tensor product of witness states for the individual runs; however, it is not too hard to show that this cannot help the prover.
This approach increases the verifier's runtime but also the required witness-size $w(n)$ by a factor $O(\log(1/\delta))$.
However, there is a beautiful and quite surprising technique (which we will not explain here) due to Marriott and Watrous~\cite{marriott&watrous:qma} that achieves the same error reduction \emph{using the same witness state}! The verifier's runtime in that amplified protocol will still go up by a factor $O(\log(1/\delta))$, but the witness-size remains $w(n)$.

\section{The local Hamiltonian problem}\label{sec:lochamproblem}

The quintessential \NP-complete problem is satisfiability (SAT): given a formula $\phi(x_1,\ldots,x_n)$ of $n$ Boolean variables $x_1,\ldots,x_n$,\footnote{To remain consistent with the literature we now use $n$ for the number of Boolean variables, not for the actual length of the input, which would be the number of bits needed to describe the instance~$\phi$, which will be larger than~$n$ now. However, the input-length should be at most polynomial in $n$.}
decide if there is an assignment of truth values to $x_1,\ldots,x_n$ that makes the formula true. The famous Cook-Levin theorem~\cite{cook:np,levin:np} says this is \NP-complete.

A special case of this is $k$-SAT, where we restrict the formula $\phi$ to be the conjunction of clauses, each of which is the disjunction of $k$ literals (a literal is a variable $x_i$ or its negation). For example, the following is a 3-SAT instance with 4 clauses on $n=5$ Boolean variables:
\begin{equation}\label{eq:3satinstance}
(x_1\vee \neg x_2\vee x_3)\wedge(\neg x_1\vee x_2\vee x_5)\wedge(\neg x_2\vee \neg x_3\vee\neg x_5)\wedge(x_3\vee x_4\vee\neg x_5).
\end{equation}
It is well-known that $k$-SAT is still \NP-complete if $k\geq 3$, while $2$-SAT is actually in \Pp.

Let us try to reformulate $k$-SAT in a way that looks ``more quantum,'' by relating it to the minimal eigenvalue of a particular Hamiltonian that is diagonal in the computational basis (recall from  Section~\ref{sec:hamiltonians} that the Hamiltonian for a physical system is the observable corresponding to total energy). For concreteness we will fix $k=3$.  Consider a clause $C=x_1\vee \neg x_2\vee x_3$. This has one non-satisfying assignment, namely $x_1=0$, $x_2=1$, $x_3=0$. With this clause let us associate the following diagonal Hamiltonian:
\[
H_C=\left(\begin{array}{cccccccc}
0 & & & & & & & \\
 & 0 & & & & & & \\
 & & 1 & & & & & \\
 & & & 0 & & & & \\
 & & & & 0 & & & \\
 & & & & & 0 & & \\
 & & & & & & 0 & \\
 & & & & & & & 0
\end{array}\right)
\]
Note that the 1 sits at the location indexed by $x_1x_2x_3=010$, the unique non-satisfying assignment for $C$. Think of $H_C$ as giving a ``penalty'' of~1 to $x$ if $x$ doesn't satisfy clause~$C$. We will implicitly treat $H_C$ as an $n$-qubit Hamiltonian, by tensoring it with identity for the other $n-3$ qubits.
That way, we have $\bra{x}H_C\ket{x}=0$ if clause $C$ is satisfied by assignment $x$, and $\bra{x}H_C\ket{x}=1$ if not. 

Now suppose we have a 3-SAT formula $\phi=C_1\wedge\cdots\wedge C_m$ that is the conjunction of $m$ clauses, each with 3 literals. To this we associate the following Hamiltonian:
\[
H_\phi=\sum_{j=1}^m H_{C_j}.
\]
Note that the eigenvalues of $H_\phi$ lie in the interval~$[0,m]$, and that $H_\phi$ is a \emph{3-local} Hamiltonian: each term involves only 3 of the qubits non-trivially (more generally, if we start with a $k$-SAT instance, then $H_\phi$ is $k$-local). Also note that the ``energy'' of assignment $x\in\01^n$ is
\[
\bra{x}H_\phi\ket{x}=\sum_{j=1}^m \bra{x}H_{C_j}\ket{x},
\]
which exactly counts the number of unsatisfied clauses under assignment~$x$. The ``minimal energy'' (lowest eigenvalue $\lambda_{\min}$) of $H_\phi$ is equal to the minimal number of unsatisfied clauses ($=m-$the maximal number of satisfied clauses).
In particular, $\phi$ is satisfiable iff $\lambda_{\min}=0$.

The above Hamiltonian is diagonal, and hence rather ``classical,'' because it is the sum of diagonal 3-local terms. If instead we allow the terms to be arbitrary 3-local (or even $k$-local) Hamiltonians, we arrive at the central problem of this chapter.

\begin{definition}
The \emph{$k$-local Hamiltonian problem} is the following: given a classical description of an $n$-qubit Hamiltonian 
\begin{equation}\label{eq:localHam}
H=\sum_{j=1}^m H_j
\end{equation}
where each $H_j$ is $k$-local (i.e., it acts nontrivially on only $k$ of the $n$ qubits) and $0\preceq H_j\preceq I$, and given parameters $a,b\in[0,m]$ with $b-a\geq 1/\poly(n)$, promised that $H$'s minimal eigenvalue $\lambda_{\min}$ is either $\leq a$ or $\geq b$, decide which is the case.
\end{definition}

Note that this is a promise problem: some $H$ of the form of Eq.~\eqref{eq:localHam} will have $\lambda_{\min}$ in $(a,b)$ and hence won't satisfy the promise. Such instances will form the set $L_*$ of this promise problem, while the ``$\leq a$'' instances will form $L_1$ and the ``$\geq b$'' instances will form $L_0$.
The input $H$ is a $2^n\times 2^n$ matrix, but because it is $k$-local we do not need to describe it literally. Instead the description of $H$ that is given as input will just consist of each of the $m$ terms as a $2^k\times 2^k$ matrix ($m\cdot 2^{2k}$ complex numbers) and $m\ceil{\log\binom{n}{k}}$ bits telling us for each of the $m$ terms on which $k$ qubits that term acts non-trivially.
Accordingly,  if $m=\poly(n)$, $k=O(\log n)$, and each complex entry is represented with $\poly(n)$ bits, then the actual input-length is $\poly(n)$ bits.

The assumption that each $H_j$ (and therefore $H$ as well) is positive semidefinite, is not essential. If we were instead to start with the weaker condition $-I\preceq H_j\preceq I$, then by defining psd matrices $H'_j=(H_j+I)/2$ we obtain an instance where $0\preceq H'_j\preceq I$ with a simple relation between the minimal eigenvalues $\lambda_{\min}$ of  $H=\sum_j H_j$ and $\lambda'_{\min}$ of $H'=\sum_j H'_j$, namely $\lambda'_{\min}=(\lambda_{\min}+m)/2$.
This allows us to also model negative energies.

The minimal eigenvalue $\lambda_{\min}$ is known as the ``ground state energy'' of the Hamiltonian: it is the energy of the state(s) we get if we cool an $n$-qubit system governed by $H$ to temperature~0.
Distinguishing $\lambda_{\min}\leq a$ or $\lambda_{\min}\geq b$ is essentially equivalent to approximating $\lambda_{\min}$ up to an additive error $O(b-a)$ (see Exercise~\ref{ex:approxlambdamin}).
Finding information about $\lambda_{\min}$, and more generally about the lowest eigenvalues of~$H$ and about the structure of the ``ground states'' (the eigenstates for eigenvalue $\lambda_{\min}$) is important for many problems in physics and chemistry, for instance in determining the properties of materials at low temperatures (including poorly-understood phenomena such as superconductivity) and the reaction speeds of chemical reactions. Much work in computational science is expended on solving such problems for particular Hamiltonians corresponding to physical systems of interest.
Such Hamiltonians are typically indeed $k$-local for small $k$, at least approximately, since particles tend to significantly influence only the particles close to them.

Unfortunately the local Hamiltonian problem is \NP-hard already for $k\geq 2$ and very large gap between $a$ and $b$. This follows from our above translation from SAT: we can convert a 2-SAT instance $\phi$ to 2-local Hamiltonian $H_\phi$ where $\lambda_{\min}=m-$the maximal number of satisfied clauses. Computing the maximal number of satisfied clauses is known as the MAX-2-SAT problem, and distinguishing between different values that are $\Omega(m)$ apart for its value is already known to be an \NP-hard problem~\cite{hastad:optimalj}. Since it is generally believed that $\NP\not\subseteq\BQP$, it is unlikely that a quantum computer can solve local Hamiltonian efficiently in general. 
Even worse, as we will see next, $k$-local Hamiltonian turns out to be complete for the complexity class \QMA\  (which is presumably larger than \NP), already for $k\geq 2$, but with polynomially small gap between $a$ and $b$.

\section{Local Hamiltonian is \QMA-complete}\label{sec:locHamQMAcomplete}

In this section we will show that $k$-local Hamiltonian is a \QMA-complete problem, proving a quantum analogue of the Cook-Levin theorem.

First, it is not too hard to see that the problem is in \QMA. The witness $\ket{\psi}$ for the instances $x\in L_1$ would be a ground state, and there are several efficient ways to approximate its energy in order to verify that it is indeed $\leq a$ and not $\geq b$ (see Exercise~\ref{eq:locHaminQMA}).

Second, we need to show that $k$-local Hamiltonian is \QMA-hard, meaning that any other problem in \QMA\ can be reduced to it. So consider an arbitrary $L=(L_1,L_0,L_*)\in\QMA$ as in Definition~\ref{def:QMA}, and fix an $n$-bit input $x\in L_1\cup L_0$. We would like to convert the circuit $C_n$ into a Hamiltonian $H$, such that $H$ has a small eigenvalue iff $C_n$ has high acceptance probability on some states $\ket{\psi}$. 
 We will assume the error probability is $\leq 1/4T$ rather than $1/3$. 
 
The circuit $C_n$ acts on $n+s+w(n)$ qubits, where the first $n$ qubits contain the fixed classical input $x$ (which we will omit below for simplicity), the circuit uses $s$ workspace qubits (which start out as $\ket{0}$), and the third register contains the purported $w(n)$-qubit witness state. 
Since $C_n$ consists of some $T=\poly(n)$ gates, we can write it as a product $C_n=U_T\cdots U_1$ where each $U_t$ is an elementary gate on 1 or 2 qubits, tensored with identity on the other qubits. For a given $w(n)$-qubit state $\ket{\psi}$, let $\ket{\psi_0}=\ket{0^s}\ket{\psi}$ and $\ket{\psi_t}=U_t\ket{\psi_{t-1}}$ for $t\in[T]$ be the initial, intermediate, and final states of the algorithm.
The output qubit of the circuit is obtained by measuring the first qubit of the final state $\ket{\psi_T}$ in the computational basis.

We will now describe a Hamiltonian $H$ that ``follows'' the state and gives ``penalties'' for every deviation from the proper sequence of states $\ket{\psi_0},\ldots,\ket{\psi_T}$, as well as penalizing a 0-output in the final measurement.
In addition to the register that $U$ acts on, we will add another register of $\log(T+1)$ qubits that acts like a ``clock,'' ranging from $0$ to $T$ (assume for simplicity that $T+1$ is a power of~2).
We will subscript the first $s+w(n)$ qubits by integers in $\{1,\ldots,s+w(n)\}$, and subscript the clock register by `$C$'.
Define
\[
\begin{array}{lll}
H_{\text{init}}  & = & \displaystyle \sum_{i=1}^s \ketbra{1}{1}_i\otimes \ketbra{0}{0}_C\\[0.3em]
H_t & = & \frac{1}{2}\left(I\otimes(\ketbra{t-1}{t-1}_C + \ketbra{t}{t}_C) - U_{t}\otimes \ketbra{t}{t-1}_C - U_{t}^*\otimes\ketbra{t-1}{t}_C\right),\text{ for }t\in[T]\\[0.5em]
H_{\text{final}} & = & \ketbra{0}{0}_1\otimes\ketbra{T}{T}_C\\ 
H & = & \displaystyle H_{\text{init}}  + \sum_{t=1}^T H_t + H_{\text{final}}. 
\end{array}
\]\label{eq:KitaevH}
The number of terms in $H$ is $m=s+T+1$.
The idea behind this Hamiltonian $H$ is that $H_{\text{init}} $ checks that the $s$ workspace qubits are all 0 in the initial state, where the clock register is~0 (giving an ``energy penalty'' if some of those workspace qubits are~1); $H_t$ checks that $U_t$ is applied properly in the $t$-th step (the factor $1/2$ is to ensure $H_t\preceq 1$); and $H_{\text{final}}$ checks that the output qubit in the final state is~1 (giving a penalty if it's 0).
Because the clock register uses $\ceil{\log(T+1)}$ bits, and each gate $U_t$ acts on at most 2 qubits, the locality of $H$ is $k=\ceil{\log(T+1)}+2=O(\log n)$. We will reduce this to a constant later. 

\subsection{Completeness and soundness}\label{sec:lochamsoundcomplete}

We now want to show that we can distinguish $x\in L_1$ and $x\in L_0$ by considering the smallest eigenvalue $\lambda_{\min}$ of the above Hamiltonian $H$. 
First, for {\bf completeness} 
we want to show that if $x\in L_1$ then there is a state with small eigenvalue.
Since $x\in L_1$, there is a $w(n)$-qubit witness state $\ket{\psi}$ that leads $U$ with initial state $\ket{\psi_0}=\ket{0^s}\ket{\psi}$ to accept (i.e., output~1) with probability $\geq 1-1/4T$.
Consider the following state:
\[
\ket{\psi'}=\frac{1}{\sqrt{T+1}}\sum_{t=0}^{T}\ket{\psi_t}\ket{t}.
\]
This state is sometimes called the ``history state'' of the circuit $U$, and you can think of it as the quantum analogue of a satisfying assignment in classical SAT. It faithfully ``follows'' the intermediate states of the computation.
This means $\ket{\psi'}$ gets penalty 0 from $H_{\text{init}}$ and from each~$H_t$. Since the probability of getting (the incorrect) measurement outcome~0 is $\leq 1/4T$, we have
\[
\lambda_{\min}\leq \bra{\psi'}H\ket{\psi'}=\frac{1}{T+1}
\bra{\psi_T}\bra{T}\,H_{\text{final}}\,\ket{\psi_T}\ket{T}=\frac{1}{T+1}
\bra{\psi_T}\left(\ketbra{0}{0}_1\otimes I\right)\ket{\psi_T}\leq \frac{1}{4T(T+1)}:=a.
\]
Second, to prove {\bf soundness} we'll show that if $x\in L_0$, then $\lambda_{\min}$ is at least $b:=2a$.
Consider any purported witness state $\ket{\psi'}$. We can write this as
\[
\ket{\psi'}=\sum_{t=0}^{T}\alpha_t\ket{\phi_t}\ket{t}
\]
for some nonnegative reals $\alpha_t$ and normalized states $\ket{\phi_t}$. 
Note that
\begin{align}
\bra{\psi'}H_t\ket{\psi'}
& =(\alpha_{t-1}\bra{\phi_{t-1}}\bra{t-1}+\alpha_{t}\bra{\phi_{t}}\bra{t})\,H_t\,(\alpha_{t-1}\ket{\phi_{t-1}}\ket{t-1}+\alpha_{t}\ket{\phi_{t}}\ket{t})\nonumber \\
& = \frac{1}{2}\left(\alpha_{t-1}^2 + \alpha_{t}^2 - \alpha_{t-1}\alpha_{t} \bra{\phi_{t}}U_t\ket{\phi_{t-1}} - \alpha_{t-1}\alpha_{t} \bra{\phi_{t-1}}U_t^*\ket{\phi_{t}}\right)\nonumber\\
& =\frac{1}{2}\norm{\alpha_{t}\ket{\phi_{t}}-\alpha_{t-1} U_t\ket{\phi_{t-1}}}^2.\label{eq:Htpenalty}
\end{align}
So intuitively, assuming $\alpha_t\approx \alpha_{t-1}$, the Hamiltonian term $H_t$ gives an energy penalty proportional to how much $\ket{\phi_{t}}$ deviates from $U_t\ket{\phi_{t-1}}$, i.e., from a correct application of the $t$-th gate of $C_n$.

For ease of presentation we will now make the following three simplifying assumptions.%
\footnote{Exercise~\ref{ex:QMAremovesimplif} shows this is without loss of generality, though the lower bound on $b-a$ becomes a worse polynomial ($1/T^6$ instead of $1/T^2$) if we drop these three assumptions and take the unary clock of the next section into account.}
\begin{itemize}
\item All $\alpha_t$ are equal to $1/\sqrt{T+1}$, as they would be in the history state. This assumption is reasonable because $\ket{\phi_t}$ and $U_t\ket{\phi_{t-1}}$ both have norm~1, so differences between $\alpha_t$ and $\alpha_{t-1}$ will only make the penalty of Eq.~\eqref{eq:Htpenalty} bigger.
\item $\ket{\phi_0}$ starts with $s$ 0s, so $\ket{\phi_0}=\ket{0^s}\ket{\psi}$ for some $w(n)$-qubit state $\ket{\psi}$. This is reasonable because if $\ket{\phi_0}$ deviates significantly from this form, then $H_{\text{init}} $ will give a large energy penalty.
\item $\ket{\phi_T}$ has acceptance probability close to~1. This is reasonable because if $\ket{\phi_T}$ has low acceptance probability, then $H_{\text{final}}$ will give a large energy penalty.
\end{itemize}
Because of the second item and the fact that $x\in L_0$, the state $C_n\ket{\phi_0}$ must have acceptance probability close to~0. Comparing with the third item, it follows that $\ket{\phi_T}$ and $C_n\ket{\phi_0}$ must be nearly orthogonal, so their distance is close to $\sqrt{2}$, and in particular at least~1. This implies
\begin{align*}
1 & \leq \norm{\ket{\phi_T}-C_n\ket{\phi_0}} 
= \norm{\sum_{t=1}^{T}U_T\cdots U_{t+1}\ket{\phi_{t}}-U_T\cdots U_{t}\ket{\phi_{t-1}}}\\
& \leq \sum_{t=1}^{T} \norm{U_T\cdots U_{t+1}\ket{\phi_{t}}-U_T\cdots U_{t}\ket{\phi_{t-1}}}
=\sum_{t=1}^{T} \norm{\ket{\phi_{t}}-U_t\ket{\phi_{t-1}}}.
\end{align*}
Here the first equality uses a telescoping sum, the second inequality is the triangle inequality, and the last equality is because the operator norm is unitarily invariant ($\norm{Uv-Uw}=\norm{v-w}$).

Using Eq.~\eqref{eq:Htpenalty} with $\alpha_t=\alpha_{t-1}=1/\sqrt{T+1}$, and Cauchy-Schwarz, we now have:
\begin{align*}
\bra{\psi'}H\ket{\psi'} & \geq \sum_{t=1}^T \bra{\psi'}H_t\ket{\psi'}
 =\frac{1}{2(T+1)}\sum_{t=1}^T \norm{\ket{\phi_{t}}-U_t\ket{\phi_{t-1}}}^2\\
& \geq \frac{1}{2T(T+1)}\left(\sum_{t=1}^{T} \norm{\ket{\phi_{t}}-U_t\ket{\phi_{t-1}}}\right)^2 \geq \frac{1}{2T(T+1)}=b.
\end{align*}
Accordingly, if $x\in L_0$ then 
$\lambda_{\min}=\min_{\ket{\psi'}}\bra{\psi'}H\ket{\psi'}\geq b$. If $x\in L_1$, then the history state witnesses $\lambda_{\min}\leq a=1/(4T(T+1))$.
We also have $b-a=a=1/(4T(T+1))\geq 1/\poly(n)$, as required.  

\subsection{Reducing the locality}\label{ssec:locahamreducing}

The above construction of $H$, with the history state as witness for $x\in L_1$, is due to Kitaev~\cite{kitaev:qnp}, who was inspired by an earlier clock construction of Feynman~\cite{feynman:qmc} (our proof of soundness is a bit different from Kitaev's). He also showed that the locality can be reduced from $O(\log n)$ to 5 by representing the clock in \emph{unary}: $t=0$ would now be represented by $\ket{0^T}_C$, $t=1$ by $\ket{10^{T-1}}_C$, $t=2$ by $\ket{110^{T-2}}_C$, etc. This now requires $T$ qubits to represent the clock instead of $\ceil{\log(T+1)}$. Denoting the $t$-th qubit of the clock by `$C_t$', for $t\in[T]$, the previous terms in $H$ now become
\[
\begin{array}{lll}
H_{\text{init}}  & = & \displaystyle\sum_{i=1}^s \ketbra{1}{1}_i\otimes \ketbra{0}{0}_{C_1}\\[0.3em]
H_t & = & \frac{1}{2}\left(I\otimes\ketbra{100}{100}_{C_{t-1},C_t,C_{t+1}} + I\otimes\ketbra{110}{110}_{C_{t-1},C_t,C_{t+1}} + \right.\\
& & ~~ - \left. U_t\otimes \ketbra{110}{100}_{C_{t-1},C_t,C_{t+1}} -
U_t^*\otimes \ketbra{100}{110}_{C_{t-1},C_t,C_{t+1}}\right)\\[0.2em]
H_{\text{final}} & = & \ketbra{0}{0}_1\otimes\ketbra{1}{1}_{C_T}\\
\end{array}
\]
($H_1$ and $H_T$ have a slightly different form than the $H_t$ written above, because the clock register starts resp.\ ends there.)
We also add the following to penalize a $T$-bit clock register that doesn't conform to the proper 1s-followed-by-0s format of a unary number:
\[
\begin{array}{lll}
\hspace*{-16.1em}H_{\text{clock}} & = & \displaystyle\sum_{t=1}^{T-1}\ketbra{01}{01}_{C_t,C_{t+1}}.
\end{array}
\]
Note that the terms in the Hamiltonian now only ``touch'' at most 5 qubits: in particular, each $H_t$ touches 1 or 2 qubits for the gate $U_t$, and 3 qubits of the clock.
This shows $k$-local Hamiltonian is \QMA-complete for $k\geq 5$.
Subsequently, Kempe, Kitaev, and Regev~\cite{KKR:localham} showed  that $k$-local Hamiltonian is \QMA-complete already for $k=2$ (in contrast to $2$-SAT, which is in \Pp).

All of these results assume a polynomially small gap $b-a$.
Intuitively, the local Hamiltonian problem becomes easier if the gap between $a$ and $b$ becomes bigger, since the 1-instances and 0-instances are further apart and should be easier to distinguish. One may ask whether $k$-local Hamiltonian is still \QMA-complete if $b-a\geq \Omega(m)$ instead of $\geq 1/\poly(n)$. We know that it is at least  \NP-hard for all $k\geq 2$ because of the connection with MAX-2-SAT mentioned at the end of Section~\ref{sec:lochamproblem}, but whether it is also \QMA-hard is a longstanding open problem in the field of Hamiltonian complexity~\cite{GHLS:hamcompl}, known as the ``quantum PCP conjecture''~\cite{aav:qpcp}.

\section{Other interesting problems in \QMA}

The main reason \NP-completeness is a prominent notion in computer science, is that very many interesting and practically important computational problems are \NP-complete: satisfiability, Traveling Salesman, scheduling problems, integer linear programming, protein folding, and many more. See~\cite{garey&johnson:npcomplete} for an already very extensive list from the late-70s.
Similarly (though not yet as extensively) there is a growing list of \QMA-complete problems, often from physics or chemistry. Below we list a few without proof of their \QMA-hardness; the fact that these problems are \emph{in} \QMA\ is usually easy to show. See~\cite{aharonov&naveh:qnp,bookatz:qma} for more.
\begin{itemize}
\item The local Hamiltonian problem for {\bf Hamiltonians corresponding to actual physical systems}, such as 2-local Ising model with 1-local transverse field and a tunable 2-local transverse coupling~\cite{biamonte&love:real}, the 2D Heisenberg Hamiltonian with local magnetic fields~\cite{schuch&vertraete:dft}, the 2D Hubbard Hamiltonian with local magnetic fields~\cite{schuch&vertraete:dft}, and the Bose-Hubbard model~\cite{CGW:BoseHubbard}.
    \item {\bf Non-Identity check}~\cite{JWB:nonidQMA}.
Given a polynomial-size quantum circuit $U$ on $n$ qubits, determine whether $U$ is not close to the identity up to some global phase:\\ 
(1) for all $\phi\in[0,2\pi)$ we have $\norm{U-e^{i\phi}I}\geq b$ or\\ 
(0) there is a $\phi\in[0,2\pi)$ such that $\norm{U-e^{i\phi}I}\leq a$,\\
promised one of these is the case, and $b-a\geq 1/\poly(n)$.
\item {\bf $k$-local density matrix consistency}~\cite{liu:consistency,LCV:NrepQMA,broadbent&grilo:zeroknowledge}.\footnote{A \emph{density matrix} is a generalization of a pure state. See Chapters~\ref{chap:qencodings} and~\ref{chap:qcrypto} for the notation used here.}
Given $m = \poly(n)$ density matrices $\rho_1,\ldots,\rho_m$ where the state of $\rho_i$ is only on qubits $C_i\subseteq[n]$ with $|C_i|\leq k$, determine whether:\\
(1) there is an $n$-qubit density matrix $\rho$ such that for all $i\in[m]$, $\Tr_{[n]\setminus C_i}(\rho)=\rho_i$, or\\ 
(0) for all $n$-qubit density matrices $\rho$ there is an $i\in[m]$ such that $\norm{\Tr_{[n]\setminus C_i}(\rho)-\rho_i}_{tr}\geq b$,\\
promised one of these is the case, and $b\geq 1/\poly(n)$.
\end{itemize}
The following problem is in \QMA, is not known to be \QMA-complete, but also not known to be in \NP\ (in contrast, group \emph{member}ship is known to be in \NP).
\begin{itemize}
\item {\bf Group non-membership}~\cite{watrous:group}.
Given a finite group $G$, subgroup $H\leq G$, and an element $g\in G$, determine whether:\\
(1) $g\not\in H$, or\\
(0) $g\in H$.
\end{itemize}
Here each group could be given for instance as a generating subset of elements, together with the multiplication tables for those elements. The witness state for case (1) is the uniform superposition over $H$. Exercise~\ref{ex:groupnonmembershipwitness} asks you to show completeness; proving soundness is a bit harder.

\section{Quantum interactive proofs}

The prover-verifier protocols in \QMA, just like \NP\ and \MA, only allow one message from the prover to the verifier. This is akin to submitting a proof to a journal, where the referee then verifies the correctness of the proof without further interaction with the prover.
One can also allow multiple rounds of interaction, formalizing the back-and-forth situation that often occurs when a mathematician (the prover) proves a complicated theorem in front of a colleague (the verifier): the verifier can raise objections or ask questions about steps of the proof, which the prover answers (hopefully to the verifier's satisfaction) before proceeding with the next steps. A ``proof'' here is a very general notion: it's any polynomial interaction that convinces the verifier of true statements and cannot convince the verifier about any false statements.
(i.e., it cannot make them accept an $x\not\in L$ with high probability). 

The complexity class \IPr\ consists of those languages that can be decided by a polynomial interaction between an unbounded prover and a polynomial-time classical verifier.
Again, if $x\in L_1$ then the prover should succeed in convincing the verifier to accept (with probability $\geq 2/3$), and if $x\in L_0$ then no matter what the prover does, the verifier should reject with probability $\geq 2/3$.
A fundamental classical complexity theory result says that \IPr=\PSPACE~\cite{LFKN:algebraicIP,shamir:IPPSPACEj,shen:IPPSPACEj}. 

One can define \emph{quantum} \IPr\  (\QIP) analogously.
The two main results known about \QIP\ are: 
\begin{enumerate}
    \item Every \QIP\ protocol can be implemented with only 3 messages, with the prover starting~\cite{watrous:constantroundPSPACE,kitaev&watrous:qip}. Roughly speaking, the 3-message protocol starts from a $\poly(n)$-message protocol for an $L\in\QIP$, using the 3 messages to check one randomly chosen one among the $\poly(n)$ messages. This results in a 3-message protocol with $1/\poly(n)$ gap between completeness and soundness parameters for the same~$L$. This small gap can then be amplified to a constant gap by repeating the protocol in parallel; this repetition increases the communication per message but not the number of messages, which remains~3.
    In contrast, it is widely believed that \IPr\ restricted to 3-message protocols is a much smaller class than the full class \IPr.
    \item \QIP=\IPr~\cite{JJUW10j}. Roughly speaking, this is proved by showing that the acceptance probability of the optimal strategy of the prover in a \QIP-protocol can be described implicitly by an exponential-size semidefinite program. Its optimal value (which is either $\geq 2/3$ or $\leq 1/3$) can then be approximated by an exponential-size but polynomial-depth circuit. Such circuits can be implemented in \PSPACE, and we already knew that \IPr=\PSPACE.
    \end{enumerate}
Accordingly, adding quantum to the model of interactive proofs does not change the class of languages that can be decided, but does reduce the required number of messages between prover and verifier from polynomial to constant. See \cite{vidick&watrous:qproofs} for much more about quantum proof systems.

\section*{Exercises}

\begin{enumerate}

\item Give a satisfying assignment for the 3-SAT instance of Eq.~\eqref{eq:3satinstance}.

\item Show that 1-local Hamiltonian is in \Pp.

\item\label{ex:approxlambdamin} (H) Suppose you had an efficient quantum algorithm for the $k$-local Hamiltonian problem for every $a,b$ that satisfy $b-a\geq 1/n$. Give an efficient quantum algorithm that approximates $\lambda_{\min}$ to within additive error $\pm 2/n$.

\item\label{eq:locHaminQMA} 
Show that $k$-local Hamiltonian is in \QMA\ in two different ways, by providing details for the following two sketches:
\begin{enumerate}
    \item Choose a $j\in[m]$ uniformly at random and measure the observable $H_j$ on the witness state. Repeat this a few times (using new witness states each time) to approximate the expected value.
    \item Apply phase estimation (Section~\ref{ssecphaseestimation}) to the unitary $U=e^{iH/m}$ with the given witness state; $U$ can be implemented using Hamiltonian simulation (Chapter~\ref{chap:hamsim}).
\end{enumerate}

\item\label{ex:QMAremovesimplif} This long  exercise completes the proof of the soundness for the Hamiltonian of Section~\ref{ssec:locahamreducing}, with the unary clock and without the three simplifying assumptions of Section~\ref{sec:lochamsoundcomplete}.
The gap $b-a$ between completeness and soundness will now be $\Omega(1/T^6)$ instead of $\Omega(1/T^2)$.
\begin{enumerate}
    \item Assume the error probability of the \QMA-protocol for $L$ is $\ll 1/T^5$. Let $\ket{\psi'''}$ be an arbitrary ground state (with energy $\lambda_{\min}$) for the Hamiltonian of  Section~\ref{ssec:locahamreducing}. Show that if $x\in L_1$, then $\lambda_{\min}\ll 1/T^6$.
    \item (H) For the remainder of the exercise assume $x\in L_0$.  Let $P_{bc}$ be the projector on the subspace of bad (i.e., non-unary) clock states. Show that $\norm{P_{bc}\ket{\psi'''}}^2\leq\lambda_{\min}$. 
    \item Show that $\norm{H}\leq O(T)$.
    \item (H) Let $\ket{\psi''}$ be the state obtained from $\ket{\psi'''}$ by removing $P_{bc}\ket{\psi'''}$ and renormalizing. Show that its energy $\lambda''$ cannot be much larger than $\lambda_{\min}$: $\lambda''\leq O(T\lambda_{\min})$.
    \item (H) Write $\ket{\psi''}=\sum_{t=0}^{T}\alpha_t\ket{\phi_t}\ket{t}$ for nonnegative reals $\alpha_t$ and normalized states $\ket{\phi_t}$. Show that $\sum_{t=1}^T |\alpha_t-\alpha_{t-1}|^2\leq 2 \lambda''$.
    \item (H) Show that for all $t,t'$ we have $|\alpha_{t'}-\alpha_t|^2\leq 2\lambda''T$.
    \item (H) Let  $\ket{\psi'}=\frac{1}{\sqrt{T+1}}\sum_{t=0}^{T}\ket{\phi_t}\ket{t}$ be the state after making all amplitudes equal in $\ket{\psi''}$.
    Show that $\norm{\ket{\psi''}-\ket{\psi'}}^2\leq O(T^2\lambda'')$.
    \item Modify $\ket{\phi_0}$ and  $\ket{\phi_T}$ to satisfy the second and third simplifying assumptions.
    \item Show that the energy $\lambda'$ of $\ket{\psi'}$ cannot be much larger than $\lambda''$: $\lambda'\leq O(T^3\lambda'')$.
    \item (H) Show that $\lambda_{\min}\geq \Omega(1/T^6)$.
\end{enumerate}

\item\label{ex:QMAwithlogwitnesses} Consider a promise problem $L=(L_1,L_0,L_*)\in\QMA$ and a protocol as in Definition~\ref{def:QMA} with witness states of $w(n)$ qubits.
\begin{enumerate}
\item (H) Show that there is a \QMA\ protocol for $L$ with witness states of $w(n)$ qubits and error probability $\leq \frac{1}{3}2^{-w(n)}$.
\item Suppose we now replace the $w(n)$ qubits of the protocol of (a) with a uniformly random $w(n)$-bit basis state. Show that if $x\in L_1$, then the acceptance probability (i.e., the probability of output~1) is $\geq \frac{2}{3}2^{-w(n)}$, while if $x\in L_0$ then it is $\leq \frac{1}{3}2^{-w(n)}$.
\item Use (b) to show that the class \QMA\ with witness states restricted to $w(n)=O(\log n)$ qubits equals the class \BQP.
\end{enumerate}

\item\label{ex:QMAinPP}
(H) Let \PP\ be the class of promise problems $L=(L_1,L_0,L_*)$ that can be decided by a polynomial-time classical algorithm with success probability $>1/2$ (meaning that for inputs $x\in L_1$ the algorithm accepts with probability $>1/2$, and for $x\in L_0$ it accepts with probability $<1/2$).
Show that $\QMA\subseteq\PP$.

\item\label{ex:locHamBQP} Consider the following computational decision problem. We are given a Hamiltonian~$H$ of the form of page~\pageref{eq:KitaevH}, with the additional property that $w(n)=0$.
We are promised that the smallest eigenvalue $\lambda_{\min}$ of $H$ is either $\leq 1/(4T(T+1))$ (``yes-instance'')
or $\geq 1/(2T(T+1))$ (``no-instance''),
and the problem is to decide which of these two cases we are in.
\begin{enumerate}
    \item (H) Show that this problem is in \BQP.
    \item Show that this problem is \BQP-hard, meaning that for every promise problem~$L=(L_1,L_0,L_*)$ in $\BQP$ there exists a classical deterministic polynomial-time algorithm that maps every $x\in L_1$ to a yes-instance of the above problem and every $x\in L_0$ to a no-instance.
\end{enumerate}

\item\label{ex:groupnonmembershipwitness} 
Suppose we are given (in some form) a finite group $G$, a subgroup $H\leq G$, and an element $g\in G$, and we can efficiently implement the unitary map $V$ corresponding to multiplication with $g$
(i.e., the map $V: \ket{h}\mapsto\ket{h\circ g}$). Let 
\[
\ket{\psi}=\frac{1}{\sqrt{|H|}}\sum_{h\in H}\ket{h}
\]
be the uniform superposition over $H$. The prover for the non-membership problem can construct this state, though not necessarily efficiently.
\begin{enumerate}
    \item Show that if $g\in H$, then $V\ket{\psi}=\ket{\psi}$.
    \item Show that if $g\not\in H$, then $V\ket{\psi}$ is orthogonal to $\ket{\psi}$.
    \item Consider the following procedure that the verifier can use to test if $g\in H$ or not:\\ 
    (1) prepare an auxiliary qubit in state $H\ket{0}$,\\ 
    (2) conditioned on that qubit apply $V$ to $\ket{\psi}$,\\ 
    (3) apply $H$ to the auxiliary qubit and measure it.\\
    Show that the probability of measurement outcome~0 is 1 if $g\in H$, and is 1/2 if~$g\not\in H$.
\end{enumerate}
\end{enumerate}

\chapter{Quantum Encodings, with a Non-Quantum Application}\label{chap:qencodings}

\section{Mixed states and general measurements}\label{sec:mixedstates}

So far, we have restricted our states to so-called \emph{pure states}: unit vectors of amplitudes.
In the classical world we often have uncertainty about the state of a system,
which can be expressed by viewing the state as a random variable that has a certain
probability distribution over the set of basis states.  Similarly we can define a
\emph{mixed} quantum state as a probability distribution (or ``mixture'') over pure states.
While pure states are written as vectors, it is most convenient to write mixed states as
\emph{density matrices}. A pure state $\ket{\phi}$ corresponds to the density matrix $\ketbra{\phi}{\phi}$,
which is the outer product of the vector $\ket{\phi}$ with itself. For example, the pure state $\ket{\phi}=\alpha\ket{0}+\beta\ket{1}$ corresponds to the density matrix 
$$
\ketbra{\phi}{\phi}
=\left(\begin{array}{c}\alpha\\ \beta\end{array}\right)\cdot\left(\alpha^*~~~\beta^*\right)
=\left(\begin{array}{cc}|\alpha|^2 & \alpha\beta^*\\ \alpha^*\beta & |\beta|^2\end{array}\right).
$$
A mixed state that is in pure states $\ket{\phi_1},\ldots,\ket{\phi_\ell}$ with probabilities $p_1,\ldots,p_\ell$,
respectively, corresponds to the density matrix $\rho=\sum_{i=1}^\ell p_i\ketbra{\phi_i}{\phi_i}$.
This $\rho$ is sometimes called a ``mixture'' of the states $\ket{\phi_1},\ldots,\ket{\phi_\ell}$.\footnote{Note that
applying the probabilities $p_i$ to the vectors $\ket{\phi_i}$ (rather than to the matrices 
$\ketbra{\phi_i}{\phi_i}$) does not make sense in general, because $\sum_{i=1}^\ell p_i\ket{\phi_i}$ need not be a unit vector.
Using square roots of the probabilities also doesn't work, because $\sum_{i=1}^\ell \sqrt{p_i}\ket{\phi_i}$ need not be a unit vector either if the $\ket{\phi_i}$ are not pairwise orthogonal.} 
The set of density matrices is exactly the set of positive semidefinite (psd) matrices of trace~1.
A mixed state is pure if, and only if, it has rank~1.

You can always write a mixed state $\rho$ as a probability distribution over \emph{orthogonal} pure states, using the diagonalization of $\rho$ (see Appendix~\ref{secsvd}) plus the observations that (1) the eigenvalues of a trace-1 psd matrix form a probability distribution, and (2) that the eigenvectors of a Hermitian matrix can be assumed to form an orthonormal set without loss of generality. But you can also write $\rho$ as a convex combination of non-orthogonal states (see Exercise~\ref{ex:samerho}.c).

Applying a unitary $U$ to a pure state $\ket{\phi}$ gives pure state $U\ket{\phi}$.
Written in terms of rank-1 density matrices, this corresponds to the map 
$$
\ketbra{\phi}{\phi}\mapsto U\ketbra{\phi}{\phi}U^*.
$$
By linearity, this actually tells us that a unitary acts on an arbitrary mixed state by conjugation:
$$
\rho\mapsto U\rho U^*.
$$
What about measurements? Recall from Section~\ref{ssecmeasurement} that an $m$-outcome projective measurement corresponds to $m$ orthogonal projectors
$P_1,\ldots,P_m$ that satisfy $\sum_{i=1}^m P_i=I$.
When applying this measurement to a mixed state $\rho$, the probability to see outcome $i$ is given by
$p_i=\Tr(P_i\rho)$. If we get outcome $i$, then the state collapses to $P_i\rho P_i/p_i$
(the division by $p_i$ renormalizes the state to have trace~1).
This may look weird, but let's recover our familiar measurement in the computational basis in this framework.
Suppose we measure a state $\ket{\phi}=\sum_{j=1}^d\alpha_j\ket{j}$ using $d$ projectors
$P_i=\ketbra{i}{i}$ (note that $\sum_i P_i$ is the identity on the $d$-dimensional space).
The probability to get outcome $i$ is given by $p_i=\Tr(P_i\ketbra{\phi}{\phi})=|\inp{i}{\phi}|^2=|\alpha_i|^2$.
If we get outcome $i$ then the state collapses to $P_i\ketbra{\phi}{\phi}P_i/p_i=\alpha_i\ketbra{i}{i}\alpha_i^*/p_i=\ketbra{i}{i}$.
This is exactly the measurement in the computational basis as we have used it until now.
Similarly, a measurement of the first register of a two-register state corresponds to
projectors $P_i=\ketbra{i}{i}\otimes I$, where $i$ goes over all basis states of the first register.

If we only care about the final probability distribution on the $m$ outcomes, 
not about the resulting state, then the most general thing we can do is
a POVM.  This is specified by $m$ positive semidefinite matrices
$E_1,\ldots,E_m$ satisfying $\sum_{i=1}^m E_i=I$. When measuring a state $\rho$, the probability of outcome $i$ is given by $\Tr(E_i\rho)$.

\section{Quantum encodings and their limits}\label{sec:qencandlimits}

Quantum information theory studies the quantum generalizations of familiar notions from classical information theory
such as Shannon entropy, mutual information, channel capacities, etc.
Here we will discuss a few quantum information-theoretic results that all have the same flavor:
they say that a low-dimensional quantum state (i.e., a small number of qubits) cannot contain too much \emph{accessible} information.

\paragraph{Holevo's Theorem:}
The mother of all such results is Holevo's theorem from 1973~\cite{holevo}, which predates the area of quantum computing by several decades.
Its proper technical statement is in terms of a quantum generalization of mutual information,
but the following consequence of it (derived by Cleve et al.~\cite{cdnt:ip}) about two communicating parties,
suffices for our purposes.

\begin{theorem}[Holevo, CDNT]\label{thholevo}
Suppose Alice wants to communicate some classical string $x$ to Bob.
\begin{itemize}
\item If Alice sends Bob $m$ qubits, and they did not share any prior entanglement, then Bob receives at most $m$ bits of information about~$x$.
\item If Alice sends Bob $m$ qubits, and they did share some prior entangled state, then Bob receives at most $2m$ bits of information about~$x$, no matter how many qubits Bob sent to Alice.
\item If Alice sends Bob $m$ classical bits, and they did share some prior entangled state, then Bob receives at most $m$ bits of information about~$x$, no matter how many qubits Bob sent to Alice.
\end{itemize}
\end{theorem}

This theorem is slightly imprecisely stated here, but the intuition should be clear: if Bob makes any measurement on his state after the communication, then the mutual information between his classical outcome and Alice's $x$, is bounded by $m$ or $2m$.
In particular, the first part of the theorem says that if we encode some classical random variable $X$ in an $m$-qubit state\footnote{Via an encoding map $x\mapsto\rho_x$; we generally use upper-case letters
like $X$ to denote random variables, lower-case letters like~$x$ to denote specific values.}, then no measurement on the quantum
state can give more than $m$ bits of information about $X$.
If we encoded the classical information in an $m$-\emph{bit} system instead of an $m$-qubit system
this would be a trivial statement, but the proof of Holevo's theorem is quite non-trivial.
Thus we see that an $m$-qubit state, despite somehow ``containing'' $2^m$ complex amplitudes,
is no better than $m$ classical bits for the purpose of storing or transmitting information.
Prior entanglement can improve this by a factor of~2 because of superdense coding (see Exercise~\ref{chap:intro}.\ref{ex:superdense}), but no more than that.

\paragraph{Low-dimensional encodings:}
Here we provide a ``poor man's version'' of Holevo's theorem due to Nayak~\cite[Theorem~2.4.2]{nayak:qfa},
which has a simple proof and often suffices for applications.
Suppose we have a classical random variable $X$, uniformly distributed over $[N]=\{1,\ldots,N\}$.\footnote{NB: unlike in most of these lecture notes, $N$ need not equal $2^n$ in this chapter!}
Let $x\mapsto\rho_x$ be some encoding of the elements of~$[N]$, where $\rho_x$ is a mixed state in a $d$-dimensional space.
Let $E_1,\ldots,E_N$ be the POVM operators applied for decoding; these sum to the $d$-dimensional identity operator.
Then the probability of correct decoding in case $X=x$, is
$$
p_x=\Tr(E_x\rho_x)\leq \Tr(E_x).
$$
The sum of these success probabilities is at most
\begin{equation}\label{nayakpibound}
\sum_{x=1}^N p_x\leq \sum_{x=1}^N \Tr(E_x)=\Tr\left(\sum_{x=1}^N E_x\right)=\Tr(I)=d.
\end{equation}
In other words, if we are encoding one of $N$ classical values in a $d$-dimensional quantum state,
then any measurement to decode the encoded classical value has average success probability at most $d/N$
(uniformly averaged over all $N$ values that we can encode).
For example, if we encode $n$ uniformly random bits into $m$ qubits, we will have $N=2^n$, $d=2^m$,
and the average success probability of decoding is at most $2^m/2^n$, which is very small unless $m$ is nearly~$n$.

\paragraph{Random access codes:}
The previous two results dealt with the situation where we encoded a classical random variable $X$ in
some quantum system, and would like to recover the original value $X$ by an appropriate
measurement on that quantum system.  However, suppose $X=X_1\ldots X_n$ is an $n$-bit random variable,
uniformly distributed and encoded by a map $x\mapsto\rho_x$, 
and it suffices for us if we are able to decode individual bits $X_i$ from this with some probability $p>1/2$.  
More precisely, for each $i\in[n]$ there should exist
a measurement $\{M_i,I-M_i\}$ allowing us to recover $x_i$. $M_i$ would correspond to output~1 and $I-M_i$ to output~0. Hence for each $x\in\01^n$
we should have $\Tr(M_i\rho_x)\geq p$ if $x_i=1$ and $\Tr(M_i\rho_x)\leq 1-p$ if $x_i=0$.
An encoding satisfying this is called a \emph{quantum random access code}, since it allows
us to choose which bit of $X$ we would like to access. Note that the measurement to recover $x_i$
can change the state $\rho_x$, so generally we may not be able to decode more than one bit of $x$
(also, we cannot copy $\rho_x$ because of the no-cloning theorem, see Exercise~\ref{chap:intro}.\ref{ex:nocloning}).

An encoding that allows us to recover (with high success probability)
an $n$-bit string requires about $n$ qubits by Holevo.
Random access codes only allow us to recover \emph{each} of the $n$ bits.
Can they be much shorter? In small cases they can be: for instance, one can encode two classical bits
into one qubit, in such a way that each of the two bits can be recovered with
success probability $85\%$ from that qubit (see Exercise~\ref{ex:2to1rac}).
However, Nayak~\cite{nayak:qfa}
proved that asymptotically quantum random access codes cannot be much shorter than classical, even if their success probability $p$ only needs to be an average over all~$x$ rather than worst-case.

\begin{theorem}[Nayak]\label{thrac}
Let $x\mapsto\rho_x$ encode $n$-bit strings~$x$ into $m$-qubit states~$\rho_x$ such that, for each $i \in [n]$, we can decode $x_i$ from $\rho_x$ with success probability $p$ (averaged over a uniform choice of $x$ and over the measurement randomness).
Then $m\geq (1-H(p))n$, where $H(p)=-p\log p - (1-p)\log(1-p)$ is the binary entropy function.
\end{theorem}

The intuition of the proof is quite simple: if the quantum state allows us to predict the bit $X_i$ with
probability $p_i$, then it reduces the ``uncertainty'' about the random variable $X_i$ from 1~bit to $H(p_i)$~bits.
Hence it contains at least $1-H(p_i)$ bits of information about $X_i$. Since all $n$ $X_i$'s
are independent, the state has to contain at least $\sum_{i=1}^n(1-H(p_i))$ bits of information about~$X$.
For more technical details see~\cite{nayak:qfa} or Appendix~B of~\cite{kerenidis&wolf:qldcj}.
The lower bound on $m$ can be achieved up to an additive $O(\log n)$ term, even by classical probabilistic encodings.

\section{Lower bounds on locally decodable codes}\label{sec:ldclowerbound}

Here we will give an application of  quantum information theory to a \emph{classical} problem.%
\footnote{There is a growing number of such applications of quantum tools to non-quantum problems. See~\cite{drucker&wolf:qproofs} for a survey.}

The development of error-correcting codes is one of the success stories of science in the second half of the 20th century.
Such codes are eminently practical, and are widely used to protect information stored on discs, communication over noisy channels, etc.
From a theoretical perspective, there exist codes that are nearly optimal in a number of different respects simultaneously:
they have constant rate, can protect against a constant noise-rate, and have linear-time encoding and decoding procedures.
We refer to Trevisan's survey~\cite{trevisan:eccsurvey} for a complexity-oriented discussion of codes and their applications.

One drawback of ordinary error-correcting codes is that we cannot efficiently decode small parts of the encoded
information. If we want to learn, say, the first bit of the encoded message then we usually still need to decode the whole encoded string.
This is relevant in situations where we have encoded a very large string (say, a library of books, or a large database),
but are only interested in recovering small pieces of it at any given time.
Dividing the data into small blocks and encoding each block separately will not work:
small chunks will be efficiently decodable but not error-correcting, since a tiny fraction of
well-placed noise could wipe out the encoding of one chunk completely.
There exist, however, error-correcting codes that are \emph{locally decodable}, in the sense that we can efficiently recover
individual bits of the encoded string. 

\begin{definition}
$C:\01^n\rightarrow\01^N$ is a \emph{$(q,\delta,\eps)$-locally decodable code}
(LDC) if there is a classical randomized decoding algorithm $A$ such that
\begin{enumerate}
\item $A$ makes at most $q$ queries to an $N$-bit string $y$.
\item For all $x\in\01^n$ and $i\in[n]$, and all $y\in\01^N$ with Hamming distance
$d(C(x),y)\leq\delta N$ we have $\Pr[A^y(i)=x_i]\geq 1/2+\eps$.
\end{enumerate}
\end{definition}

Here $\delta$ is an upper bound on the fraction of bits of the codeword that may have been corrupted (by some noise process, or by our worst enemy), and $\eps$ is a lower bound on the advantage we have compared to just randomly guessing the value of the bit~$x_i$. 
The notation $A^y(i)$ reflects that the decoder $A$ has two different types of input.
On the one hand there is the (possibly corrupted) codeword $y$, to which the decoder has oracle access and from which it can read at most $q$
bits of its choice.  On the other hand there is the index $i$ of the bit that needs to be recovered, and which is known fully to the decoder.

The main question about LDCs is the tradeoff between the codelength $N$ and the number of queries $q$ (which is a proxy for the decoding-time).
This tradeoff is still not very well understood.
The only case where we know the answer is the case of $q=2$ queries.\footnote{For $q=1$, LDCs don't 
exist once $n$ is sufficiently large~\cite{katz&trevisan:ldc}. For $q=3$, the best upper bound known on the codelength~$N$ is roughly $2^{2^{\sqrt{\log n}}}$~\cite{yekhanin:3ldcj,efremenko:ldc}, while the best known lower bound is roughly $n^3$~\cite{kerenidis&wolf:qldc,wehner&wolf:improvedldc,woodruff:3querylower,Alrabiah2023NearCubic}. An exponential lower bound was recently proved for the special case of \emph{linear} 3-query LDCs~\cite{KothariManohar:3LDC}.}
For $q=2$ there is the Hadamard code: given $x\in\01^n$, define a codeword of length $N=2^n$ by writing down
the bits $x\cdot z$ mod 2, for all $z\in\01^n$, with the $z$'s ordered in lexicographic order. For example for $n=2$ and $x=10$, the codeword would be
\[
C(x)=(x\cdot 00,x\cdot 01,x\cdot 10,x\cdot 11)=0011.
\]
One can decode $x_i$ with 2 queries as follows: choose $z\in\01^n$ uniformly at random, query the (possibly corrupted) codeword at indices $z$ and $z\oplus e_i$ (where the latter denotes the string obtained from $z$ by flipping its $i$-th bit), and output the sum of the two returned bits modulo~2.
Individually, each of these two indices $z$ and $z\oplus e_i$ is uniformly
distributed.  Hence for each of them, the probability that the returned bit is corrupted is at most $\delta$. By the union bound, with probability
at least $1-2\delta$, both queries return the uncorrupted values. Adding these two bits mod 2 gives the correct answer:
$$
C(x)_z\oplus C(x)_{z\oplus e_i}=(x\cdot z)\oplus (x\cdot (z\oplus e_i))=x\cdot e_i=x_i.
$$
Thus the Hadamard code is a $(2,\delta,1/2-2\delta)$-LDC of exponential length.

The only superpolynomial \emph{lower bound} known on the length of general LDCs is for the case of 2 queries:
there one needs an exponential codelength and hence the Hadamard code is essentially optimal. 
This is shown via a \emph{quantum} argument~\cite{kerenidis&wolf:qldcj}---despite the fact that
the result is a purely classical result, about classical codes and classical decoders.
The easiest way to present this argument is to assume the following fact, which states a kind of
``normal form'' for the decoder.

\begin{fact}[Katz \&\ Trevisan~\cite{katz&trevisan:ldc} + folklore]\label{ldcfact}
For every $(q,\delta,\eps)$-LDC $C:\01^n\rightarrow\01^N$, and for each $i\in[n]$, there exists a set ${\cal M}_i$ of $\Omega(\delta\eps N/q^2)$ disjoint tuples,
each of at most $q$ indices from $[N]$, and a bit $a_{i,t}$ for each tuple $t\in{\cal M}_i$, such that the following holds:
\begin{equation} \label{eqldcfact}
\Pr_{x\in\01^n}\left[x_i = a_{i,t}\oplus\sum_{j\in t}C(x)_j\right]\geq 1/2+\Omega(\eps/2^q),
\end{equation}
where the probability is taken uniformly over $x$.
Hence to decode $x_i$ from $C(x)$, the decoder can just query the indices in a uniformly-randomly chosen tuple $t$ from ${\cal M}_i$,
outputting the sum of those $q$ bits and $a_{i,t}$.
\end{fact}

\noindent
Note that the above decoder for the Hadamard code is already of this form, with ${\cal M}_i$ consisting of the $2^{n-1}$ pairs $\{z,z\oplus e_i\}$. We omit the fairly easy proof of Fact~\ref{ldcfact}, which uses purely classical ideas.

Now suppose $C:\01^n\rightarrow\01^N$ is a $(2,\delta,\eps)$-LDC.  We want to show that the codelength $N$ must be exponentially large in $n$.
Our strategy is to show that the following $N$-dimensional quantum encoding is a quantum random access code for $x$
(with some success probability $p>1/2$):
$$
x\mapsto \ket{\phi_x}=\frac{1}{\sqrt{N}} \sum_{j=1}^{N} (-1)^{C(x)_j}\ket{j}.
$$
Theorem~\ref{thrac} then implies that the number of qubits of this state (which is $\ceil{\log N}$) is at least $(1-H(p))n=\Omega(n)$, and we are done.

Suppose we want to recover $x_i$ from $\ket{\phi_x}$.
We'll do this by a sequence of two measurements, as follows.
We turn each ${\cal M}_i$ from Fact~\ref{ldcfact} into a projective measurement:
for each pair $(j,k)\in {\cal M}_i$ form the projector $P_{jk}=\ketbra{j}{j}+\ketbra{k}{k}$,
and let $P_{rest}=\sum_{j\not\in\cup_{t\in {\cal M}_i}t}\ketbra{j}{j}$ be the projector on the remaining indices.
These $|{\cal M}_i|+1$ projectors sum to the $N$-dimensional identity matrix, so they form a valid projective measurement.
Applying this to $\ket{\phi_x}$ gives outcome $(j,k)$ with probability $\norm{P_{jk}\ket{\phi_x}}^2=2/N$ for each $(j,k)\in {\cal M}_i$.
There are $|{\cal M}_i|=\Omega(\delta\eps N)$ different $(j,k)$-pairs in ${\cal M}_i$, so the probability to see one of those 
as outcome of the measurement, is $|{\cal M}_i|\cdot 2/N=\Omega(\delta\eps)$.
With the remaining probability $r=1-\Omega(\delta\eps)$, we'll get ``rest'' as outcome of the measurement.
In the latter case we didn't get anything useful from the measurement, so we'll just output a fair coin flip as our guess for $x_i$
(then the output will equal $x_i$ with probability exactly 1/2).  
In case we got one of the $(j,k)$ as measurement outcome, the state has collapsed to the following useful superposition:
$$
\frac{1}{\sqrt{2}}\left((-1)^{C(x)_j}\ket{j}+(-1)^{C(x)_k}\ket{k}\right)=\frac{(-1)^{C(x)_j}}{\sqrt{2}}\left(\ket{j}+(-1)^{C(x)_j\oplus C(x)_k}\ket{k}\right)
$$
We know what $j$ and $k$ are, because it is the outcome of the measurement on $\ket{\phi_x}$.
Now do a 2-outcome projective measurement with projectors $P_0$ and $P_1$ corresponding to the two vectors $\frac{1}{\sqrt{2}}(\ket{j}+\ket{k})$ and $\frac{1}{\sqrt{2}}(\ket{j}-\ket{k})$, respectively.
The measurement outcome equals the value $C(x)_j\oplus C(x)_k$ with probability~1.
By Eq.~(\ref{eqldcfact}), if we add the bit $a_{i,(j,k)}$ to this, we get $x_i$ with probability at least $1/2+\Omega(\eps)$.
The success probability of recovering $x_i$, averaged over all $x$, is
$$
p\geq \frac{1}{2}r + \left(\frac{1}{2}+\Omega(\eps)\right)(1-r)=\frac{1}{2}+\Omega(\delta\eps^2).
$$
Thus we have constructed a random access code that encodes $n$ bits into $\log N$ qubits,
and has success probability at least $p$.
Applying Theorem~\ref{thrac} and using that
\begin{equation}
1-H(1/2+\eta)=\Theta(\eta^2)\mbox{ for }\eta\in[0,1/2],
\end{equation}
we obtain the following:

\begin{theorem}
If $C:\01^n\rightarrow\01^N$ is a $(2,\delta,\eps)$-locally decodable code, then $N\geq 2^{\Omega(\delta^2\eps^4 n)}$.
\end{theorem}

\section*{Exercises}

\begin{enumerate}

\item\label{ex:samerho}
Suppose we have a qubit whose density matrix is
$\rho$.
\begin{enumerate}
\item Show that there exist real numbers $r_0,r_1,r_2,r_3$ such that 
$\rho=\frac{r_0}{2} I + \frac{r_1}{2}X + \frac{r_2}{2}Y + \frac{r_3}{2}Z$, where $I,X,Y,Z$ are the Pauli matrices (see Appendix~\ref{app:pauli}).
\item Show that $r_0=1$.
\item Show that $\Tr(\rho^2)=\frac{1}{2}(r_0^2+r_1^2+r_2^2+r_3^2)$.
\item Show that $r_1^2+r_2^2+r_3^2\leq 1$. 
\item Show that $r_1^2+r_2^2+r_3^2=1$ iff $\rho$ is a pure state. 
\end{enumerate}
{\footnotesize Comment: One can represent a qubit by the corresponding vector $(r_1,r_2,r_3)\in\mathbb{R}^3$. By part (e) the pure states are exactly the ones that are on the boundary of the 3-dimensional sphere of radius~1. The mixed states are in the interior of the sphere, and the maximally mixed state $\rho=I/2$ is at the origin~$(0,0,0)$. This geometric picture is called the \emph{Bloch-sphere} representation of a qubit, and is very useful in physics. For example, single-qubit gates correspond to rotations on this sphere. Unfortunately this picture does not generalize cleanly to more than one qubit.}

\item\label{ex:2to1rac}
\begin{enumerate}
\item (H) Give a quantum random access code that encodes 2 classical bits into 1 qubit, such that each of the two classical
bits can be recovered from the quantum encoding with success probability $p\geq 0.85$.
\item Prove an upper bound of $1/2+O(1/\sqrt{n})$ on the success probability $p$ for every random access code that encodes $n$ classical bits into 1 qubit.
\end{enumerate}

\item\label{ex:teleportopt} (H)
Teleportation transfers an arbitrary unknown qubit from Alice to Bob,
using 1 EPR-pair and 2 classical bits of communication from Alice to Bob (see Section~\ref{sec:teleport}).
Prove that these 2 bits of communication are necessary, i.e., you cannot teleport an arbitrary unknown qubit using 1 EPR-pair and only 1 classical bit of communication.

\item Suppose $n+1=2^k$ for some integer $k$.
For $\ell\in\{0,\ldots,n\}$ define $n$-qubit state $$\ket{\psi_\ell}=\frac{1}{\sqrt{\binom{n}{\ell}}}\sum_{x\in\01^n: |x|=\ell}\ket{x},
$$
where $|x|$ denotes the Hamming weight (number of 1s) in $x$.
\begin{enumerate}
\item Show that $\inp{\psi_\ell}{\psi_{\ell'}}$ equals 1 if $\ell=\ell'$, and equals 0 otherwise.
\item 
Consider a qubit $\ket{\phi}=\alpha\ket{0}+\beta\ket{1}$.
Show that the $n$-qubit state $\ket{\phi}^{\otimes n}$ can be written as a linear combination of the states $\ket{\psi_\ell}$. Say explicitly what the coefficients of this linear combination are.
\item Give a unitary $V$, independent of $\alpha,\beta$, that encodes $\ket{\phi}^{\otimes n}$ into a $k$-qubit state $\ket{\psi}$ in the sense that
$$
V:\ket{\phi}^{\otimes n}\mapsto\ket{\psi}\otimes\ket{0^{n-k}}.
$$
Say explicitly what your state $\ket{\psi}$ is  and how it depends on $\alpha$ and $\beta$ (you're not required to write out circuits).
\end{enumerate}

\item Consider the Hadamard code $C$ that encodes $n=2$ bits $x_1x_2$ into a codeword of $N=4$ bits.
\begin{enumerate}
\item Give the 4-bit codeword $C(11)$.
\item What are the states $\ket{\phi_x}$ that arise as quantum random access code when we apply the LDC lower bound proof of Section~\ref{sec:ldclowerbound} to~$C$? Give the 4 states, not one general formula.
\item What is the measurement used for recovering $x_2$ from $\ket{\phi_x}$ at the end of that proof? You may either describe this as a sequence of two projective measurements, or as one (combined) projective measurement.
\end{enumerate}

\item\label{ex:qftcommunication} 
\begin{enumerate}
\item Let $x\in\01^n$. Suppose we apply the $2n$-qubit Fourier transform $F_{2^{2n}}$ on the $2n$-bit basis state $\ket{x}\ket{0^n}$, followed by $F_{2^n}^{-1}$ on the last $n$ qubits (and identity on the first $n$ qubits).
Show that we end up with the $2n$-qubit state $\ket{+}^{\otimes n}\ket{x}$. 
\item (H) Consider a circuit $C$ that implements $F_{2^{2n}}$ in some way using arbitrary 1-qubit and 2-qubit gates ($C$ can do anything, it need not be one of the specific QFT circuits from the lecture notes).
Show that there must be $\Omega(n)$ 2-qubit gates in $C$ where the control bit lies in the first $n$ qubits of the state and the target qubit lies in the second $n$ qubits (or vice versa).
\end{enumerate}

\item\label{ex:dist2states} Suppose there are two classically-known mixed states $\rho_0$ and $\rho_1$, and we are given one copy of quantum state $\rho_b$ for a uniformly random $b\in\01$. We want to learn $b$ using some 2-outcome projective measurement with operators $P_0$ and $P_1$, which we can choose ourselves depending on what $\rho_0$ and $\rho_1$ are. The success probability of such a measurement is $\frac{1}{2}(\Tr(P_0\rho_0)+\Tr(P_1\rho_1))$.
\begin{enumerate}
\item (H) Give a projective measurement with success probability $\geq\frac{1}{2}+\frac{1}{4}\norm{\rho_0-\rho_1}_1$, where the norm $\norm{A}_1$ of a matrix $A$ is defined as the sum of $A$'s singular values.
\item Show that every 2-outcome projective measurement $P_0,P_1$ has a success probability that is $\leq\frac{1}{2}+\frac{1}{4}\norm{\rho_0-\rho_1}_1$.
\end{enumerate}

\end{enumerate}

\chapter{Quantum Communication Complexity}\label{chap:qcomcompl}

Communication complexity was first introduced by Yao~\cite{yao:distributive}, and has been studied extensively in the area of
theoretical computer science and has deep connections
with seemingly unrelated areas, such as VLSI design, circuit lower
bounds, lower bounds on branching programs, sizes of data structures, and bounds on the length
of logical proof systems, to name just a few.

\section{Classical communication complexity}\label{sec:setting}

\begin{figure}[htb]
\centering
\includegraphics{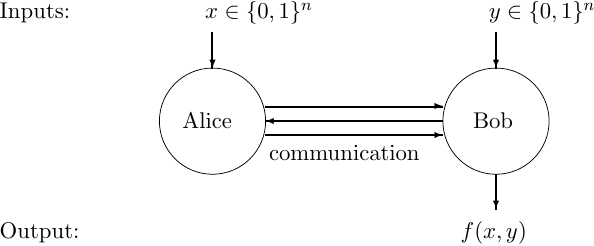}
\caption{Alice and Bob solving a communication complexity problem}\label{fig:cc}
\end{figure}

First we sketch the setting for classical communication complexity.
Alice and Bob want to compute some function $f: \mathcal{D}\rightarrow\01$,
where $\mathcal{D}\subseteq X\times Y$.%
\footnote{If the domain $\mathcal{D}$ equals $X\times Y$ then $f$ is called
a \emph{total} function, otherwise it is called a \emph{partial} or \emph{promise} function.}
Alice receives input $x\in X$, Bob receives input $y\in Y$, with
$(x,y)\in\mathcal{D}$.
A typical situation, illustrated in Fig.~\ref{fig:cc}, 
is where $X=Y=\01^n$, so both Alice and Bob receive an $n$-bit input string.
As the value $f(x,y)$ will generally depend on both $x$ and~$y$,
some communication between Alice and Bob is required in order
for them to be able to compute $f(x,y)$.
We are interested in the \emph{minimal} amount of communication they need.

A communication \emph{protocol} is a distributed algorithm where
first Alice does some individual computation, and then sends a message
(of one or more bits) to Bob, then Bob does some computation and sends
a message to Alice, etc. Each message is called a \emph{round}.
After one or more rounds the protocol terminates and one of the parties (let's say Bob) 
outputs some value that should be $f(x,y)$.
The \emph{cost} of a protocol is the total number of bits communicated on the worst-case input.
A \emph{deterministic} protocol for $f$ always has to output
the right value $f(x,y)$ for all $(x,y)\in\mathcal{D}$.
In a \emph{bounded-error} protocol, Alice and Bob may
flip coins and  the protocol has to output the right
value $f(x,y)$ with probability $\geq 2/3$ for all $(x,y)\in\mathcal{D}$.
We could either allow Alice and Bob to toss coins
individually (local randomness, or ``private coin'') or jointly (shared randomness, or ``public coin'').
A public coin can simulate a private coin and is potentially more powerful.
However, Newman's theorem~\cite{newman:random} says that having a public coin can save
at most $O(\log n)$ bits of communication, compared to a protocol with a private coin.

To illustrate the power of randomness, let us give a simple yet efficient
bounded-error protocol for the equality problem, where
the goal for Alice is to determine whether her $n$-bit input is the same as Bob's or not: 
$f(x,y)=1$ if $x=y$, and $f(x,y)=0$ otherwise.
Alice and Bob jointly toss a random string $r\in\01^n$.
Alice sends the bit $a=x\cdot r$ to Bob
(where `$\cdot$' is inner product mod~2).
Bob computes $b=y\cdot r$ and compares this with $a$.
If $x=y$ then $a=b$, but if $x\neq y$ then $a\neq b$ with probability~1/2.
Repeating this a few times, Alice and Bob can decide equality with small error probability
using $O(n)$  public coin flips and a constant amount of communication.
This protocol uses public coins, but note that Newman's theorem implies that there exists an $O(\log n)$-bit protocol that uses a private coin (see Exercise~\ref{ex:qfpfrompolys} for an explicit protocol).
Note that the correct output of the equality function depends on all $n$ bits of~$x$, but Bob does not need to learn all $n$ bits of $x$ in order to be able to decide equality with high success probability.
In contrast, one can show that \emph{deterministic} protocols for the equality problem need $n$ bits of communication, so then Alice might as well just send $x$ to Bob.

\section{The quantum question}

Now what happens if we give Alice and Bob a quantum computer
and allow them to send each other qubits and/or to make use of EPR-pairs
that they share at the start of the protocol?

Formally speaking, we can model a quantum protocol as follows.
The total state consists of 3 parts: Alice's private space,
the channel, and Bob's private space.
The starting state is $\ket{x}\ket{0}\ket{y}$:
Alice gets $x$, the channel is initialized to 0, and Bob gets $y$.
Now Alice applies a unitary transformation to her space and
the channel. This corresponds to her private computation as well
as to putting a message on the channel (the length of this message
is the number of channel-qubits affected by Alice's operation).
Then Bob applies a unitary transformation to his space and the channel, etc.
At the end of the protocol Alice or Bob makes a measurement
to determine the output of the protocol.
This model was introduced by Yao~\cite{yao:qcircuit}.

In the second model, introduced by Cleve and
Buhrman~\cite{cleve&buhrman:subs}, Alice and Bob share an unlimited
number of EPR-pairs at the start of the protocol,
but now they communicate via a \emph{classical} channel:
the channel has to be in a classical state throughout the protocol.
We only count the communication, not the number of EPR-pairs used.
Protocols of this kind can simulate protocols of the first kind with only a factor 2 overhead:
using teleportation, the parties can send each other a qubit using an EPR-pair
and two classical bits of communication. Hence the qubit-protocols that we describe below also
immediately yield protocols that work with entanglement and a classical channel.
Note that an EPR-pair can simulate a public coin toss:
if Alice and Bob each measure their half of the pair of qubits, they get the same random bit.

The third variant combines the strengths of the other two:
here Alice and Bob start out with an unlimited number of EPR-pairs
\emph{and} they are allowed to communicate qubits.
This third kind of communication complexity is in fact equivalent to the
second, up to a factor of 2, again by teleportation.

Before continuing to study this model, we first have to face
an important question: \emph{is there anything to be gained here?}
At first sight, the following argument seems to rule out any significant gain.
Suppose that in the classical world $k$ bits
have to be communicated in order to compute $f$.
Since Holevo's theorem says that $k$ qubits cannot contain more
information than $k$ classical bits, it seems that the quantum
communication complexity should be roughly $k$ qubits as well
(maybe $k/2$ to account for superdense coding, but not less).
Surprisingly (and fortunately for us), this argument is false, and quantum
communication can sometimes be much less than classical communication
complexity. The information-theoretic argument via Holevo's theorem
fails, because Alice and Bob do not need to communicate the information
in the $k$ bits of the classical protocol;
they are only interested in the value $f(x,y)$, which is just 1 bit.
Below we will go over four of the main examples that have so far been found of
differences between quantum and classical communication complexity.
These are large \emph{provable}
 quantum advantages in a \emph{realistic} model of computation, in contrast to Shor's factoring algorithm (which is in a realistic model of computation but where we are unable to prove strong classical lower bounds) and the advantages we know in query complexity (which are provable but rely on the somewhat unrealistic idea of black-box access to the input bits).

\section{Example~1: Distributed Deutsch-Jozsa}\label{secdistributeddj}

The first impressively large gaps between quantum and classical
communication complexity were exhibited by Buhrman, Cleve, and Wigderson~\cite{BuhrmanCleveWigderson98}.
Their protocols are distributed versions of known quantum
query algorithms, like the Deutsch-Jozsa and Grover algorithms.
Let us start with the first one. It is actually explained most easily in a direct way,
without reference to the Deutsch-Jozsa algorithm, though that is where the idea came from.
The problem is a promise version of the equality problem.  Suppose the $n$-bit
inputs $x$ and $y$ are restricted to the following case:
\begin{quote}
{Distributed Deutsch-Jozsa:}
either $x=y$, or $x$ and $y$ differ in exactly $n/2$ positions
\end{quote}
Note that this promise only makes sense if $n$ is an even number, otherwise $n/2$ would not be integer.
In fact it will be convenient to assume $n$ is a power of~2.
Here is a simple quantum protocol to solve this promise version of equality using only $\log n$ qubits of communication:
\begin{enumerate}
\item Alice sends Bob the $\log n$-qubit state $\frac{1}{\sqrt{n}}\sum_{i=0}^{n-1}(-1)^{x_i}\ket{i}$,
which she can prepare unitarily from $x$ and $\log n$ $\ket{0}$-qubits.
\item Bob applies the unitary map $\ket{i}\mapsto(-1)^{y_i}\ket{i}$ to the state, applies a Hadamard transform
to each qubit (for this it is convenient to view $i$ as a $\log n$-bit string), and measures the resulting $\log n$-qubit state.
\item Bob outputs 1 if the measurement gave $\ket{0^{\log n}}$ and outputs 0 otherwise.
\end{enumerate}
It is clear that this protocol only communicates $\log n$ qubits, but why does it work?
Note that the state that Bob measures is
$$
H^{\otimes\log n}\left(\frac{1}{\sqrt{n}}\sum_{i=0}^{n-1}(-1)^{x_i+y_i}\ket{i}\right)=
\frac{1}{n}\sum_{i=0}^{n-1} (-1)^{x_i+y_i}\sum_{j\in\01^{\log n}}(-1)^{i\cdot j}\ket{j}.
$$
This superposition looks rather unwieldy, but consider the amplitude of the $\ket{0^{\log n}}$ basis state.
It is $\frac{1}{n}\sum_{i=0}^{n-1} (-1)^{x_i+y_i}$, which
is 1 if $x=y$ and $0$ otherwise because the promise now guarantees
that $x$ and $y$ differ in exactly $n/2$ of the bits!  Hence Bob will always give the correct answer.

What about efficient \emph{classical} protocols (without entanglement) for this problem?
Proving lower bounds on communication complexity often requires a very technical combinatorial
analysis.  Buhrman, Cleve, and Wigderson used a deep combinatorial result from~\cite{frankl&rodl:forbidden}
to prove that every classical errorless protocol for this problem needs to send at least $0.007 n$ bits.

This $\log n$-qubits-vs-$0.007 n$-bits example was the first exponentially large
separation of quantum and classical communication complexity.
Notice, however, that the difference disappears if we move to the \emph{bounded-error} setting,
allowing the protocol to have some small error probability. We can
 use the randomized protocol for equality discussed above or even simpler:
Alice can just send a few $(i,x_i)$ pairs to Bob, who then compares the $x_i$'s with his $y_i$'s.
If $x=y$ he will not see a difference, but if $x$ and $y$ differ in $n/2$ positions,
then Bob will probably detect this. Hence $O(\log n)$ classical bits of communication
suffice in the bounded-error setting, in sharp contrast to the errorless setting.

\section{Example~2: The Intersection problem}\label{sec:qccintersection}

Now consider the Intersection function, which is 1 if $x_i=y_i=1$ for at least one $i$.
Buhrman, Cleve, and Wigderson~\cite{BuhrmanCleveWigderson98} also presented an efficient quantum protocol for this,
based on Grover's search algorithm (Chapter~\ref{chap:grover}).
We can solve Intersection if we can solve the following search problem: 
find some $i$ such that $x_i=y_i=1$, if such an $i$ exists.\footnote{This is sometimes called
the \emph{appointment-scheduling problem}: view $x$ and $y$ as Alice's and Bob's agendas, respectively,
with a~1 at the $i$-th bit indicating that timeslot~$i$ is available. Then the goal is to find a timeslot where Alice and Bob are both available, so they can schedule an appointment.}
We want to find a solution to the search problem on 
the string $z=x\wedge y$ (which is the bit-wise AND of $x$ and $y$),
since  $z_i =1$ whenever both $x_i=1$ and $y_i=1$. The idea is now to let Alice run
Grover's algorithm to search for such a solution. Clearly, she can prepare the uniform starting state herself.
She can also apply the unitaries $H$ and $R$ herself.  The only thing where she needs Bob's help, is in implementing the phase-query 
$O_{z,\pm}$ (which she needs to do $O(\sqrt{n})$ times, because that's how many queries Grover makes). Alice and Bob can together implement a phase-query as follows. Whenever Alice wants to apply $O_{z,\pm}$ to a state
$$
\ket{\phi}=\sum_{i=1}^n\alpha_i\ket{i},
$$
she tags on her $x_i$'s in an extra qubit (which she can do by the unitary map $\ket{i}\ket{0}\mapsto\ket{i}\ket{x_i}$) and sends Bob the state
$$
\sum_{i=1}^n\alpha_i\ket{i}\ket{x_i}.
$$
Bob applies the unitary map
$$
\ket{i}\ket{x_i}\mapsto(-1)^{x_i\wedge y_i}\ket{i}\ket{x_i}
$$
and sends back the result. Alice sets the last qubit back to $\ket{0}$ (which she
can do unitarily because she has $x$), and now she has the state $O_{z,\pm}\ket{\phi}$!
Thus we can simulate $O_{z,\pm}$ using 2 messages of $\log(n)+1$ qubits each.
Thus Alice and Bob can run Grover's algorithm to find an intersection, using
$O(\sqrt{n})$ messages of $O(\log n)$ qubits each, for total communication of $O(\sqrt{n}\log n)$ qubits.
Later Aaronson and Ambainis~\cite{aaronson&ambainis:searchj} gave a more complicated protocol
that uses $O(\sqrt{n})$ qubits of communication.

What about lower bounds? It is a well-known result of classical communication complexity that
classical bounded-error protocols for the Intersection problem need about $n$ bits of communication.
Thus we have a quadratic quantum-classical separation for this problem.
Could there be a quantum protocol that uses much less than $\sqrt{n}$ qubits of communication?
This question was open for quite a few
years after~\cite{BuhrmanCleveWigderson98} appeared, until finally Razborov~\cite{razborov:qdisj}
showed that any bounded-error quantum protocol for Intersection needs to communicate
about $\sqrt{n}$ qubits.

\section{Example~3: The vector-in-subspace problem}\label{sec:ranraz}

Notice the contrast between the examples of the last two sections.
For the Distributed Deutsch-Jozsa problem we get an \emph{exponential}
quantum-classical separation, but the separation only holds
if we require the classical protocol to be errorless.
On the other hand, the gap for the disjointness function is only
\emph{quadratic}, but it holds even if we allow classical protocols
to have some error probability.

Here is a function where the quantum-classical
separation has both features: the quantum protocol is exponentially better than
the classical protocol, even if the latter is allowed some error:
\begin{quote}
Alice receives a unit vector $v\in\mathbb{R}^m$\\
Bob receives two $m$-dimensional projectors $P_0$ and $P_1$ such that $P_0+P_1=I$\\
Promise: either $P_0v=v$ or $P_1v=v$.\\
Question: which of the two?
\end{quote}
As stated, this is a problem with continuous input,
but it can be discretized in a natural way by approximating
each real number by $O(\log m)$ bits. Alice and Bob's input
is now $n=O(m^2\log m)$ bits long.
There is a simple yet efficient 1-round quantum protocol for this problem:
Alice views $v$ as a $\log m$-qubit state and sends this to Bob;
Bob measures with operators $P_0$ and $P_1$, and outputs the measurement result (0 or 1).
This takes only $\log m=O(\log n)$ qubits of communication, and Bob's output is correct with probability~1 thanks to the promise on the inputs.

The efficiency of this protocol comes from the fact that an
$m$-dimensional unit vector can be ``compressed'' or ``represented''
as a $\log m$-qubit state.
Similar compression is not possible with classical bits, which
suggests that any classical protocol will have to send
the vector $v$ more or less literally and hence will require a lot of communication. This turns out to be true, but the proof is quite hard~\cite{klartag&regev:lowerbound}. It shows that any bounded-error protocol needs to send $\Omega(m^{1/3})$ bits.

\section{Example~4: Quantum fingerprinting}\label{sec:qfingerprint}

The examples of the previous section were either exponential quantum improvements 
for promise problems (Deutsch-Jozsa and vector-in-subspace) or polynomial improvements
for total problems (disjointness).
We will now give an exponential improvement for the total problem of equality-testing, but in a restricted setting called the \emph{simultaneous message passing} (SMP) model.
Alice and Bob receive $n$-bit input $x$ and $y$, respectively.
They do not have any shared resources like shared
randomness or an entangled state, but they do have local randomness.
They don't communicate with each other directly, but instead
send a single message to a third party, called the Referee.
The Referee, upon receiving message $m_A$ from Alice and $m_B$ from Bob,
should output the value $f(x,y)$.
The goal is to compute $f(x,y)$ with a minimal amount of communication
from Alice and Bob to the Referee.

We will see that for the equality problem there is an exponential savings
in communication when qubits are used instead of classical bits.
Classically, the problem of the bounded-error communication complexity of
equality in the SMP model was first raised by Yao~\cite{yao:distributive}, and was open for almost twenty years until Newman and Szegedy~\cite{newman&szegedy:1round} exhibited a lower bound of $\Omega(\sqrt{n})$ bits.
This is tight, since Ambainis~\cite{ambainis:3computer} constructed a bounded-error protocol for this problem where the messages are
$O(\sqrt{n})$ bits long (see Exercise~\ref{ex:ambainissmpeq}).
In contrast, in the quantum setting this problem can be solved with very little communication: only $O(\log{n})$ qubits suffice~\cite{bcww:fp}.

The quantum trick is to associate each $x\in\01^n$ with a short quantum state $\ket{\phi_x}$,
called the \emph{quantum fingerprint} of~$x$. Just like with physical fingerprints, the idea is
that a quantum fingerprint is a small object that doesn't contain very much information about the object~$x$, but that suffices for testing if the fingerprinted object equals some other fingerprinted object. As we will see below, we can do such testing if the fingerprints are pairwise almost orthogonal.
More precisely, an \emph{$(n,m,\eps)$-quantum fingerprinting scheme} maps $n$-bit string~$x$ to $m$-qubit state $\ket{\phi_x}$ with the property that for all distinct $x,y\in\01^n$, we have $|\inp{\phi_x}{\phi_y}|\leq\eps$.

We will now show how to obtain a specific $(n,m,0.02)$-quantum fingerprinting scheme from an error-correcting code $C:\01^n\rightarrow\01^N$ where $m=\log N\approx\log n$. There exist codes where $N=O(n)$ and any two codewords $C(x)$ and $C(y)$ have Hamming distance close to $N/2$,
say $d(C(x),C(y))\in[0.49N,0.51N]$ (we won't prove this here, but for instance a random linear code will work).
Define the quantum fingerprint of $x$ as follows:
$$
\ket{\phi_x}=\frac{1}{\sqrt{N}}\sum_{j=1}^N (-1)^{C(x)_j}\ket{j}.
$$
This is a unit vector in an $N$-dimensional space, so it corresponds to only $\ceil{\log N}=\log n +O(1)$ qubits.
For distinct $x$ and $y$, the corresponding fingerprints will have small inner product:
$$
\inp{\phi_x}{\phi_y}=\frac{1}{N}\sum_{j=1}^N (-1)^{C(x)_j+C(y)_j}=\frac{N-2d(C(x),C(y))}{N}\in[-0.02,0.02].
$$

\begin{figure}[hbt]
\centering
\includegraphics{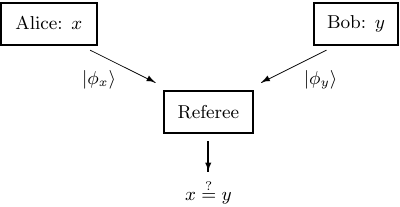}
\caption{Quantum fingerprinting protocol for the equality problem}\label{fig:fingerprint}
\end{figure}

\begin{figure}[hbt]
\centering
\includegraphics{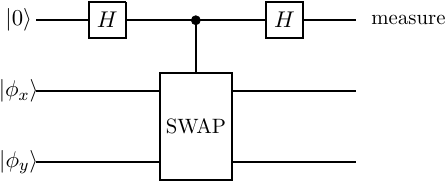}
\caption{Quantum circuit to test if $\ket{\phi_x}=\ket{\phi_y}$ or
$|\inp{\phi_x}{\phi_y}|$ is small}\label{fig:swaptest}
\end{figure}

The quantum protocol is very simple (see Figure~\ref{fig:fingerprint}): 
Alice and Bob send quantum fingerprints of $x$ and $y$ to the Referee, respectively. 
The referee now has to determine whether $x=y$ (which corresponds to $\inp{\phi_x}{\phi_y}=1$)
or $x\neq y$ (which corresponds to $\inp{\phi_x}{\phi_y}\in[-0.02,0.02]$).
The following test (Figure~\ref{fig:swaptest}), sometimes called
the \emph{SWAP-test}~\cite{bbdejm:stab,bcww:fp}, accomplishes this with small error probability.

This circuit first applies a Hadamard transform to a qubit that is initially $\ket{0}$, then SWAPs
the other two registers conditioned on the value of the first qubit being $\ket{1}$,
then applies another Hadamard transform to the first qubit and measures it.
Here SWAP is the operation that swaps the two registers: 
$\ket{\phi_x}\ket{\phi_y} \mapsto \ket{\phi_y}\ket{\phi_x}$. The
Referee receives $\ket{\phi_x}$ from Alice and $\ket{\phi_y}$ from Bob 
and applies the test to these two states. An easy calculation reveals that the outcome of the
measurement is $1$ with probability $(1-|\inp{\phi_x}{\phi_y}|^2)/2$.
Hence if $\ket{\phi_x}=\ket{\phi_y}$ then we observe a~1 with probability~0,
but if $|\inp{\phi_x}{\phi_y}|$ is close to~0 then we observe a~1 with probability close to~1/2. 
Repeating this procedure with several individual
fingerprints can make the error probability arbitrarily close to $0$.

\section*{Exercises}

\begin{enumerate}

\item\label{ex:detEQ} (H) Prove that classical deterministic protocols with one message (from Alice to Bob), need to send $n$ bits to solve the equality problem.

\item\label{ex:QeEQ}
\begin{enumerate}
\item (H) Show that if $\ket{\phi}$ and $\ket{\psi}$ are non-orthogonal states (i.e., $\inp{\phi}{\psi}\neq 0$),
then there is no two-outcome projective measurement that perfectly distinguishes these two states,
in the sense that applying the measurement on $\ket{\phi}$ always gives a different outcome from applying the same measurement to $\ket{\psi}$.
\item Prove that quantum protocols with one message (from Alice to Bob), need to send at least $n$ qubits to solve the equality problem (on $n$-bit inputs)
with success probability~1 on every input.
Assume for simplicity that Bob does a projective measurement rather than a general POVM.
\item (H) Prove that quantum protocols with one message (from Alice to Bob), need to send at least $\log n$ qubits to solve the distributed Deutsch-Jozsa problem
with success probability~1 on every input.
Again assume for simplicity that Bob does a projective measurement rather than a general POVM.
\end{enumerate}

\item\label{ex:oneroundqcc} (H)
Consider \emph{one-round} quantum communication complexity.
Alice gets input $x\in\01^n$, Bob gets input $y\in\01^n$, and they want
to compute some Boolean function $f(x,y)$ of their inputs.
Assume that all rows of the communication matrix are different,
i.e., for all $x$ and~$x'$ there is a $y$ such that $f(x,y)\neq f(x',y)$.
They are allowed only one round of communication:
Alice sends a quantum message to Bob and Bob must then
be able to give the right answer with probability 1.
Prove that Alice needs to send $n$ qubits to Bob for this.
You may assume that Alice's messages are pure states (this is without loss of generality).

\item  Suppose Alice and Bob each have $n$-bit agendas, and they know that for exactly 25\%\ of the timeslots they are both free.
Give a quantum protocol that finds such a timeslot with probability~1, using only $O(\log n)$ qubits of communication.

\item\label{ex:DISJonewayLB} (H)
The \emph{disjointness problem} of communication complexity is the following decision version of the intersection problem: Alice receives an $x\in\01^n$, Bob receives $y\in\01^n$, and $f(x,y)=0$ if there is an $i$ such that $x_i=y_i=1$, and $f(x,y)=1$ otherwise (i.e., $f$ says whether $x$ and $y$ represent disjoint subsets of $[n]$).
Suppose there exists an $m$-qubit one-way protocol that solves this problem, so where Alice sends Bob $m$ qubits and then Bob outputs $f(x,y)$ with probability at least $2/3$. Prove the lower bound $m=\Omega(n)$ on the number of qubits sent.

\item\label{ex:intersectionrrounds}
(H) Consider the intersection problem:
Alice has input $x\in\01^n$, Bob has input $y\in\01^n$, and they want to find (with success probability $\geq 2/3$) an $i$ such that $x_i=y_i=1$, if such an $i$ exists. We know that using $r=O(\sqrt{n})$ messages between Alice and Bob, they can solve the intersection problem with $O(\sqrt{n}\log n)$ qubits of communication (see Section~\ref{sec:qccintersection}). We also know that with only $r=1$ message (i.e., one-way communication) $\Theta(n)$ qubits of communication are necessary and sufficient (see Exercise~\ref{ex:DISJonewayLB}).
Now suppose we limit them to some $r\in\{1,\ldots,\sqrt{n}\}$ messages. This $r$ is known to Alice and Bob. Describe a communication protocol by means of which Alice and Bob can solve the intersection problem with at most $r$ messages, and $O((n/r)\log n)$ qubits of communication in total.

\item 
\begin{enumerate}
\item Consider the following variant of the search problem: we are given query access to a string $x\in\01^n$, and we know a set $S\subseteq[n]$ of $k<n$ elements such that $x_i=0$ for all $i\not\in S$.
Show that there is a quantum algorithm that can find a solution for this search problem (i.e., an $i$ such that $x_i=1$, if there is one) with success probability $\geq 2/3$, using $O(\sqrt{k})$ queries to $x$.
\item Consider the following variant of the intersection problem of communication complexity: Alice holds a string $x\in\01^n$ of Hamming weight $k$, and Bob holds a string $y\in\01^n$ of Hamming weight $k$. Give a quantum communication protocol that finds an $i$ such that $x_i=y_i=1$ (if such an $i$ exists) with success probability $\geq 2/3$, using $O(\sqrt{k}\log n)$ qubits of communication.
\end{enumerate}

\item\label{ex:ambainissmpeq}
 Consider an error-correcting code $C:\01^n\rightarrow\01^N$ where $N=O(n)$, $N$ is a square, and any two distinct codewords are
at Hamming distance $d(C(x),C(y))\in[0.49N,0.51N]$ (such codes exist, but you don't have to prove that).
\begin{enumerate}
\item  View the codeword $C(x)$ as a $\sqrt{N}\times\sqrt{N}$ matrix.
Show that if you choose a row index  uniformly at random and choose a column index uniformly at random, then the unique index~$i$ where these row and column intersect, is uniformly distributed over~$i\in\{1,\ldots,N\}$.
\item (H) Give a classical bounded-error SMP-protocol for the equality problem where Alice and Bob each send $O(\sqrt{n})$ bits to the Referee.
\end{enumerate}

\item\label{ex:qfpfrompolys}
Alice and Bob want to solve the equality problem on $n$-bit inputs $x$ and $y$ (i.e., decide whether $x=y$). They do not share randomness or entanglement but can use local (private) randomness.
\begin{enumerate}
\item (H) Fix a prime number $p\in[3n,6n]$, then the set $\mathbb{F}_p$ of integers modulo $p$ is a finite field (i.e., it has a well-defined addition and multiplication).
For $x=(x_0,\ldots,x_{n-1})\in\01^n$, define the univariate polynomial $P_x:\mathbb{F}_p\to\mathbb{F}_p$ of degree $<n$ as $P_x(t)=\sum_{i=0}^{n-1} x_i t^i$ (note that the $n$ bits of $x$ are used as coefficients here, not as the argument of the polynomial). Show that for distinct $n$-bit strings $x$ and $y$, we have
$\Pr_{t\in\mathbb{F}_p}[P_x(t)=P_y(t)]\leq 1/3$, where the probability is taken over a uniformly random $t\in\mathbb{F}_p$.
\item Use (a) to give a classical communication protocol where Alice sends an $O(\log n)$-bit message to Bob, and Bob can decide whether $x=y$ with success probability $\geq 2/3$.
\item Use (a) to give a quantum fingerprinting scheme $x\mapsto\ket{\phi_x}$, where quantum state $\ket{\phi_x}$ has $O(\log n)$ qubits, and $|\inp{\phi_x}{\phi_y}|\in[0,1/3]$ for all distinct $n$-bit strings $x$ and $y$ (prove the latter property explicitly, it's not enough to write down only the states).
\end{enumerate}

\item\label{ex:IPLB} 
The inner product problem in communication complexity is the function $f:\01^n\times\01^n\rightarrow\01$ defined by $f(x,y)=\sum_{i=1}^n x_iy_i$ mod 2.  Suppose there exists a quantum protocol~$P$ for Alice and Bob that uses $q$ qubits of communication (possibly using multiple messages between Alice and Bob) and computes the inner product function with success probability~1 (on every possible inputs $x,y$). The protocol does not assume any shared entangled state at the start.
\begin{enumerate}
\item Give a quantum protocol that uses $2q$ qubits of communication and implements the $2n$-qubit map $\ket{x}_A\ket{y}_B\mapsto(-1)^{x\cdot y}\ket{x}_A\ket{y}_B$ (possibly with some auxiliary qubits for each of Alice and Bob; these should start and end in state $\ket{0}$).
\item (H) Give a protocol where Alice transmits $x$ to Bob using $2q$ qubits of communication. 
\item Derive a lower bound on $q$ from (b) and Holevo's theorem (Theorem~\ref{thholevo} of Chapter~\ref{chap:qencodings}; be specific about which part of the theorem you invoke).
\end{enumerate}

\item\label{ex:vecinsubspace} 
Consider the following problem in communication complexity.  Alice's input has two parts: a unit vector $v\in\mathbb{R}^m$ and two orthogonal projectors $P_0$ and $P_1$.  Bob's input is an $m\times m$ unitary $U$. They are promised that the vector $Uv$ either lies in the subspace corresponding to $P_0$ (i.e., $P_0 Uv=Uv$) or in the subspace corresponding to $P_1$ (i.e., $P_1 Uv=Uv$), and the problem for Alice and Bob is to find out which of these two cases holds.
\begin{enumerate}
\item Give a quantum protocol that uses two messages of $O(\log m)$ qubits (one message from Alice to Bob and one from Bob to Alice) to solve this problem with success probability~1.
\item (H) Show that there exists a constant $c>0$ such that classical protocols need to send $\Omega(m^{c})$ bits of communication to solve this problem with error probability $\leq 1/3$, even when they are allowed to send many messages.
\end{enumerate}

\item\label{ex:HidMat} (H)
  Consider the following communication complexity problem, called the ``Hidden Matching Problem.''
Alice's input is some $x\in\01^n$. Bob's input is a matching $M$, i.e., a partition of $\{1,\ldots,n\}$ into $n/2$ disjoint unordered pairs (assume $n$ is a power of~2 for simplicity).
Their goal is that Bob outputs a pair $\{i,j\}\in M$ together with the parity $x_i\oplus x_j$ of the two bits indexed by that pair. It doesn't matter which pair $\{i,j\}\in M$ Bob outputs, as long as the additional bit of output equals the parity of the two indexed bits of~$x$. Show that they can solve this problem with success probability~1 using only a message of $\log n$ qubits from Alice to Bob (and no communication from Bob to Alice).\\
{\footnotesize Comment: One can show that classical one-way protocols need $\Omega(\sqrt{n})$ bits of communication to solve this problem with small error probability.}

\item\label{ex:HtestHSP}
\begin{enumerate}
\item Suppose you have a state $\frac{1}{\sqrt{2}}(\ket{0}\ket{\phi}+\ket{1}\ket{\psi})$, where $\ket{\phi}$ and $\ket{\psi}$ are quantum states with real amplitudes. Suppose you apply a Hadamard gate to its first qubit and then measure that first qubit. Show that the probability of measurement outcome~0 is $\frac{1}{2}(1+\inp{\phi}{\psi})$.
\item Suppose $H$ is a subgroup of a finite group $G$, and $g\in G$ some element. Show (1) if $g\in H$ then the cosets $g\circ H$ and $H$ are equal\\
and (2) if $g\not\in H$ then the cosets $g\circ H$ and $H$ are disjoint.
\item Suppose you are given quantum state $\ket{\psi_H}=\frac{1}{\sqrt{H}}\sum_{h\in H}\ket{h}$ (for an unknown $H\leq G$), and an element $g\in G$. You may assume you have a unitary $A$ available that implements the group operation, $A:\ket{g,h}\mapsto\ket{g,g\circ h}$, and you may also apply a controlled version of~$A$. 
Give an algorithm that acts on $\ket{\psi_H}$ and possibly some auxiliary qubits, and that outputs 0 with probability 1 if $g\in H$, and outputs 0 with probability $\leq 1/2$ if $g\not\in H$.
\item (H) Consider the following communication complexity problem. Alice and Bob both know a finite group $G$, Alice gets as input some subgroup $H\leq G$ (for instance in the form of a generating set for $H$) and Bob gets input $g\in G$. Give a one-way quantum protocol where Alice sends to Bob a message of $O(\log|G|)$ qubits, and then Bob decides with success probability $\geq 2/3$ whether $g\in H$.
\end{enumerate}

\end{enumerate}

\chapter{Entanglement and Non-Locality}\label{chap:nonloc}

\section{Quantum non-locality}

Entangled pure states are those that cannot be written as a tensor product of separate states.  
The most famous one is the EPR-pair:
$$
\frac{1}{\sqrt{2}}(\ket{00}+\ket{11}).
$$
Suppose Alice has the first qubit of the pair, and Bob has the second.
If Alice measures her qubit in the computational basis and gets outcome $b\in\01$,
then the state collapses to $\ket{bb}$. Similarly, if Alice measures her qubit in some other basis, 
this will collapse the joint state (including Bob's qubit) to 
some state that depends on her measurement basis as well as its outcome. Somehow Alice's action seems to have
an \emph{instantaneous} effect on Bob's side---even if the two qubits are light-years apart!
This was a great bother to Einstein, whose theory of relativity posits that information
and causation cannot travel faster than the speed of light.
Einstein called such effects of entanglement ``spooky action at a distance'' 
(in German: ``spukhafte Fernwirkungen''), and viewed it as a fundamental problem for quantum mechanics~\cite{epr}.
In his view, quantum mechanics should be replaced by some ``local realist'' physical
theory that would still have the same predictive power as quantum mechanics.
Here ``local'' means that information and causation act locally, not faster than light,
and ``realist'' means that physical systems have definite, well-defined properties 
(even if those properties may be unknown to us).

Note that the above experiment where Alice measures her half of the EPR-pair doesn't actually violate
locality: no information is transferred from Alice to Bob.  From Bob's perspective there is no
difference between the situation where Alice measured and the situation where she didn't.%
\footnote{In fact, one can show that entanglement cannot replace communication, see for example Exercise~\ref{chap:qcrypto}.\ref{ex:unitaryinv}.}
For this experiment, a shared coin flip between Alice and Bob is a local realist physical model that has exactly 
the same observable consequences as measuring the qubits of the EPR-pair in the computational basis:
a 50-50 distribution on outcomes $\ket{00}$ and $\ket{11}$.
This shared-coin-flip model is \emph{local} because no information is transferred between Alice and Bob, and it's \emph{realist} 
because the coin flip has a definite outcome (even if that outcome is unknown to Alice and Bob 
before they measure). 

Given this example, one might hope (and Einstein expected) that any kind of behavior that comes
from entangled states can be replaced by some local realist physical model.
This way, quantum mechanics could be replaced by an alternative physical theory with
less counter-intuitive behavior.
Surprisingly, in the 1960s, John Bell~\cite{bell:epr} devised entanglement-based experiments whose
behavior \emph{cannot be reproduced by any local realist theory}.
In other words, we can let Alice and Bob do certain measurements on an entangled state,
and the resulting distributions on their outputs predicted by quantum mechanics,
\emph{cannot be obtained} from any local realist theory.
This phenomenon is known as ``quantum non-locality.''
It could of course be that the quantum mechanical predictions of the resulting correlations are just wrong.
However, in the early 1980s, such experiments were actually done by Aspect and others~\cite{AspectGR81:Bell}, and they gave the outcomes 
that quantum mechanics predicted.\footnote{Modulo some technical ``loopholes'' due to imperfect photon sources, measurement devices, Alice and Bob not being sufficiently far apart etc.  These are still hotly debated in some corners, but most people accept that Aspect's and later experiments are convincing, and kill any reasonable hope of a complete local-realist explanation of nature. \cite{hansonea:loopholefreebell} did an experiment that simultaneously closed the two most important loopholes.}
Note that such experiments don't \emph{prove} quantum mechanics, but they \emph{disprove} 
any local realist physical theory.\footnote{Despite its name, non-locality doesn't disprove locality, 
but rather disproves the conjunction of locality and realism---at least one of the two assumptions has to fail.}  

Such experiments, which realize correlations that are \emph{provably impossible} to realize with local
realist models, are among the deepest and most philosophical results of 20th century physics:
the commonsense idea of local realism is most probably false!
Since Bell's seminal work, the concept of quantum non-locality has been extensively studied, by physicists, philosophers, and more recently by computer scientists. 

\begin{figure}[h]
\centering
\includegraphics{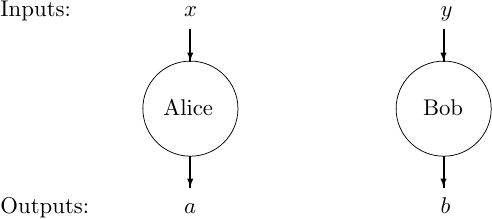}
\caption{The non-locality scenario involving two parties:
Alice and Bob receive inputs $x$ and~$y$, respectively,
and are required to produce outputs $a$ and $b$ that satisfy certain conditions.
Once the inputs are received, no communication is permitted between the parties.
}\label{fig:chsh}
\end{figure}

In the next sections we review some interesting examples.
The two-party setting of these examples is illustrated in Fig.~\ref{fig:chsh}:
Alice receives input $x$ and Bob receives input $y$, and they produce outputs $a$ and $b$, respectively,
that have to be correlated in a certain way (which depends on the game).
They are not allowed to communicate. In physics language, we could assume they are ``space-like separated,''
which means that they are so far apart that they cannot influence each other during the course of the experiment
(assuming information doesn't travel faster than the speed of light).
In the classical scenario they are allowed to share a random variable.  Physicists would call this 
the ``local hidden variable'' that gives properties their definite value (that value may be unknown to 
the experimenter).  This setting captures all local realist models.
In the quantum model Alice and Bob are allowed to share entangled states, such as EPR-pairs.
The goal is to show that entanglement-based strategies can do things that
local realist strategies cannot.

\section{CHSH: Clauser-Horne-Shimony-Holt}\label{sec:CHSH}

In the CHSH game~\cite{chsh} Alice and Bob receive input bits $x$ and $y$, and
their goal is to output bits $a$ and $b$, respectively, such that
\begin{equation}\label{eq:chsh}
a \oplus b = x \wedge y,
\end{equation}
(`$\wedge$' is logical AND; `$\oplus$' is parity, i.e.~addition mod 2)
or, failing that, to satisfy this condition with as high a probability as possible.

First consider the case of classical \emph{deterministic} strategies, so without any randomness.
For these, Alice's output bit depends solely on her input bit $x$, and similarly for Bob.
Let $a_0$ be the bit that Alice outputs if her input is $x=0$, and $a_1$ the bit she outputs if $x=1$.
Let $b_0, b_1$ be the outputs Bob gives on inputs $y=0$ and $y=1$, respectively.
These four bits completely characterize any deterministic strategy.
Condition~(\ref{eq:chsh}) becomes
\begin{eqnarray}\label{eq:eqchsh}
a_0 \oplus b_0 & = & 0, \nonumber \\
a_0 \oplus b_1 & = & 0, \nonumber \\
a_1 \oplus b_0 & = & 0, \nonumber \\
a_1 \oplus b_1 & = & 1.
\end{eqnarray}
It is impossible to satisfy all four equations simultaneously, since summing
them modulo 2 yields $0 = 1$.
Therefore it is impossible to satisfy Condition~(\ref{eq:chsh}) perfectly.
Since a probabilistic strategy (where Alice and Bob share randomness)
is a probability distribution over deterministic strategies, it follows
that no probabilistic strategy can have success probability better than 3/4 on every possible input
(the 3/4 can be achieved simultaneously for every input, see Exercise~\ref{ex:classchsh}).%
\footnote{Such statements, upper bounding the optimal success probability of classical strategies for a specific game,
are known as \emph{Bell inequalities}. This specific one is called the \emph{CHSH inequality}.}

Now consider the same problem but where Alice and Bob are supplied with
a shared 2-qubit system initialized to the entangled state
\begin{equation*}
\textstyle{\frac{1}{\sqrt 2}}(\ket{00} - \ket{11}).
\end{equation*}
Such a state can easily be obtained from an EPR-pair by local operations, for instance if Alice applies a $Z$-gate to her qubit.
Now the parties can produce outputs that satisfy
Condition~(\ref{eq:chsh}) with probability $\cos(\pi/8)^2 \approx 0.85$
(higher than what is possible in the classical case), as follows.
Recall the unitary operation that rotates the qubit by angle $\theta$:
$R(\theta)=\left(
\begin{array}{lr}
\cos\theta & -\sin\theta \\
\sin\theta & \cos\theta
\end{array}
\right)$. 
If $x=0$ then Alice applies $R(-\pi/16)$ to her qubit; and if $x=1$ she applies $R(3\pi/16)$.
Then Alice measures her qubit in the computational basis and outputs the resulting bit~$a$.
Bob's procedure is the same, depending on his input bit $y$.
It is straightforward to calculate that if Alice rotates by $\theta_A$ and
Bob rotates by $\theta_B$, the state becomes
$$
\frac{1}{\sqrt{2}}
\left(\cos(\theta_A + \theta_B)(\ket{00} - \ket{11})
+ \sin(\theta_A + \theta_B)(\ket{01} + \ket{10})\right).
$$
After the measurements, the probability that $a \oplus b = 0$ is $\cos(\theta_A + \theta_B)^2$.
Note that if $x\wedge y=0$ then $\theta_A + \theta_B=\pm\pi/8$, while if $x\wedge y=1$ then $\theta_A + \theta_B=3\pi/8$.
Hence Condition~\ref{eq:chsh} is satisfied with probability $\cos(\pi/8)^2$ for all four input possibilities, showing that quantum entanglement allows Alice and Bob to win the game with a probability that's higher than what the best classical strategy can achieve. Tsirelson~\cite{tsirelson80} showed that $\cos(\pi/8)^2$ is the best that quantum strategies can do for CHSH,
even if they are allowed to use much more entanglement than one EPR-pair (see Exercise~\ref{ex:tsirelsonbound}).

\section{Magic square game}

Is there a game where the quantum protocol \emph{always}
succeeds, while the best classical success probability is bounded below~$1$?
A particularly elegant example is the following \emph{magic square game}~\cite{aravind:magicsquare}.
Consider the problem of labeling the entries of
a $3 \times 3$ matrix with bits so that the parity of each row is
even, whereas the parity of each column is odd.
This is clearly impossible:
if the parity of each row is even then the sum of the 9 bits is 0 mod~2,
but if the parity of each column is odd then the sum of the 9 bits is 1 mod~2.
The two matrices
\begin{center}
\begin{tabular}{|c|c|c|}
\hline
0 & 0 & 0 \\
\hline
0 & 0 & 0 \\
\hline
1 & 1 & 0 \\
\hline
\end{tabular}
\hspace*{40mm}
\begin{tabular}{|c|c|c|}
\hline
0 & 0 & 0 \\
\hline
0 & 0 & 0 \\
\hline
1 & 1 & 1 \\
\hline
\end{tabular}
\end{center}
each satisfy five out of the six constraints.
For the first matrix, all rows have even parity,
but only the first two columns have odd parity.
For the second matrix, the first two rows have even
parity, and all columns have odd parity.

Consider the game where Alice receives
$x \in \{1,2,3\}$ as input (specifying the number of a row), and Bob receives
$y \in \{1,2,3\}$ as input (specifying the number of a column).
Their goal is to each produce $3$-bit outputs, $a_1a_2a_3$ for Alice
and $b_1b_2b_3$ for Bob, such that
\begin{enumerate}
\item They satisfy the row/column parity constraints:
$a_1 \oplus a_2 \oplus a_3 = 0$ and $b_1 \oplus b_2 \oplus b_3 = 1$.
\item They are consistent where the row intersects the column: $a_y = b_x$.
\end{enumerate}
As usual, Alice and Bob are forbidden from communicating once the
game starts, so Alice does not know $y$ and Bob does not know $x$.
We shall show the best classical strategy has success probability $8/9$, 
while there is a quantum strategy that always succeeds.

An example of a deterministic strategy that attains success probability $8/9$
(when the input $xy$ is uniformly distributed) is
where Alice plays according to the rows of the first matrix above
and Bob plays according the columns of the second matrix above.
This succeeds in all cases, except where $x = y = 3$.
To see why this is optimal, note that for any other
classical strategy, it is possible to represent it as two matrices
as above but with different entries.
Alice plays according to the rows of the first matrix and Bob
plays according to the columns of the second matrix.
We can assume that the rows of Alice's matrix all have even
parity; if she outputs a row with odd parity then they immediately
lose, regardless of Bob's output.
Similarly, we can assume that all columns of Bob's matrix have
odd parity.%
\footnote{In fact, the game can be simplified so that Alice and Bob
each output just two bits, since the parity constraint determines
the third bit.}
Considering such a pair of matrices, the players lose at each entry where they differ.
There must be such an entry, since otherwise it would be possible
to have all rows even and all columns odd with one matrix.
Thus, when the input $xy$ is chosen uniformly from
$\{1,2,3\}\times\{1,2,3\}$, the success probability of the best-possible  classical strategy is at most~$8/9$.

We now give the \emph{quantum} strategy for this game.
Let $I$, $X$, $Y$, $Z$ be the $2 \times 2$ Pauli matrices from Appendix~\ref{app:pauli}.
Each is a 1-qubit \emph{observable} with eigenvalues in $\{+1,-1\}$.\footnote{See Section~\ref{ssecmeasurement}. In particular, a $\pm 1$-valued observable $A$ can be written as $A=P-Q$, where $P$ and $Q$ are projectors on two orthogonal subspaces such that $P+Q=I$. This corresponds to a two-outcome measurement specified by projectors $P$ and $Q$ with outcomes $+1$ and $-1$, respectively.}
That is, each can be written as $P_{+} - P_{-}$ where $P_{+}$ and $P_{-}$ are orthogonal
projectors that sum to identity, and hence define a two-outcome measurement with outcomes $+1$ and $-1$.
For example, $Z=\ketbra{0}{0}-\ketbra{1}{1}$, corresponding to a measurement in
the computational basis (with state $\ket{b}$ giving rise to measurement outcome $(-1)^b$).
And $X=\ketbra{+}{+}-\ketbra{-}{-}$, corresponding to a measurement in the Hadamard basis.
The Pauli matrices are self-inverse, distinct non-identity Paulis anti-commute (e.g., $XY=-YX$), and we have 
$X=iZY$, $Y=iXZ$, $Z=iYX$.
Consider the following table, where each entry is a tensor product of two Paulis:
\begin{center}
\begin{tabular}{|c|c|c|}
\hline
$X \otimes X$ & $Y \otimes Z$ & $Z \otimes Y$ \\
\hline
$Y \otimes Y$ & $Z \otimes X$ & $X \otimes Z$ \\
\hline
$Z \otimes Z$ & $X \otimes Y$ & $Y \otimes X$ \\
\hline
\end{tabular}
\end{center}
Because $(P_+ - P_-)\otimes(Q_+ - Q_-)=(P_+\otimes Q_+ + P_-\otimes Q_-)-(P_+\otimes Q_- + P_-\otimes Q_+)$, 
each such product is itself a $\{+1,-1\}$-valued observable.
Hence each product of Pauli matrices corresponds to a measurement 
on a 2-qubit space, with outcomes $+1$ and $-1$.

Note that the observables along each row commute and their product is
$I \otimes I$, and the observables along each column
commute and their product is $- I \otimes I$.
This implies that for any 2-qubit state, performing the
three measurements along any row results in three
$\{+1,-1\}$-valued bits whose product is $+1$.
Also, performing the three measurements along any column
results in three $\{+1,-1\}$-valued bits whose product is $-1$.

We can now describe the quantum protocol.
It uses two pairs of entangled qubits, each of which is in initial state
\begin{equation*}\label{eq:singlet}
\frac{1}{\sqrt 2}(\ket{01}-\ket{10})
\end{equation*} 
(again, such states can be obtained from EPR-pairs by local operations).
Alice, on input $x$, applies three 2-qubit measurements
corresponding to the observables in row~$x$ of the above table.
For each measurement, if the result is $+1$ then she outputs~0,
and if the result is $-1$ then she outputs~1.
Similarly, Bob, on input $y$, applies the measurements
corresponding to the observables in column~$y$, and
converts the $\pm 1$-outcomes into bits.

We have already established that Alice and Bob's output bits
satisfy the required parity constraints.
It remains to show that Alice and Bob's output bits agree at the point where the row meets the column.
For that measurement, Alice and Bob are measuring with respect
to the same observable in the above table. Because all the observables
in each row and in each column commute, we may assume that the place
where they intersect is the first observable applied.
Those bits are obtained by Alice and Bob each measuring
$\frac{1}{2}(\ket{01}-\ket{10})(\ket{01}-\ket{10})$
with respect to the observable in entry $(x,y)$ of the table.
To show that their measurements will agree for all cases of $xy$,
we consider the individual Pauli measurements on the individual
entangled pairs of the form $\frac{1}{\sqrt 2}(\ket{01}-\ket{10})$.
Let $a'$ and $b'$ denote the $0/1$-valued outcomes of the first measurement,
and $a''$ and $b''$ denote the outcomes of the second.
The measurement associated with the tensor product of two
observables gives the same distribution over outcomes as measuring each individual observable and then taking the product of the two results. Hence we have
$a_y = a' \oplus a''$ and $b_x = b' \oplus b''$.
It is straightforward to verify that if the same measurement from $\{X,Y,Z\}$ is applied to each qubit of
$\frac{1}{\sqrt 2}(\ket{01}-\ket{10})$ then the outcomes will be distinct: $a' \oplus b' = 1$ and $a'' \oplus b'' = 1$.
We now have $a_y = b_x$, because 
\begin{equation}
a_y \oplus b_x  =  (a' \oplus a'') \oplus (b' \oplus b'')
 = (a' \oplus b') \oplus (a'' \oplus b'')
 = 1 \oplus 1
 = 0.
\end{equation}

\section{A non-local version of distributed Deutsch-Jozsa}

The previous two examples used small amounts of entanglement:
one EPR-pair for CHSH, two EPR-pairs for magic square.
In both cases we could show that classical protocols need at least \emph{some}
communication if they want to achieve the same as what entanglement-based protocols can achieve without any communication.
We will now give a non-locality game that's parametrized by a number $n$,
and where Alice and Bob's quantum strategy uses $\log n$ EPR-pairs~\cite{bct:simulating}.
The advantage is that we can show that classical protocols for this game  actually need \emph{much} classical
communication rather than at least some nonzero amount.
\begin{quote}
{\bf Non-local DJ problem:}
Alice and Bob receive $n$-bit inputs $x$ and $y$ that satisfy the DJ promise:
either $x=y$, or $x$ and $y$ differ in exactly $n/2$ positions. The task is for Alice and Bob to provide outputs $a,b\in \{0,1\}^{\log n}$ such that
if $x=y$ then $a=b$, and if $x$ and $y$ differ in exactly $n/2$ positions then $a\neq b$.
\end{quote}
They achieve this as follows
\begin{enumerate}
\item Alice and Bob share $\log n$ EPR-pairs, i.e., the maximally entangled state
$\frac{1}{\sqrt{n}}\sum_{i=0}^{n-1}\ket{i}\ket{i}$.\footnote{Note that $k$ EPR-pairs $\left(\frac{1}{\sqrt{2}}(\ket{0}_A\ket{0}_B+\ket{1}_A\ket{1}_B)\right)^{\otimes k}$ can also be written as $\displaystyle\frac{1}{\sqrt{2^k}}\sum_{i\in\01^k}\ket{i}_A\ket{i}_B$ if we reorder the qubits, 
putting Alice's $k$ qubits on the left and Bob's on the right. While these two ways of writing the state strictly speaking correspond
to two different vectors of amplitudes, they still represent the same bipartite physical state, and we will typically
view them as equal.}
\item They both apply locally a conditional phase to obtain:
$\frac{1}{\sqrt{n}}\sum_{i=0}^{n-1}(-1)^{x_i}\ket{i}(-1)^{y_i}\ket{i}$.
\item They both apply a Hadamard transform, obtaining
\begin{align*}
\frac{1}{n\sqrt{n}}\sum_{i=0}^{n-1} (-1)^{x_i+y_i}\sum_{a\in\01^{\log n}}(-1)^{i\cdot a}\ket{a}\sum_{b\in\01^{\log n}}(-1)^{i\cdot b}\ket{b}
\\
=\frac{1}{n\sqrt{n}}\sum_{a,b\in\01^{\log n}}\left (\sum_{i=0}^{n-1} (-1)^{x_i + y_i + i\cdot(a\oplus b)}\right)\ket{a}\ket{b}.
\end{align*}
\item They measure in the computational basis and output the results $a$ and $b$, respectively.
\end{enumerate}
For every $a$, the amplitude of the basis state $\ket{a}\ket{a}$ in the state before the measurement is
\[
\frac{1}{n\sqrt{n}}\sum_{i=0}^{n-1} (-1)^{x_i + y_i}.
\]
This equals $1/\sqrt{n}$ if $x=y$, and equals~0 otherwise. There are $n$ possible values $a$, so the probability that Bob's measurement outcome equals Alice's measurement outcome $n\cdot (1/\sqrt{n})^2=1$ if $x=y$, and equals~0 otherwise. This solves the problem perfectly, using an $n$-dimensional entangled state at the start (namely $\log n$ EPR-pairs).

What about classical protocols? Suppose there is a classical protocol that uses $C$ bits of communication, and that wins the \emph{non-local}  Deutsch-Jozsa problem with success probability~1.
If Alice and Bob ran this protocol, and then Alice communicated her output $a$ to Bob (using an additional $\log n$ bits), 
then they could solve the \emph{distributed} Deutsch-Jozsa problem since Bob could then check whether $a=b$ or $a\neq b$. 
But we know from Section~\ref{secdistributeddj} that solving the distributed Deutsch-Jozsa problem requires at least $0.007n$ bits of communication.
Hence $C+\log n\geq 0.007 n$, so $C\geq 0.007n - \log n=\Omega(n)$.
Thus we have a non-locality problem that can be solved perfectly without any communication if Alice and Bob share $\log n$ EPR-pairs,
while classically it needs not just \emph{some} communication, but actually a \emph{lot} of communication if we want to solve it perfectly.

\section*{Exercises}

\begin{enumerate}

\item\label{ex:EPRUT} Suppose Alice and Bob share an EPR-pair $\frac{1}{\sqrt{2}}(\ket{00}+\ket{11})$.
\begin{enumerate}
\item Let $U$ be a 1-qubit unitary. Show that the following two states are the same:
(1) the state obtained if Alice applies $U$ to her qubit of the EPR-pair;\\
(2) the state obtained if Bob applies the transpose $U^T$ to his qubit of the EPR-pair.
\item (H) What state do you get if each of Alice and Bob applies a Hadamard transform to their qubit of the EPR-pair?
\end{enumerate}

\item\label{ex:EPRXX}
Alice and Bob share an EPR-pair, $\frac{1}{\sqrt{2}}(\ket{00}+\ket{11})$.
Suppose they each measure their qubit with an $X$-observable (which corresponds to a particular projective measurement with possible outcomes $+1,-1$).
\begin{enumerate}
\item Show that Alice's measurement outcome is uniformly distributed, so 50\%\ probability of outcome $+1$ and 50\%\ probability of outcome $-1$.
\item (H) Show that Alice's and Bob's measurement outcomes are always equal.
\item Suppose we view $X\otimes X$ as one 2-qubit observable (with possible outcomes $+1,-1$) instead of two 1-qubit observables. What is the probability distribution on the two possible outcomes?
\end{enumerate}

\item 
Alice and Bob share $n$ EPR-pairs. Call their shared $2n$-qubit state $\ket{\psi}_{AB}$.
\begin{enumerate}
\item Let $U$ be an arbitrary $n$-qubit unitary and $\bar{U}$ be $U$ after conjugating its entries (without transposing). Prove that $(U\otimes\bar{U})\ket{\psi}_{AB}=\ket{\psi}_{AB}$.
\item Suppose Alice receives some input $x$, and she does an $n$-qubit unitary $U_x$ on her part of the state and then measures in the computational basis, obtaining a classical outcome $a\in\01^n$. What is the probability distribution over Alice's measurement outcomes, and why?
\item Suppose Bob receives the same input $x$ as Alice already received. How can he learn Alice's measurement outcome $a$ from part (b) without communication?
(you may assume Bob knows the map $x\mapsto U_x$)
\end{enumerate}

\item\label{ex:classchsh} (H)
Give a classical strategy using shared randomness for the CHSH game, such that Alice and Bob
win the game with probability at least 3/4 for every possible input $x,y$ (note the order of quantification:
the same strategy has to work for every $x,y$).

\item\label{ex:mermin} 
``Mermin's game'' (a.k.a.\ the GHZ game) is the following.
Consider three space-like separated players: Alice, Bob, and Charlie.
Alice receives input bit $x$, Bob receives input bit $y$, and Charlie receives input bit $z$.
The input satisfies the promise that $x\oplus y\oplus z=0$.
The goal of the players is to output bits $a,b,c$, respectively, such that $a\oplus b\oplus c=\mbox{OR}(x,y,z)$.
In other words, the outputs should sum to~0 (mod~2) if $x=y=z=0$, and should sum to~1 (mod~2) if $x+y+z=2$.
\begin{enumerate}
\item Show that every classical deterministic strategy will fail on at least one of the 4 allowed inputs.\item Show that every classical randomized strategy has success probability at most 3/4 under the uniform distribution on the four allowed inputs $xyz$.
\item Suppose the players share the following entangled 3-qubit state:
$$
\frac{1}{2}(\ket{000}-\ket{011}-\ket{101}-\ket{110}).
$$
Suppose each player does the following: if his/her input bit is~1, apply $H$ to his/her qubit, otherwise do nothing.
Describe the resulting 3-qubit superposition.
\item Using (c), give a quantum strategy that wins the above game with probability~1 on every input that satisfies the promise.
\end{enumerate}

\item 
Suppose Alice, Bob, and Charlie share the 3-qubit ``GHZ-state'' $\frac{1}{\sqrt{2}}(\ket{000}+\ket{111})$, each of them holding one qubit.
\begin{enumerate}
\item What is the local density matrix of Charlie's qubit (viewing the 2 qubits of Alice and Bob as one party)?
\item What is the local density matrix of the 2-qubit state held by Alice and Bob? 
\item Give a protocol where Charlie measures his qubit in some way, communicates the outcome (one classical bit) to Bob, who then applies a local operation so that Alice and Bob end up with an EPR-pair $\frac{1}{\sqrt{2}}(\ket{00}+\ket{11})$.
\end{enumerate}

\item\label{ex:tsirelsonbound} (H)
  This question examines how well the best quantum protocol can do for CHSH (resulting in the so-called ``Tsirelson bound'').
  Consider a protocol where Alice and Bob share a $2k$-qubit state $\ket{\psi}=\ket{\psi}_{AB}$ with $k$ qubits for Alice and $k$ for Bob (the state can be arbitrary and need not consist of EPR-pairs). Alice has two possible $\pm 1$-valued observables
  $A_0$ and $A_1$, and Bob has two possible $\pm 1$-valued observables $B_0$ and $B_1$. Each of these observables acts on $k$ qubits. On inputs $x\in\01$ and $y\in\01$, respectively, Alice measures her half of $\ket{\psi}$ with $A_x$ and outputs the resulting sign $a\in\{+1,-1\}$, and Bob measures his half of $\ket{\psi}$ with $B_y$ and outputs the resulting sign~$b$.  Note that we treat the output bits as signs instead of 0/1 now.  However, the winning condition is the same: the AND of the input bits should equal the parity (XOR) of the output bits. So Alice and Bob \emph{win} the game if $(-1)^{xy}=ab$. 
\begin{enumerate}
\item Show that the expected value of the product $ab$ on inputs $x,y$ is $\bra{\psi}A_x\otimes B_y\ket{\psi}$ (this is the same as $\Tr\left[(A_x\otimes B_y)\ketbra{\psi}{\psi}\right]$).
\item Define $2k$-qubit operator $C=A_0\otimes B_0 + A_0\otimes B_1 + A_1\otimes B_0 - A_1\otimes B_1$.
Show that the winning probability of the protocol (averaged over all 4 inputs pairs $x,y$) is $\frac{1}{2}+\frac{1}{8}\bra{\psi}C\ket{\psi}$.
\item Show that $C^2=4I+(A_0A_1-A_1A_0)\otimes(B_1B_0-B_0B_1)$, where $I$ is the $2k$-qubit identity matrix.
\item Show that $\bra{\psi}C\ket{\psi}\leq \sqrt{8}$. 
\item What can you conclude about the best-possible winning probability among all possible quantum protocols for CHSH?
\end{enumerate}

\end{enumerate}

\chapter{Quantum Cryptography}\label{chap:qcrypto}

\section{Saving cryptography from Shor}

Most classical public-key cryptography in use today can be broken by a large quantum computer. In particular, the RSA system relies on the hardness of factoring integers and hence is broken by Shor's factoring algorithm (see Exercise~\ref{chap:factoring}.\ref{ex:rsa});
and Diffie-Hellman key exchange relies on the hardness of the discrete logarithm problem which was also broken by Shor (see Exercise~\ref{chap:hsp}.\ref{ex:diffie}).
This could clearly become a huge problem for society if and when a large quantum computer is realized: if we cannot securely send messages, make payments or sign transactions online anymore, then much of our economy and society breaks down, or will at least need to be heavily reconfigured.

There are two ways to address this problem.
On the one hand we can try to design other classical cryptographic systems, not based on factoring or discrete log. This part of classical cryptography is---slightly confusingly---called \emph{post-quantum cryptography} (PQC)~\cite{bernstein&lange:pqc}. It relies on the assumed hardness (even for quantum computers) of certain specific computational problems, for instance related to finding short vectors in integer lattices, such as ``learning with errors'' (LWE)~\cite{regev:lwe}.
PQC is by now a big subfield, and we won't cover it here.

On the other hand, we can also try to design cryptographic systems that explicitly rely on quantum effects.
This area is called \emph{quantum cryptography} and is the topic of this chapter.
Compared to post-quantum cryptography, this has the disadvantage that even the honest users of the scheme need to have a (simple) quantum computer at their disposal, but it has the advantage that the security against adversaries in some cases is information-theoretic, not predicated on the assumed but unproven hardness of some computational problems.

\section{Quantum key distribution}
One of the most basic tasks of cryptography is to allow Alice to send a message to Bob (whom she trusts)
over a public channel, without allowing a third party Eve (for ``eavesdropper'') 
to get any information about $M$ from tapping the channel.
Suppose Alice wants to send message $M\in\01^n$ to Bob.
The goal here is not minimal communication, but \emph{secrecy}.
This is often done by public-key cryptography such as RSA.
Such schemes, however, are only computationally secure, not information-theoretically secure: all the information about
the private key can be computed from the public key, it just appears to take a lot of time to compute it---assuming of course that problems like factoring are classically hard, and that nobody builds a quantum computer\ldots

In contrast, the following ``one-time pad'' scheme is information-theoretically secure.
If Alice and Bob share a \emph{secret key} $K\in\01^n$ then Alice can send $C=M\oplus K$ over the channel.
By adding $K$ to what he received, Bob learns $M$.
On the other hand, if Eve didn't know anything about $K$ then she learns nothing about $M$
from tapping the message $M\oplus K$ that goes over the channel.
How can we make Alice and Bob share a secret key, about which Eve will have no or negligibly little information?
In the classical world this is impossible, but with quantum communication it can be done!

Below we describe the famous BB84 quantum key distribution (QKD) protocol of Bennett and Brassard~\cite{bb84}.
Consider two possible bases: basis~0 is the computational basis $\{\ket{0},\ket{1}\}$,
and basis~1 is the Hadamard basis $\{\ket{+},\ket{-}\}$.
The main property of quantum mechanics that we'll use, is that if a bit $b$ is encoded in an unknown basis,
then Eve cannot get information about $b$ without disturbing the state, and the latter can
be detected by Alice and Bob.\footnote{Quantum key distribution might in fact better be called ``quantum eavesdropper detection.'' There is another important assumption underlying BB84 that should be made explicit: we assume that the classical channel used
in steps~3--5 is ``authenticated,'' meaning that Alice and Bob know they are talking to each other, and Eve can listen but not change the bits sent over the classical channel (in contrast to the qubits sent during step~1 of the protocol, which Eve
is allowed to manipulate in any way she wants). One can authenticate a classical communication channel by using some shared secret key; if this is used, then one may think of QKD as something that allows to grow an initial shared secret key, rather than as something that conjures up a shared random key out of nothing.}
\begin{enumerate}
\item
Alice chooses $n$ random bits $a_1,\ldots,a_n$ and $n$ random bases $b_1,\ldots,b_n$.
She sends $a_i$ to Bob in basis $b_i$ over the public quantum channel. For example, if $a_i=0$ and $b_i=1$ then the 
$i$-th qubit that she sends is in state $\ket{+}$.
\item
Bob chooses random bases $b'_1,\ldots,b'_n$ and measures the qubits he received in those bases, 
yielding bits $a'_1,\ldots,a'_n$.
\item Bob sends Alice all $b_i'$ (this also signals to Alice that Bob has measured the qubits he received),
  and Alice sends Bob all $b_i$. 
Note that for roughly $n/2$ of the $i$s, Alice and Bob used the same basis $b_i=b'_i$.
For those $i$ Bob should have $a'_i=a_i$ (if there was no noise and Eve didn't tamper with the $i$-th qubit on the channel). 
Both Alice and Bob know for which $i$s this holds.  Let's call these roughly $n/2$ positions the ``shared string.''
\item Alice randomly selects $n/4$ locations in the shared string, and sends Bob those locations as well as
the values~$a_i$ at those locations.  Bob then checks whether they have the same bits in those positions. 
If the fraction of errors is significantly  higher than the error rate induced by the natural noise in the channel, then they suspect some eavesdropper was
tampering with the channel, and they abort.
\item If the test is passed, then they discard the $n/4$ test-bits, and have roughly $n/4$ bits left in their shared string.
This is called the ``raw key.''
Now they do some classical postprocessing on the raw key:
``information reconciliation'' to ensure they end up with exactly the same shared string, and ``privacy amplification'' to ensure that Eve has negligible information about that shared string.\footnote{This can be done for instance by something called the ``leftover hash lemma.''}
\end{enumerate}
The communication is $n$ qubits in step~1, $2n$ bits in step~3, $O(n)$ bits in step~4, and $O(n)$ bits in step~5.
So the required amount of communication is linear in the length of the shared secret key that Alice and Bob end up with.

It's quite hard to formally \emph{prove} that this protocol yields (with high probability) a shared key about which Eve has negligible information.  
In fact it took more than 12~years before BB84 was finally proven secure~\cite{mayers:security,lo&chau:security}.
The main reason it works is that when the qubits that encode $a_1,\ldots,a_n$ are going
over the public channel, Eve doesn't know yet in which bases $b_1,\ldots,b_n$ these are encoded
(she will learn the $b_i$ later from tapping the classical communication in step~3, 
but at that point this information is not of much use to her anymore).  
She could try to get as much information as she can about $a_1,\ldots,a_n$ by some measurement, 
but there's an \emph{information-vs-disturbance tradeoff}:
the more information Eve learns about $a_1,\ldots,a_n$ by measuring the qubits, the more she will disturb the state,
and the more likely it is that Alice and Bob will detect her presence in step~4.

\begin{figure}[hbt]
\centering
\includegraphics{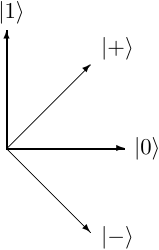}
\caption{The four possible states in BB84 encoding:  
$\ket{0}$ and $\ket{+}$ are two different encodings of~0, and 
$\ket{1}$ and $\ket{-}$ are two different encodings of~1.}\label{fig:bb84states}
\end{figure}

We won't go into the full proof details here, just illustrate the information-disturbance tradeoff for the case where Eve individually attacks the qubits encoding each bit in step~1 of the protocol.\footnote{The more complicated situation where Eve does 
an $n$-qubit measurement on all qubits of step~1 simultaneously can be reduced to the case of individual-qubit 
measurements by something called the \emph{quantum De Finetti theorem}, but we won't go into the details here.}
In Fig.~\ref{fig:bb84states} we give the four possible states for one BB84-qubit.
If Alice wants to send $a_i=0$, then she sends a uniform mixture of $\ket{0}$ and $\ket{+}$ across the channel;
if Alice wants to send $a_i=1$ she sends a uniform mixture of $\ket{1}$ and $\ket{-}$.
Suppose Eve tries to learn $a_i$ from the qubit on the channel.
The best way for her to do this is to measure in the orthonormal basis corresponding to state $\cos(\pi/8)\ket{0}+\sin(\pi/8)\ket{1}$
and $-\sin(\pi/8)\ket{0}+\cos(\pi/8)\ket{1}$.  Note that the first state is halfway between the two encodings of~0,
and the second state is halfway between the two encodings of~1 (remember that $\ket{-}$ and $-\ket{-}$ are physically indistinguishable because they only differ by a global phase).
This will give her the value of~$a_i$ with probability $\cos(\pi/8)^2\approx 0.85$ 
(remember the 2-to-1 quantum random access code from Exercise~\ref{ex:2to1rac} 
of Chapter~\ref{chap:qencodings}).
However, this measurement will change the state of the qubit by an angle of at least $\pi/8$, 
so if Bob now measures the qubit he receives in the same basis as Alice, then his probability 
of recovering the incorrect value of $a_i$ is at least $\sin(\pi/8)^2\approx 0.15$
(if Bob measured in a different basis than Alice, then the result will be discarded anyway).
If this $i$ is among the test-bits Alice and Bob use in step~4 of the protocol (which happens with probability 1/2), 
then they will detect an error. 
Eve can of course try a less disturbing measurement to reduce the probability of being detected, 
but such a measurement will also have lower probability of telling her $a_i$.

\section{Reduced density matrices and the Schmidt decomposition}

Suppose Alice and Bob share some pure state $\ket{\phi}$.
If this state is entangled, it cannot be written as a tensor product $\ket{\phi_A}\otimes\ket{\phi_B}$
of separate pure states for Alice and Bob. Still, there is a way to describe Alice's local state
as a mixed state, by \emph{tracing out} Bob's part.
Formally, if $C\otimes D$ is a tensor product matrix then $\Tr_B(C\otimes D)=C\cdot \Tr(D)$.
By extending this linearly to matrices that are not of product form, the operation $\Tr_B$ is
well-defined on all mixed states. Note that $\Tr_B$ removes Bob's part of the state, leaving just Alice's part of the state. 
If $\rho_{AB}$ is some bipartite state (mixed or pure, entangled or not),
then $\rho_A=\Tr_B(\rho_{AB})$ is Alice's local density matrix.  This describes all the information she has.
For example, for an EPR-pair $\ket{\phi}=\frac{1}{\sqrt{2}}(\ket{00}+\ket{11})$, the corresponding
density matrix is 
\begin{eqnarray*}
\rho_{AB} & = & \frac{1}{2}(\ketbra{00}{00}+\ketbra{00}{11}+\ketbra{11}{00}+\ketbra{11}{11})\\
& = & \frac{1}{2}(\ketbra{0}{0}\otimes\ketbra{0}{0}+\ketbra{0}{1}\otimes\ketbra{0}{1}+\ketbra{1}{0}\otimes\ketbra{1}{0}+\ketbra{1}{1}\otimes\ketbra{1}{1}),
\end{eqnarray*}
and since $\Tr(\ketbra{a}{b})=1$ if $a=b$ and $\Tr(\ketbra{a}{b})=0$ if $\ket{a}$ and~$\ket{b}$ are orthogonal, we have
$$
\rho_A=\Tr_B(\rho_{AB})=\frac{1}{2}(\ketbra{0}{0}+\ketbra{1}{1}).
$$
In other words, Alice's local state is the same as a random coin flip!
Similarly we can compute Bob's local state by tracing out Alice's part of the space:  $\rho_B=\Tr_A(\rho_{AB})$.
Note that the original 2-qubit density matrix $\rho_{AB}$ is \emph{not} equal to $\rho_A\otimes\rho_B$,
because the tracing-out operation has ``removed'' the entanglement between the two qubits.

The Schmidt decomposition is a very useful way to write bipartite pure states, and allows us to easily
calculate the local density matrices of Alice and Bob.
It says the following: for every bipartite pure state $\ket{\phi}$ there is a unique integer $d$ (called the \emph{Schmidt rank} of $\ket{\phi}$), an orthonormal set of states $\ket{a_1},\ldots,\ket{a_d}$ for Alice's space,
an orthonormal set of states $\ket{b_1},\ldots,\ket{b_d}$ for Bob's space,
and positive reals $\lambda_1,\ldots,\lambda_d$ whose squares sum to~1, such that
\begin{equation}\label{eqschmidtdecom}
\ket{\phi}=\sum_{i=1}^d\lambda_i\ket{a_i}\ket{b_i}.
\end{equation}
For example, an EPR-pair has Schmidt coefficients $\lambda_1=\lambda_2=1/\sqrt{2}$ and hence has Schmidt rank~2.
The Schmidt rank and the Schmidt coefficients of a state $\ket{\phi}$ are unique, but there is some freedom in the choice of bases if the $\lambda_j$ are not all distinct. For example
$$
\frac{1}{\sqrt{2}}(\ket{00}+\ket{11})=\frac{1}{\sqrt{2}}(\ket{++}+\ket{--})
$$
are two distinct Schmidt decompositions of the EPR-pair.

The existence of the Schmidt decomposition is shown as follows.
Let $\rho_A=\Tr_B(\ketbra{\phi}{\phi})$ be Alice's local density matrix.
This is Hermitian, so it has a spectral decomposition $\rho_A=\sum_{i=1}^d\mu_i\ket{a_i}\bra{a_i}$ 
with orthonormal eigenvectors $\ket{a_i}$ and positive real eigenvalues $\mu_i$.
Note that $d$ is the rank of $\rho_A$, and $\sum_i\mu_i=\Tr(\rho_A)=1$.
Then there are $c_{ij}$ such that
$$
\ket{\phi}=\sum_{i=1}^d\sum_{j=1}^D\sqrt{\mu_i}c_{ij}\ket{a_i}\ket{j},
$$
where the $\ket{j}$ are the computational basis states for Bob's $D$-dimensional overall space.
Define $\lambda_i=\sqrt{\mu_i}$ and $\ket{b_i}=\sum_{j=1}^D c_{ij}\ket{j}$.
This gives the decomposition of $\ket{\phi}$ of Eq.~(\ref{eqschmidtdecom}).
It only remains to show that $\{\ket{b_i}\}$ is an orthonormal set, which we do as follows.
The density matrix version of Eq.~(\ref{eqschmidtdecom}) is
$$
\ketbra{\phi}{\phi}=\sum_{i,j=1}^d\lambda_i\lambda_j\ketbra{a_i}{a_j}\otimes\ketbra{b_i}{b_j}.
$$
We know that if we trace out the $B$-part from $\ketbra{\phi}{\phi}$, then we should get $\rho_A=\sum_i\lambda_i^2\ketbra{a_i}{a_i}$,
but that can only happen if $\inp{b_j}{b_i}=\Tr(\ketbra{b_i}{b_j})=1$ for $i=j$ and $\inp{b_j}{b_i}=0$ for $i\neq j$.
Hence the $\ket{b_i}$ form an orthonormal set.
Note that from Eq.~(\ref{eqschmidtdecom}) it easily follows that Bob's local density matrix is $\rho_B=\sum_i\lambda_i^2\ketbra{b_i}{b_i}$.

\section{The impossibility of perfect bit commitment}

Key distribution is just one of the many tasks cryptographers would like to solve.
Another important primitive is \emph{bit commitment}.
In this scenario there is no eavesdropper, but Alice and Bob don't trust each other.
Suppose Alice has a bit $b$ which for the time being she doesn't want to reveal to Bob,
though she would like to somehow convince Bob that she has already made up her mind about $b$ and won't change its value later.
A protocol for bit commitment comes in two stages, each of which may involve several rounds of communication:
\begin{enumerate}
\item In the ``commit'' phase Alice gives Bob a state which is supposed to commit her to the value of $b$ 
(without informing Bob about the value of $b$).
\item In the ``reveal'' phase Alice sends $b$ to Bob, and possibly some other information to allow him to check that this is indeed the same value $b$ that Alice committed to before.
\end{enumerate}
A protocol is \emph{binding} if Alice can't change her mind, meaning she can't get Bob to ``open'' $1-b$.
A protocol is \emph{concealing} if Bob cannot get any information about $b$ before the ``reveal phase.''%
\footnote{A good metaphor to think about this: in the commit phase Alice locks $b$ inside a safe which she sends to Bob. This commits her to the value of $b$, since the safe is no longer in her hands.
During the reveal phase she sends Bob the key to the safe, who can then open it and learn $b$.}

A good protocol for bit commitment would be a very useful building block for many other cryptographic applications.  
For instance, it would allow Alice and Bob (who still don't trust each other) to jointly flip a fair coin.  
Maybe they're going through a divorce, and need to decide who gets to keep their joint car.
Alice can't just flip the coin by herself because Bob doesn't trust her to do this honestly, and vice versa.
Instead, Alice would pick a random coin $b$ and commit to it.
Bob would then pick a random coin $c$ and send it to Alice.
Alice then reveals $b$, and the outcome of the coin flip is defined to be $b\oplus c$.
As long as at least one of the two parties follows this protocol, the result will be a fair coin flip.

Perfect coin flipping (and hence also perfect bit commitment) are known to be impossible in the classical world.
After BB84 there was some hope that perfect bit commitment (and hence also perfect coin flipping) would be possible in the quantum world,
and there were some seemingly information-theoretically secure proposals for quantum protocols to achieve this.
Unfortunately it turns out that there is \emph{no} quantum protocol for bit commitment that is both
perfectly binding and perfectly concealing.

To show that a protocol for perfect bit commitment is impossible,
consider the joint pure state $\ket{\phi_b}$ that Alice and Bob would have if Alice wants to commit to bit-value $b$,
and they both honestly followed the protocol.\footnote{The assumption that the state is pure rather than mixed is without loss of generality.}
If the protocol is perfectly concealing, then the
reduced density matrix on Bob's side should be independent of $b$, i.e., 
$\Tr_A(\ketbra{\phi_0}{\phi_0})=\Tr_A(\ketbra{\phi_1}{\phi_1})$.
The way we constructed the Schmidt decomposition in the previous section now 
implies that there exist Schmidt decompositions of $\ket{\phi_0}$ and $\ket{\phi_1}$ 
with the same $\lambda_i$'s and the same $b_i$'s: there exist orthonormal bases $\{a_i\}$ and $\{a'_i\}$ such that
$$
\ket{\phi_0}=\sum_{i=1}^d\lambda_i\ket{a_i}\ket{b_i}\mbox{ \ and \ }
\ket{\phi_1}=\sum_{i=1}^d\lambda_i\ket{a'_i}\ket{b_i}
$$
Now Alice can locally switch from $\ket{\phi_0}$ to $\ket{\phi_1}$ by just applying on her part of the state
the map $\ket{a_i}\mapsto\ket{a'_i}$. 
Alice's map is unitary because it takes one orthonormal basis to another orthonormal basis.
But then the protocol is not binding at all:
Alice can still freely change her mind about the value of $b$ after the ``commit'' phase is over!
Accordingly, if a quantum protocol for bit commitment is perfectly concealing, it cannot be binding at all.

\section{More quantum cryptography}

Quantum cryptography is by now a pretty large subset of the area of quantum information and computation.
Here we just briefly mention a few other topics in quantum crypto (see~\cite{broadbent&schaffner:beyondQKD}):
\begin{itemize}
\item There are quantum protocols for bit commitment that are partially concealing and partially
binding---something which is still impossible in the classical world.
A primitive called ``weak coin flipping'' can be implemented almost perfectly in the quantum world, 
and cannot be implemented at all in the classical world.
\item Under assumptions on the fraction of dishonest players among a set of $k$ parties, it is possible 
to implement \emph{secure multi-party quantum computation}.  This is a primitive that allows the players
to compute any function of their $k$ inputs, without revealing more information to player~$i$ 
than can be inferred from $i$'s input plus the function value.
\item One can actually do nearly perfect bit commitment, coin flipping, etc., assuming
the dishonest party has \emph{bounded quantum storage}, meaning that it can't keep large quantum states coherent for longer times.  
At the present state of quantum technology this is a very reasonable assumption, 
though a breakthrough in physical realization of quantum computers could wipe out this approach.
\item In \emph{device-independent} cryptography, Alice and Bob want to solve certain cryptographic tasks like key distribution or randomness generation without trusting their own devices (for instance because they don't trust the vendor of their apparatuses).  Roughly speaking, the idea here is to use Bell-inequality violations to prove the presence of entanglement, and then use this entanglement for cryptographic purposes.  Even if Alice or Bob's apparatuses have been tampered with, they can still only violate things like the CHSH inequality if they actually share an entangled state.
\item Experimentally it is much easier to realize quantum key distribution than general quantum computation,
because you basically just need to prepare qubits (usually photons) in either the computational or the Hadamard basis,
send them across a channel (usually an optical fibre, but sometimes free space), and measure them in either the computational or the Hadamard basis.
Many sophisticated experiments have already been done. 
Somewhat surprisingly, you can already commercially buy quantum key distribution machinery.  Unfortunately the implementations are typically not perfect (for instance, we don't have perfect photon sources or perfect photon detectors), and once in a while another loophole is exposed in the implementation, which the vendor then tries to patch, etc.
\end{itemize}

\section*{Exercises}

\begin{enumerate}

\item Here we will consider in more detail the information-disturbance tradeoff for measuring a qubit in one of the four BB84 states
(each of which occurs with probability 25\%).
\begin{enumerate}
\item Suppose Eve measures the qubit in the orthonormal basis given by $\cos(\theta)\ket{0}+\sin(\theta)\ket{1}$ and  $\sin(\theta)\ket{0}-\cos(\theta)\ket{1}$,
  for some parameter $\theta\in[0,\pi/4]$. The first basis vector corresponds to output~0, the second to output~1.
  For each of the four possible BB84 states, give the probabilities of outcome~0 and outcome~1
(so your answer should consist of 8~numbers, each of which is a function of $\theta$).
\item What is the average probability that Eve's measurement outcome equals the encoded bit $a_i$, as a function of $\theta$?
(average taken both over the uniform distribution over the four BB84 states, and over the probabilities calculated in part (a))
\item By what angle does the state change in each of the 8 cases of~(a)?
\end{enumerate}

\item
\begin{enumerate}
\item What is the Schmidt rank of the state $\frac{1}{2}(\ket{00}+\ket{01}+\ket{10}+\ket{11})$?
\item Suppose Alice and Bob share $k$ EPR-pairs.  What is the Schmidt rank of their joint state?
\item Prove that a pure state $\ket{\phi}$ is entangled if, and only if, its Schmidt rank is greater than~1.
\end{enumerate}
   
\item Give the Schmidt decomposition of the state $\frac{1}{2}(\ket{0}_A\ket{0}_B+\ket{0}_A\ket{1}_B+\ket{1}_A\ket{1}_B+\ket{1}_A\ket{2}_B)$. Here Alice's space has dimension~2, and Bob's space has dimension~3. It suffices if you write down your Schmidt decomposition, being explicit about the values of the $\lambda_i$'s and what are the states $\ket{a_i}$ and $\ket{b_i}$. You can add your calculation (involving local density matrices etc.) as a justification, but you don't have to.

\item 
Consider a density matrix $\rho$ on Alice's Hilbert space. A bipartite pure state $\ket{\psi}_{AB}$ is called a \emph{purification} of $\rho$, if $\rho=\Tr_B(\ketbra{\psi}{\psi})$. The $B$-register in $\ket{\psi}_{AB}$ is called the \emph{purifying register}.
\begin{enumerate}
\item Show that an EPR-pair is a purification of the 1-qubit mixed state $\rho=I/2$.
\item Show that if $\rho$ is a density matrix of rank~$r$, then there exists a purification of $\rho$ where the purifying register has at most $\ceil{\log r}$ qubits.
\item Show that if $\ket{\psi}_{AB}$ and $\ket{\psi'}_{AB}$ are purifications of the same $\rho$, then there exists a unitary $U$ on Bob's space such that $\ket{\psi'}_{AB}=(I\otimes U)\ket{\psi}_{AB}$.
\end{enumerate}

\item\label{ex:qonetimepad} Suppose Alice has a 1-qubit state $\rho$.
\begin{enumerate}
\item Suppose Alice chooses a uniformly random Pauli matrix (see Appendix~\ref{app:pauli}) and applies it to $\rho$. What is the resulting density matrix, averaged over the four cases?
\item Suppose Alice and Bob shared a uniformly distributed secret 2-bit string $ab$, which is unknown to Eve. How can Alice send $\rho$ to Bob over a public quantum channel, without leaking any information to Eve (i.e., the quantum state sent over the channel should by itself be independent of $\rho$), in such a way that Bob can recover $\rho$?\\
{\footnotesize Comment: This may be called a ``quantum one-time pad'': we use two classical bits of secret key to hide one qubit, while the classical one-time pad uses one classical bit of secret key to hide one classical bit.}
\end{enumerate}

\item\label{ex:secretsharing} (H) Suppose we have a qubit in mixed state $\rho$ that we want to hide from Alice and Bob individually, but in such a way that if Alice and Bob cooperate, then they can recover $\rho$.

Describe how we can change $\rho$ into some other 1-qubit state $\rho'$, what secret keys we give to Alice and Bob, why individually they can get no information about $\rho$ from the qubit $\rho'$, and why jointly they can fully recover the qubit in state $\rho$ from $\rho'$. The keys should be classical.
  
\item\label{ex:unitaryinv} (H)
Prove that Alice cannot give information to Bob by doing a unitary operation on her part of an entangled pure state.

\item Suppose Alice sends \emph{two} $n$-bit messages $M_1$ and $M_2$ with the one-time pad scheme, reusing the \emph{same} $n$-bit key $K$.
Show that Eve can now get some information about $M_1,M_2$ from tapping the classical channel.

\item\label{ex:decentpurestate}
\begin{enumerate}
    \item (H) Consider a bipartite pure state shared between Alice and Bob, where Alice and Bob's local spaces have dimension $d$ each:
    $$
    \sum_{i,j\in\{1,\ldots,d\}}\alpha_{ij}\ket{i}_A\ket{j}_B.
    $$
    The state is given to you classically, as a list of $d^2$ amplitudes, each described by $O(d)$ bits. Give a classical polynomial-time algorithm to find the Schmidt coefficients and to find Alice and Bob's basis for a Schmidt decomposition.
    \item Give a classical polynomial-time algorithm that decides whether a given bipartite pure state (given as in (a)) is entangled or not.\\
 {\footnotesize Comment: If the given state were mixed instead of pure, this decision problem is known to be NP-hard and hence probably not polynomial-time solvable.}
\end{enumerate}

\end{enumerate}

\chapter{Quantum Machine Learning}\label{chap:qml}

\section{Introduction}

Machine learning tries to extract patterns and regularities from given data for the purposes of prediction and understanding. In a slogan, one could say:
ML $=$ data $+$ optimization.
The data is what you learn from; the optimization finds a good model or hypothesis for the given data, which hopefully has some generalization power.
ML has gone through several ups and downs over the years, but currently is booming thanks to the success of so-called ``deep learning,'' based on neural networks.\footnote{Machine learning based on neural networks has been studied for decades but quite suddenly became much more successful starting around 2012, due to the availability of more data, stronger computing hardware (incl.\ special-purpose GPUs for fast parallel matrix-vector calculations), and better software to do the training.} 
ML is often subdivided into three subareas, depending on the data one has:
\begin{enumerate}
    \item In \emph{supervised learning} we are given labelled data, for instance pictures of animals annotated with the kind of animal that's on the picture, and we want to learn how to predict the label.
    \item In \emph{unsupervised learning} we are just given unlabeled data, and need to find patterns in it. The canonical example is the \emph{clustering} problem, where we are given unlabelled data items that we want to group into ``similar'' subsets. For example, it could be that our data consists of pictures of different kinds of animals (not labeled with the type of animal), and we somehow want to cluster the cat-pictures together, the wolf-pictures together,  etc. We may or may not know in advance what the number of clusters should be.
    \item In \emph{reinforcement learning} the learner actually interacts with the environment, receiving rewards or penalties for desirable or undesirable behavior, and tries to learn from this interactive data to behave more successfully in the environment. This is roughly how a child learns.\footnote{It's also how a computer can learn to play games. One of the big breakthroughs of machine learning was in 2016 when the AlphaGo program learned to play the game of Go so well that it beat one of the world's best human Go players. Computers have been better than humans at chess already since the late 1990s (the IBM program ``Deep Blue'' beat Kasparov in 1997), but Go was viewed as a much more complicated game than chess.}
\end{enumerate}
It is a very interesting question to see how quantum computing changes and helps machine learning. Here the learner would be a quantum computer, and the data may be classical or quantum. 
Quantum ML is by now a rather large area, and in this chapter we will go over a few representative results and methods for supervised and unsupervised learning, mostly with classical output. See~\cite{DTB:advqreinforcement}
for quantum applications to reinforcement learning, and \cite{biamonteea:qml,arunachalam:quantumlearningsurvey,schuld&petruccione:qmlbook} for much more.

\section{Supervised learning from quantum data}

\subsection{The PAC model of learning}

Let us first describe a mathematical model of what it means to learn from labeled data. This is Valiant's PAC model~\cite{valiant:paclearning}, for ``probably approximately correct'' learning (see \cite{shwartz&david:learningbook,MRT:foundationsML} for more).

Assume for simplicity that the labels are just binary: 0 or 1. 
Our goal is to learn a Boolean function $f:{\cal X}\to\01$ from examples of the form $(x,f(x))$, where $x\in{\cal X}$. 
A typical case would be ${\cal X}=\01^n$.
The last bit $f(x)$ of the example is called the \emph{label}. 
Think for instance about the case where we are given $1000\times 1000$-pixel black-and-white pictures ($n=1000,000$) whose labels $f(x)$ indicate whether $x$ is the picture of a wolf or not. We would like to learn $f$, or some good approximation of it, to be able to recognize pictures of wolves in the future. Some $x$'s are more important and more likely to appear as examples than others: many $1000\times 1000$-grids don't depict anything. The assumption in PAC learning is that the examples are generated (independent and identically distributed) according to some distribution $D$ on $\cal X$. The idea is that this $D$ represents ``the world'' or ``Nature,'' which provides us with examples. 
We assume $f$ cannot be completely arbitrary (in that case there would be an $f$ consistent with every possible sequence of labeled examples) but comes from some known ``concept class'' $\cal C$ of Boolean functions. For instance, $\cal C$ could be a set of small logical formulas $f$ on $n$ Boolean variables, or a set of small-depth or small-size decision trees on $n$ input bits, or neural networks with a restricted number of nodes or depth. 

A learning algorithm should generate a ``hypothesis'' $h:{\cal X}\to\01$ that has small error compared to the unknown $f$ that we're trying to learn, \emph{measured under the same distribution $D$ that generated the data}.\footnote{It is important to be taught and tested according to the same distribution $D$. Imagine a quantum-computing course whose lectures focused on the mathematics of quantum algorithms, but with an exam that focuses on physics questions about how to implement qubits and gates---that would clearly be very unreasonable.} The \emph{generalization error} of $h$ w.r.t.\ the target function $f$ is defined as
\[
err_D(f,h)=\Pr_{x\sim D}[f(x)\neq h(x)].
\]
This error measures how well we've generalized the examples, and how well we can predict the labels of future examples.
We say that $h$ is ``approximately correct'' if this error is small, at most some specified~$\eps$.
The goal in PAC learning is to output an $h$ that is probably approximately correct:

\begin{definition}
An $(\eps,\delta)$-PAC learner for a concept class $\cal C$ w.r.t.\ distribution $D$ on $\cal X$, is an algorithm that receives $m$ labeled examples $(x_1,f(x_1)),\ldots,(x_m,f(x_m))$ for a \emph{target function} $f\in{\cal C}$, where each $x_i\sim D$, and that outputs a hypothesis $h$ such that
$$
\Pr[err_D(f,h)\leq\eps]\geq 1-\delta.
$$
The learning algorithm has to satisfy the above for every possible target function $f\in{\cal C}$, and the probability is over both the choice of the examples and over the internal randomness of the algorithm.

An $(\eps,\delta)$-PAC learner for a concept class $\cal C$ is an algorithm that is an $(\eps,\delta)$-PAC learner for $\cal C$ w.r.t.\ every possible distribution $D$. 
\end{definition}

Note that the first part of the  definition is about learners that are only required to work correctly for one specific distribution $D$ (for instance, the uniform distribution over $\cal X$), while the second part is ``distribution-independent'': here we want a learner that works well irrespective of what (unknown) distribution $D$ generates the data.
This is in keeping with the usual attitude towards algorithms in computer science: these should work well even for a worst-case input. We don't require the class $\cal H$ of possible hypotheses~$h$ to equal the class $\cal C$ of possible target functions~$f$ (if we add this requirement, then it's called \emph{proper} PAC learning). This allows us for instance to use neural networks to learn target functions that come from some other class~$\cal C$, say logical formulas. 

The number of examples $m$ that a particular learning algorithm uses is called its ``sample complexity,'' and the overall time or number of elementary operations it takes to output $h$ is its ``time complexity.''
Clearly the latter upper bounds the former, since we need at least one operation to process one example. 
The sample complexity of a concept class~$\cal C$ (as a function of $\eps,\delta$) is the minimal sample complexity among all PAC learners for $\cal C$.
Ideally, a good learner for $\cal C$ has both small sample complexity and small time complexity (say, polynomial in $n$). For some concept classes $\cal C$ efficient distribution-independent PAC learners exist, for example the class of logical formulas in $k$-Conjunctive Normal Form (i.e., each $f$ would be the AND of several ORs, each of at most $k$ variables or negated variables) or the class of regular languages (with the added help of so-called ``membership queries''), but there are also many $\cal C$ that are not efficiently learnable.

\subsection{Learning from quantum examples under the uniform distribution}

There are different ways to define learning from \emph{quantum} data. One natural way, due to Bshouty and Jackson~\cite{bshouty:quantumpac}, is to replace each classical random example $(x,f(x))$, with $x\sim D$, by a superposition. Focusing on the typical case ${\cal X}=\01^n$, a \emph{quantum example} would be the $(n+1)$-qubit state
\[
\sum_{x\in\01^n}\sqrt{D(x)}\ket{x,f(x)}.
\]
Of course, the world doesn't usually present us with quantum examples, in contrast to the abundance of classical data for machine learning. So this model is only relevant in special cases, for example if we have a physical experiment producing such states.

One thing we could do with such a quantum example is measure it in the computational basis, but that would just give us back a classical example $(x,f(x))$ with $x\sim D$.
A more clever thing we can do is \emph{Fourier sampling}.
Suppose $D$ is the uniform distribution.
Exercise~\ref{ex:labelinphase} shows how to convert a quantum example (with probability 1/2) into an $n$-qubit state where the labels are $\pm 1$-phases:
\[
\frac{1}{\sqrt{2^n}}\sum_{x\in\01^n}(-1)^{f(x)}\ket{x}.
\]
If we apply $n$ Hadamard gates to this state, then we get 
\[
\sum_{s\in\01^n}
\frac{1}{2^n}\sum_{x\in\01^n}(-1)^{x\cdot s}(-1)^{f(x)}\ket{s}=\sum_{s\in\01^n}\alpha_s\ket{s}.
\]
If we measure this state, then we'll see outcome $s\in\01^n$ with probability $\alpha_s^2$.
The amplitudes $\alpha_s$ are called the \emph{Fourier coefficients} of the function $(-1)^{f(x)}$, whence the name ``Fourier sampling.''
In some cases Fourier sampling gives a lot of information about the $f$ we're trying to learn. 

\paragraph{Learning linear functions.}
A perfect illustration of Fourier sampling is for the following class:
\[
{\cal C}=\{f_a\mid a\in\01^n, \forall x: f_a(x)=a\cdot x\mbox{ mod 2}=\sum_{i=1}^n a_ix_i\mbox{ mod 2}\},
\]
these are the linear functions modulo 2.
It is easy to calculate that if we do Fourier sampling on a quantum example for function $f_a$, then $\alpha_a=1$ and $\alpha_{s}=0$ for all $s\neq a$. So one Fourier sample already tells us what $a$ is! Hence we can learn $f_a$ \emph{exactly} (i.e., with $\eps=0$), with high probability, using $O(1)$ examples and $O(n)$ elementary gates. In contrast, learning linear functions from classical examples under the uniform distribution requires $\Theta(n)$ examples (see Exercise~\ref{ex:claslinearfns}).

\paragraph{Learning DNF.}
A richer concept class that can be learned efficiently from uniform quantum examples is the class of $s$-term Disjunctive Normal Form (DNF) formulas on $n$ Boolean variables. These are formulas of the form $f(x)=(x_1\wedge\neg x_3)\vee(x_2\wedge x_3\wedge x_5)$,
i.e., an OR of up to $s$ different ANDs of variables or negations of variables. The concept class~$\cal C$ of $s$-term DNF is not known to be efficiently PAC learnable w.r.t.\ the uniform distribution $D$ classically.
However, Bshouty and Jackson~\cite{bshouty:quantumpac} showed that $s$-term DNF \emph{can} be learned in polynomial time (in $s$ and $n$) from uniform \emph{quantum} examples. Roughly speaking, they use Fourier sampling to produce a linear function that is weakly correlated with the target DNF function $f$, and then use a classical ``boosting'' algorithm to combine multiple such weak hypotheses into one good hypothesis~$h$. We'll skip the details here.

\subsection{Learning from quantum examples under all distributions}

We saw a few cases where quantum examples reduce the sample and/or time complexity of learning algorithms w.r.t.\ a fixed data-generating distribution~$D$, namely uniform~$D$. But in the PAC model we ideally want a \emph{distribution-independent} learner that works well for every possible distribution~$D$. Can allowing quantum instead of classical examples significantly reduce the sample complexity of learning a class $\cal C$ in the distribution-independent setting?
It turns out the answer is `no'.

Classically, the number of examples that is necessary and sufficient for ($\eps,\delta)$-PAC learning a concept class $\cal C$ is known to be~\cite{blumer:optimalpacupper,hanneke:optimalpac} 
\begin{equation}\label{eq:PACsamplecomplexity}
m=\Theta\left(\frac{VCdim({\cal C})}{\eps} + \frac{\log(1/\delta)}{\eps}\right),
\end{equation}
where $VCdim({\cal C})$ is the so-called \emph{VC-dimension} of $\cal C$, named after Vapnik and Chervonenkis~\cite{vapnik:vcdimension} and defined as follows.  We say that a set $S\subseteq\01^n$ is \emph{shattered} by $\cal C$ if for each of the $2^{|S|}$ possible labelings $\ell:S\to\01$, there is a function $f\in{\cal C}$ that has the same labeling of $S$ (i.e., $f_{|S}=\ell$). $VCdim({\cal C})$ is the size of a largest $S$ shattered by~$\cal C$. Intuitively, larger VC-dimension corresponds to a more complex or ``richer'' (and hence harder to learn) concept class. 
We won't prove the characterization of Eq.~\eqref{eq:PACsamplecomplexity} here, but Exercises~\ref{ex:VCdimUB} and~\ref{ex:VCdimLB} go most of the way towards the claimed upper and lower bounds on~$m$, respectively.

It was proven in~\cite{arunachalam:optimalpaclearning} that in fact the same formula Eq.~\eqref{eq:PACsamplecomplexity} determines the number of \emph{quantum} examples that are necessary and sufficient for learning $\cal C$.
The sufficiency is trivial: just measure the quantum examples  and run the best classical PAC learner.
The necessity was proved by reducing a quantum measurement problem to the problem of PAC learning $\cal C$ from quantum examples, and showing that the number of copies of the example-state required to solve that measurement problem is at least the expression of Eq.~\eqref{eq:PACsamplecomplexity}.
So, up to constant factors, quantum examples are not more useful than classical examples for distribution-independent PAC learning.

\subsection{Learning quantum states from classical data}

One can generalize PAC learning from Boolean-valued to \emph{real-valued} target functions $f:{\cal X}\to[0,1]$, and then consider a hypothesis $h:{\cal X}\to[0,1]$ to be approximately correct (for some small $\gamma$) if
\[
err_{D,\gamma}(f,h)=\Pr_{x\sim D}[|f(x)-h(x)|>\gamma]\leq\eps.
\]
So now a good hypothesis $h$ is supposed to be close to $f$ (rather than equal) for most~$x$.

An interesting example is the problem of learning an unknown $n$-qubit quantum state $\rho$ from measurement data.
Let ${\cal X}$ be the set of measurement elements, i.e., psd matrices $M$ with $\norm{M}\leq 1$.
If we measure $\rho$ with some POVM of which $M$ is one element, then the probability to get the outcome corresponding to $M$, is $\Tr(M\rho)$. Accordingly, we can define $f:{\cal X}\to[0,1]$ as $f(M)=\Tr(M\rho)$ and consider the class $\cal C$ of all such functions (one $f$ for each possible $\rho$, so this class is uncountable).
Aaronson~\cite{aaronson:qlearnability} showed that this $\cal C$ is classically PAC learnable from $O(n)$ examples of the form $(x,f(x))$ (with some polynomial dependence of the sample complexity on $\gamma,\eps$, and exponential time complexity).
Note that we are not really learning $\rho$ itself, but rather learn to predict the measurement probabilities.
In contrast, learning a good approximation of $\rho$ itself (with small error in trace distance) requires a number of copies of $\rho$ that is exponential in~$n$~\cite{haah16,odonnell16}.
Some positive results for learning specific classes of quantum states can be found in~\cite{low:clifford,AAKW:learnHam,LC:qlearnTgate}.

\section{Unsupervised learning from quantum data}\label{sec:unsupML}

In this section we will look at an example of \emph{unsupervised} learning from quantum data: dimension-reduction via Principal Component Analysis.
Suppose we are given $m$ vectors $v_1,\ldots,v_m\in\mathbb{R}^d$, say unit vectors for simplicity.
Let's say the dimension $d$ of the data-vectors is very large, and we would like to reduce it to some  much smaller $k$, say at most $k=\polylog(d)$.
Many machine learning tasks, for example clustering, become much easier if we can significantly reduce this dimension.

One way to achieve this dimension-reduction is to find $k$ suitable unit vectors $c_1,\ldots,c_k\in\R^d$ (which may or may not be in the set  $\{v_i\}$ themselves), such that the  projection $P_Sv_i$ of the $v_i$'s on the $k$-dimensional space $S=span\{c_1,\ldots,c_k\}$ typically doesn't lose much, i.e., $P_S v_i$ is close to $v_i$ for most $i\in[m]$.
Then we can replace each $v_i$ by the $k$-dimensional vector $P_S v_i=\sum_{j=1}^k \alpha_j c_j$, expressed as the vector of coefficients $(\alpha_j)\in\R^k$ (note that $\alpha_j=\inp{c_j}{v_i}$). 
How to find those $k$ ``directions'' $c_1,\ldots,c_k$?
One method that often (though not always) works well is to find the $k$ eigenvectors corresponding to the $k$ largest eigenvalues of the following $d\times d$ ``correlation matrix'':
\[
A=\sum_{i=1}^m v_iv_i^T.
\]
Those $k$ eigenvectors are called the $k$ ``principal components'' of~$A$. They intuitively correspond to the $k$ most important directions in the data, and we can choose them 
for dimension-reduction.

Classically, we can find those $k$ eigenvectors by diagonalizing $A$, which takes time polynomial in~$d$.
In the quantum case we can do something very different, under the (very strong) assumption that we can efficiently, say in time $\polylog(d)$, prepare the $\ceil{\log(d)}$-qubit quantum states $\ket{v_i}$ corresponding to the vectors~$v_i$. By choosing $i\in[m]$ uniformly at random and preparing $\ket{v_i}$, we prepare the following $\ceil{\log(d)}$-qubit mixed state, which is proportional to the correlation matrix:
\[
\rho=\frac{1}{m}\sum_{i=1}^m \ketbra{v_i}{v_i}=\frac{1}{m}A.
\]
Let's say this has (unknown) spectral decomposition $\rho=\sum_{j=1}^d \lambda_j\ketbra{c_j}{c_j}$ with $\lambda_1\geq \cdots\geq\lambda_d\geq 0$.

We would now like to find the top-$k$ eigenstates $\ket{c_1},\ldots,\ket{c_k}$ of this~$\rho$, with the associated eigenvalues.
We will sketch the main idea here. Assume the first $k$ eigenvalues sum to something close to~1, and are not too close together, at least $1/\poly(k)$ apart.\footnote{These assumptions make this principal-component analysis more like a \emph{heuristic} method for clustering, not something that provably always works. It is anyway not clear in this case what a correct or optimal output would be: because this is unlabeled data, we do not have a clear standard for correctness.} 
Note that the unitary $U=e^{i\rho}$ has the same eigenstates as $\rho$ itself, with every eigenvalue $\lambda_j$ of $\rho$ translating into eigenvalue $e^{i\lambda_j}$ of~$U$.
Lloyd, Mohseni, and Rebentrost~\cite{lmr:pca} (with more precise analysis and matching lower bound in~\cite{kimmelea:hamsim}) showed that we can actually implement the power $U^t$ up to error~$\eps$  using $O(t^2/\eps)$ copies of the state~$\rho$ (see Exercise~\ref{ex:hamsimrho}).
We now use phase estimation with the unitary $U$ \emph{on a copy of $\rho$ itself}, with additive error $\delta=1/\poly(k)$.
By Section~\ref{ssecphaseestimation}, phase estimation with additive error $\delta$ corresponds to running controlled versions of $U^t$ for $t$ up to $O(1/\delta)$.
Under our earlier assumptions, this only takes $\poly(k)=\polylog(d)$ time.
Ignoring for simplicity the small errors ($\leq\delta$) that phase estimation makes in estimating the values $\lambda_j$, phase estimation transforms the copy of $\rho$ and a few auxiliary $\ket{0}$-qubits into the state 
\[
\sum_{j=1}^d\lambda_j \ketbra{c_j}{c_j}\otimes\ketbra{\lambda_j}{\lambda_j}.
\]
If we measure the second register, then we obtain state $\ket{c_j}\otimes\ket{\lambda_j}$ with probability $\lambda_j$.\footnote{Actually, in the second register we will see (with high probability) a $\delta$-close approximation of $\lambda_j$ rather than $\lambda_j$ itself, but if we assume $\delta$ is much smaller than the spacing between the eigenvalues $\lambda_1,\ldots,\lambda_k$, then such a $\delta$-approximation is good enough to ``recognize'' the state in the first register as the $j$-th principal component.} 
Doing this $\poly(k)$ many times, we learn the $k$ largest values $\lambda_1,\ldots,\lambda_k$, and for each of those $\lambda_j$'s we'll have a number of copies of the eigenstate $\ket{c_j}$.
This is a quantum form of Principal Component Analysis.

This collection of eigenstates determines a $k$-dimensional subspace on which we could re-express the $v_i$'s (approximately), but it is not very explicit: we only have the $k$ basis vectors of this space as quantum states! Suppose we want to express some unit vector $v$ (which again we assume we can prepare efficiently as a state~$\ket{v}$) as a linear combination of the $c_j$'s. 
One thing we can do is use a few copies of each $\ket{c_j}$ to approximate $|\inp{c_j}{v}|^2$ for each $i$ using the SWAP-test,
which gives us at least partial information about the coefficients $\inp{c_j}{v}$ (see Exercise~\ref{ex:estdistance}).

``Quantum PCA'' has a lot of drawbacks, but at least it shows some genuinely quantum tricks that we can use under the assumption that our input vectors can be efficiently prepared as quantum states. 
There have also been some quantum approaches for the prominent unsupervised learning problem of clustering, but we will not describe those here (see for instance~\cite{lmp:supunsup,KLLP:qmeans}). 

\section{Optimization}

In the previous two sections we assumed quantum data: either the data is already given as a superposition, or we can efficiently put given classical data in superposition. However, in most real-world applications of machine learning we have classical data without the means to efficiently make this quantum.
Remembering the slogan ML $=$ data $+$ optimization, if there's any room left for quantum improvements when data is classical, it would be in the optimization to find a well-fitting model for the data. We'll look at some examples where quantum computing might help.

\subsection{Variational quantum algorithms}

One approach that has received a lot of attention is to optimize over \emph{parametrized circuits}. Suppose we have a quantum circuit $U(\theta)$ with a vector~$\theta$ of parameters. This could for instance be a circuit where CNOTs and single-qubit rotations are already in place, but the angles of the single-qubit gates are parameters that we can tweak. This $U(\theta)$ is then applied to a fixed starting state, say~$\ket{0}$, yielding a final state $\ket{\psi(\theta)}=U(\theta)\ket{0}$. The goal is now to minimize the expected value of some observable $M$, i.e., to find a $\theta$ to minimize the function $f(\theta)=\bra{\psi(\theta)}M\ket{\psi(\theta)}$. In the case of supervised learning applications, $U(\theta)$ could for instance represent some hypothesis (i.e., a way to predict labels of $x$'s), $M$ could incorporate the given labeled examples $(x,f(x))$, and $f(\theta)$ could be the ``empirical error'': the fraction of mis-predicted labels among the given examples. 

Note that $f(\theta)$ can be computed approximately (for classically given $\theta$) on a quantum computer by repeatedly preparing $\ket{\psi(\theta)}$ and measuring the observable $M$.
If the circuits $U(\theta)$ are relatively simple (say, few qubits, few gates, low depth) and $M$ is relatively easy to measure (say, a sum of a few $n$-qubit Pauli matrices with few non-identity terms) then this could already be done on a relatively small and simple quantum computer.
Variational quantum algorithms (VQAs) are typically hybrid classical-quantum algorithms: the minimization over $\theta$ is usually done by a classical outer loop that iteratively improves $\theta$. Using the ability to approximately compute $f$ we can for instance try to do approximate gradient descent (move $\theta$ by some step-size in the direction of steepest descent of $f$) or some other method. This is analogous to the iterative way the weights in neural networks are optimized, and these variational quantum approaches are sometimes 
(with a keen sense for marketing) called ``quantum neural networks'' or ``quantum deep learning.''
For combinatorial optimization, a very structured version of the variational approach is the Quantum Approximate Optimization Algorithm (QAOA)~\cite{FGG:QAOA}. See \cite{VQAsurvey} for a general overview of VQAs.

One interesting application of this variational idea is in trying to find the smallest eigenvalue of a given Hamiltonian~$H$.
For example, $H$ could describe the energy of a chemical system as a function of the locations of the particles (nuclei and electrons) of the system; the smallest eigenvalue of $H$ would be the ``ground-state energy'' of the system, which is an important quantity in chemistry.
We know from Chapter~\ref{ch:QMAlocham} that in general this problem of determining or even well-approximating this ground state energy is \QMA-hard, even in the special case where $H$ is a sum of 2-local terms, so in general this shouldn't be efficiently solvable on a quantum computer. However, suppose that from some general physics or chemistry intuition we have a rough idea of what the ground state of our particular Hamiltonian~$H$ should look like, something we can prepare using a simple parametrized circuit $U(\theta)$. The set of states $\ket{\psi(\theta)}=U(\theta)\ket{0}$ that we are limiting ourselves to, is called an ``Ansatz'' (German for ``approach'' or ``attempt''). We can now try to optimize the parameters $\theta$ in order to minimize the expected value $f(\theta)=\bra{\psi(\theta)}H\ket{\psi(\theta)}$, i.e., the energy of the state $\ket{\psi(\theta)}$.
This approach is called the ``variational quantum eigensolver'' (VQE)~\cite{VQE}, and is one of the best hopes for applying smallish, near-term quantum computers to problems in chemistry.

\subsection{Some provable quantum speed-ups for optimization}

The variational approach is rather heuristic: it very much depends on how good the ``Ansatz'' (the choice of the class of parametrized circuits $U(\theta)$) happens to be for the particular problem at hand. Here we mention some other approaches, which yield provable (albeit usually only polynomial) quantum speed-ups under some assumptions on how the input is given.
\begin{itemize}
\item There are many quantum speed-ups for optimization problems on graphs, typically using Grover search (Section~\ref{sec:grover}), Grover-based minimum-finding (Exercise~\ref{chap:grover}.\ref{ex:minfind}), amplitude amplification (Section~\ref{sec:amplamplif}), or amplitude estimation (Exercise~\ref{chap:grover}.\ref{ex:amplitudeestimation}) as a subroutine. Examples are finding shortest paths~\cite{dhhm:graphproblemsj} and approximating minimum cuts or graph sparsification~\cite{AW20}.
\item Solving linear systems and other basic linear algebra is ubiquitous in classical optimization algorithms. Since quantum states are vectors and quantum operations are matrices, one can try to improve such classical algorithms using quantum algorithms. Examples are phase estimation (Section~\ref{ssecphaseestimation}),  the block-encoding approach (Section~\ref{sec:blockenc} and \cite[Section~3.2.4]{gilyen:thesis}), and HHL (Chapter~\ref{chap:hhl}). The trouble with this approach is that it often assumes the input is a quantum state (which is not always practical) and/or that it produces the output as a quantum state (which is not always useful). For example, HHL and quantum PCA have both features. See~\cite{aaronson:fineprint} for more discussion, and \cite{CGW:topeigenvectors} for a small (but optimal) polynomial speed-up for PCA with classical input and output.

One interesting application of ``quantum linear algebra'' (with classical inputs and outputs!) is the quantum recommendation system of Kerenidis and Prakash~\cite{KP16}, which can generate recommendations of type ``you might also like'' to a user of systems like Amazon or Netflix, based on the user's and other users' earlier behavior. Initially \cite{KP16} was believed to give an exponential speed-up over classical recommendation systems, until Tang showed how to ``dequantize'' their quantum algorithm under similar classical access assumptions~\cite{tang:RC,chiaetal:dequant}. 
\item In \emph{convex optimization} we minimize a convex function $f:\R^n\to\R$, either over all $x\in\R^n$ or over all $x$ that are constrained to lie in some convex domain ${\cal X}\subset\R^n$. This covers a big part of continuous optimization. Convexity ensures that the only local minima are also global minima, but such methods often still work to find good local minima for non-convex problems (such as training neural networks).
\emph{First-order} methods such as gradient descent use the gradient of $f$ at a given point, which in some cases can be computed more efficiently 
by quantum algorithms~\cite{jordan:gradient,GAW:gradient,cornelissen:mscthesis}  (see Exercise~\ref{ex:Jordangradient}).
\emph{Second-order} methods often solve a linear system involving the Hessian (the $n\times n$ matrix of partial second derivatives at a given point), and we can try to use quantum linear algebra for that as well. If the matrix is symmetric and diagonally dominant and the output needs to be classical, then we could use the linear-system  solver of~\cite{AW20}.

Quantum algorithms are known for the specific cases of linear programming (LPs) and semidefinite programming (SDPs)~\cite{brandao2016QSDPSpeedup,vAGGdW17,brandao2017QSDPSpeedupsLearning,vAG19,AG19zerosumgame}, for learning support vector machines (SVMs)~\cite{rebentrost2014QSVM,schuld&killoran:feature,saeedi&arodz:qsvm,allcock&hsieh:qsvm,SPA:semi},
for least-squares linear regression with an $\ell_1$-regularizer~\cite{Chen&Wolf:lasso}, and for PCA with classical output~\cite{CGW:topeigenvectors} (in contrast to the quantum output of Section~\ref{sec:unsupML}).
\end{itemize}

\section*{Exercises}

\begin{enumerate}
  
\item\label{ex:labelinphase} Suppose that for some unknown Boolean function $f:\01^n\to\01$ and amplitude-vector $(\alpha_x)_{x\in\01^n}$, you are given one copy of the $(n+1)$-qubit state 
\[
\sum_{x\in\01^n}\alpha_x\ket{x}\ket{f(x)}. 
\]
Show how you can convert this into state
\[
\sum_{x\in\01^n}\alpha_x(-1)^{f(x)}\ket{x}\ket{1} 
\]
with success probability 1/2, in such a way that you know when you succeeded.

\item\label{ex:claslinearfns}
Consider again the concept class $\cal C$ of linear functions mod~2.
\begin{enumerate}
    \item Give a classical learning algorithm to learn a linear function exactly with high success probability ($\eps=0,\delta=1/3$) using $O(n)$ uniform random examples and $O(n^3)$ time.
    \item Argue that every classical PAC learner for $\cal C$ under uniform $D$, with $\eps<1/4$, needs $\Omega(n)$ examples.
\end{enumerate}

\item In the model of \emph{exact learning with membership queries}, the goal is to exactly learn a target function $f\in{\cal C}$ from queries to $f$ (so there are no examples in this setting, or rather the learner can choose their own examples).

Show that if $\cal C$ is the concept class of linear functions, then a target function $f\in\cal C$ can be learned with 1 quantum membership query, but requires $\Omega(n)$ classical membership queries.

\item\label{ex:VCdimUB} Consider a concept class $\cal C$ of functions $f:{\cal X}\to\01$, with $|{\cal X}|=N$ and $VCdim({\cal C})=d$.
\begin{enumerate}
    \item Consider the following simple (and probably not very time-efficient) learning algorithm:\\[1mm]
    \hspace*{0.6em}Draw $m$ examples for target function $f$; output an $h\in{\cal C}$ consistent with these examples.\\[2mm]
    Let $h\in{\cal C}$ be a function with $err_D(f,h)>\eps$.
    Show that at the end of this algorithm, the probability that this particular $h$ is still consistent with the $m$ examples is $<(1-\eps)^m$.
    \item Set $m=\ceil{\log(3|{\cal C}|)/\log(1/(1-\eps))}$. Show that with probability $\geq 2/3$, the only $h$ that are consistent with the $m$  examples have $err_D(f,h)\leq\eps$.
    \item 
Derive an upper bound $m=O(d\log(N)/\eps)$ on the classical sample complexity of ($\eps,1/3)$-PAC learning the class $\cal C$ using Sauer's lemma, which says that $|{\cal C}|\leq \sum_{i=0}^d \binom{N}{i}$.
\end{enumerate}

\item\label{ex:VCdimLB} Suppose the set $S=\{x_1,\ldots,x_d\}\subseteq{\cal X}$ is shattered by concept class $\cal C$.
Consider a distribution $D$ that puts $1-4\eps$ probability on $x_1$ and $4\eps/(d-1)$ probability on each of $x_2,\ldots,x_d$.
\begin{enumerate}
\item Let $f\in{\cal C}$ be the target function. Show that you need $\Omega((d-1)/\eps)$ examples $\sim D$ to see (with probability $\geq 2/3$) $(x_i,f(x_i))$ for at least 50\%\ of the $i\in \{2,\ldots,d\}$.
\item Show that the sample complexity of every $(\eps,1/3)$-PAC learner for the class $\cal C$ is at least $\Omega((d-1)/\eps)$.
\end{enumerate}

\item\label{ex:Jordangradient} This exercise is about efficiently finding the gradient $\nabla f(z)$ of a function $f:\R^d\to\R$ at a point $z\in\R^d$. The gradient is the $d$-dimensional real vector of the $d$ partial derivatives $\partial f/\partial x_i$, evaluated at the point~$z$.
\begin{enumerate}
    \item (H) Let $f(x)=a+bx$ be a linear (strictly speaking, affine-linear) function from $\R$ to $\R$, where the real number $b\in[0,1)$ can be written with $n$ bits of precision. Suppose we have a unitary $O_f$ that maps $\ket{x,0}\to\ket{x,f(x)}$ (assume we have enough qubits to write down $x$ and $f(x)$).
    Give a quantum algorithm to compute $b$ using one application of $O_f$ and one application of $O_f^{-1}$, and some  unitaries that do not depend on~$f$.
   \item Let $f(x_1,\ldots,x_d)=a+b_1x_1+\cdots+b_dx_d$ be a linear function from $\R^d$ to $\R$, where $a,b_1,\ldots,b_d\in\R$. Show that the gradient $\nabla f(z)$ is equal to $(b_1,\ldots,b_d)$ for every~$z\in\R^d$.
   \item Assume that for the function $f$ in (b), each coefficient $b_k$ is $\in[0,1)$ and can be written with $n$ bits of precision. Suppose we have a unitary $O_f$ that maps $\ket{x_1,\ldots,x_d,0}\to\ket{x_1,\ldots,x_d,f(x_1,\ldots,x_d)}$. Give a quantum algorithm that computes the gradient $\nabla f(z)$ using one application of $O_f$ and $O_f^{-1}$, and some  unitaries that do not depend on~$f$.
\end{enumerate}
{\footnotesize Comment: The quantum algorithm of (c) is somewhat reminiscent of the Bernstein-Vazirani algorithm (Section~\ref{ssec:bernsteinvazirani}), though that one is for functions over $\mathbb{F}_2^d$ rather than $\R^d$. Variants of the algorithm of (c) can also efficiently approximate the gradient of a sufficiently smooth \emph{non-linear} function $f:\R^d\to\R$ at a given point $z\in\R^d$, since a smooth $f$ can be well-approximated around $z$ by a linear function whose coefficients are the entries of~$\nabla f(z)$.
The algorithm is due to Jordan~\cite{jordan:gradient} and was further analyzed by Gily\'en et al.~\cite{GAW:gradient,cornelissen:mscthesis}.}

\item\label{ex:hamsimrho}
(H) Let $\sigma$ and $\rho$ be $k$-qubit mixed states, $\eps>0$ small, $t\geq 0$, and $U=e^{i\rho}$ be a unitary. Our goal in this exercise is to apply $U^t$ to $\sigma$ at the expense of using some copies of the state $\rho$. The final state should have error $\leq\eps$ in trace norm compared to the intended state $U^t\sigma U^{-t}$.
\begin{enumerate}
    \item Show that for small $\eta\geq 0$, $U^\eta\sigma U^{-\eta}=\sigma+i\eta(\rho\sigma-\sigma\rho)+E$, where $\norm{E}_1=O(\eta^2)$.
    \item Let $W$ be the $2k$-qubit unitary that swaps the first $k$ qubits with the last $k$ qubits, i.e., it maps $\ket{a}\ket{b}\to\ket{b}\ket{a}$ for all $a,b\in\01^k$. Let $\sigma'$ be the $k$-qubit local state of the first register after applying the unitary $e^{iW\eta}$ to the $2k$-qubit state $\sigma\otimes\rho$.  Show that $\sigma'=\sigma+i\eta(\rho\sigma-\sigma\rho)+E'$, where $\norm{E'}_1=O(\eta^2)$.
    \item Show that $\norm{\sigma'-U^\eta\sigma U^{-\eta}}_1=O(\eta^2)$.   
    \item Show you can implement $U^t$ on $\sigma$ with error $\eps$ in trace norm, using $O(t^2/\eps)$ copies of $\rho$.
\end{enumerate}
{\footnotesize Comment: $O(kt^2/\eps)$ elementary gates suffice to implement this, but you don't have to show that.}

\item\label{ex:estdistance} 
Suppose $\ket{\phi}$ and $\ket{\psi}$ are unknown $n$-qubit pure states. 
\begin{enumerate}
\item (H) Show how a quantum computer can estimate the overlap $|\inp{\phi}{\psi}|$ (in absolute value) up to additive error 1/100 using $O(1)$ given copies of $\ket{\phi}$ and $\ket{\psi}$, and $O(n)$ elementary gates.  
\item Assume the inner product $\inp{\phi}{\psi}$ is a real number. Show that $\norm{\ket{\phi}-\ket{\psi}}^2=2-2\inp{\phi}{\psi}$.
\item Assume $\inp{\phi}{\psi}$ is real and positive. Show how a quantum computer can estimate the distance $\norm{\ket{\phi}-\ket{\psi}}$ up to additive error 1/100 using $O(1)$ copies of $\ket{\phi}$ and $\ket{\psi}$, and $O(n)$ gates.
\item Can a quantum computer detect the difference between the two cases $\ket{\psi}=\ket{\phi}$ and $\ket{\psi}=-\ket{\phi}$, given arbitrarily many copies of these two states? Explain your answer.
\end{enumerate}

\end{enumerate}

\chapter{Stabilizer States, Clifford Circuits, and One-Way Quantum Computing}\label{chap:stabilizer}

\section{Stabilizer states}

In this chapter we will look at a useful and important subset of the set of all $n$-qubit pure states, namely the \emph{stabilizer states}.
They make appearances everywhere in quantum computing, in particular in circuit-design and in quantum error-correction.
Consider the 1-qubit Paulis $I,X,Y,Z$ (Appendix~\ref{app:pauli}). These matrices are unitary and Hermitian, and hence self-inverse, so their eigenvalues are $\pm 1$. Non-identity Paulis anti-commute ($XZ=-ZX$ etc.) and $Y=iXZ$.
Let 
$
{\cal P}_n=\{+1,-1,+i,-i\}\cdot\{I,X,Y,Z\}^{\otimes n}
$
be the set of $n$-qubit Paulis with phases, which has  $4^{n+1}$ elements; the phases ensure that ${\cal P}_n$ is a \emph{group} under multiplication.\footnote{Section~\ref{sec:grouptheory} gives an introduction to group theory but isn't necessary for understanding this chapter.}  

If an $n$-qubit state $\ket{\psi}$ is an eigenstate of $P\in{\cal P}_n\backslash\{\pm I\}$ with eigenvalue $+1$ (i.e., $\ket{\psi}=P\ket{\psi}$) then we say that $\ket{\psi}$ is \emph{stabilized} by $P$.
One can think of the equation $\ket{\psi}=P\ket{\psi}$ as cutting the $2^n$-dimensional space of all pure states in half: the
set $\{\ket{\psi}:P\ket{\psi}=\ket{\psi}\}$ is a subspace of dimension $2^{n-1}$. If we had another equation $P'\ket{\psi}=\ket{\psi}$, where $P$ and $P'$ are independent\footnote{$S\subseteq{\cal P}_n$ is ``independent'' if we can't write one $P\in S$ as a product of some other elements $\in S$, times $\pm 1,\pm i$.}, then the remaining subspace $\{\ket{\psi}:P\ket{\psi}=P'\ket{\psi}=\ket{\psi}\}$ that is stabilized by both $P$ and $P'$, has dimension $2^{n-2}$. Continuing with this, if we know $\ket{\psi}$ is stabilized by a set $P^{(1)},\ldots,P^{(n)}$ of $n$ independent Paulis, then that determines a subspace of dimension~$2^{n-n}=1$, so it determines the state $\ket{\psi}$ up to an irrelevant global phase.\footnote{This global phase wouldn't even be there if we took the density matrix view $\rho=\ketbra{\psi}{\psi}$.}
Such a state that is determined by $n$ independent elements of ${\cal P}_n$ is called a \emph{stabilizer state}.
For example, for $n=1$, the basis state $\ket{0}$ is stabilized by $Z$, and $\ket{1}$ is stabilized by $-Z$.
For $n=2$, the basis state $\ket{01}$ is stabilized by $Z\otimes I$ and $-I\otimes Z$. In general, every basis state is a stabilizer state.
Superpositions can also be stabilizer states. For example, $\ket{+}$ is stabilized by $X$, and $\ket{-}$ is stabilized by $-X$.
Stabilizer states can be entangled, for instance an EPR-pair (Exercise~\ref{ex:EPRstabilizer}) and tensor products of $n$ EPR-pairs (Exercise~\ref{ex:productstabilizer}) are stabilizer states.

It is easy to see that the $n$ independent stabilizers must be pairwise commuting because two anti-commuting matrices can't have the same $+1$-eigenstate, and that $-I$ cannot be written as a product of some of them. If some $P\in\{I,X,Y,Z\}^{\otimes n}$ commutes with all $n$ of those stabilizers, then that $P$ is a product of some of the $P^{(1)}$, possibly with a phase. Hence $\ket{\psi}$ will be an eigenstate of that $P$ as well (with eigenvalue $+1$ or $-1$, because $P$ is Hermitian).

\section{Clifford circuits}

An $n$-qubit \emph{Clifford circuit} $U$ consists of Pauli gates, Hadamard gates ($H$), $S=\left(\begin{array}{cc}1 & 0\\
0 & i
\end{array}\right)$-gates,
and CNOT gates.
These are exactly the unitaries that preserve the Pauli group: they map (by conjugation) all elements of ${\cal P}_n$ to (possibly other) elements of ${\cal P}_n$.
For example, conjugation by $H$ maps 1-qubit Paulis to 1-qubit Paulis as in Figure~\ref{fig:HPaulis}: $HXH=Z$, $HYH=-Y$ etc.
\begin{figure}[hbt]
    \centering 
$I\longleftrightarrow I$, $X\longleftrightarrow Z$, $Z\longleftrightarrow X$, $Y\longleftrightarrow -Y$
    \caption{The way conjugation by $H$ maps the 4 1-qubit Paulis}\label{fig:HPaulis}
    \end{figure}
 Conjugation by CNOT maps the 2-qubit Paulis as depicted in Figure~\ref{fig:CNOTPaulis} (adapted from~\cite{kruw:noisethresholdj}).
\begin{figure}[hbt]
    \centering \includegraphics[width=0.28\linewidth]{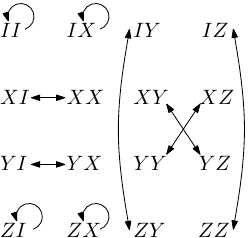}
    \caption{The way that conjugation by CNOT maps the 16 2-qubit Paulis (tensor product omitted and phases $\pm 1,\pm i $ omitted)}\label{fig:CNOTPaulis}
    \end{figure}

See Exercise~\ref{ex:SCliff} for how the $S$-gate acts on Paulis.

Since CNOT, $H$, and $S$ map Paulis to Paulis, and since every Clifford circuit consists of those 3 types of gates, every $n$-qubit Clifford circuit~$C$ maps $n$-qubit Paulis to $n$-qubit Paulis.%
\footnote{Conversely, one can show that every unitary $C$ that has this ``Pauli-preserving'' property can be written as a Clifford circuit of $O(n^2)$ gates (see Exercise~\ref{ex:paulipreserving}).}
This also implies that Clifford circuits map stabilizer states to stabilizer states, as follows.
Suppose $U$ is an $n$-qubit unitary which is one Clifford gate (CNOT, $H$, or $S$) tensored with identities on the other qubits. Suppose we apply this to a state $\ket{\psi}$ stabilized by $P^1,\ldots,P^n\in{\cal P}_n$.
Then it is easy to see that the new state $U\ket{\psi}$ is stabilized by the $n$ unitaries
$UP^1U^*,\ldots,UP^nU^*$. Because the gate is a Clifford, the latter are still elements of ${\cal P}_n$ and they are easy to compute using the rules from the previous section about how conjugation by CNOT, $H$, or $S$ changes a Pauli.
For example, if 2-qubit state $\ket{\psi}$ is stabilized by $P^1=X\otimes X$ and $P^2=Z\otimes Z$, and we conjugate by the 2-qubit Clifford circuit  $U=H\otimes I$, then $U\ket{\psi}$ is stabilized by  
$P'^1=Z\otimes X$ and $P'^2=X\otimes Z$.

In addition to unitary gates, we may also add the ability to measure a Pauli $M\in\{I,X,Y,Z\}^{\otimes n}$ during the computation, i.e., doing the projective measurement with operators $M_+$ and $M_-$ with outcomes $+1$ and $-1$ corresponding to the positive and negative eigenspaces of~$M$.
For example, measuring the first qubit in the computational basis would correspond to $M=Z\otimes I_{2^{n-1}}$, with $M_+=\ketbra{0}{0}\otimes I_{2^{n-1}}$
and $M_-=\ketbra{1}{1}\otimes I_{2^{n-1}}$.
We can then let the next gate depend on the earlier measurement outcomes, using some polynomial-time classical (possibly randomized) algorithm to compute what the next gate or measurement would be. This way the algorithm does not deterministically generate one specific stabilizer state at the end (as a circuit made up only of Clifford gates would), but rather it samples from a probability distribution over stabilizer states.
To distinguish such potentially more powerful circuits from the purely unitary Clifford circuits, we will use the term \emph{measurement-controlled} Clifford circuits for such circuits that have intermediate Pauli measurements and classical control.
Note that strictly speaking we'd have to talk about \emph{families} of circuits $\{C_n\}$ here, one for each number of qubits $n$, otherwise it doesn't make sense to speak of ``polynomial time'' for the classical control computations, but that's a technicality.

\section{Efficient simulation of Cliffords: Gottesman-Knill theorem}

We saw in Section~\ref{sec:bqpinpspace} that we can simulate a polynomial-size quantum circuit classically in polynomial space and exponential time. The polynomial space is nice (and initially surprising, since writing down the amplitude-vector of an arbitrary $n$-qubit state takes exponential space). The exponential time is terrible, but it is probably also unavoidable to a large extent: if the factoring problem does not have a polynomial-time classical algorithm (which is so widely believed that it underlies much of our current public-key cryptography) then there cannot be a time-efficient classical simulation of Shor's quantum factoring algorithm.

While we do not expect all quantum circuits to be efficiently simulable, it is still interesting and useful to try to delineate where the boundary for efficient simulability lies, because beyond that boundary lies the power of quantum computing. In other words, we would like to identify special classes of quantum circuits that \emph{can} be efficiently simulated.
One prominent example of such a special case is the class of Clifford circuits.
This was proved by Gottesman and Knill~\cite{gottesman&knill} (see \cite{aaronson&gottesman:improvedstabilizer} for more efficient simulation). 

\begin{theorem}[Gottesman-Knill]
There exists a classical polynomial-time algorithm that, given an $n$-qubit measurement-controlled Clifford circuit $C$ and an $n$-qubit initial basis state $\ket{x}$, samples from the same final probability distribution over stabilizer states as $C$. The classical simulator will output a stabilizer description of the sampled final state. 
\end{theorem}

The key idea is to keep track of the state via $n$ stabilizers (which can be written down with $n\times( 2n+2)=O(n^2)$ bits) rather than via the state's exponentially large vector of amplitudes.
Suppose our quantum algorithm starts with a particular stabilizer state, for instance a computational basis state $\ket{x}$, whose stabilizer description is easy to write down. Then it applies a (possibly measurement-controlled) Clifford circuit. We can efficiently update the stabilizer representation of the state step-by-step, for each of the Clifford gates and Pauli measurements that are applied. 
We already saw in the previous section how to update when a unitary Clifford-gate ($H$, $S$, CNOT) is applied, but we also have to deal with the intermediate Pauli measurements.

\paragraph{Updating the stabilizer representation after a measurement.}
Now suppose that at some point in our circuit we have an $n$-qubit state $\ket{\psi}$ stabilized by independent pairwise-commuting Paulis $P^{(1)},\ldots,P^{(n)}\in{\cal P}_n$, and suppose we measure a Pauli $M\in\{I,X,Y,Z\}^{\otimes n}$.
This is a projective measurement with outcome $+1$ and $-1$, and associated projectors $M_+=(I+M)/2$ and $M_-=(I-M)/2$, which correspond to the $+1$-eigenspace and $-1$-eigenspace of $M$, respectively.
There are now two 
possibilities.

{\bf Case~1.} 
$M$ commutes with all $P^{(j)}$. Because the $P^{(j)}$ form a maximal set of pairwise-commuting Paulis, $M$ can be written as a product of some of the $P^{(j)}$s times some phase $m\in \{\pm 1\}$ (the phase can't be $\pm i$, because $M$ is Hermitian and thus has real eigenvalues). Hence $M\ket{\psi}=m\ket{\psi}$.
Accordingly, one of the two possible measurement outcomes ($+1$ or $-1$) has probability~1. We can compute what this deterministic measurement outcome~$m$ is, and the state doesn't change after the measurement so its stabilizer description remains $P^{(1)},\ldots,P^{(n)}$.
The computed measurement outcome may affect which gates we will do further on, but that is not a problem for the simulation.

{\bf Case~2.}
$M$ anti-commutes with at least one of the $P^{(j)}$s, say $MP^{(1)}=-P^{(1)}M$. By multiplying all other $P^{(j)}$ that anti-commute with $M$ by $P^{(1)}$, we may assume $P^{(1)}$ is the only one that anti-commutes with $M$.
Using the fact that $P^{(1)}$ stabilizes $\ket{\psi}$, and the cyclic property of the trace, the measurement outcome $m\in\{\pm 1\}$ satisfies
\begin{eqnarray*}
\Pr[m=1]
& = &\Tr\left(\frac{(I+M)}{2}\ketbra{\psi}{\psi}\right)
=\Tr\left(\frac{(I+M)}{2}P^{(1)}\ketbra{\psi}{\psi}\right)
 = \Tr\left(P^{(1)}\frac{(I-M)}{2}\ketbra{\psi}{\psi}\right)\\
& = & \Tr\left(\frac{(I-M)}{2}\ketbra{\psi}{\psi}P^{(1)}\right)
=\Tr\left(\frac{(I-M)}{2}\ketbra{\psi}{\psi}\right)
=\Pr[m=-1].
\end{eqnarray*}
Hence the measurement outcome $m$ is distributed uniformly: both probabilities are equal to 1/2. We can simulate this simply by picking $m$ by a fair coin flip. With that measurement outcome, the post-measurement state $\ket{\psi'}=\frac{1}{\sqrt{2}}(I+mM)\ket{\psi}$ has collapsed to the eigenspace of $M$ that has eigenvalue~$m$, so it will be stabilized by $mM$.
Note that every stabilizer of $\ket{\psi}$ that commutes with~$M$ (that is, all except $P^{(1)}$), also stabilizes $\ket{\psi'}$.
We update the stabilizer description by replacing~$P^{(1)}$ by $mM$, resulting in $n$ independent and pairwise-commuting stabilizers of $\ket{\psi'}$.

Note that computing the next operation (Clifford gate or Pauli measurement) from the earlier measurement outcomes was assumed to be classical polynomial-time computable in the measurement-controlled Clifford circuit that we're simulating, so we can do the same computation during our simulation. Putting everything together, we end up with exactly the same probability distribution over the final states as the quantum circuit, except that the final output is now a classical description of the final state in the form of $n$ stabilizers, rather than an actual $n$-qubit state.

\section{$T$-gates and magic states}\label{sec:magicstate}

 The Clifford gates do not form a universal set in the sense of Section~\ref{sec:universalgatesets}, and in fact Clifford circuits on classical basis states can be efficiently simulated classically, as we just saw in the Gottesman-Knill theorem. 
One can show that adding just one non-Clifford gate to the available gate set $\{H,S,\mbox{CNOT}\}$ already gives us a universal model of computation. The most common choice of a gate to add is the $T=\left(\begin{array}{cc}1 & 0\\
0 & e^{i\pi/4}
\end{array}\right)$-gate, often one sees circuits described as ``Clifford$+T$'' for that reason. To see that $T$ is not Clifford, observe for instance that $TXT^*=\left(\begin{array}{cc}0 & e^{-i\pi/4}\\
e^{i\pi/4} & 0
\end{array}\right)\not\in{\cal P}_1$, so conjugation-by-$T$ does not send Paulis to Paulis. 

The $T$-gate looks rather straightforward, it just adds a relatively simple phase $e^{i\pi/4}$ to $\ket{1}$ similarly to the Clifford-gates $Z$ (which adds $-1$) and $S$ (which adds $i$). However, it turns out that in the setting of quantum error-correction (Chapter~\ref{chap:qec}), Clifford gates are usually easy to do and non-Clifford gates (such as $T$) are often hard. Instead of actually doing a $T$-gate, there is a very nice trick that allows us to simulate a $T$-gate by a Clifford circuit (incl.\ an intermediate measurement used to control a subsequent gate) starting from one copy of the state 
\[
\ket{T}=T\ket{+}=\frac{1}{\sqrt{2}}(\ket{0}+e^{\pi i/4}\ket{1}).
\]
See Exercise~\ref{ex:Tgate} for how this is done.
Accordingly, we can still do universal quantum computation with only measurement-controlled Clifford circuits if we start with a number of copies of $\ket{T}$ equal to the number of $T$-gates that we'd like to simulate. Somehow the $\ket{T}$-states ``magically'' add universality to Clifford circuits; $\ket{T}$ is called a ``magic state'' for that reason. We can think of the $\ket{T}$-states as a resource that we can consume during the computation, somewhat similar to how shared EPR-pairs are a bipartite resource that can be consumed when doing teleportation or non-locality experiments. We can thus move the cost of doing $T$-gates from the actual circuit to a pre-processing phase where we produce many copies of $\ket{T}$. Alternatively, the $\ket{T}$-states might be produced on-demand by some side-process during the computation, called a ``magic-state factory.''

\section{Efficient simulation of circuits with few $T$-gates}

A quantum algorithm will typically need to do a lot of $T$-gates, and in general we don't believe we can simulate such an algorithm efficiently.
But what if (besides Clifford operations) we only need a small number~$t$ of $T$-gates or $\ket{T}$-states?
Intuitively such circuits are ``not very far'' from Cliffords, and one might hope that efficient simulation is still possible. This indeed turns out to be the case.
The trick is to take the non-stabilizer resource state $\ket{R}=\ket{T}^{\otimes t}$, and to write it as a linear combination of some $r$ stabilizer states 
\[
\ket{R}=\sum_{j=1}^r\alpha_j\ket{S_j}.
\]
Suppose we want to run an $n$-qubit measurement-controlled Clifford circuit with $t$ $T$-gates on an initial $n$-qubit basis state $\ket{x}$.
We can equivalently run a measurement-controlled Clifford circuit without $T$-gates on an initial $(n+t)$-qubit state $\ket{x}\ket{R}$.
We can now run a variant of the Gottesman-Knill algorithm $r$ times in parallel, once for each $\ket{x}\ket{S_j}$, keeping track of the $r$ stabilizer descriptions of the $(n+t)$-qubit states $\ket{S'_j}$ that evolved starting from $\ket{x}\ket{S_j}$. We have to be careful here about the intermediate measurements. Suppose we want to measure a Pauli $M$ on intermediate state $\ket{\psi}=\sum_{j=1}^r\alpha'_j\ket{S'_j}$, where we have a classical description of the $\alpha'_j$s and stabilizer descriptions of the $\ket{S'_j}$. We can compute the expectation value 
\[
v=\bra{\psi}M\ket{\psi}
=\sum_{j,k=1}^r
\alpha_j'^*\alpha_k'\bra{S_j'}M\ket{S_k'} 
\]
by computing all $r^2$ values $\bra{S'_j}M\ket{S'_k}$.
This, however, would take time linear in~$r^2$.
There is a better way to calculate this $v$ in time proportional to $r$ that we will not describe here, see~\cite{bravyi&gosset:improved}.
The $v$ that we calculated is the difference between measurement outcomes $m=1$ and $m=-1$. We can now sample an $m\in\{\pm 1\}$ according to that distribution, and for all $j$ update the stabilizer description of $\ket{S'_j}$ according to that measurement outcome~$m$ ($mM$ will be one of the new stabilizers for each of the $r$ states), and update the coefficient $\alpha'_j$ depending on the probability of outcome $m$ for that particular stabilizer state (some $\alpha'_j$ may even become 0). 
This gives a classical simulation of a measurement-controlled Clifford circuit that has $t$ additional $T$-gates, in time $r$ times a polynomial in the circuit size. 

\paragraph{Stabilizer rank.}
Clearly, the smaller we can make the number of terms $r$ in the decomposition of~$\ket{R}$, the faster the above simulation will be.
The minimal~$r$ possible (as a function of $t$) is called the ``stabilizer rank'' of the state~$\ket{R}$, and a fair amount of research has gone into proving upper and lower bounds on it. Tight bounds are not known, but the best known upper bound on the stabilizer rank of $\ket{R}$ is $3^{t/4}\approx 2^{0.3963 t}$, which comes from writing $\ket{T}^{\otimes 4}$ as the sum of 3 stabilizer states~\cite{QPG:improved}. The best stabilizer rank of a good \emph{approximation} of $\ket{R}$ (which is still good enough for classical simulation with small error) is roughly $2^{0.228t}$~\cite{bravyi&gosset:improved};
see Exercise~\ref{ex:stabrank} for a weaker but already non-trivial upper bound of $2^{0.61t}$.
This is still exponential in~$t$, but better than the trivial $2^t$, and it does imply that Clifford$+T$ circuits that have only $O(\log n)$ $T$-gates, or measurement-controlled Clifford circuits that start from $t=O(\log n)$ copies of the magic state $\ket{T}$, can be classically simulated in polynomial time.

\section{One-way quantum computing}\label{sec:1wqc}

We know that 1-qubit and 2-qubit gates suffice for universal quantum computing. If we start with a basis state, or any other product state, then the 2-qubit gates are essential: without them we would never get away from product states, and those are easy to describe classically ($2n$ complex numbers suffice rather than $2^n$).
Also in Clifford+$T$ circuits where we replaced the $T$-gates by little measurement-controlled Clifford-computations on magic states, we still need to do the 2-qubit CNOT gates.
2-qubit gates are much harder to implement in practice than 1-qubit gates, because they require us to interact two different physical systems (two electrons, two photons, two atoms, or whatever system is used to realize qubits).

There is an interesting way to avoid doing any 2-qubit operations at all during the computation, called \emph{one-way quantum computing} (1WQC). This was introduced by Raussendorf and Briegel~\cite{raussendorf&briegel:oneway1,raussendorf&briegel:oneway2}, and is the most prominent example of ``measurement-based'' quantum computing (MBQC). In 1WQC the computation consists only of single-qubit measurements in bases that are controlled by (i.e., that may depend on) the classical outcomes of earlier single-qubit measurements. The trick that makes this possible is that we start the computation in a rather large entangled resource state, called a ``cluster state,'' which we will explain below. 
If we have some good way to build a cluster state, either fully ready before the start of the computation, or on the fly during the computation, then we can avoid having to do 2-qubit gates during the rest of the computation (similar to how preparing magic states beforehand avoids having to do $T$-gates during the rest of the computation).

How does this work?
Suppose we have an $n$-qubit circuit~$C$ of depth~$d$ that we would like to convert to the one-way model.
Suppose the initial state on which we want to apply~$C$ is $\ket{+}^{\otimes n}$; this is  without loss of generality, since an initial product state, including any basis state, can be obtained from $\ket{+}^{\otimes n}$ by single-qubit unitaries. Suppose the circuit~$C$ uses only  controlled-$Z$s ($\CZ$s) and single-qubit gates of the form 
\[J(\alpha)=H\left(\begin{array}{cc}
1 & 0\\
0 & e^{i\alpha}
\end{array}\right),
\]
for $\alpha\in[0,2\pi)$.
This maps $\ket{0}$ to $\ket{+}$, and $\ket{1}$ to $e^{i\alpha}\ket{-}$.
Note that $J(0)=H$, $J(0)J(\pi/2)=S$, $J(0)J(\pi/4)=T$, so this gate-set can implement Clifford$+T$ and hence is universal in the sense of Section~\ref{sec:universalgatesets}: it can do any kind of quantum computation up to arbitrarily small error. 

\paragraph{Implementing 1-qubit $J(\alpha)$-gates by measurement.}
Let's start with the simple case of one qubit on which we want to apply one gate $J(\alpha)$, so $n=1$ and $d=1$. The 2-qubit cluster state we would start from in this case is 
\[
\CZ\ket{+}\ket{+}=\frac{1}{\sqrt{2}}(\ket{0}\ket{+}+\ket{1}\ket{-}).
\]
In order to apply $J(\alpha)$, we measure the first ($=$ leftmost) qubit in the basis 
\[
B(\alpha)=\left\{\ket{0}_\alpha=J^{-1}(\alpha)\ket{0}=\frac{1}{\sqrt{2}}(\ket{0}+e^{-i\alpha}\ket{1}), \, \ket{1}_\alpha=J^{-1}(\alpha)\ket{1}=\frac{1}{\sqrt{2}}(\ket{0}-e^{-i\alpha}\ket{1})\right\}.
\]
If the measurement outcome is $b\in\01$, then the state has collapsed to 
\[
\ket{b}_\alpha\otimes \frac{1}{\sqrt{2}}(\ket{+}+(-1)^b e^{i\alpha}\ket{-})
=\ket{b}_\alpha\otimes X^bJ(\alpha)\ket{+}. 
\]
So it is as if we applied the intended gate $J(\alpha)$ to the \emph{second} $\ket{+}$, with an additional Pauli~$X$ if we saw measurement outcome~$b=1$. We will discard the first qubit, and keep track of the  additional ``Pauli corrections'' until the end of the computation.

Now suppose our circuit consists of two single-qubit gates, first $J(\alpha)$ and then $J(\beta)$ ($n=1$ and $d=2$).
Now the initial three-qubit cluster 
state is 
\[
\CZ_{23}\CZ_{12}\ket{+}\ket{+}\ket{+}.
\]
The subscripts on the CZ-gates indicate the qubits they act on; note that the CZ-gates commute, so the order in which we apply them to create the cluster state from $\ket{+}\ket{+}\ket{+}$ doesn't matter. There is also no difference between which qubit is the control and which is the target for a $\CZ$, since the only nontrivial thing $\CZ$ does, is put a minus in front of $\ket{11}$. We execute the two gates by measurement, as follows.
We start by measuring the leftmost qubit in the basis 
$B(\alpha)$ like before. With measurement outcome $b\in\01$, the state has now collapsed to
\[
\ket{b}_\alpha \otimes \CZ_{23}(X^bJ(\alpha)\ket{+}\ket{+}).
\]
Note that the $\CZ_{23}$ commutes with what we just did, since we only acted on the first qubit.
We now would like to measure in basis $B(\beta)$ in order to apply the gate $J(\beta)$, but we have to adapt if $b=1$, because we don't want to end up applying $J(\beta)X J(\alpha)$.
Note that $J(-\beta)X=ZJ(\beta)$. Accordingly, we measure the second qubit in the basis $B(\beta)$ if $b=0$ but in the basis $B(-\beta)$ if $b=1$ (here we see how the earlier measurement result $b$ influences the next measurement basis).
Let's say we got measurement outcome $c\in\01$ from this. We discard the second qubit, and the state of the third qubit is now $X^cZ^b J(\beta)J(\alpha)\ket{+}$ (see Exercise~\ref{ex:writeoutJbJa}), so we have applied the intended $J(\beta)J(\alpha)$ to the initial state $\ket{+}$, up to a ``Pauli correction'' $X^cZ^b$ that we know.%
\footnote{We should also be able to do single-qubit \emph{identity} gates. These can be implemented by means of a measurement in the $B(0)$ basis. This is like applying $H$, not $I$, but this $H$ can be absorbed in the next single-qubit gate.}

\paragraph{Implementing a 2-qubit $\CZ$-gate by measurement.}
Now suppose somewhere in the computation we would like to do a $\CZ$-gate on the state $\ket{\psi}$ on the first two qubits. The relevant part of the half-consumed cluster state is now  
\[
\CZ_{12}\CZ_{13}\CZ_{24}\ket{\psi}_{12}\ket{+}_3\ket{+}_4
\]
So not only do we have $\CZ$s between the two qubits of $\ket{\psi}$ and their horizontal neighbors on the right (numbered 3 and 4 above), but we also have a ``vertical'' $\CZ$ between those two qubits (numbered 1 and 2) themselves.
We now measure the first two qubits in the $J(0)$-basis, say with measurement outcomes $b_1,b_2\in\01$. A small calculation (Exercise~\ref{ex:writeoutCZ1wqc}) shows that the state on qubits~3 and~4 is now
\[
(X^{b_1}H\otimes X^{b_2}H)\CZ\ket{\psi}.
\]
This is the intended 2-qubit $\CZ$-gate applied on $\ket{\psi}$ (implemented by means of \emph{single-qubit} measurements on the current cluster state, made possible by the $\CZ$s that were used earlier to make the initial cluster state),
up to Pauli corrections that we keep track of, and Hadamards that can be absorbed in the next single-qubit gates.

\paragraph{One-way quantum computing.}
By combining these two types of steps we can implement all gates of the original circuit step-by-step by means of single-qubit measurements, starting from a $n\times (d+1)$ rectangular quantum state where all qubits start as $\ket{+}$ and $\CZ$s are applied beforehand on each horizontal edge, and on each vertical edge that corresponds to a $\CZ$ in the original circuit. The idea is that the first ``column'' of $n$ qubits corresponds to the initial state, and each subsequent column corresponds to the qubits after the next layer of gates.%
\footnote{Note that the cluster state is a stabilizer state, and has a lot of entanglement. Because $\CZ$s commute and because each qubit in the rectangular grid of the cluster state has only $O(1)$ neighbors, we can prepare it with a constant-depth circuit.
Of course, we need 2-qubit $\CZ$ gates to prepare the cluster state, but the selling point of 1WQC is that this initial state can be pre-prepared and stored somewhere before we need it for the computation. This shifts some of the hard work of doing a quantum computation to a pre-processing phase that could be done at our leisure.
As explained here, 1WQC needs a lot more than the $n$ qubits used in the standard circuit model because the full cluster state contains $n(d+1)$ qubits, which we will ``consume'' from left to right with our measurements. However, it is not necessary to have the full depth-$(d+1)$ cluster state present at the start. One may start instead with a small $O(1)$-depth cluster state, start the one-way computation, and recycle the qubits that were measured on the left side to grow additional layers to the right side of the cluster state on the fly, in some sort of ``conveyor belt'' architecture~\cite{browne&rudolph:linopt}. This way, only $O(n)$ qubits are present in the system at any one time. 
This conveyor-belt idea is useful when doing 1WQC with photons because we can recover from failed attempts to add qubits to the cluster state, and try again (in contrast, a failed gate in the middle of a circuit can irretrievably destroy the whole computation).} 

The Paulis $X$ and $Z$ that we accumulate for each of the $n$ qubits during the computation only affect a final measurement in the computational basis in trivial ways: the $Z$-gate only changes the signs of amplitudes so it doesn't affect the measurement probabilities at all, and the $X$-gate on some qubit is easy to correct for by just flipping the binary measurement outcome on that qubit.

\section*{Exercises}

\begin{enumerate}

\item\label{ex:EPRstabilizer} Show that an EPR-pair $\frac{1}{\sqrt{2}}(\ket{00}+\ket{11})$ is a stabilizer state by exhibiting two independent 2-qubit Paulis that stabilize that state.

\item\label{ex:productstabilizer} Show that if $\ket{\phi}$ is an $n$-qubit stabilizer state and $\ket{\psi}$ is an $m$-qubit stabilizer state, then their tensor product $\ket{\phi}\otimes\ket{\psi}$ is an $(n+m)$-qubit stabilizer state.

\item\label{ex:SCliff} Show how conjugation by the $S$-gate changes the 4 Pauli matrices, i.e.\ compute $SXS^*$, $SYS^*$, and $SZS^*$.

\item\label{ex:cliffordcircuitmatrix}
\begin{enumerate}
\item (H) Show that every $n$-qubit Clifford circuit corresponds to an invertible linear map $C:\mathbb{F}_2^{2n}\to\mathbb{F}_2^{2n}$, if we represent Paulis as binary strings via $I=00,X=10,Y=11,Z=01$ and ignore the phases $\pm 1,\pm i$.
\item Show that the total number of $n$-qubit unitaries that can be written as a Clifford circuit, is $2^{O(n^2)}$.
\end{enumerate}

\item\label{ex:paulipreserving}(H) Show that every $n$-qubit unitary circuit that maps (by conjugation) elements from ${\cal P}_n$ to elements of ${\cal P}_n$, can be written as a circuit of $O(n^2)$ Clifford gates. 

\item\label{ex:Tgate}
Suppose we have a qubit $\ket{\phi}=\alpha\ket{0}+\beta\ket{1}$ to which we would like to apply a
 $T$-gate, but for some reason we cannot.
However, we have a second qubit available in state $\ket{T}=\frac{1}{\sqrt{2}}(\ket{0}+e^{i\pi/4}\ket{1})$, and we can apply a CNOT-gate and an
$S$-gate.
\begin{enumerate}
\item What state do we get if we apply a CNOT to the first and second qubit (with, as usual, the first qubit as the control)?
\item Suppose we then measure the second qubit in the computational basis. What are the probabilities of outcomes 0 and~1, respectively?
\item Suppose the measurement of part~(b) gives outcome~0. Show how we can get $T\ket{\phi}$ in the first qubit.
\item Suppose the measurement of part~(b) gives outcome~1. Show how we can get $T\ket{\phi}$ in the first qubit, up to an (irrelevant) global phase.
\end{enumerate}
{\footnotesize Comment: This method still works in the same way if the qubit to which we want to apply the $T$-gate is entangled with other qubits. This way of implementing the $T$-gate is very helpful in fault-tolerant computing (Chapter~\ref{chap:qec}), where often CNOT and $S$ are easy to do on encoded states but $T$ is not. This exercise shows that we can prepare (encodings of) the ``magic state'' $\ket{T}=\frac{1}{\sqrt{2}}(\ket{0}+e^{i\pi/4}\ket{1})$ beforehand, and use those to indirectly implement a $T$-gate using only CNOT and $S$-gates.}

\item Algorithms like Deutsch-Jozsa (Chapter~\ref{chap:qircuitdj}) and Simon (Chapter~\ref{chap:simon}) 
achieve exponential speed-up over the best possible classical algorithms, and seem to do so using only Clifford gates. Why does this not contradict the Gottesman-Knill theorem?

\item\label{ex:stabrank}
This exercise gives a non-trivial upper bound on the stabilizer rank of good approximations of $\ket{T}^{\otimes t}$, which is $t$ copies of the  magic state $\ket{T}$.
\begin{enumerate}
\item Show that $e^{-i\pi/8}H\ket{T}=\cos(\pi/8)\ket{0}-i\sin(\pi/8)\ket{1}$.
\item (H) Suppose you flip $t$ independent coins, each with probability $p=\sin(\pi/8)^2$ of coming up~1. The expected number of 1s in the resulting $t$-bit string is exactly $pt$. Let $\eps$ be a small positive constant. Show that the probability that the resulting $t$-bit string has more than $(p+\eps)t$ 1s in it, is exponentially small in $t$, where the constant in the exponent may depend on~$\eps$.
\item How many $t$-bit strings are there with at most $(p+\eps)t$ 1s? Think of $\eps>0$ as an arbitrarily small constant.
\item Show that there exists a $t$-qubit state that is a linear combination of at most $2^{0.61t}$ basis states, and that is exponentially (in $t$) close to $(\cos(\pi/8)\ket{0}-i\sin(\pi/8)\ket{1})^{\otimes t}$.
\item What can you conclude about the stabilizer rank of good approximations of $\ket{T}^{\otimes t}$? 
\end{enumerate}

\item\label{ex:writeoutJbJa} Write out the sequence of 3-qubit states that you get from applying $J(\alpha),J(\beta)$ on the 3-qubit cluster state for $n=1,d=2$, following Section~\ref{sec:1wqc}, and assuming measurement outcomes $b=c=1$.

\item\label{ex:writeoutCZ1wqc} Write out the sequence of 4-qubit states that you get when implementing a $\CZ$ on 2-qubit state $\ket{\psi}$ as sketched at the end of Section~\ref{sec:1wqc}, assuming measurement outcomes $b_1=0,b_2=1$.

\end{enumerate}

\chapter{Error-Correction and Fault-Tolerance}\label{chap:qec}

\section{Introduction}

When Shor's algorithm had just appeared in 1994, most people (especially experimental physicists, who were very aware of the difficulties in manipulating subatomic particles) were
extremely skeptical about the prospects of actually building a quantum computer.
In their view, it would be impossible to avoid errors when manipulating small quantum systems,
and such errors would very quickly overwhelm the computation, rendering it no more useful than classical computation.
However, in the few years that followed, the theory of quantum error-correction and fault-tolerant computation 
was developed.  This shows, roughly speaking, that if the error-rate per operation can be brought down
to something reasonably small (say 1\%), and the errors between different qubits are not very correlated,
then we can actually do near-perfect quantum computing for as long as we want.
Below we give a succinct and somewhat sketchy introduction to this important but complex area, 
just explaining the main ideas. See the surveys by Gottesman~\cite{gottesman:eccintro} and Terhal~\cite{terhal:qmemories} for more (in particular the latter for the important ``surface code,'' which we won't cover here).

\section{Classical error-correction}\label{secclasserrcor}
In the early days of classical computing, errors were all over the place:
memory-errors, errors in bits sent over a channel, incorrectly applied instructions, little animals creeping into the machine, etc.
Nowadays hardware is much more reliable, but we also have much better ``software solutions'' for errors,
in particular error-correcting codes.  Such codes take a string of data and encode it in a larger
string (the ``codeword''), adding a lot of redundancy so that a small fraction of errors on the codeword won't be able to reduce
the information about the encoded data.

The simplest example is of course the repetition code.
If we want to protect a bit $b$, we could repeat it three times:
$$
b\mapsto bbb.
$$
If we want to decode the encoded bit $b$ from the (possibly corrupted) 3-bit codeword, we just take the majority value of the 3~bits.

Consider a very simple noise model: every bit is flipped (independently of the other bits)
with probability~$p$.  Then initially, before applying the code, $b$ has probability $p$ to be flipped.
But if we apply the repetition code, the probability that the majority-value of the three bits
is different from $b$, is the probability of 2 or 3 bitflips, which is $3p^2(1-p)+p^3<3p^2$.
Hence the error-rate has been reduced from $p$ to less than $3p^2$. If the initial error-rate 
$p_0$ was $<1/3$, then the new error-rate $p_1<3p_0^2$ is less than $p_0$ 
and we have made progress: the error-rate on the encoded bit is smaller than the error-rate on the unencoded bits.  
If we'd like it to be even smaller, we could concatenate the code with itself, i.e., 
repeat each of the three bits in the code three times, so the codelength becomes~9.
This would give error-rate $p_2=3p_1^2(1-p_1)+p_1^3<3p_1^2<27p_0^4$, giving a further improvement.
As we can see, as long as the initial error-rate $p$ was at most 1/3, we can reduce the error-rate
to whatever we want: $k$ levels of concatenation encode one ``logical bit'' into $3^k$ ``physical bits,''
but the error-rate for each logical bit has been reduced to $\frac{1}{3}(3p_0)^{2^k}$.\footnote{With a bit more work we can show that this even works if the initial error-rate $p$ is only strictly less than 1/2 rather than $\leq 1/3$. See Exercise~\ref{ex:treshold1/2}.}
This is a very good thing: if the initial error is below~$1/3$, then $k$ levels
of concatenation increase the number of bits exponentially (in $k$) but reduce the error-rate \emph{double-exponentially fast}!

Typically, already a small choice of $k$ gets the error-rate down to negligible levels.
For example, suppose we want to protect some polynomial (in some $n$) number of bits for some polynomial number of time-steps, 
and our physical error-rate is some fixed $p_0<1/3$.  
Choosing $k=2\log\log n$ levels of concatenation already suffices for this, 
because then $p_k\leq\frac{1}{3}(3p_0)^{2^k}\sim 2^{-(\log n)^2}=n^{-\log n}$ goes to~0 faster than any polynomial.
In that case, by the union bound, even the probability that there exists an error anywhere among our polynomially many logical bits in polynomially many time-steps, will be negligibly small.
With this choice of $k$, each logical bit would be encoded in $3^k=(\log n)^{2\log(3)}$ physical bits, 
so we only increase the number of bits by a polylogarithmic factor.

\section{Quantum errors}
The need for error-correction is far greater for quantum computers than for classical computers,
because ``quantum hardware'' is much more fragile than classical hardware.
Unfortunately, error-correction is also substantially more difficult in the quantum world, for several reasons:
\begin{itemize}
\item The classical solution of just repeating a state is not available in general in the quantum world, 
because of the no-cloning theorem.
\item The classical world has basically only bitflip-errors, while the quantum world is continuous
and hence has infinitely many different possible errors.
\item Measurements that test whether a state is correct can collapse the state, losing information.
\end{itemize}
Depending on the specific model of errors that one adopts, it is possible to deal with all of these issues.
We will consider the following simple error model.
Consider quantum circuits with $S$ qubits, and $T$ time-steps; in each time-step, 
several gates on disjoint sets of qubits may be applied in parallel.
After each time-step, at each qubit, independently from the other qubits, some unitary error hits that qubit with probability~$p$.
Note that we assume the gates themselves to operate perfectly;  
a perfect gate followed by the unitary errors of our model may be viewed as an imperfect gate, though this doesn't fully cover every way in which a gate could fail.

Let's investigate what kind of (unitary) errors we could get on one qubit.
Consider the four Pauli matrices from Appendix~\ref{app:pauli}:
$$
I  =  \left(\begin{array}{cc}1 & 0\\ 0 & 1\end{array}\right), \
X  =  \left(\begin{array}{cc}0 & 1\\ 1 & 0\end{array}\right), \
Y  =  \left(\begin{array}{cc}0 & -i\\ i & 0\end{array}\right), \
Z  =  \left(\begin{array}{cc}1 & 0\\ 0 & -1\end{array}\right).
$$ 
These have an interpretation as possible errors:
$I$ corresponds to no-error, $X$ is a bitflip-error, $Z$ is a phaseflip-error,
and $Y=iXZ$ is a phaseflip-error followed by a bitflip-error (and a global phase of $i$, which doesn't matter).
These four matrices span the space of all possible $2\times 2$ matrices,
so every possible error-operation $E$ on a qubit is some linear combination $E=\alpha_0I+\alpha_1X+\alpha_2Y+\alpha_3Z$ of the 4 Pauli matrices. More generally, every $2^k\times 2^k$ matrix can be written uniquely as a linear combination of matrices
that each are the tensor product of $k$ Pauli matrices.

Consider for example the error which puts a small phase $\phi$ on $\ket{1}$:
$$
E=\left(\begin{array}{cc}1 & 0\\ 0 & e^{i\phi}\end{array}\right)=e^{i\phi/2}\cos(\phi/2)I - ie^{i\phi/2}\sin(\phi/2)Z. 
$$
Note that for small $\phi$ most of the weight in this linear combination sits on $I$, which corresponds to the fact that $E$ is close to $I$.
The sum of squared moduli of the two coefficients is~1 in this case. That's not a coincidence: 
whenever we write a unitary as a linear combination of Pauli matrices, the sum of squares of the coefficients will be~1 (see Exercise~\ref{eq:paulidecomp}).

The fact that all 1-qubit errors are linear combinations of $I,X,Y,Z$, together with the linearity of quantum mechanics, 
implies that if we can correct bitflip-errors ($X$), phaseflip-errors ($Z$), and their product ($Y$) on a qubit, then we can correct \emph{all possible} unitary errors on that qubit.\footnote{We can even correct the \emph{non-unitary} 
errors that arise from undesired interaction between qubits of our circuit with the environment, but we won't talk about such errors here.}
So typically, quantum error-correcting codes are designed to correct bitflip and phaseflip-errors (their product is
then typically also correctable), and all other possible errors are then also handled without further work.

Our noise model does not explicitly consider errors on multiple qubits that are not a product of errors on individual qubits.
However, even such a joint error on, say, $k$ qubits simultaneously can still be written as a linear combination
of products of $k$ Pauli matrices.  So also here the main observation applies: if we can just correct bitflip and phaseflip-errors
on individual qubits, then we can correct all possible errors!

\section{Quantum error-correcting codes}
Quantum error-correcting codes encode a number of ``logical qubits'' into a larger number of ``physical qubits,''
in such a way that errors on some number of its physical qubits can be corrected.
The first and simplest is Peter Shor's 9-qubit code~\cite{shor:scheme}, 
which encodes 1 logical qubit into 9 physical qubits, and can correct an
error on any one of the 9~physical qubits. Here are the codewords for the two logical basis states:
$$
\ket{0}\mapsto\ket{\overline{0}}=\frac{1}{\sqrt{8}}(\ket{000}+\ket{111})(\ket{000}+\ket{111})(\ket{000}+\ket{111})
$$
$$
\ket{1}\mapsto\ket{\overline{1}}=\frac{1}{\sqrt{8}}(\ket{000}-\ket{111})(\ket{000}-\ket{111})(\ket{000}-\ket{111})
$$
These two quantum codewords $\ket{\overline{0}}$ and $\ket{\overline{1}}$ span a 2-dimensional space $\{\alpha\ket{\overline{0}}+\beta\ket{\overline{1}}\}$. This 2-dimensional subspace of the overall $2^9$-dimensional space is called the ``codespace''
and corresponds to one logical qubit.

Suppose an error happens on one of these 9~qubits.
We would like to have a procedure that maps the resulting state back to the codespace.
By linearity, it suffices if we can do this for the basis states $\ket{\overline{0}}$ and $\ket{\overline{1}}$.
First consider bitflip and phaseflip-errors.

\paragraph{Detecting a bitflip-error.}
If a bitflip-error occurs on one the first 3 qubits, then we can detect its location by noting which of the 3 positions is the minority bit.
We can do this for each of the three 3-qubit blocks.
Hence there is a unitary that writes down in 4~auxiliary qubits (which are all initially $\ket{0}$)
a number $e_{b}\in\{0,1,\ldots,9\}$.  Here $e_{b}=0$ means that no bitflip-error was detected,
and $e_{b}\in\{1,\ldots,9\}$ means that a bitflip-error was detected on qubit number~$e_{b}$.
Note that we don't specify what should happen if more than one bitflip-error occurred.

\paragraph{Detecting a phaseflip-error.}
To detect a phaseflip-error, we can consider the relative phase for each of the three blocks $\ket{000}\pm\ket{111}$,
and if they are not all the same, unitarily write down in 2 more auxiliary qubits (again, initially~$\ket{0}$)
a number $e_{p}\in\{0,1,2,3\}$.  Here $e_{p}=0$ means that no phaseflip-error was detected,
and $e_{p}\in\{1,2,3\}$ means that a phaseflip-error was detected in the $e_{p}$-th block. Note that we are not discovering exactly on which of the 9 qubits the phaseflip-error happened (in contrast to the case of bitflips), but that's not needed anyway: we can correct the phaseflip-error by applying $Z$ to any one of the 3 qubits in the block where the affected qubit sits.

\bigskip

\noindent
Together the above two procedures form one unitary $U$ (i.e., one circuit) that acts on $9+4+2=15$ qubits,
and that ``writes down'' $e_{b}$ in 4 auxiliary qubits and $e_{p}$ in 2 auxiliary qubits.
For example, suppose we have the state $\ket{\overline{0}}$.
If $X_i$ denotes a bitflip-error on the $i$-th qubit ($i\in[9]$) and $Z_j$ denotes a phaseflip-error on the $j$-th qubit
(let $j'\in[3]$ denote the number of the block in which qubit $j$ lies).
Then after these errors our state is $X_iZ_j\ket{\overline{0}}$.
After fresh auxiliary qubits $\ket{0^4}\ket{0^2}$ are added, $U$ maps
$$
X_iZ_j\ket{\overline{0}}\ket{0^4}\ket{0^2}\mapsto X_iZ_j\ket{\overline{0}}\ket{i}\ket{j'}.
$$
Together, $e_{b}=i$ and $e_{p}=j'$ form the ``error syndrome''; this tells us which error occurred where.
The error-correction procedure can now measure this syndrome in the computational basis,
and take corrective action depending on the classical outcomes $e_{b}$ and $e_{p}$: 
apply an $X$ to qubit $e_{b}$ (or no $X$ if $e_{b}=0$), and apply a $Z$ to one qubit in the $e_{p}$-th block
(or no $Z$ if $e_{p}=0$). The case of a $Y$-error on the $i$-th qubit corresponds to the case where $i=j$ (i.e., the $i$-th qubit is
hit by both a phaseflip and a bitflip); our procedure still works in this case.
Hence we can perfectly correct one Pauli-error on any one of the 9~codeword qubits.

As we argued before, the ability to correct Pauli-errors on a qubit suffices to correct all possible errors on that qubit.
Let's see in more detail how this works.
Consider for instance some 9-qubit unitary error $E$ that is a linear combination of $X/Z$ errors on at most one of the 9 qubits. Assume it can be decomposed as
$$
E=(\alpha_0I+\sum_{i=1}^9\alpha_iX_i)(\beta_0I+\sum_{j=1}^9\beta_jZ_j).
$$
Suppose this error~$E$ occurs on $\ket{\overline{0}}$:
$$
E\ket{\overline{0}}=(\alpha_0I+\sum_{i=1}^9\alpha_iX_i)(\beta_0I+\sum_{j=1}^9\beta_jZ_j)\ket{\overline{0}}=\sum_{i,j=0}^9\alpha_i\beta_jX_iZ_j\ket{\overline{0}},
$$
where we denote $X_0=Z_0=I$.

If we now add auxiliary qubits $\ket{0^4}\ket{0^2}$ and apply the above unitary $U$, then we go into a superposition of error syndromes:
$$
U(E\otimes I^{\otimes 6})\ket{\overline{0}}\ket{0^4}\ket{0^2}=\sum_{i,j=0}^9\alpha_i\beta_jX_iZ_j\ket{\overline{0}}\ket{i}\ket{j'}.
$$
Measuring the 6 auxiliary qubits will now probabilistically give us one of the syndromes $\ket{i}\ket{j'}$, with $i\in\{0,1,\ldots,9\}$ and $j'\in\{0,1,2,3\}$, and it will collapse the state to 
$$
X_iZ_j\ket{\overline{0}}\ket{i}\ket{j'}.
$$
In a way, this measurement of the syndrome ``discretizes'' the continuously many possible errors 
to the finite set of Pauli-errors.  Once the syndrome has been measured, 
we can apply a corrective $X$ and/or $Z$ to the first 9 qubits to undo the specific error corresponding to the specific 
syndrome we got as outcome of our measurement. It is also possible that the measurement outcome is $0^4,0^2$; in that case the state has collapsed to $\ket{\overline{0}}\ket{0^4}\ket{0^2}$, so the syndrome measurement itself already removed the error!

So now we can correct an error on one qubit.
To achieve this, however, we have substantially increased the number 
of locations where such an error could occur: the number of qubits has gone from~1 to~9 (even to~15 if we also count the 6 auxiliary qubits used for the syndrome measurements), 
and we need a number of time-steps to compute and measure the syndrome, and to correct a detected error.
Hence this procedure only gains us something if the error-rate $p$ is so small that the probability 
of 2 or more errors on the larger encoded system is smaller than the probability of~1 error 
in the unencoded qubit.
We will get back to this issue below, when talking about the threshold theorem.
Note also that each new application of the correction-procedure needs a new, fresh 6-qubit register initialized to $\ket{0^4}\ket{0^2}$.
After one run of the error-correction procedure these auxiliary qubits will contain the measured error syndrome,
and we can just discard this.  In a way, error correction acts like a refrigerator: a fridge pumps heat out
of its system and dumps it into the environment, and error-correction pumps noise out of its system and dumps it in the environment
in the form of the discarded auxiliary qubits.

The above 9-qubit code is just one example of a quantum error-correcting code.
Better codes exist, and a lot of work has gone into simultaneously optimizing the different
parameters: we want to encode a large number of logical qubits into a not-much-larger number of
physical qubits, while being able to correct as many errors as possible.  
The shortest code that encodes one logical qubit and protects against one~error, has five physical qubits.
There are also ``asymptotically good'' quantum error-correcting codes; these encode $k$ logical qubits into $O(k)$ physical qubits and can correct errors on a constant fraction of the physical qubits (rather than just an error on one qubit).

\section{Fault-tolerant quantum computation}
Encoding a quantum state in a quantum error-correcting code to protect it against noise is 
good, but not enough: we also need to be able to do \emph{operations} on the encoded qubits  (Hadamards, CNOTs, etc.).
One way is to decode the logical qubits, do the operation on them, and then re-encode them.
This, however, is a recipe for disaster: if an error occurs in the interval between the decoding and subsequent encoding, then we're unprotected and we cannot detect (let alone undo) errors happening during that interval. Accordingly, we need to be able to do operations on the logical qubits \emph{while they are encoded}.
Additionally, we need operations for regular stages of error-correction, i.e., measuring the syndrome and then 
correcting errors based on the outcomes of those measurements. These operations may also introduce errors, and the big worry is that error-correction steps may themselves introduce more errors than they correct.\footnote{It's like being inside a leaky boat 
on the open seas, using a leaky bucket to scoop out water all the time to prevent the boat from filling 
up with water and sinking.  It's doable if the holes in the boat and in the bucket aren't too big, but it's not easy.}

There is a 7-qubit code due to Steane~\cite{steane:errcor} (see Exercise~\ref{ex:steanecode}) which has some nice properties:
a Hadamard on the logical qubit corresponds to $H^{\otimes 7}$ on the physical qubits, and a CNOT between 
two logical qubits corresponds to applying CNOTs between the 7 pairs of the two blocks of physical qubits
(i.e., between the 1st qubit of one block and the 1st qubit of the other block, etc.). Such implementations are called \emph{transversal}.
Adding the $T$-gate ($\ket{b}\mapsto e^{ib\pi/4}\ket{b}$) to $H$ and CNOT would yield a gate-set that suffices for universal quantum computation.
Unfortunately, implementing the $T$-gate fault-tolerantly takes more work, and we won't go into detail here (see Exercise~\ref{chap:stabilizer}.\ref{ex:Tgate} for the main idea of using (encoded) ``magic states'').

When designing schemes for fault-tolerant computing, 
it is very important to ensure that errors do not spread too quickly.  Consider for instance
a logical CNOT: if its control-bit is erroneous but its target bit is not, then after doing the CNOT both bits will be erroneous. 
The trick is to keep the errors on the physical qubits under control in such a way that regular stages of error-correction don't get overwhelmed by the errors.
For example, suppose we have a code that is able to correct up to one error in each encoded block (logical qubit); then the implementation of a logical CNOT may convert two encoded blocks where only one physical qubit has an error, into two blocks each of which has a single error, but not to multiple errors within one block, because our code will be able to handle two blocks with one error each but not one block with two errors (this is why the transversal implementation of the CNOT for Steane's code is nice).
In addition, we need to be able to fault-tolerantly prepare states, and measure logical
qubits in the computational basis.  We won't go into the many further details of fault-tolerant quantum computing here.

\section{Concatenated codes and the threshold theorem}

The idea to concatenate a code with itself, described at the end of Section~\ref{secclasserrcor} for classical codes,
also applies to quantum codes as we will sketch now.  Suppose we have some code that encodes one qubit into $C$ qubits, 
suppose that it can correct one error on any one of its $C$ qubits, and uses
$D$ time-steps per stage of error-correcting (each time-step may involve a number of elementary gates in parallel).
Instead of only~1, we now have $CD$ locations where an error could occur!
Assuming error-rate $p$ per-qubit-per-time-step, the probability for the code to fail on a specific logical qubit at a specific time
(i.e., to have \emph{more than~1} physical error on its $CD$ locations) is $p'=\sum_{i=2}^{CD}\binom{CD}{i}p^i(1-p)^{CD-i}$.
If $p$ is a sufficiently small constant, then this sum is dominated by the term for $i=2$, and we have $p'\approx (CD)^2p^2$.
Accordingly, if the initial error-rate $p$ is below some magical constant $\approx 1/(CD)^2$, 
then $p'<p$ and hence each level of error-correction reduces the error-rate by a constant factor.

More generally, suppose we concatenate this code $k$ times with itself.
Then every ``logical qubit'' gets encoded into $C^k$ qubits, but (by the same calculation as in Section~\ref{secclasserrcor}) 
the error-rate for each logical qubit gets reduced to $O((CDp)^{2^k})$.
Suppose we want to be able to ``survive'' $T=\poly(n)$ time-steps without any error on the logical qubits;
that is what we would need to run an efficient quantum algorithm on faulty quantum hardware.
Then it suffices if we reduce the error rate to $\ll 1/T$, for which
$k=O(\log\log T)$ levels of concatenation are enough.  
These layers of error-correction increase the number of qubits and the computation time by a factor which is exponential in $k$,
but that is still only a polylogarithmic overhead, since $2^{O(\log\log T)}=(\log T)^{O(1)}$.\footnote{One can even bring the overhead down to $O(1)$~\cite{fawzi2018constant}.}

The above sketch (when implemented precisely) gives us the famous ``threshold theorem''~\cite{aharonov&benor:faulttolj,klz:treshold}:
if the initial error-rate of the quantum hardware can be brought down below some constant 
(known as the ``fault-tolerance threshold,'' though it depends on the assumed noise model, the allowed gate set, and the code used, it is not one number), then we can use software-solutions like quantum error-correcting codes
and fault-tolerant computing to ensure that we can quantum compute for long periods of time without serious errors.

Much research has gone into finding the best value for this fault-tolerance threshold, under realistic assumptions on the noise.  
The more efficient our basic quantum error-correcting codes are (i.e., the smaller $C$ and $D$),
the higher (= better) the value of the threshold is.
Currently the best rigorous lower bounds for the threshold are around $0.1\%$ for reasonable assumptions on the noise, 
but there is numerical evidence that even a few percent might be tolerable
(there are also some largish \emph{upper} bounds known on the threshold value, e.g.~\cite{kruw:noisethresholdj}).
The existence of a constant threshold is actually one of the most important results in the area of quantum computing,
and is the main answer to the quantum-computing skeptics mentioned at the start of this chapter: as long as 
experimentalists manage to implement basic operations with an error below that threshold, in a scalable way, then we should be able to build large-scale quantum computers.%
\footnote{This is of course assuming our simple model of independent noise on each physical qubit is not too far off;
if the noise can be correlated in devious ways, then it becomes much harder (though often still possible) to protect against.}
This is starting to happen~\cite{GoogleQuantumAI2025:QECC}, and as far as we know there is no fundamental reason why we shouldn't be able to do this at scale. It is, however, an extremely hard engineering problem.

\section*{Exercises}

\begin{enumerate}

\item\label{eq:paulidecomp} (H)
Let $E$ be an arbitrary 1-qubit unitary.
We know that it can be written as
$$
E=\alpha_0 I + \alpha_1 X +\alpha_2 Y +\alpha_3 Z,
$$
for some complex coefficients $\alpha_i$.  Show that $\sum_{i=0}^3|\alpha_i|^2=1$.

\item
\begin{enumerate}
\item Write the 1-qubit Hadamard transform $H$ as a linear combination of the four Pauli matrices.
\item Suppose an $H$-error happens on the first qubit of $\alpha\ket{\overline{0}}+\beta\ket{\overline{1}}$ using the 9-qubit code.
Give the various steps in the error-correction procedure that corrects this error.
\end{enumerate}

\item
Give a quantum circuit for the encoding of Shor's 9-qubit code, i.e., a circuit that maps
$\ket{00^8}\mapsto\ket{\overline{0}}$ and
$\ket{10^8}\mapsto\ket{\overline{1}}$.
Explain why the circuit works.

\item
Shor's 9-qubit code allows to \emph{correct} a bit flip and/or 
a phase flip on one of its 9 qubits.  Below we give a 4-qubit code
which allows to \emph{detect} a bitflip and/or a phaseflip.
By this we mean that after the detection procedure we either have the 
original uncorrupted state back, or we know that an error occurred 
(though we do not know which one). The logical 0 and 1 are encoded as:
\begin{quote}
$\ket{\overline{0}} = \frac{1}{2}(\ket{00}+\ket{11})\otimes(\ket{00}+\ket{11})$\\[2mm]
$\ket{\overline{1}} = \frac{1}{2}(\ket{00}-\ket{11})\otimes(\ket{00}-\ket{11})$
\end{quote}
\begin{enumerate}
\item Give a procedure (either as a circuit or as sufficiently-detailed pseudo-code) that detects a bitflip error on one of the 4 qubits of $\alpha\ket{\overline{0}}+\beta\ket{\overline{1}}$.
\item Give a procedure (either as a circuit or as sufficiently-detailed pseudo-code) that detects a phaseflip error on one of the 4 qubits of $\alpha\ket{\overline{0}}+\beta\ket{\overline{1}}$.
\item Does that mean that we can now detect \emph{any} unitary 1-qubit error on one of the 4 qubits? Explain your answer.
\end{enumerate}

\item\label{ex:steanecode}
This exercise is about the Steane code, which (as you'll show) encodes 1 logical qubit into 7 physical qubits, and can correct an error on any one of the 7 physical qubits.
\begin{enumerate}
\item Consider the subspace $C\subseteq\01^7$ spanned by the 3 elements $0001111,0110011,1010101$. Write out a list of the 8 elements of $C$.
\item 
Let $C'=\{c\oplus 1^7\mid c\in C\}$ be the set of binary complements of $C$.
It is easy to verify that the minimal Hamming distance between any two elements of $C\cup C'$ is at least 3 (you may assume this without proof in your answer).
Show that there exists a procedure that, for every $c\in C\cup C'$, maps $X_j\ket{c}\ket{0}\mapsto X_j\ket{c}\ket{j}$, where $j\in\{1,\ldots,7\}$ (so the bitflip-error acts on the $j$-th qubit, and the procedure finds out the location $j$) and the second register consists of a number of initially-$\ket{0}$ qubits. You don't have to write out the details, just argue clearly why such a procedure exists.
\item The Steane code is defined by the two logical basis states
\[
\ket{\overline{0}}=\frac{1}{\sqrt{8}}\sum_{c\in C}\ket{c}\mbox{~~~and~~~}\ket{\overline{1}}=\frac{1}{\sqrt{8}}\sum_{c'\in C'}\ket{c'}
\]
What is the 7-qubit unitary $\overline{X}$ on the physical qubits that corresponds to the logical $X$-gate? In other words, what simple 7-qubit unitary maps $\ket{\overline{b}}\mapsto\ket{\overline{1-b}}$?
\item Show that there exists a quantum procedure (which may involve measurements in the computational basis) that identifies and corrects a bitflip-error on any one of the 7 qubits, i.e., that maps $X_j(\alpha\ket{\overline{0}}+\beta\ket{\overline{1}})\ket{0}\mapsto(\alpha\ket{\overline{0}}+\beta\ket{\overline{1}})\ket{j}$. 
\item (H) Show that there exists a quantum procedure (which may involve measurements in the computational basis) that identifies and corrects a phaseflip-error on any one of the 7 qubits, i.e., that maps $Z_j(\alpha\ket{\overline{0}}+\beta\ket{\overline{1}})\ket{0}\mapsto (\alpha\ket{\overline{0}}+\beta\ket{\overline{1}})\ket{j}$.
\end{enumerate}

\item\label{ex:2kqecc} (H)
Show that there cannot be a quantum code that encodes one logical qubit into $2k$ physical qubits while being able to correct errors on up to $k$ of the physical qubits. 

\item Suppose we have a qubit whose density matrix is
$\rho=\alpha_0 I + \alpha_1 X + \alpha_2 Y + \alpha_3 Z$,
where $\alpha_0,\alpha_1,\alpha_2,\alpha_3$ are real coefficients
and $I,X,Y,Z$ are the Pauli matrices.
\begin{enumerate}
\item Show that $\alpha_0=1/2$.
\item Depolarizing noise (of strength $p\in[0,1]$) acts on a qubit as follows:
with probability $1-p$ nothing happens to the qubit, and with probability $p$ the qubit is replaced
by the ``completely mixed state'' of a qubit, whose density matrix is $I/2$.

Show that depolarizing noise on the above qubit doesn't change the coefficient $\alpha_0$, but shrinks each of $\alpha_1,\alpha_2,\alpha_3$ by a factor of $1-p$.
\end{enumerate}

\item 
Consider a quantum-error correcting code that encodes $k$ qubits (and $n-k$ $\ket{0}$s) into an $n$-qubit codeword state, via the unitary encoding map
$$
U:\ket{x,0^{n-k}}\mapsto \ket{C(x)}, \mbox{ where }x\in\01^k\mbox{, and $\ket{C(x)}$ need not be a basis state.}
$$
A ``weight-$w$ Pauli error'' is the tensor product of $n$ Pauli matrices, of which at most $w$ are not identity (e.g., something like $X\otimes I \otimes Z\otimes I\otimes I$ if $w=2$ and $n=5$). Suppose that there is a unitary map $S$ on $3n$ qubits that can identify every weight-$w$ Pauli error $E$ on a codeword, by writing the name of $E$ (the ``error syndrome,'' which we can think of as a $2n$-bit string $"E"$, for example writing $00$ for $I$, $10$ for $X$, $01$ for $Z$, $11$ for $Y$) in a second register that's initially~$0^{2n}$. In other words, for every $x\in\01^k$ and weight-$w$ Pauli error $E$, this $S$ maps
$$
S:(E\ket{C(x)})\ket{0^{2n}}\mapsto (E\ket{C(x)})\ket{"E"}.
$$
\begin{enumerate}
\item
  Show that if $x$ and $y$ are $k$-bit strings, and $E$ and $F$ are weight-$w$ Pauli errors, then the $n$-qubit states $E\ket{C(x)}$ and $F\ket{C(y)}$ are orthogonal unless both $x=y$ and $E=F$.
\item Prove the inequality $\displaystyle
  2^k\sum_{i=0}^w \binom{n}{i} 3^i\leq 2^n$.\\
{\footnotesize Comment: This inequality implies a lower bound on the required number of qubits $n$, in terms of the number of encoded qubits $k$ and the weight $w$ of errors that you can correct, but you don't need to derive that consequence.}
\end{enumerate}

\item\label{ex:treshold1/2}
In this exercise we will see that the argument about concatenating the classical 3-bit code at the end of Section~\ref{secclasserrcor} still works if the initial error rate is $p\leq 1/2-\eps$ for some $\eps>0$.
\begin{enumerate}
\item Define the function $f$ as $f(p)=3p^2(1-p)+p^3$. Show that if $p=1/2-\eps$ for some $\eps\in[0,1/6]$, then $f(p)\leq 1/2-(13/9)\eps$.
\item Show that there is an $m=O(\log(1/\eps))$ such that $m$ levels of concatenation reduce the error to $p_m<1/3$.
\item\label{ex:mklevels} Show that $m+k$ levels of concatenation reduce the error to something exponentially small in $2^k$.
\item How many bits are used to encode one logical bit in the scheme from (c)?
\end{enumerate}

\end{enumerate}

\appendix

\chapter{Some Useful Linear Algebra}\label{chaplinalg}

In this appendix we quickly introduce the basic elements of linear algebra,\index{linear algebra}
most of which will be used somewhere or other in these notes.

\section{Vector spaces} 
A \emph{vector space} $V$ over a field $\mathbb{F}$ is a set of objects (called vectors) satisfying that if $v,w \in V$, then $cv+dw\in V$ for all $c,d\in\mathbb{F}$. In other words, $V$ is closed under addition and scalar multiplication. A (linear) \emph{subspace} $W$ is a subset $W\subseteq V$ which is itself a vector space (i.e., closed under addition and scalar multiplication).
For example, $V={\mathbb C}^d$ is the $d$-dimensional complex vector
space, which is the set of all column vectors of $d$ complex numbers.
The set $W\subseteq V$ of vectors whose first two entries are~0 is a subspace of~$V$.
As another example, the set $V=\01^d$ of $d$-bit vectors, with entrywise addition modulo~2, is a linear space. The field here is $\mathbb{F}_2=\01$. The set $W\subseteq V$ of vectors whose first two entries are equal is a subspace of~$V$.

A set of vectors $v_1,\ldots,v_m\in V$ is 
\emph{linearly independent}\index{linear independence} if the only way to get
$\sum_{i=1}^m a_i v_i$ equal to the zero-vector $0$, is to set $a_1=\cdots=a_m=0$.
The \emph{span} (over field $\mathbb{F}$) of a set of vectors $S=\{v_1,\ldots,v_m\}\subseteq V$ is the set $\mbox{span}(S)$ of vectors that can be written as a linear combination $\sum_{i=1}^m a_i v_i$ (with coefficients $a_1,\ldots,a_m\in\mathbb{F}$).
A \emph{basis}\index{basis of vector space} for $V$ is a linearly independent set $S$ of vectors such that $\mbox{span}(S)=V$. One can show that all bases of $V$ have the same size; this size is called the \emph{dimension} of $V$.
If we fix an ordered basis $S=(v_1,\ldots,v_m)$, then every $w\in V$ can be written uniquely as a linear combination $\sum_{i=1}^m w_iv_i$ with coefficients $w_i$, and can also be written (in that basis) as $(w_1,\ldots,w_m)$.
The \emph{support} of such a $w$ is the set $\{i\mid w_i\neq 0\}$ of locations where $w$ is nonzero. For example, the support of $w=(1,5,0,4)$ is $\{1,2,4\}$ if we number the locations starting from~1 (we will also sometimes number positions from~0 in these notes).

\section{Matrices}
Matrices represent linear maps between two vector spaces with particular bases.
We assume familiarity with the basic rules of matrix addition and multiplication.
We use $A_{ij}$ for the $(i,j)$-entry of a matrix $A$,
and $A^T$ for its \emph{transpose},\index{matrix!transpose} 
which has $A^T_{ij}=A_{ji}$.
We use $I_d$ to denote the $d\times d$ identity matrix, 
which has 1s on its diagonal and 0s elsewhere;
we usually omit the subscript~$d$ when the dimension is clear from context.
If $A$ is square and there is a matrix $B$ such that $AB=BA=I$, 
then we use $A^{-1}$ to denote this $B$, which is called the \emph{inverse}\index{matrix!inverse} 
of $A$ (and is unique if it exists).
Note that $(AB)^{-1}=B^{-1}A^{-1}$.

In the remainder of this appendix we will mostly consider the complex field.
If $A$ is a matrix (not necessarily square), then $A^*$ denotes its 
\emph{conjugate transpose} (or \emph{adjoint}):\index{matrix!conjugate transpose} 
the matrix obtained by transposing $A$ and taking the complex conjugates of all entries. Note that $(AB)^*=B^*A^*$.
Physicists usually write $A^\dagger$ (pronounced ``$A$-dagger'') instead of $A^*$, but in these notes we will stick with the $A^*$~notation that is common in mathematics.

\section{Inner product and norms for vectors}
For vectors $v,w$ with real or complex entries, we use $\inp{v}{w}=v^*w=\sum_i v_i^*w_i$ 
for their \emph{inner product}.\footnote{Here we follow a physics convention by conjugating the entries of the first vector; mathematicians usually conjugate the entries of the second vector, defining $\inp{v}{w}=\sum_i v_iw_i^*$.}\index{inner product} 
The combination of the vector space $V$ with this inner product 
is called a \emph{Hilbert space}.\index{Hilbert space}
Two vectors $v,w$ are \emph{orthogonal} if $\inp{v}{w}=0$.\index{orthogonal vectors}
A set $\{v_i\}$ of vectors is called \emph{orthogonal} if 
all vectors are pairwise orthogonal: $\inp{v_i}{v_j}=0$ if $i\neq j$.
If additionally the vectors all have norm~1, then the set is called \emph{orthonormal}.

The inner product induces a vector 
\emph{norm} $\norm{v}=\sqrt{\inp{v}{v}}=\sqrt{\sum_i|v_i|^2}$.
This is the usual Euclidean norm (or ``length'' of the vector).
The norm in turn induces a \emph{distance} $\norm{v-w}$ between vectors $v$ and~$w$.
Note that distance and inner product are closely related: 
$$
\norm{v-w}^2=\inp{v-w}{v-w}=\norm{v}^2+\norm{w}^2-\inp{v}{w}-\inp{w}{v}.
$$
In particular, for unit vectors $v$ and $w$ the real part of their inner product equals $1-\frac{1}{2}\norm{v-w}^2$.
Hence unit vectors that are close together have an inner product close to~1, and vice versa.
The \emph{Cauchy-Schwarz inequality}\index{Cauchy-Schwarz inequality} 
gives $|\inp{v}{w}|\leq \norm{v}\cdot\norm{w}$ (see also Appendix~\ref{appusefulmath}).

The \emph{outer product}\index{outer product} of $v$ and $w$ is the matrix $vw^*$.

\section{Unitary matrices}

Below we will restrict attention to square matrices, unless explicitly mentioned otherwise.

A matrix $A$ is \emph{unitary}\index{matrix!unitary} if $A^{-1}=A^*$.
The following conditions are equivalent:
\begin{enumerate}
\item $A$ is unitary
\item $A$ preserves inner product: $\inp{Av}{Aw}=\inp{v}{w}$ for all $v,w$
\item $A$ preserves norm: $\norm{Av}=\norm{v}$ for all $v$
\item $\norm{Av}=1$ if $\norm{v}=1$
\end{enumerate}
(1) implies (2) because if $A$ is unitary then $A^*A=I$, 
and hence $\inp{Av}{Aw}=(v^*A^*)Aw=\inp{v}{w}$.
(2) implies (1) as follows: if $A$ is not unitary then $A^*A\neq I$, 
so then there is a $w$ such that $A^*Aw\neq w$ and, hence, a $v$ such 
that $\inp{v}{w}\neq\inp{v}{A^*Aw}=\inp{Av}{Aw}$, contradicting (2).
Clearly (2) implies (3). 
Moreover, (3) implies that $A$ preserves the real part of the inner product via 
the following identity:
$$
\norm{v+w}^2=\norm{v}^2+\norm{w}^2+\inp{v}{w}+\inp{w}{v}=\norm{v}^2+\norm{w}^2+2\mathbb{R}(\inp{v}{w}).
$$
Similarly it preserves the imaginary part of the inner product, so (3) implies (2).
The equivalence of (3) and (4) is obvious.

Note that by (4), the eigenvalues of a unitary matrix have absolute value~1.

\section{Diagonalization and singular values}\label{secsvd}

The complex number $\lambda$ is an \emph{eigenvalue}\index{eigenvalue} 
of (square) matrix $A$ if there is some nonzero vector $v$ (called an \emph{eigenvector}\index{eigenvector}) such that $Av=\lambda v$.
The set of such eigenvalues is called the \emph{spectrum} of~$A$.

Matrices $A$ and $B$ are \emph{similar}\index{matrix!similarity} 
if there is an invertible matrix $S$ such that $A=SBS^{-1}$.
Note that if $Av=\lambda v$, then $BS^{-1}v=\lambda S^{-1}v$,
so similar matrices have the same eigenvalues.
Schur's triangularization theorem
states that every matrix $A$ is similar to an
upper triangular matrix: $A=UTU^{-1}$ for some unitary $U$ 
and upper triangular $T$. 
Since similar matrices have the same eigenvalues and the eigenvalues 
of an upper triangular matrix are exactly its diagonal entries, the diagonal of $T$ has the eigenvalues of~$A$.

A matrix $D$ is \emph{diagonal}\index{matrix!diagonal} 
if $D_{ij}=0$ whenever $i\neq j$.
Let $S$ be some matrix satisfying $AS=SD$ for
some diagonal matrix $D$. Let $v_i$ be the $i$-th column of $S$ and
$\lambda_i$ be the $i$-th entry on the diagonal of $D$, then
$$
\underbrace{\left(\begin{array}{ccc} \vdots & & \vdots\\
                   Av_1 & \cdots & Av_d\\
                    \vdots & & \vdots\end{array}\right)}_{AS}=
\underbrace{\left(\begin{array}{ccc} \vdots & & \vdots\\
                   \lambda_1 v_1 & \cdots & \lambda_d v_d\\
                    \vdots & & \vdots\end{array}\right)}_{SD},
$$
and we see that $v_i$ is an eigenvector of $A$ associated
with eigenvalue $\lambda_i$. Conversely, if $v_1,\ldots,v_d$
are eigenvectors of $A$ with eigenvalues $\lambda_1,\ldots,\lambda_d$, 
then we have $AS=SD$, where $S$ has the $v_i$ as columns and 
$D$ is the diagonal matrix of $\lambda_i$.
We call a square matrix $A$ \emph{diagonalizable} if it is similar 
to some diagonal matrix $D$: $A=SDS^{-1}$.
This $D$ then has $A$'s eigenvalues $\lambda_i$ on its diagonal, 
some of which may be zero.
Note that $A$ is diagonalizable iff it has a linearly
independent set of $d$ eigenvectors.
These eigenvectors will form the columns of $S$, giving $AS=SD$, 
and linear independence ensures that $S$ has an inverse, 
giving $A=SDS^{-1}$.
A matrix $A$ is \emph{unitarily} diagonalizable iff it can be
diagonalized via a unitary matrix $U$: $A=UDU^{-1}$.
If the columns of $U$ are the vectors $u_i$, and the diagonal entries of $D$ are $\lambda_i$, then we can also write $A=\sum_i \lambda_i u_i u_i^*$; this is sometimes called the \emph{spectral decomposition} of~$A$.
By the same argument as before, $A$ will be unitarily diagonalizable
iff it has an orthonormal set of $d$ eigenvectors.

A matrix $A$ is \emph{normal}\index{matrix!normal} 
if it commutes with its conjugate transpose ($A^*A=AA^*$). 
For example, unitary matrices are normal.
If $A$ is normal and $A=UTU^{-1}$ for some upper triangular $T$ 
(which must exist because of Schur's triangularization theorem), 
then $T=U^{-1}AU$ and $T^*=U^{-1}A^*U$, so $TT^*=U^{-1}AA^*U=U^{-1}A^*AU=T^*T$.
Hence $T$ is normal and upper triangular, 
which implies (with a little work) that $T$ is diagonal.
This shows that normal matrices are unitarily diagonalizable.
Conversely, if $A$ is unitarily diagonalizable as $UDU^{-1}$, then
$AA^*=UDD^*U^*=UD^*DU^{-1}=A^*A$, so then $A$ is normal.
Thus a matrix is normal iff it is unitarily diagonalizable.
If $A$ is not normal, it may still be diagonalizable
via a non-unitary $S$, for example:
$$
\underbrace{\left(\begin{array}{rr}1 & 1\\ 0 & 2\end{array}\right)}_A=
\underbrace{\left(\begin{array}{rr}1 & 1\\ 0 & 1\end{array}\right)}_S\cdot
\underbrace{\left(\begin{array}{rr}1 & 0\\ 0 & 2\end{array}\right)}_D\cdot
\underbrace{\left(\begin{array}{rr}1 & -1\\ 0 & 1\end{array}\right)}_{S^{-1}}.
$$
If $A=UDU^{-1}$ then $A^*=UD^*U^{-1}$, so the eigenvalues of~$A^*$ 
are the complex conjugates of the eigenvalues of~$A$. 

An important class of normal (and hence unitarily diagonalizable)
matrices are the \emph{Hermitian} matrices,\index{matrix!Hermitian}
which are the ones satisfying $A=A^*$. Note that the last line of the previous paragraph
implies that the eigenvalues of Hermitian matrices are real.

A Hermitian matrix~$A$ is called \emph{positive semidefinite} (resp.~\emph{positive definite}), if all its eigenvalues are nonnegative (resp.~positive). 
An equivalent definition is that $A$ is positive semidefinite (psd) iff there exists a matrix $C$ such that $A=C^*C$ (in other words, there exist vectors $c_i$ such that for all $i,j$, we have $A_{ij}=\inp{c_i}{c_j}$).
A useful and easy to prove property is that $A$ is psd iff $\Tr(AB)\geq 0$ for all psd matrices~$B$.
By defining $A\succeq B$ iff $A-B$ is positive semidefinite, we obtain a partial ordering on the set of all Hermitian matrices.
If all eigenvalues are 0 or 1, then $A$ is called a \emph{projection}\index{matrix!projection}
(or \emph{projection matrix} or \emph{projector}). This is equivalent to requiring $A^2=A$.

Not all matrices are diagonalizable, 
for instance $A=\left(\begin{array}{cc}0 & 1\\ 0 & 0\end{array}\right)$ is not.
However, every matrix has a \emph{singular value decomposition} (SVD), which is almost as useful as a diagonalization.
We will derive this below for an arbitrary (not necessarily square) matrix~$A$.
Since $A^*A$ is psd, we can write  $A^*A=VDV^{-1}$ 
for some unitary $V$ whose columns $v_i$ are the orthonormal eigenvectors of $A^*A$, and some nonnegative diagonal matrix~$D$ with the corresponding eigenvalues. 
The entries $\sigma_i$ of the matrix $\Sigma=\sqrt{D}$ are called the \emph{singular values}\index{singular value} of~$A$ (some of which may be zero).
Define vectors $u_i=Av_i/\sigma_i$. Note that the $u_i$ form an orthonormal system because $u_i^*u_j=v_i^*(A^*Av_j)/\sigma_i\sigma_j=v_i^*(\sigma_j^2 v_j)/\sigma_i\sigma_j$, which is 1 if $i=j$ and 0 if $i\neq j$. Hence the matrix $U$ that has these $u_i$'s as columns is unitary. We have $U=AV\Sigma^{-1}$, so we can write  $A=U\Sigma V^{-1}$, which is the SVD of~$A$. Equivalently, we can write  $A=\sum_i\sigma_i u_iv_i^*$.
This derivation of the SVD $A=U\Sigma V^{-1}$ can easily be extended to arbitrary $m\times n$ matrices~$A$; $U$ will be an $m\times m$ unitary, $\Sigma$ will be $m\times n$ (padded with 0s if the rank of $A$ is $<m,n$), and $V$ will be an $n\times n$ unitary.


\section{Tensor products}\label{app:tensor}

If $A=(A_{ij})$ is an $m\times n$ matrix and $B$ an $m'\times n'$ matrix,
then their \emph{tensor} product (a.k.a.\ \emph{Kronecker} product) is the $mm'\times nn'$ matrix
$$
A\otimes B=\left(\begin{array}{rrr}A_{11}B & \cdots & A_{1n}B\\
                            A_{21}B & \cdots & A_{2n}B\\
                                    & \ddots & \\
                            A_{m1}B & \cdots & A_{mn}B
\end{array}
\right).
$$
For example:
$$
\left(\begin{array}{rr}\frac{1}{\sqrt{2}} & \frac{1}{\sqrt{2}}\\
                       \frac{1}{\sqrt{2}} & -\frac{1}{\sqrt{2}}
\end{array}\right)
\otimes
\left(\begin{array}{rr}0 & 1\\
                       -1 & 0
      \end{array}\right)
    =
\left(\begin{array}{rrrr} 0 & \frac{1}{\sqrt{2}} & 0 & \frac{1}{\sqrt{2}}\\
                         -\frac{1}{\sqrt{2}} & 0 & -\frac{1}{\sqrt{2}} & 0\\
                         0 & \frac{1}{\sqrt{2}} & 0 & -\frac{1}{\sqrt{2}}\\
                         -\frac{1}{\sqrt{2}} & 0 & \frac{1}{\sqrt{2}} & 0
      \end{array}\right).
    $$
Note that the tensor product of two numbers (i.e., $1\times 1$ matrices) is itself just a number, and the tensor product of two column vectors is itself a column vector.

The following properties of the tensor product are easily verified:
\begin{itemize}
\item $c(A\otimes B)=(cA)\otimes B=A\otimes(cB)$ for all scalars $c$
\item $(A\otimes B)^*=A^*\otimes B^*$, and similarly for inverse and transpose (note that the order of the tensor factors doesn't change).
\item $A\otimes(B+C)=(A\otimes B)+(A\otimes C)$
\item $A\otimes(B\otimes C)=(A\otimes B)\otimes C$
\item $(A\otimes B)(C\otimes D)=(AC)\otimes (BD)$
\end{itemize}
Different vector spaces can also be combined using tensor products.
If $V$ and $V'$ are vector spaces of dimension $d$ and $d'$
with basis $\{v_1,\ldots,v_d\}$ and $\{v'_1,\ldots,v'_{d'}\}$, 
respectively, then their tensor product space is the 
$d\cdot d'$-dimensional space $W=V\otimes V'$
spanned by $\{v_i\otimes v'_j\mid 1\leq i\leq d, 1\leq j\leq d'\}$.
Applying a linear operation $A$ to $V$ and $B$ to $V'$ corresponds to 
applying the tensor product $A\otimes B$ to the tensor product space $W$.

\section{Trace}

The \emph{trace}\index{matrix!trace} of a square matrix $A$ is the sum of its 
diagonal entries: $\Tr(A)=\sum_i A_{ii}$.
Some important and easily verified properties of $\Tr(A)$ are:
\begin{itemize}
\item $\Tr(A+B)=\Tr(A)+\Tr(B)$
\item $\Tr(AB)=\Tr(BA)$, which is known as the ``cyclic property'' or ``cyclicity'' of the trace.\\
For example, $\Tr(Avv^*)=v^*Av$.
\item $\Tr(A)$ is the sum of the eigenvalues of $A$.\footnote{The \emph{product} of the eigenvalues is the \emph{determinant} of $A$, but we will rarely use that in these notes.}\\
(This follows from Schur and the previous item: 
$\Tr(A)=\Tr(UTU^{-1})=\Tr(U^{-1}UT)=\Tr(T)=\sum_i\lambda_i$)
\item $\Tr(A\otimes B)=\Tr(A)\Tr(B)$
\end{itemize}

\section{Rank}

The \emph{rank}\index{matrix!rank} of a matrix $A$ (over a field $\mathbb{F}$) is 
the size of a largest linearly independent set of rows of $A$ 
(linear independence taken over $\mathbb{F}$).
Unless mentioned otherwise, we take $\mathbb{F}$ to be the field of complex numbers.
We say that $A$ has \emph{full rank} if its rank equals its dimension.
The following properties are all easy to show:
\begin{itemize}
\item $\rank(A)=\rank(A^*)$
\item $\rank(A)$ equals the number of nonzero singular values of $A$
(counting multiplicity)
\item $\rank(A+B)\leq \rank(A)+\rank(B)$
\item $\rank(AB)\leq\min\{\rank(A),\rank(B)\}$
\item $\rank(A\otimes B)=\rank(A)\cdot \rank(B)$
\item $A$ has an inverse iff $A$ has full rank
\end{itemize}

\section{The Pauli matrices}\label{app:pauli}

The four \emph{Pauli matrices} are:
\begin{equation*}
I = \left(\begin{array}{rr}1 & 0\\ 0&1\end{array}\right), \
X = \left(\begin{array}{rr}0&1\\ 1&0\end{array}\right), \
Y = \left(\begin{array}{rr}0&-i\\i&0\end{array}\right), \mbox{ and }
Z = \left(\begin{array}{rr}1&0\\ 0&-1\end{array}\right).
\end{equation*}
Note that each Pauli matrix~$P$ is both unitary and Hermitian, and hence self-inverse: $P^{-1}=P$. This implies that their eigenvalues are in $\{-1,1\}$.
Distinct non-identity Paulis anti-commute: if $P,Q\in\{X,Y,Z\}$ are distinct then $PQ=-QP$. Note that $Y=iXZ$. Also, products of two distinct Pauli matrices have trace~0.

Define the \emph{Hilbert-Schmidt} inner product on the space of $d\times d$ matrices as $\inpc{A}{B}=\frac{1}{d}\Tr(A^* B)$. With respect to this inner product (for $d=2$), the four Pauli matrices form an orthonormal set. This implies that every complex $2\times 2$ matrix $A$ can be written as a linear combination of the Pauli matrices:
$$
A=\alpha_0 I + \alpha_1 X + \alpha_2 Y + \alpha_3 Z,
$$
with complex coefficients $\alpha_i$. If $A$ is Hermitian, then these coefficients will be real. 

We can also consider the \emph{$n$-qubit} Paulis, which are $n$-fold tensor products of the above $2\times 2$ Paulis. For example $X\otimes Z\otimes I\otimes Y\otimes Z$ is a 5-qubit Pauli. There are $4^n$ $n$-qubit Paulis, since we have 4 possibilities for each of the $n$ tensor factors, and these $4^n$ matrices form an orthonormal set w.r.t.\ Hilbert-Schmidt inner product. Accordingly, every $2^n\times 2^n$ matrix~$A$ can be written uniquely as a linear combination of the $4^n$ $n$-qubit Paulis. Again, if $A$ is Hermitian, then the $4^n$ coefficients will be real. 

\section{Dirac notation}

Physicists often write their linear algebra in \emph{Dirac notation},
and we will follow that custom for denoting quantum states.\index{Dirac notation}
In this notation we write $\ket{v}=v$ and $\bra{v}=v^*$.
The first is called a \emph{ket}, the second a \emph{bra}.
Some points about this notation:
\begin{itemize}
\item $\inp{v}{w}=\bra{v}\ket{w}$: inner products are bra-ket (``bracket'') products.
\item If matrix $A$ is unitarily diagonalizable, then $A=\sum_i\lambda_i\ket{v_i}\bra{v_i}$ for some orthonormal set of eigenvectors $\{v_i\}$.
\item $\ketbra{v}{v}\otimes\ketbra{w}{w}=(\ket{v}\otimes\ket{w})(\bra{v}\otimes\bra{w})$, the latter is often abbreviated to $\ket{v}\otimes\ket{w}\bra{v}\otimes\bra{w}$. Abbreviating the latter further by omitting the tensor product leads to dangerous ambiguity, though sometimes it's still clear from context.
\item $(U\ket{v})^*=\bra{v}U^*$ and $(\ket{u}\otimes\ket{v})^*=\bra{u}\otimes\bra{v}$ (the ordering of tensor factors doesn't change).
\item Don't write kets inside of other kets or bras: the notation $\inp{u}{(\alpha\ket{v}+\beta\ket{w})}$ doesn't really make sense.
\end{itemize}

\chapter{Some other Useful Math and CS}\label{appusefulmath}

Here we collect various basic but useful facts and definitions needed in parts of the lecture notes.

\section{Some notation, equalities and inequalities}\label{appsec:eqs}

\begin{itemize}
\item We use $[n]$ to denote the set $\{1,\ldots,n\}$. If $P$ is a statement which can be true or false, then $[P]\in\01$ denotes its truth value. We write $\delta_{a,b}=[a=b]$, indicating whether $a=b$ or not.
Logarithms will always be to base~2 unless stated otherwise.
\item A \emph{complex number} is of the form $c=a+bi$, where $a,b\in\mathbb{R}$,
and $i$ is the imaginary unit, which satisfies $i^2=-1$.  Such a $c$ can also be written as $c=r e^{i\phi}$
where $r=|c|=\sqrt{a^2+b^2}$ is the \emph{magnitude} (a.k.a.\ \emph{modulus} or \emph{norm}) of $c$, and $\phi\in[0,2\pi)$ is the angle that
$c$ makes with the positive horizontal axis when we view it as a point $(a,b)$ in the plane.
Note that complex numbers of magnitude~1 lie on the unit circle in this plane.
We can also write those as $e^{i\phi}=\cos(\phi)+i\sin(\phi)$.
The complex conjugate $c^*$ is $a-ib$, equivalently $c^*=r e^{-i\phi}$.
\item The Cauchy-Schwarz inequality: for $a=(a_1,\ldots,a_n)\in\mathbb{C}^n$ and $b=(b_1,\ldots,b_n)\in\mathbb{C}^n$
$$
\left|\sum_{i=1}^n a^*_i b_i\right| \leq \sqrt{\sum_{i=1}^n |a_i|^2}\sqrt{\sum_{i=1}^n |b_i|^2}.
$$
Equivalently, written in terms of inner products and norms of vectors: $|\inp{a}{b}|\leq\norm{a}\cdot\norm{b}$. {\footnotesize Proof for the case with real entries:
for every real $\lambda$ we have $0\leq \inp{a-\lambda b}{a-\lambda b}=\norm{a}^2+\lambda^2\norm{b}^2-2\lambda\inp{a}{b}$. Now set $\lambda=\norm{a}/\norm{b}$ and rearrange (a slightly more complicated proof works if $a,b\in\mathbb{C}^n$).}
\item Geometric sum: $\displaystyle\sum_{j=0}^{m-1} z^j =\left\{\begin{array}{ll}
m & \mbox{if }z=1\\
\frac{1-z^m}{1-z} & \mbox{if }z\neq 1
\end{array}\right.$\\
{\footnotesize Proof: The case $z=1$ is obvious; for the case $z\neq 1$, observe $(1-z)(\sum_{j=0}^{m-1} z^j)=\sum_{j=0}^{m-1} z^j - \sum_{j=1}^m z^j=1-z^m$.}\\[0.8mm]
For example, if $z=e^{2\pi i r/N}$ is a root of unity, with $r$ an integer in $\{1,\ldots,N-1\}$, then
$\sum_{j=0}^{N-1} z^j=\frac{1-e^{2\pi i r}}{1-e^{2\pi i r/N}}=0$.
\item The ratio in the previous line can be rewritten using the identity $|1-e^{i\theta}|=2|\sin(\theta/2)|$;
this identity can be seen by drawing the numbers 1 and $e^{i\theta}$ as vectors from the origin 
in the complex plane, and dividing their angle $\theta$ in two.
Some other useful trigonometric identities: $\cos(\theta)^2+\sin(\theta)^2=1$, $\sin(2\theta)=2\sin(\theta)\cos(\theta)$.
\item $1+x\leq e^x$ for all real numbers $x$ (positive as well as negative).\\[-1mm]
\item If $\eps_j\in[0,1]$ for all $j\in[k]$, then
$\displaystyle
1-\sum_{j=1}^k\eps_j\leq\prod_{j=1}^k(1-\eps_j)\leq e^{-\sum_{j=1}^k\eps_j}.$\\
{\footnotesize Proof: The upper bound comes from the preceding item.
  The lower bound follows easily by induction on~$k$, using the fact that $(1-\eps_1)(1-\eps_2)=1-\eps_1-\eps_2+\eps_1\eps_2\geq 1-\eps_1-\eps_2$.}
\end{itemize}

\section{Algorithms and probabilities}\label{appusefulCS}

\begin{itemize}
\item When we don't care about constant factors, we'll often use big-O notation: $T(n)=O(f(n))$ means there exist constants $c,n_0\geq 0$
such that for all integers $n\geq n_0$, we have $T(n)\leq cf(n)$.  Similarly, big-Omega notation is used for lower bounds: 
$T(n)=\Omega(f(n))$ means there exist constants $c,n_0\geq 0$ such that $T(n)\geq cf(n)$ for all~$n\geq n_0$. $T(n)=\Theta(f(n))$ means that simultaneously $T(n)=O(f(n))$ and $T(n)=\Omega(f(n))$. Such notation is often used to write upper and/or lower bounds on running time of algorithms as a function of input-length~$n$.
\item For $N=2^n$, we can identify the integers $\{0,\ldots,N-1\}$ with their $n$-bit binary representations as follows: the bitstring $x=x_{n-1}\ldots x_1x_0\in\01^n$ corresponds to the integer $\sum_{i=0}^{n-1} x_i2^i$. The leftmost bit $x_{n-1}$ is called the \emph{most significant bit} of~$x$ (since it corresponds to the largest power of two, $2^{n-1}$), and the rightmost bit $x_0$ is its \emph{least significant} bit (it corresponds to $2^0=1$, so determines whether $x$ is an even or odd integer). For example, if $n=3$ then the bitstring $x=x_2x_1x_0=101$ corresponds to the integer $x_2\cdot 4 + x_1\cdot 2 + x_0\cdot 1=4+1=5$. The integer~$0$ corresponds to the bitstring $0^n$ (if we use $0$ to denote a bitstring of $0$s, then the value of $n$ should be clear from context).

We can also use binary notation for non-integral numbers, with the bits to the right of the decimal dot corresponding to negative powers of two ($1/2$, $1/4$, $1/8$, etc.). For example, 0.1 denotes $1/2$ and 10.101 denotes $2+1/2+1/8=21/8$. Note that multiplying by two corresponds to shifting the dot to the right, and dividing corresponds to shifting to the left.
\item The \emph{union bound} says that the probability of the union (or logical ``or'') of two events is at most the sum of their probabilities:
$\Pr[A\vee B]\leq \Pr[A]+\Pr[B]$. This inequality should be obvious from drawing a Venn diagram. More generally, if we have $T$ events $A_1,\ldots,A_T$, then $\Pr[A_1\vee\cdots \vee A_T]\leq \sum_{i=1}^T\Pr[A_i]$.
\item A (discrete) \emph{random variable} $X$ is an object that takes value $x_i$ with probability $p_i$. Its \emph{expected value} (or \emph{expectation}) is $\mu=\Exp[X]=\sum_i p_ix_i$. Its \emph{variance} is $\sigma^2=\Var[X]=\Exp[(X-\Exp[X])^2]=\Exp[X^2]-\Exp[X]^2$. Its \emph{standard deviation} is $\sigma$.
\item \emph{Linearity of expectation} says that, for values $a_1,\ldots,a_m$ and random variables $X_1,\ldots,X_m$, we have  $\Exp[\sum_{j=1}^m a_jX_j]=\sum_{j=1}^m a_j\Exp[X_j]$ (easily verified using the definition of expectation).
\item Random variable $X$ is \emph{independent} from random variable $Y$, if the value of $Y$ does not affect the probability distribution of $X$, i.e., $\Pr[X=x\mbox{ and }Y=y]=\Pr[X=x]\cdot\Pr[Y=y]$ for all possible values $x,y$. If $X$ and $Y$ are independent, then $\Exp[X\cdot Y]=\Exp[X]\cdot\Exp[Y]$.
\item Three basic upper bounds on the tails of probability distributions:\\[0.2em]
{\bf Markov}: if $X$ is a nonnegative random variable with expectation $\mu$, then $\Pr[X\geq k\mu]\leq 1/k$. 
{\footnotesize Proof: Since $X$ is nonnegative, $\mu\geq \Pr[X\geq k\mu]\cdot k\mu$.}\\[0.2em]
{\bf Chebyshev}: if $X$ is a random variable with expectation $\mu$ and standard deviation $\sigma$, then $\Pr[|X-\mu|\geq k\sigma]\leq 1/k^2$.\\
{\footnotesize Proof: Apply Markov to the random variable $|X-\mu|^2$, whose expectation is $\sigma^2$.}\\[0.2em]
{\bf Chernoff/Hoeffding}: if $X=\sum_{i=1}^n X_i$ is the sum of $n$ independent, identically distributed random variables $X_i\in\{0,1\}$, each with expectation $\Pr[X_i=1]=p$, then $X$ has expectation $\mu=np$, and exponentially decreasing tail bound $\Pr[|X-\mu|\geq\alpha n]\leq 2e^{-2\alpha^2 n}$.\\
{\footnotesize Proof idea: For all parameters $\lambda\geq 0$, we have $\Pr[X-\mu\geq t]=\Pr[e^{\lambda X}\geq e^{\lambda(t+\mu)}]$. Upper bound the latter probability by applying Markov to random variable $e^{\lambda X}$. This is a product of $n$ independent random variables~$e^{\lambda X_i}$, so its expectation is the product of $n$ simpler expectations. Choose~$\lambda$ to minimize the bound.}
\item A \emph{randomized} algorithm is a classical algorithm that can flip random coins during its operation, meaning its behavior is partially determined by chance and its output is a random variable that depends on its input, rather than a deterministic function of its input. One can think of a randomized algorithm as a probability distribution over deterministic algorithms (one deterministic algorithm for each setting of the coins).
\item When we say a (randomized or quantum) algorithm has error probability $\leq 1/3$, this typically means \emph{in the worst case}: for every possible input, the algorithm produces the correct answer with probability $\geq 2/3$, where probability is taken over the random coin flips and/or quantum measurements during its operation. Such statements do not refer to ``most'' inputs under some distribution on the inputs, unless stated explicitly.
\item If a (randomized or quantum) algorithm produces the correct answer in \emph{expected} running time~$T$ (meaning \emph{for each} input its expected running time is $\leq T$), then we can convert that into an algorithm with \emph{worst-case} running time $3T$ and error probability $\leq 1/3$, as follows. Run the original algorithm for $3T$ steps, and just cut it off if it hasn't terminated by itself. The probability of non-termination within $3T$ steps is at most $1/3$ by Markov's inequality. Hence with probability $\geq 2/3$ we will have received the correct answer from the algorithm.
\item The union bound is very useful for analyzing an algorithm with $T$ randomized subroutines, each of which has its own small failure probability $\leq 1/(3T)$. If these failure events are independent, then the probability that none of the $T$ subroutines fails is at least $(1-1/(3T))^T\geq 1-T/(3T)=2/3$ (using the last bullet of \ref{appsec:eqs}). But what if these events are dependent on each other, which may well happen if the subroutines depend on what happens elsewhere in the algorithm? Then by the union bound, the probability that one or more of the subroutines fails is $\leq T\cdot 1/(3T)\leq 1/3$. In other words, the probability that none of the $T$ subroutines fails is still~$\geq 2/3$.
\item If a (randomized or quantum) algorithm with 0/1-outputs has error probability $\leq 1/3$, then we can cheaply reduce this error probability to small $\eps>0$, as follows. Choose odd $n=O(\log(1/\eps))$ such that $2e^{-2\alpha^2 n}\leq\eps$ for $\alpha=1/6$. Run the original algorithm $n$ times, each time with newly-chosen randomness, and output the majority among the $n$ output bits. The probability that this majority is wrong (i.e., that the number of correct output bits is more than $\alpha n$ below its expectation), is at most $\eps$ by the Chernoff bound. 
\end{itemize}

\chapter{Hints for Exercises}\label{app:hints}

\subsection*{Chapter~\ref{chap:intro}}
\ref{ex:povmvsprojective}. Find the maximal $p$ such that $I-2p\ketbra{-}{-}-2p\ketbra{1}{1}$ is still psd.\\
\ref{ex:nocloning}. Consider what $U$ has to do when $\ket{\phi}=\ket{0}$, when $\ket{\phi}=\ket{1}$, and when $\ket{\phi}$ is a superposition of these two.\\
\ref{ex:TrC}.b. Use the facts that $\Tr(D\ketbra{\psi}{\psi})=\bra{\psi}D\ket{\psi}$ for every $D$, and that products of 2 distinct Paulis have trace~0 because at least one them has to be different from $I$ (i.e., $X$, $Y$, or $Z$). This exercise is just superdense coding in disguise.

\subsection*{Chapter~\ref{chap:qircuitdj}}
\ref{ex:controlledcircuit}. Use Exercise~\ref{ex:controlledgate}.\\
\ref{ex:intermediatemeas}. Instead of measuring the qubit, apply a CNOT that ``copies'' it to a new $\ket{0}$-qubit, which is then left alone until the end of the computation.  Analyze what happens.\\
\ref{ex:parityquery}. Use the  Bernstein-Vazirani algorithm.

\subsection*{Chapter~\ref{chap:simon}}
\ref{ex:oracleinter}.c. Approximate the state of part (a) using the subroutine of part (b), and see what happens if you apply Hadamards to the approximate state. Use the fact that $\frac{1}{2^N}\sum_{w=0}^{N/2+2\sqrt{N}}\binom{N}{w}$ is nearly~1, because this is the probability that if you flip $N$ fair coins, $N/2-2\sqrt{N}$ or more of them come up ``heads.''

\subsection*{Chapter~\ref{chap:ft}}
\ref{ex:tvddistance}. Use $|\alpha_i^2-\beta_i^2|=|\alpha_i-\beta_i|\cdot|\alpha_i+\beta_i|$ and the Cauchy-Schwarz inequality.\\
\ref{ex:approxqft}.e. Use triangle inequality.\\
\ref{ex:approxqft}.f. Drop all phase-gates with small angles $\phi<1/n^3$ from the $O(n^2)$-gate circuit for $F_{2^n}$ explained in Section~\ref{sec:qftcircuit}. Upper bound the number of gates that are left in the new circuit, and analyze the distance between the unitaries corresponding to the new circuit and the original circuit.

\subsection*{Chapter~\ref{chap:factoring}}
\ref{ex:modexp}.a. You may invoke here (without proof) the Sch\"onhage-Strassen
algorithm for fast multiplication~\cite{schonhage&strassen,knuth:acp2}\index{Knuth, D.}. This allows you to multiply two $n$-bit integers mod $N$ using $O(n\log(n)\log\log(n))$ steps (where $n=\lceil{\log N\rceil}$).\footnote{In his original paper, Shor used the Sch\"onhage-Strassen algorithm for fast integer multiplication. We could also invoke the more recent improvement of Harvey and van der Hoeven~\cite{harvey&hoeven:fastermultj}, who remove the $\log\log n$ factor.}\\
\ref{ex:rsa}.a. The prime number theorem implies that $\Omega(N/\ln N)$ of the numbers between 1 and $N$ are prime; also there is an efficient classical algorithm to test if a given number is prime~\cite{AKS:primesinP}. You may use these facts, but be explicit in how many bits your primes $p$ and $q$ will have.
You may also invoke the extended Euclidean algorithm without providing detail of how/why it works.\\
\ref{ex:rsa}.b. Use the result of Exercise~\ref{ex:modexp} (no need to rederive that here).\\
\ref{ex:rsa}.c. The set of all possible messages forms a group of size $\phi(N)$. Euler's Theorem says that in any group $G$, we have $a^{|G|}=1$ for all $a\in G$ (here `$1$' is the identity element in the group).

\subsection*{Chapter~\ref{chap:hsp}}
\ref{ex:povm}.b. For $M=\sum_{i=1}^K E_i$, show that $\norm{M}^2=\norm{M^2}\leq\frac{2}{3}\norm{M}\,+$ a small constant.\\
\ref{ex:GIswap}. You could use the SWAP-test from Section~\ref{sec:qfingerprint}.

\subsection*{Chapter~\ref{chap:grover}}
\ref{ex:findallsols}.b. Recall that if there are $i>0$ solutions, then one variant of Grover's algorithm finds a solution using an expected number of $O(\sqrt{N/i})$ queries.\\
\ref{ex:exactgrover}.e. Choose $\gamma$ in (d) such that applying $\ceil{\tilde{k}}$ rounds of amplitude amplification to $\cal A$ results in a solution for $y$ with probability~1.\\
\ref{ex:log1overeps}.a. Try running the exact version of Grover (see end of Section~\ref{sec:grover}) with different guesses for what the actual $t$ is.\\
\ref{ex:approxcount}.a. Run the basic Grover search with a cleverly chosen number of iterations.\\
\ref{ex:approxcount}.b. Use binary search on top of~(a).\\
\ref{ex:amplitudeestimation}.d. The eigenvalues of a 2-dimensional rotation matrix over angle $\lambda$ are $e^{i\lambda}$ and $e^{-i\lambda}$ (you don't need to prove this). You may also assume $a\ll 1$, so that the differences between  $\sqrt{a},\sin\sqrt{a},\arcsin\sqrt{a}$ are negligible.
You may refer to the lecture notes for phase estimation without further proof, incl.\ the fact that phase estimation gives a good $n$-bit approximation with high probability even if the phase cannot be represented exactly with $n$ bits of precision.\\
\ref{ex:amplitudeestimation}.e. Define an $A$ that involves one query to $x$ and where $a=t/N$. Invoke (d) with $\eps$ proportional to $1/\sqrt{N}$. You can use $|\tilde{a}-a|=|\sqrt{\tilde{a}}-\sqrt{a}|\cdot|\sqrt{\tilde{a}}+\sqrt{a}|$ in your analysis of the approximation error.\\
\ref{ex:grovergates}.b. Combine amplitude amplification with the algorithm of (a), for a smart choice of $k$. Your answer may refer to the lecture notes for the details of amplitude amplification.\\
\ref{ex:minfind}. Start with $m=x_i$ for a random $i$, and repeatedly use Grover's algorithm to
find an index $j$ such that $x_j<m$ and update $m=x_j$.  Continue this until you can find no element smaller than $m$,
and analyze the number of queries of this algorithm. You are allowed to argue about this algorithm on a high level
(i.e., things like ``use Grover to search for a $j$ such that\ldots'' are OK), no need to write out complete circuits.
You do, however, have to take into account that the various runs of Grover each have their own error probability\\
\ref{ex:collisionfind}.b. What is the probability in (a) if you set $s$ to roughly $\sqrt{N}$?\\
\ref{ex:collisionfind}.c. Choose a set $S$ of size $s=O(N^{1/3})$, and classically query all its elements. First check if $S$ contains a collision. If yes, then you're done. If not, then use Grover to find a $j\not\in S$ that collides with an $i\in S$.\\
\\
\ref{ex:exactsimon}.a.
Amplitude amplification can be made exact if we know the success probability~$p$, similar to Exercise~\ref{ex:exactgrover}; you may use this here without proof.

\subsection*{Chapter~\ref{chap:qwalks}}
\ref{ex:matrixtest}.a. Choose a uniformly random vector $v\in\01^n$, calculate $ABv$ and $Cv$, and check whether these two vectors are the same.\\
\ref{ex:matrixtest}.b. Consider the case where $A$ is the all-0 matrix.\\
\ref{ex:matrixtest}.c. Modify the algorithm for collision-finding: use a quantum walk on the Johnson graph $J(n,r)$, where each vertex corresponds to a set $R\subseteq[n]$, and that vertex is marked if there are $i,j\in R$ such that $(AB)_{i,j}\neq C_{i,j}$. Optimize over $r$.\\
\ref{ex:3sat}.b. There's no need to use the $C,U,S$-framework of the chapter here; the answer is much simpler. View the $3n$-step random walk algorithm as a deterministic algorithm with an additional input $r\in\01^n\times \{1,2,3\}^{3n}$, where the first $n$ bits determine $x$, and the last $3n$ entries determine which variable of the leftmost false clauses will be flipped in the $3n$ steps of the random walk.
Use Grover search or amplitude amplification (no need to write out complete circuits here).

\subsection*{Chapter~\ref{chap:hamsim}}
\ref{ex:TrotterE}. Use induction on $m$, and the fact that there exists a constant $c$ such that for $A,B$ of small norm we have $e^{A+B}=e^A e^B +E$ for some $E$ of norm $\norm{E}\leq c\norm{A}\cdot\norm{B}$.\\
\ref{ex:improvertrotter}. The sequence of inequalities is very similar to the analysis of basic Trotter of Section~\ref{sec:trotter}.\\
\ref{ex:justifyMpsi}. Calculate the subnormalized second-register state $(\bra{0}\otimes I)(W^{-1}\otimes I)V(W\otimes I)\ket{0}\ket{\psi}$.\\
\ref{ex:oblivamplampl}.d. Like in the analysis of Grover's algorithm and regular amplitude amplification (Chapter~\ref{chap:grover}), the product of two reflections on $\cal S$ is a rotation of $\cal S$.\\
\ref{ex:errorTaylor}. Use triangle inequality, $\norm{H}\leq 1$, and the fact that $k!\geq (k/e)^k$.\\
\ref{ex:sparseblock}.b. $W_2$ just implements a rotation on the first qubit, by an angle that depends on $A_{kj}$. If you have a basis state $\ket{0}\ket{a}$ where $a\in[-1,1]$ is some real number written in some fixed finite number of bits, then you can rotate the first qubit to $a\ket{0}+\sqrt{1-a^2}\ket{1}$ by a small circuit that does some single-qubit gates on the first qubit, conditioned on the number $a$ in the $\ket{a}$-part. That circuit is the same for all values of $a$, so it's independent of the particular $\ket{0}\ket{a}$ you're acting on. You may just assume you can do this circuit, without writing out its details.\\
\ref{ex:simPaulimatrix}.d. Note that the computational basis states $\ket{x}$ are the eigenstates of $P$ and hence also of~$U$, so the only thing you need to do is multiply with the right phases for them.\\
\ref{ex:simPaulimatrix}.e. By conjugating with single-qubit gates you can change non-$Z$ Paulis to $Z$s, in order to reduce to the case of~(d).

\subsection*{Chapter~\ref{chap:hhl}}
No hints for this chapter, sorry!

\subsection*{Chapter~\ref{chap:lowerb}}
\ref{ex:paritylower}.a. Use Exercise~\ref{ex:parityin2}.\\
\ref{ex:paritylower}.b. Show that the symmetrized approximate polynomial $r$ induced by the algorithm has degree at least $N$.\\
\ref{ex:allequal}.c. Use the result of Exercise~\ref{ex:ANDinN} for $N=2$.\\
\ref{ex:maj}.b. When defining the relation $R$, consider that the hardest task for this algorithm is to distinguish inputs of weight $N/2$ from inputs of weight $N/2+1$.\\
\ref{ex:LBinvpermutation}.b. Consider the Boolean-valued problem of distinguishing the inputs where 0 sits at an odd location~$i$ in the string~$x$ from those where 0 sits at an even location.\\
\ref{ex:sortLB}. Show how you can use sorting to solve the Majority-problem
and then use the lower bound from Exercise~\ref{ex:maj} to get an $\Omega(N)$ lower bound on sorting.
(It is actually known that sorting takes $\Omega(N\log N)$ comparisons even on a quantum computer, but you don't have to show that.)\\
\ref{ex:6thpower}.a. Reduce the $bs(f)$-bit OR function (restricted to inputs of weight~0 or~1) to~$f$ and invoke the lower bound that we know for OR.\\
\ref{ex:hybridmethod}.b. Use induction on $T$ and triangle inequality.\\
\ref{ex:hybridmethod}.d. Add up the inequalities of (b) and (c) over all $i$, and use the Cauchy-Schwarz inequality.\\
\ref{ex:kwiseindep}.b. Compare the expected value of a monomial of degree $\leq 2T$ under distributions $U$ and $D$, and then use the fact that a polynomial is a sum of monomials.

\subsection*{Chapter~\ref{chap:adversary}}
\ref{ex:advST}.a. Use that $I=P_0P_0+P_1P_1$.\\
\ref{ex:advST}.b. Note that $t=T$ here, so we're considering the final states of an algorithm that,  for every input~$x$, outputs the wrong value $1-f(x)$ with probability $\leq\eps$.\\
\ref{ex:advST}.c. Use (a), the fact that $P_{f(x)}P_{1-f(y)}+P_{f(y)}P_{1-f(x)}=I$ whenever $f(x)\neq f(y)$, and the fact that $\Gamma_{xy}=0$ whenever $f(x)=f(y)$.\\
\ref{ex:advST}.d. Cauchy-Schwarz and the definition of operator norm imply $|\bra{\phi}M\ket{\psi}|\leq \norm{\ket{\phi}}\cdot\norm{M\ket{\psi}}\leq \norm{\ket{\phi}}\cdot\norm{M}\cdot\norm{\ket{\psi}}$ for all matrices $M$ and vectors $\ket{\phi},\ket{\psi}$.\\
\ref{ex:advDelta}.b. Note that the $\ket{\phi_i^t}$'s are pairwise orthogonal for different $i,i'$ due to having a different basis state in their query register.\\
\ref{ex:advDelta}.c.
Show and use that $I-O_{x,\pm}O_{y,\pm}=2\sum_{i:x_i\neq y_i}P_i$. Also use that $(\Gamma_i)_{xy}=0$ whenever $x_i=y_i$.\\
\ref{ex:advDelta}.d. Cauchy-Schwarz and the definition of operator norm imply $|\bra{\phi_i^t}(I\otimes \Gamma_i)\ket{\phi_i^t}|\leq\norm{I\otimes \Gamma_i}\cdot\norm{\ket{\phi_i^t}}^2$.\\
\ref{ex:weakduality}. 
Take feasible solutions for primal and dual. Use that $C-\sum_{i=1}^m y_i A_i$ is psd.

\subsection*{Chapter~\ref{chap:qcomplexitytheory}}
\ref{ex:factoringdec}. Use binary search, running the algorithm with different choices of $k$ to ``zoom in'' on the largest prime factor.\\
\ref{ex:bqptobqp}.a. Use the last item of Appendix~\ref{appusefulCS} to make the error probability exponentially small.\\
\ref{ex:bqptobqp}.c. Use the error analysis of Exercise~\ref{chap:ft}.\ref{ex:approxqft}.\\
\ref{eq:BQPinPP}.a. Write $\ket{\psi_x}=\alpha\ket{0}\ket{\phi_0}+\beta\ket{1}\ket{\phi_1}$, and consider the inner product between $(Z\otimes I)\ket{\psi_x}$ and $\ket{\psi_x}$.\\
\ref{eq:BQPinPP}.b. Use part (a). Analyze the amplitude of $\ket{x,0^{S-n}}$ in the final state $\ket{\psi_x}$, using ideas from the proof of $\BQP\subseteq\PSPACE$ in Section~\ref{sec:bqpinpspace}. Note that in contrast to that proof, you cannot use more than polynomial time for this exercise.\\
\ref{ex:entdetection}. Recall the satisfiability problem: given a logical formula $\phi$ with $m$ Boolean variables, decide if there is a setting of those $m$ variables that makes the formula true. We know this problem is $\NP$-complete, meaning that every problem in $\NP$ can be efficiently reduced to satisfiability (you can just assume this here, no need to prove it). Show how satisfiability can be efficiently solved if we have an efficient algorithm for the entanglement-detection problem.  Be explicit about what the circuit $C$ induced by an satisfiability-instance~$\phi$ does (without writing out the circuit fully). If you claim some state is entangled, then you have to prove that.

\subsection*{Chapter~\ref{ch:QMAlocham}}
\ref{ex:approxlambdamin}. Use binary search with different values of $a,b$ to zoom in on the right value.\\
\ref{ex:QMAremovesimplif}.b. Argue about the penalty given by $H_{\text{clock}}$, which can't be larger than $\lambda_{\min}$.\\
\ref{ex:QMAremovesimplif}.d. Use $\bra{\psi''}H\ket{\psi''}=\norm{\sqrt{H}\ket{\psi''}}^2$ and triangle inequality.\\
\ref{ex:QMAremovesimplif}.e. Use Eq.~\eqref{eq:Htpenalty}.\\
\ref{ex:QMAremovesimplif}.f. Use (e) and  Cauchy-Schwarz.\\
\ref{ex:QMAremovesimplif}.g. Sum (f) over all $(T+1)^2$ pairs $t,t'$ to lower bound $\inp{\psi'}{\psi''}$.\\
\ref{ex:QMAremovesimplif}.j. Once all three simplifying assumptions have been satisfied, we can invoke the energy lower bound of $\Omega(1/T^2)$ proved in Section~\ref{sec:lochamsoundcomplete}.\\
\ref{ex:QMAwithlogwitnesses}.a. Invoke the Marriott-Watrous result mentioned at the end of Section~\ref{sec:qmadef} (without proving it).\\
\ref{ex:QMAinPP}. Combine ideas from Exercises \ref{ex:QMAwithlogwitnesses}.b and \ref{chap:qcomplexitytheory}.\ref{eq:BQPinPP}.\\
\ref{ex:locHamBQP}.a. Note that from the description of $H$ you can infer the circuit $C_n=U_T\cdots U_1$ for $n$-bit instances, and you can then run $U_t\cdots U_1$ in a controlled manner, for every $t\in\{0,\ldots,T\}$ of your choice.

\subsection*{Chapter~\ref{chap:qencodings}}
\ref{ex:2to1rac}.a. It suffices to use pure states with real amplitudes as encoding. Try to ``spread out'' the 4 encodings
$\ket{\phi_{00}}$, $\ket{\phi_{01}}$, $\ket{\phi_{10}}$, $\ket{\phi_{11}}$ in the 2-dimensional real plane as well as possible.\\
\ref{ex:teleportopt}. Use the fact that 1 classical bit of communication can only send 1 bit of information, no matter how much entanglement Alice and Bob share. Combine this fact with superdense coding.\\
\ref{ex:qftcommunication}.b. Think of the first $n$ qubits as Alice and the last $n$ qubits as Bob; use Holevo's theorem.\\
\ref{ex:dist2states}.a. Consider the positive and negative eigenvalues in the spectral decomposition of $\rho_0-\rho_1$, and analyze the success probability minus the error probability.

\subsection*{Chapter~\ref{chap:qcomcompl}}
\ref{ex:detEQ}. Argue that if Alice sends the same message for distinct inputs $x$ and $x'$, then Bob doesn't know what to output if his input is $y=x$.\\
\ref{ex:QeEQ}.a. Argue that if $P$ is a projector then we can't have both $P\ket{\phi}=\ket{\phi}$ and $P\ket{\psi}=0$.\\
\ref{ex:QeEQ}.c. Observe that among Alice's possible $n$-bit inputs are the $n$ codewords of the Hadamard code that encodes $\log n$ bits (see Section~\ref{sec:ldclowerbound}); each pair of distinct Hadamard codewords is at Hamming distance exactly $n/2$.
Now use part (b), applied to equality protocols with $n'=\log n$ inputs bits.\\
\ref{ex:oneroundqcc}. Use the fact that 2 non-orthogonal states cannot be distinguished perfectly 
(Exercise~\ref{ex:QeEQ}), and that a set of $2^n$ vectors that are pairwise orthogonal must have dimension $2^n$.\\
\ref{ex:DISJonewayLB}. Invoke the quantum random access lower bound, Theorem~\ref{thrac} of Section~\ref{sec:qencandlimits}.\\
\ref{ex:intersectionrrounds}. Partition the $n$ positions into disjoint sets and run (in parallel) a separate $r$-message intersection protocol for each of these sets.\\
\ref{ex:ambainissmpeq}.b. Let Alice send a random row of $C(x)$ (with the row-index) and let Bob send a random column of $C(y)$ (with the column-index).\\
\ref{ex:qfpfrompolys}.a. Two distinct polynomials, each of degree $\leq d$, are equal on at most $d$ points of the domain~$\mathbb{F}_p$.\\
\ref{ex:IPLB}.b. Run the protocol of part (a) on an initial state where Bob has a well-chosen superposition over many~$\ket{y}$.\\
\ref{ex:vecinsubspace}.b. You can derive this from one of the communication lower bounds mentioned in this chapter, you don't need to prove this from scratch.\\
\ref{ex:HidMat}. The matching $M$ induces a projective measurement that Bob can do on the message he receives.\\
\ref{ex:HtestHSP}.d. Alice could send a uniform superposition over all $h\in H$.

\subsection*{Chapter~\ref{chap:nonloc}}
\ref{ex:EPRUT}.b. You could write this out, but you can also get the answer almost immediately from part (a) and the fact that $H^T=H^{-1}$.\\
\ref{ex:EPRXX}.b. It's helpful here to write the EPR-pair in the basis $\ket{+}=\frac{1}{\sqrt{2}}(\ket{0}+\ket{1})$, $\ket{-}=\frac{1}{\sqrt{2}}(\ket{0}-\ket{1})$.\\
\ref{ex:classchsh}. For every fixed input $x,y$, there is a classical strategy that gives a wrong output
only on that input, and that gives a correct output on all other possible inputs.  Use the shared randomness
to randomly choose one of those deterministic strategies.\\
\ref{ex:tsirelsonbound}.b. Argue that $\frac{1}{4}\bra{\psi}C\ket{\psi}=\Pr[\mbox{win}]-\Pr[\mbox{lose}]$.\\
\ref{ex:tsirelsonbound}.c. Use that $A_x^2$ and $B_y^2$ are the $k$-qubit identity matrix.\\
\ref{ex:tsirelsonbound}.d. Use Cauchy-Schwarz to show $(\bra{\psi}C\ket{\psi})^2\leq \bra{\psi}C^2\ket{\psi}$, and then upper bound the latter.\\
\ref{ex:tsirelsonbound}.e. $\cos(\pi/8)^2=\frac{1}{2}+\frac{1}{\sqrt{8}}$.

\subsection*{Chapter~\ref{chap:qcrypto}}
\ref{ex:secretsharing}. Use the encoding of Exercise~\ref{ex:qonetimepad}, so that Alice and Bob need to cooperate to learn the key used to change $\rho$.\\
\ref{ex:unitaryinv}. Show that a unitary on Alice's side of the state won't change Bob's local density matrix $\rho_B$.\\
\ref{ex:decentpurestate}.a. The singular value decomposition (see end of Appendix~\ref{secsvd}) of the $d\times d$ matrix $M$ whose entries are $M_{ij}=\alpha_{ij}$ can be computed in polynomial time, you may assume this without proof.
    
\subsection*{Chapter~\ref{chap:qml}}

\ref{ex:Jordangradient}.a. Start with a uniform superposition over all $x\in\01^n$ and end with an inverse QFT. You're allowed to use a unitary like $\ket{c}\mapsto e^{2\pi i c}\ket{c}$ since it does not depend on $f$.\\
\ref{ex:hamsimrho}.a. It's helpful to write $e^{i\rho\eta}=I+i\rho\eta+F$ for some matrix $F$ with $\norm{F}_1=O(\eta^2)$ whose precise form doesn't matter. This follows from Taylor series and the fact that $\rho$ has trace~1, you don't need to prove this. Here the trace norm $\norm{A}_1$ is defined as the sum of $A$'s singular values.\\
\ref{ex:hamsimrho}.b. You can first prove this for the case where $\sigma=\ketbra{a}{a}$ and $\rho=\ketbra{b}{b}$ are pure states, and then extend to general mixed states by linearity.\\
\ref{ex:hamsimrho}.d. Apply part~(b) some $r$ times with small $\eta$, choosing $r,\eta$ such that $r\eta=t$ (and hence $(U^\eta)^r=U^t$) and you can upper bound the overall error using triangle inequality.\\
\ref{ex:estdistance}.a. Use the SWAP-test from Section~\ref{sec:qfingerprint}. ``$O(1)$ given copies'' means you are allowed to use any number of copies of $\ket{\phi}$ and $\ket{\psi}$, as long as that number is independent of $n$. You may count the 3-qubit gate which is the controlled SWAP of a pair of qubits as an elementary gate here. 
    
\subsection*{Chapter~\ref{chap:stabilizer}}
\ref{ex:cliffordcircuitmatrix}.a. Use induction on the number of gates, and see how each $S,H,\CNOT$ gates changes the $2n$-bit string representing a Pauli.\\
\ref{ex:paulipreserving}. Use Exercise~\ref{ex:cliffordcircuitmatrix}, argue how all such matrices can be generated by the Clifford gates.\\
\ref{ex:stabrank}.b. Use the Chernoff bound from Appendix~\ref{appusefulCS}.

\subsection*{Chapter~\ref{chap:qec}}
\ref{eq:paulidecomp}. Compute the trace $\Tr(E^*E)$ in two ways, and use the fact that $\Tr(AB)=0$ if $A$ and $B$ are distinct Paulis, and $\Tr(AB)=\Tr(I)=2$ if $A$ and $B$ are the same Pauli.\\
\ref{ex:steanecode}.e. 
Consider the ``dual code'' $C^\perp=\{z\mid z\cdot c=0~({\rm mod}~2)~\forall c\in C\}$. The set $C^\perp\cup(C^\perp\oplus 1^7)$ also has minimal distance 3 (you may assume this without proof, though it's easy to see). Use that $H^{\otimes 7}$ maps a uniform superposition over $C$ to a uniform superposition over $C^\perp$ (this was part of Exercise \ref{chap:simon}.\ref{ex:subspacesimon}, you may also assume this without proof). Also note that $HZ=XH$, so in some sense the Hadamard gate converts phaseflip-errors into bitflip-errors.\\
\ref{ex:2kqecc}. Given an unknown qubit $\alpha\ket{0}+\beta\ket{1}$ encoded using this code, you could split the $2k$ qubits into two sets of $k$ qubits each, and use each to recover a copy of the unknown qubit. 

\bibliographystyle{plain}
\bibliography{qc}

\begin{thebibliography}{100}

\bibitem{aaronson:qlearnability}
S.~Aaronson.
\newblock The learnability of quantum states.
\newblock {\em Proceedings of the Royal Society of London}, 463(2088), 2007.
\newblock quant-ph/0608142.

\bibitem{aaronson:fineprint}
S.~Aaronson.
\newblock Quantum machine learning algorithms: Read the fine print.
\newblock {\em Nature Physics}, 11(4):291--293, April 2015.

\bibitem{aaronson&ambainis:searchj}
S.~Aaronson and A.~Ambainis.
\newblock Quantum search of spatial regions.
\newblock {\em Theory of Computing}, 1(1):47--79, 2005.
\newblock Earlier version in FOCS'03. quant-ph/0303041.

\bibitem{aaronson&gottesman:improvedstabilizer}
S.~Aaronson and D.~Gottesman.
\newblock Improved simulation of stabilizer circuits.
\newblock {\em Physical Review A}, 70(052328), 2004.
\newblock quant-ph/0406196.

\bibitem{aaronson&shi:collision}
S.~Aaronson and Y.~Shi.
\newblock Quantum lower bounds for the collision and the element distinctness
  problems.
\newblock {\em Journal of the ACM}, 51(4):595--605, 2004.

\bibitem{AKS:primesinP}
M.~Agrawal, N.~Kayal, and N.~Saxena.
\newblock {PRIMES} is in {P}.
\newblock {\em Annals of Mathematics}, 160(2):781--793, 2004.

\bibitem{aav:qpcp}
D.~Aharonov, I.~Arad, and T.~Vidick.
\newblock Guest column: The quantum {PCP} conjecture.
\newblock {\em ACM SIGACT News}, 44(2):47--79, 2013.
\newblock arXiv:1309.7495.

\bibitem{aharonov&benor:faulttolj}
D.~Aharonov and M.~{Ben-Or}.
\newblock Fault tolerant quantum computation with constant error rate.
\newblock {\em SIAM Journal on Computing}, 38(4):1207--1282, 2008.
\newblock Earlier version in STOC'97. quant-ph/9611025.

\bibitem{aharonov&naveh:qnp}
D.~Aharonov and T.~Naveh.
\newblock Quantum {NP} - a survey, 2002.
\newblock quant-ph/0210077.

\bibitem{allcock&hsieh:qsvm}
J.~Allcock and C-Y. Hsieh.
\newblock A quantum extension of {SVM-perf} for training nonlinear {SVM}s in
  almost linear time.
\newblock {\em Quantum}, 4:342, 2020.
\newblock arXiv:2006.10299.

\bibitem{Alrabiah2023NearCubic}
O.~Alrabiah, V.~Guruswami, P.~Kothari, and P.~Manohar.
\newblock A near-cubic lower bound for 3-query locally decodable codes from
  semirandom csp refutation.
\newblock In {\em Proceedings of 55th ACM STOC}, pages 1438--1448, 2023.

\bibitem{ambainis:3computer}
A.~Ambainis.
\newblock Communication complexity in a 3-computer model.
\newblock {\em Algorithmica}, 16(3):298--301, 1996.

\bibitem{ambainis:lowerboundsj}
A.~Ambainis.
\newblock Quantum lower bounds by quantum arguments.
\newblock {\em Journal of Computer and System Sciences}, 64(4):750--767, 2002.
\newblock Earlier version in STOC'00. quant-ph/0002066.

\bibitem{ambainis:degreevsqueryj}
A.~Ambainis.
\newblock Polynomial degree {vs.} quantum query complexity.
\newblock {\em Journal of Computer and System Sciences}, 72(2):220--238, 2006.
\newblock Earlier version in FOCS'03. quant-ph/0305028.

\bibitem{ambainis:edj}
A.~Ambainis.
\newblock Quantum walk algorithm for element distinctness.
\newblock {\em SIAM Journal on Computing}, 37(1):210--239, 2007.
\newblock Earlier version in FOCS'04. quant-ph/0311001.

\bibitem{ambainis:variabletime}
A.~Ambainis.
\newblock Quantum search with variable times.
\newblock In {\em Proceedings of 25th Annual Symposium on Theoretical Aspects
  of Computer Science (STACS'08)}, pages 49--61, 2008.
\newblock arXiv:1010.4458.

\bibitem{ambainis2019quantum}
A.~Ambainis, K.~Balodis, J.~Iraids, M.~Kokainis, K.~Pr{\=u}sis, and J.~Vihrovs.
\newblock Quantum speedups for exponential-time dynamic programming algorithms.
\newblock In {\em Proceedings of 30th ACM-SIAM SODA}, pages 1783--1793, 2019.
\newblock arXiv:1807.05209.

\bibitem{abrw:qjunta}
A.~Ambainis, A.~Belovs, O.~Regev, and R.~{de} Wolf.
\newblock Efficient quantum algorithms for (gapped) group testing and junta
  testing.
\newblock In {\em Proceedings of 27th ACM-SIAM SODA}, pages 903--922, 2016.
\newblock arXiv:1507.03126.

\bibitem{acrsz:andorj}
A.~Ambainis, A.~Childs, B.~Reichardt, R.~{\v{S}}palek, and S.~Zhang.
\newblock Any {AND-OR} formula of size {$N$} can be evaluated in time
  {$N^{1/2+o(1)}$} on a quantum computer.
\newblock {\em SIAM Journal on Computing}, 39(6):2513--2530, 2010.
\newblock Earlier version in FOCS'07.

\bibitem{amtw:pqc}
A.~Ambainis, M.~Mosca, A.~Tapp, and R.~{de} Wolf.
\newblock Private quantum channels.
\newblock In {\em Proceedings of 41st IEEE FOCS}, pages 547--553, 2000.
\newblock quant-ph/0003101.

\bibitem{AAKW:learnHam}
A.~Anshu, S.~Arunachalam, T.~Kuwahara, and M.~Soleimanifar.
\newblock Sample-efficient learning of quantum many-body systems.
\newblock {\em Nature Physics}, 17:931--–935, 2021.
\newblock Earlier version in FOCS'20. arXiv:2004.07266.

\bibitem{AG19zerosumgame}
{J. van} Apeldoorn and A.~Gily{\'e}n.
\newblock Quantum algorithms for zero-sum games.
\newblock arXiv:1904.03180, 2019.

\bibitem{vAGGdW17}
{J. van} Apeldoorn, A.~Gily{\'{e}}n, S.~Gribling, and {R. de} Wolf.
\newblock Quantum {SDP}-solvers: better upper and lower bounds.
\newblock {\em Quantum}, 4:230, 2020.
\newblock Earlier version in FOCS'17. arXiv:1705.01843.

\bibitem{vAG19}
{J. van} Apeldoorn and A.~Gilyén.
\newblock Improvements in quantum {SDP}-solving with applications.
\newblock In {\em Proceedings of 46th International Colloquium on Automata,
  Languages, and Programming}, volume 132 of {\em Leibniz International
  Proceedings in Informatics}, pages 99:1--99:15, 2019.
\newblock arXiv:1804.05058.

\bibitem{AW20}
S.~Apers and {R. de} Wolf.
\newblock Quantum speedup for graph sparsification, cut approximation and
  {L}aplacian solving.
\newblock In {\em Proceedings of 61st {IEEE} Annual Symposium on Foundations of
  Computer Science}, pages 637--648, 2020.
\newblock arXiv:1911.07306.

\bibitem{aravind:magicsquare}
P.~K. Aravind.
\newblock A simple demonstration of {Bell's} theorem involving two observers
  and no probabilities or inequalities.
\newblock quant-ph/0206070, 2002.

\bibitem{ABP:cbforms}
S.~Arunachalam, J.~Bri{\"{e}}t, and C.~Palazuelos.
\newblock Quantum query algorithms are completely bounded forms.
\newblock {\em SIAM Journal on Computing}, 48(3):903--925, 2019.
\newblock Earlier version in ITCS'18. arXiv:1711.07285.

\bibitem{arunachalam:quantumlearningsurvey}
S.~Arunachalam and R.~{de} Wolf.
\newblock Guest column: {A} survey of quantum learning theory.
\newblock {\em {SIGACT} News}, 48(2):41--67, 2017.
\newblock arXiv:1701.06806.

\bibitem{arunachalam&wolf:grovergate}
S.~Arunachalam and R.~{de} Wolf.
\newblock Optimizing the number of gates in quantum search.
\newblock {\em Quantum Information and Computation}, 17(4):251--261, 2017.
\newblock arXiv:1512.07550.

\bibitem{arunachalam:optimalpaclearning}
S.~Arunachalam and R.~{de} Wolf.
\newblock Optimal quantum sample complexity of learning algorithms.
\newblock {\em Journal of Machine Learning Research}, 19, 2018.
\newblock Earlier version in CCC'17. arXiv:1607.00932.

\bibitem{google:supremacy}
F.~Arute, ..., and J.~Martinis.
\newblock Quantum supremacy using a programmable superconducting processor.
\newblock {\em Nature}, 574:505--510, 2019.
\newblock arXiv:1910.11333.

\bibitem{AspectGR81:Bell}
A.~Aspect, Ph. Grangier, and G.~Roger.
\newblock Experimental tests of realistic local theories via {B}ell's theorem.
\newblock {\em Physical Review Letters}, 47:460, 1981.

\bibitem{babai:quasipolygi}
L.~Babai.
\newblock Graph isomorphism in quasipolynomial time.
\newblock In {\em Proceedings of 48th ACM STOC}, pages 684--697, 2016.
\newblock arXiv:1512.03547.

\bibitem{babai&luks:gi}
L.~Babai and E.~M. Luks.
\newblock Canonical labeling of graphs.
\newblock In {\em Proceedings of 15th ACM STOC}, pages 171--183, 1983.

\bibitem{babai&moran:am}
L.~Babai and S.~Moran.
\newblock Arthur-{M}erlin games: a randomized proof system, and a hierarchy of
  complexity classes.
\newblock {\em Journal of Computer and System Sciences}, 36(2):254--276, 1988.

\bibitem{bjk:q1wayj}
Z.~{Bar-Yossef}, T.~S. Jayram, and I.~Kerenidis.
\newblock Exponential separation of quantum and classical one-way communication
  complexity.
\newblock {\em SIAM Journal on Computing}, 38(1):366--384, 2008.
\newblock Earlier version in STOC'04.

\bibitem{bbdejm:stab}
A.~Barenco, A.~Berthiaume, D.~Deutsch, A.~Ekert, R.~Jozsa, and C.~Macchiavello.
\newblock Stabilization of quantum computations by symmetrization.
\newblock {\em SIAM Journal on Computing}, 26(5):1541--1557, 1997.
\newblock quant-ph/9604028.

\bibitem{beals:fourier}
R.~Beals.
\newblock Quantum computation of {Fourier} transforms over symmetric groups.
\newblock In {\em Proceedings of 29th ACM STOC}, pages 48--53, 1997.

\bibitem{bbcmw:polynomialsj}
R.~Beals, H.~Buhrman, R.~Cleve, M.~Mosca, and R.~{de} Wolf.
\newblock Quantum lower bounds by polynomials.
\newblock {\em Journal of the ACM}, 48(4):778--797, 2001.
\newblock Earlier version in FOCS'98. quant-ph/9802049.

\bibitem{beauregard:shor}
S.~Beauregard.
\newblock Circuit for {S}hor's algorithm using {$2n+3$} qubits.
\newblock {\em Quantum Information and Computation}, 3(2):175--185, 2003.
\newblock quant-ph/0205095.

\bibitem{bell:epr}
J.~S. Bell.
\newblock On the {E}instein-{P}odolsky-{R}osen paradox.
\newblock {\em Physics}, 1:195--200, 1964.

\bibitem{belovs:kdistinctness}
A.~Belovs.
\newblock Learning-graph-based quantum algorithm for k-distinctness.
\newblock In {\em Proceedings of 53rd IEEE FOCS}, pages 207--216, 2012.
\newblock arXiv:1205.1534.

\bibitem{belovs:learninggraphs}
A.~Belovs.
\newblock Span programs for functions with constant-sized 1-certificates.
\newblock In {\em Proceedings of 43rd ACM STOC}, pages 77--84, 2012.
\newblock arXiv:1105.4024.

\bibitem{belovs:phd}
A.~Belovs.
\newblock {\em Applications of the Adversary Method in Quantum Query
  Algorithms}.
\newblock PhD thesis, University of Latvia, 2014.
\newblock 1402.3858.

\bibitem{belovs:learningjuntas}
A.~Belovs.
\newblock Quantum algorithms for learning symmetric juntas via adversary bound.
\newblock {\em Computational Complexity}, 24(2):255--293, 2015.
\newblock Earlier version in Complexity'14. arXiv:1311.6777.

\bibitem{belovs:variations}
A.~Belovs.
\newblock Variations on quantum adversary, 27 Apr 2015.
\newblock arXiv:1504.06943.

\bibitem{BL20}
A.~Belovs and T.~Lee.
\newblock The quantum query complexity of composition with a relation.
\newblock arXiv:2004.06439, 2020.

\bibitem{belovs&reichardt:conn}
A.~Belovs and B.~Reichardt.
\newblock Span programs and quantum algorithms for st-connectivity and claw
  detection.
\newblock In {\em Proceedings of 20th European Symposium on Algorithms
  (ESA'12)}, pages 193--204, 2012.
\newblock arXiv:1203.2603.

\bibitem{benioff:hamiltonian}
P.~A. Benioff.
\newblock Quantum mechanical {H}amiltonian models of {T}uring machines.
\newblock {\em Journal of Statistical Physics}, 29(3):515--546, 1982.

\bibitem{teleporting}
C.~Bennett, G.~Brassard, C.~Cr{\'e}peau, R.~Jozsa, A.~Peres, and W.~Wootters.
\newblock Teleporting an unknown quantum state via dual classical and
  {Einstein-Podolsky-Rosen} channels.
\newblock {\em Physical Review Letters}, 70:1895--1899, 1993.

\bibitem{superdense}
C.~Bennett and S.~Wiesner.
\newblock Communication via one- and two-particle operators on
  {Einstein-Podolsky-Rosen} states.
\newblock {\em Physical Review Letters}, 69:2881--2884, 1992.

\bibitem{bbbv:str&weak}
C.~H. Bennett, E.~Bernstein, G.~Brassard, and U.~Vazirani.
\newblock Strengths and weaknesses of quantum computing.
\newblock {\em SIAM Journal on Computing}, 26(5):1510--1523, 1997.
\newblock quant-ph/9701001.

\bibitem{bb84}
C.~H. Bennett and G.~Brassard.
\newblock Quantum cryptography: Public key distribution and coin tossing.
\newblock In {\em Proceedings of the IEEE International Conference on
  Computers, Systems and Signal Processing}, pages 175--179, 1984.

\bibitem{bernstein&lange:pqc}
D.~Bernstein and T.~Lange.
\newblock Post-quantum cryptography.
\newblock {\em Nature}, 549(6):188--194, 2017.

\bibitem{bernstein&vazirani:qcomplexity}
E.~Bernstein and U.~Vazirani.
\newblock Quantum complexity theory.
\newblock {\em SIAM Journal on Computing}, 26(5):1411--1473, 1997.
\newblock Earlier version in STOC'93.

\bibitem{Berryetal:expimpr}
D.~Berry, A.~Childs, R.~Cleve, R.~Kothari, and R.~Somma.
\newblock Exponential improvement in precision for simulating sparse
  {H}amiltonians.
\newblock In {\em Proceedings of 46th ACM STOC}, pages 283--292, 2014.
\newblock arXiv:1312.1414.

\bibitem{BerryChilds:hamsim}
D.~Berry, A.~Childs, R.~Cleve, R.~Kothari, and R.~Somma.
\newblock Simulating {H}amiltonian dynamics with a truncated {T}aylor series.
\newblock {\em Physical Review Letters}, 114:090502, 2015.
\newblock arXiv:1412.4687.

\bibitem{BCK:nearlyoptimal}
D.~Berry, A.~Childs, and R.~Kothari.
\newblock Hamiltonian simulation with nearly optimal dependence on all
  parameters.
\newblock In {\em Proceedings of 56th IEEE FOCS}, pages 792--809, 2015.
\newblock arXiv:1501.01715.

\bibitem{biamonteea:qml}
J.~Biamonte, P.~Wittek, N.~Pancotti, P.~Rebentrost, N.~Wiebe, and S.~Lloyd.
\newblock Quantum machine learning.
\newblock {\em Nature}, 549(7671), 2017.
\newblock arXiv:1611.09347.

\bibitem{biamonte&love:real}
J.~D. Biamonte and P.~J. Love.
\newblock Realizable {H}amiltonians for universal adiabatic quantum computers.
\newblock {\em Physical Review A}, 78(1)(012352), 2008.
\newblock arXiv:0704.1287.

\bibitem{biasse&song:princideal}
J-F. Biasse and F.~Song.
\newblock Efficient quantum algorithms for computing class groups and solving
  the principal ideal problem in arbitrary degree number fields.
\newblock In {\em Proceedings of 27th ACM-SIAM SODA}, pages 893--902, 2016.

\bibitem{blumer:optimalpacupper}
A.~Blumer, A.~Ehrenfeucht, D.~Haussler, and M.~K. Warmuth.
\newblock Learnability and the {V}apnik-{C}hervonenkis dimension.
\newblock {\em Journal of the ACM}, 36(4):929--965, 1989.

\bibitem{bookatz:qma}
A.~Bookatz.
\newblock {QMA}-complete problems.
\newblock {\em Quantum Information and Computation}, 14(5--6):361--383, 2014.
\newblock arXiv:1212.6312.

\bibitem{bbht:bounds}
M.~Boyer, G.~Brassard, P.~H{\o}yer, and A.~Tapp.
\newblock Tight bounds on quantum searching.
\newblock {\em Fortschritte der Physik}, 46(4--5):493--505, 1998.
\newblock Earlier version in Physcomp'96. quant-ph/9605034.

\bibitem{brandao2017QSDPSpeedupsLearning}
F.~Brand{\~a}o, A.~Kalev, T.~Li, C.~Lin, K.~Svore, and X.~Wu.
\newblock Quantum {SDP} solvers: Large speed-ups, optimality, and applications
  to quantum learning.
\newblock In {\em Proceedings of 46th International Colloquium on Automata,
  Languages, and Programming}, volume 132 of {\em Leibniz International
  Proceedings in Informatics}, pages 27:1--27:14, 2019.
\newblock arXiv:1710.02581.

\bibitem{brandao2016QSDPSpeedup}
F.~Brand{\~a}o and K.~Svore.
\newblock Quantum speed-ups for solving semidefinite programs.
\newblock In {\em Proceedings of 58th {IEEE} {FOCS}}, pages 415--426, 2017.
\newblock arXiv:1609.05537.

\bibitem{bct:simulating}
G.~Brassard, R.~Cleve, and A.~Tapp.
\newblock The cost of exactly simulating quantum entanglement with classical
  communication.
\newblock {\em Physical Review Letters}, 83(9):1874--1877, 1999.
\newblock quant-ph/9901035.

\bibitem{bhmt:countingj}
G.~Brassard, P.~H{\o}yer, M.~Mosca, and A.~Tapp.
\newblock Quantum amplitude amplification and estimation.
\newblock In {\em Quantum Computation and Quantum Information: A Millennium
  Volume}, volume 305 of {\em AMS Contemporary Mathematics Series}, pages
  53--74. 2002.
\newblock quant-ph/0005055.

\bibitem{bht:collision}
G.~Brassard, P.~H{\o}yer, and A.~Tapp.
\newblock Quantum algorithm for the collision problem.
\newblock {\em ACM SIGACT News (Cryptology Column)}, 28:14--19, 1997.
\newblock quant-ph/9705002.

\bibitem{bravyi&gosset:improved}
S.~Bravyi and D.~Gosset.
\newblock Improved classical simulation of quantum circuits dominated by
  {C}lifford gates.
\newblock {\em Physical Review Letters}, 116(25):250501, 2016.
\newblock arXiv:1601.07601.

\bibitem{broadbent&grilo:zeroknowledge}
A.~Broadbent and A.~B. Grilo.
\newblock {QMA}-hardness of consistency of local density matrices with
  applications to quantum zero-knowledge.
\newblock In {\em Proceedings of 61st IEEE FOCS}, pages 196--205, 2020.
\newblock arXiv:1911.07782.

\bibitem{broadbent&schaffner:beyondQKD}
A.~Broadbent and C.~Schaffner.
\newblock Quantum cryptography beyond quantum key distribution.
\newblock {\em Designs, Codes and Cryptography}, 78(1):351--382, 2016.
\newblock arXiv:1510.06120.

\bibitem{brouwer&haemers:spectra}
A.~E. Brouwer and W.~H. Haemers.
\newblock {\em Spectra of Graphs}.
\newblock Springer, 2012.

\bibitem{browne&rudolph:linopt}
D.~E. Browne and T.~Rudolph.
\newblock Resource-efficient linear optical quantum computation.
\newblock {\em Physical Review Letters}, 95(1):010501, 2005.
\newblock quant-ph/0405157.

\bibitem{bshouty:quantumpac}
N.~H. Bshouty and J.~C. Jackson.
\newblock Learning {DNF} over the uniform distribution using a quantum example
  oracle.
\newblock {\em SIAM Journal on Computing}, 28(3):1136–--1153, 1999.
\newblock Earlier version in COLT'95.

\bibitem{bcmw:survey}
H.~Buhrman, R.~Cleve, S.~Massar, and R.~{de} Wolf.
\newblock Non-locality and communication complexity.
\newblock {\em Reviews of Modern Physics}, 82:665--698, 2010.
\newblock arXiv:0907.3584.

\bibitem{bcww:fp}
H.~Buhrman, R.~Cleve, J.~Watrous, and R.~{de} Wolf.
\newblock Quantum fingerprinting.
\newblock {\em Physical Review Letters}, 87(16), September 26, 2001.
\newblock quant-ph/0102001.

\bibitem{BuhrmanCleveWigderson98}
H.~Buhrman, R.~Cleve, and A.~Wigderson.
\newblock Quantum vs.~classical communication and computation.
\newblock In {\em Proceedings of 30th ACM STOC}, pages 63--68, 1998.
\newblock quant-ph/9802040.

\bibitem{bs:matrix}
H.~Buhrman and R.~{\v{S}}palek.
\newblock Quantum verification of matrix products.
\newblock In {\em Proceedings of 17th ACM-SIAM SODA}, pages 880--889, 2006.
\newblock quant-ph/0409035.

\bibitem{bun&thaler:approxdeg}
M.~Bun and J.~Thaler.
\newblock Dual lower bounds for approximate degree and {Markov-Bernstein}
  inequalities.
\newblock In {\em Proceedings of 40th ICALP}, volume 7965 of {\em Lecture Notes
  in Computer Science}, pages 303--314, 2013.

\bibitem{caoetal:qchemistry}
Y.~Cao, J.~Romero, J.~Olson, M.~Degroote, P.~Johnson, M.~Kieferov\'{a},
  I.~Kivlichan, T.~Menke, B.~Peropadre, N.~Sawaya, S.~Sim, L.~Veis, and
  A.~Aspuru-Guzik.
\newblock Quantum chemistry in the age of quantum computing.
\newblock {\em Chemical Reviews}, 119(19):10856--11091, 2019.
\newblock arXiv:1812.09976.

\bibitem{VQAsurvey}
M.~Cerezo, A.~Arrasmith, R.~Babbush, S.~Benjamin, S.~Endo, K.~Fujii,
  J.~McClean, K.~Mitarai, X.~Yuan, L.~Cincio, and P.~Coles.
\newblock Variational quantum algorithms.
\newblock {\em Nature Reviews Physics}, 1, 2021.
\newblock arXiv:2012.09265.

\bibitem{CGJ:blockenc}
S.~Chakraborty, A.~Gily{\'e}n, and S.~Jeffery.
\newblock The power of block-encoded matrix powers: improved regression
  techniques via faster {H}amiltonian simulation.
\newblock In {\em Proceedings of 46th International Colloquium on Automata,
  Languages, and Programming}, volume 132 of {\em Leibniz International
  Proceedings in Informatics}, pages 33:1--33:14, 2019.
\newblock arXiv:1804.01973.

\bibitem{CGW:topeigenvectors}
Y.~Chen, A.~Gily{\'e}n, and {R. de} Wolf.
\newblock A quantum speed-up for approximating the top eigenvectors of a
  matrix.
\newblock In {\em Proceedings of 36th ACM-SIAM SODA}, pages 994--1036, 2024.
\newblock arXiv:2405.14765.

\bibitem{Chen&Wolf:lasso}
Y.~Chen and {R. de} Wolf.
\newblock Quantum algorithms and lower bounds for linear regression with norm
  constraints.
\newblock In {\em Proceedings of 50th ICALP}, volume 261 of {\em Leibniz
  International Proceedings in Informatics (LIPIcs)}, pages 38:1--38:21, 2023.
\newblock arXiv:2110.13086.

\bibitem{chiaetal:dequant}
N-H. Chia, A.~Gily{\'e}n, T.~Li, H-H. Lin, E.~Tang, and C.~Wang.
\newblock Sampling-based sublinear low-rank matrix arithmetic framework for
  dequantizing quantum machine learning.
\newblock In {\em Proceedings of 52nd ACM STOC}, pages 387--400, 2020.
\newblock arXiv:1910.06151.

\bibitem{childs:qalgonotes}
A.~Childs.
\newblock Lecture notes on quantum algorithms, 2017.
\newblock Available at \url{https://cs.umd.edu/~amchilds/qa/}.

\bibitem{CKS:optimalHHL}
A.~Childs, R.~Kothari, and R.~Somma.
\newblock Quantum algorithm for systems of linear equations with exponentially
  improved dependence on precision.
\newblock {\em SIAM Journal on Computing}, 46(6):1920--1950, 2017.
\newblock arXiv:1511.02306.

\bibitem{CGW:BoseHubbard}
A.~M. Childs, D.~Gosset, and Z.~Webb.
\newblock The {Bose-Hubbard} model is {QMA}-complete.
\newblock {\em Theory of Computing}, 11(20):491--603, 2015.
\newblock arXiv:1311.3297.

\bibitem{childsetal:trotter}
A.~M. Childs, Y.~Su, M.~C. Tran, N.~Wiebe, and S.~Zhu.
\newblock A theory of {T}rotter error.
\newblock arXiv:1912.08854, 18 Dec 2019.

\bibitem{tsirelson80}
B.~S. Cirel'son.
\newblock Quantum generalizations of {B}ell's inequality.
\newblock {\em Letters in Mathematical Physics}, 4(2):93--100, 1980.

\bibitem{chsh}
J.~F. Clauser, M.~A. Horne, A.~Shimony, and R.~A. Holt.
\newblock Proposed experiment to test local hidden-variable theories.
\newblock {\em Physical Review Letters}, 23(15):880--884, 1969.

\bibitem{cleve:orderfinding}
R.~Cleve.
\newblock The query complexity of order-finding.
\newblock In {\em Proceedings of 15th IEEE Conference on Computational
  Complexity}, pages 54--59, 2000.
\newblock quant-ph/9911124.

\bibitem{cleve&buhrman:subs}
R.~Cleve and H.~Buhrman.
\newblock Substituting quantum entanglement for communication.
\newblock {\em Physical Review A}, 56(2):1201--1204, 1997.
\newblock quant-ph/9704026.

\bibitem{cdnt:ip}
R.~Cleve, W.~{van} Dam, M.~Nielsen, and A.~Tapp.
\newblock Quantum entanglement and the communication complexity of the inner
  product function.
\newblock In {\em Proceedings of 1st NASA QCQC conference}, volume 1509 of {\em
  Lecture Notes in Computer Science}, pages 61--74. Springer, 1998.
\newblock quant-ph/9708019.

\bibitem{cemm:revisited}
R.~Cleve, A.~Ekert, C.~Macchiavello, and M.~Mosca.
\newblock Quantum algorithms revisited.
\newblock In {\em Proceedings of the Royal Society of London}, volume A454,
  pages 339--354, 1998.
\newblock quant-ph/9708016.

\bibitem{cook:np}
S.~Cook.
\newblock The complexity of theorem-proving procedures.
\newblock In {\em Proceedings of 3rd ACM STOC}, pages 151--158, 1971.

\bibitem{cooley&tukey:fft}
J.~W. Cooley and J.~W. Tukey.
\newblock An algorithm for the machine calculation of complex {F}ourier series.
\newblock {\em Mathematics of Computation}, 19(90):297--301, 1965.

\bibitem{coppersmith:fourier}
D.~Coppersmith.
\newblock An approximate {F}ourier transform useful in quantum factoring.
\newblock IBM Research Report No.~RC19642, quant-ph/0201067, 1994.

\bibitem{cornelissen:mscthesis}
A.~Cornelissen.
\newblock Quantum gradient estimation and its application to quantum
  reinforcement learning.
\newblock Master's thesis, Delft University, 2018.

\bibitem{dam:oracle}
W.~{van} Dam.
\newblock Quantum oracle interrogation: Getting all information for almost half
  the price.
\newblock In {\em Proceedings of 39th IEEE FOCS}, pages 362--367, 1998.
\newblock quant-ph/9805006.

\bibitem{deutsch:uqc}
D.~Deutsch.
\newblock Quantum theory, the {Church-Turing} principle, and the universal
  quantum {T}uring machine.
\newblock In {\em Proceedings of the Royal Society of London}, volume A400,
  pages 97--117, 1985.

\bibitem{deutsch:networks}
D.~Deutsch.
\newblock Quantum computational networks.
\newblock In {\em Proceedings of the Royal Society of London}, volume A425,
  1989.

\bibitem{deutsch&jozsa}
D.~Deutsch and R.~Jozsa.
\newblock Rapid solution of problems by quantum computation.
\newblock In {\em Proceedings of the Royal Society of London}, volume A439,
  pages 553--558, 1992.

\bibitem{drucker&wolf:qproofs}
A.~Drucker and R.~{de} Wolf.
\newblock Quantum proofs for classical theorems.
\newblock {\em Theory of Computing}, 2011.
\newblock ToC Library, Graduate Surveys 2. arXiv:0910.3376.

\bibitem{DTB:advqreinforcement}
V.~Dunjko, J.~Taylor, and H.~Briegel.
\newblock Advances in quantum reinforcement learning.
\newblock {\em IEEE SMC}, pages 282--287, 2017.
\newblock arXiv:1811.08676.

\bibitem{dhhm:graphproblemsj}
C.~D{\"u}rr, M.~Heiligman, P.~H{\o}yer, and M.~Mhalla.
\newblock Quantum query complexity of some graph problems.
\newblock {\em SIAM Journal on Computing}, 35(6):1310--1328, 2006.
\newblock Earlier version in ICALP'04. quant-ph/0401091.

\bibitem{durr&hoyer:minimum}
C.~D{\"u}rr and P.~H{\o}yer.
\newblock A quantum algorithm for finding the minimum.
\newblock quant-ph/9607014, 18 Jul 1996.

\bibitem{efremenko:ldc}
K.~Efremenko.
\newblock 3-query locally decodable codes of subexponential length.
\newblock In {\em Proceedings of 41st ACM STOC}, pages 39--44, 2009.

\bibitem{ehlich&zeller:schwankung}
H.~Ehlich and K.~Zeller.
\newblock Schwankung von {P}olynomen zwischen {G}itterpunkten.
\newblock {\em Mathematische Zeitschrift}, 86:41--44, 1964.

\bibitem{epr}
A.~Einstein, B.~Podolsky, and N.~Rosen.
\newblock Can quantum-mechanical description of physical reality be considered
  complete?
\newblock {\em Physical Review}, 47:777--780, 1935.

\bibitem{emdeboas:mm}
P.~van Emde~Boas.
\newblock Machine models and simulations.
\newblock In van Leeuwen \cite{handbookA}, pages 1--66.

\bibitem{ehk:hsppoly}
M.~Ettinger, P.~H{\o}yer, and M.~Knill.
\newblock The quantum query complexity of the hidden subgroup problem is
  polynomial.
\newblock {\em Information Processing Letters}, 91(1):43--48, 2004.
\newblock quant-ph/0401083.

\bibitem{FGG:QAOA}
E.~Farhi, J.~Goldstone, and S.~Gutmann.
\newblock A quantum approximate optimization algorithm.
\newblock arXiv:1411.4028, 2014.

\bibitem{fawzi2018constant}
O.~Fawzi, A.~Grospellier, and A.~Leverrier.
\newblock Constant overhead quantum fault-tolerance with quantum expander
  codes.
\newblock In {\em Proceedings of 59th IEEE FOCS}, pages 743--754, 2018.
\newblock arXiv:1808.03821.

\bibitem{feynman:simulating}
R.~Feynman.
\newblock Simulating physics with computers.
\newblock {\em International Journal of Theoretical Physics}, 21(6/7):467--488,
  1982.

\bibitem{feynman:qmc}
R.~Feynman.
\newblock Quantum mechanical computers.
\newblock {\em Optics News}, 11:11--20, 1985.

\bibitem{fortnow&rogers:limitations}
L.~Fortnow and J.~Rogers.
\newblock Complexity limitations on quantum computation.
\newblock {\em Journal of Computer and System Sciences}, 59(2):240--252, 1999.
\newblock Earlier version in Complexity'98. Also cs.CC/9811023.

\bibitem{frankl&rodl:forbidden}
P.~Frankl and V.~R{\"o}dl.
\newblock Forbidden intersections.
\newblock {\em Transactions of the American Mathematical Society},
  300(1):259--286, 1987.

\bibitem{freivalds:matrixmult}
R.~Freivalds.
\newblock Probabilistic machines can use less running time.
\newblock In {\em Proceedings of 7th IFIP Congress}, pages 839--842, 1977.

\bibitem{furer:fastermultj}
M.~F{\"u}rer.
\newblock Faster integer multiplication.
\newblock {\em SIAM Journal on Computing}, 39(3):979--1005, 2009.
\newblock Earlier version in STOC'07.

\bibitem{garey&johnson:npcomplete}
M.~Garey and D.~Johnson.
\newblock {\em Computers and Intractability : A Guide to the Theory of
  {NP}-completeness}.
\newblock W.~H.\ Freeman and Company, 1979.

\bibitem{gkkrw:1wayj}
D.~Gavinsky, J.~Kempe, I.~Kerenidis, R.~Raz, and R.~{de} Wolf.
\newblock Exponential separation for one-way quantum communication complexity,
  with applications to cryptography.
\newblock {\em SIAM Journal on Computing}, 38(5):1695--1708, 2008.
\newblock Earlier version in STOC'07. quant-ph/0611209.

\bibitem{GHLS:hamcompl}
S.~Gharibian, Y.~Huang, Z.~Landau, and S.~W. Shin.
\newblock Quantum hamiltonian complexity.
\newblock {\em Foundations and Trends in Theoretical Computer Science},
  10(3):159--282, 2015.

\bibitem{gilyen:thesis}
A.~Gily{\'e}n.
\newblock {\em Quantum Singular Value Transformation \&\ Its Algorithmic
  Applications}.
\newblock PhD thesis, University of Amsterdam, 2019.

\bibitem{GAW:gradient}
A.~Gily{\'e}n, S.~Arunachalam, and N.~Wiebe.
\newblock Optimizing quantum optimization algorithms via faster quantum
  gradient computation.
\newblock In {\em Proceedings of 30th ACM-SIAM SODA}, pages 1425--1444, 2019.
\newblock arXiv:1711.00465.

\bibitem{gilyenea:svtrans}
A.~Gily{\'e}n, Y.~Su, G.~H. Low, and N.~Wiebe.
\newblock Quantum singular value transformation and beyond: exponential
  improvements for quantum matrix arithmetics.
\newblock In {\em Proceedings of 51st ACM STOC}, pages 193--204, 2019.
\newblock arXiv:1806.01838.

\bibitem{GoogleQuantumAI2025:QECC}
{Google Quantum AI and Collaborators}.
\newblock Quantum error correction below the surface code threshold.
\newblock {\em Nature}, 638(8052):920--926, 2025.
\newblock arXiv:2408.13687.

\bibitem{gottesman&knill}
D.~Gottesman.
\newblock The {H}eisenberg representation of quantum computers.
\newblock In {\em Proceedings of the XXII International Colloquium on Group
  Theoretical Methods in Physics}, pages 32--43, 2009.
\newblock quant-ph/9807006.

\bibitem{gottesman:eccintro}
D.~Gottesman.
\newblock An introduction to quantum error correction and fault-tolerant
  quantum computation.
\newblock In {\em Quantum Information Science and Its Contributions to
  Mathematics, Proceedings of Symposia in Applied Mathematics}, volume~68,
  pages 13--58, 2010.
\newblock arXiv:0904.2557.

\bibitem{grigni:qhsp}
M.~Grigni, L.~Schulman, M.~Vazirani, and U.~Vazirani.
\newblock Quantum mechanical algorithms for the nonabelian hidden subgroup
  problem.
\newblock {\em Combinatorica}, 24(1):137--154, 2004.
\newblock Earlier version in STOC'01.

\bibitem{grover:search}
L.~K. Grover.
\newblock A fast quantum mechanical algorithm for database search.
\newblock In {\em Proceedings of 28th ACM STOC}, pages 212--219, 1996.
\newblock quant-ph/9605043.

\bibitem{haah16}
J.~Haah, A.~W. Harrow, Z.~Ji, X.~Wu, and N.~Yi.
\newblock Sample-optimal tomography of quantum states.
\newblock In {\em Proceedings of 48th ACM STOC}, pages 913--925, 2016.
\newblock arXiv:1508.01797.

\bibitem{hales&hallgren:improvedfourier}
L.~Hales and S.~Hallgren.
\newblock An improved quantum {F}ourier transform algorithm and applications.
\newblock In {\em Proceedings of 41st IEEE FOCS}, pages 515--525, 2000.

\bibitem{hallgren:pellj}
S.~Hallgren.
\newblock Polynomial-time quantum algorithms for {P}ell's equation and the
  principal ideal problem.
\newblock {\em Journal of the ACM}, 54(1):653--658, 2007.
\newblock Earlier version in STOC'02.

\bibitem{hallgren:limcoset}
S.~Hallgren, C.~Moore, M.~Roetteler, A.~Russell, and P.~Sen.
\newblock Limitations of quantum coset states for graph isomorphism.
\newblock {\em Journal of the ACM}, 57(6):34, 2010.
\newblock Earlier version in STOC'06.

\bibitem{hrt:normal}
S.~Hallgren, A.~Russell, and A.~{Ta-Shma}.
\newblock The hidden subgroup problem and quantum computation using group
  representations.
\newblock {\em SIAM Journal on Computing}, 32(4):916--934, 2003.
\newblock Earlier version in STOC'00.

\bibitem{hallgren:phd}
S.~J. Hallgren.
\newblock {\em Quantum {F}ourier Sampling, the Hidden Subgroup Problem, and
  Beyond}.
\newblock PhD thesis, University of California, Berkeley, 2000.

\bibitem{hanneke:optimalpac}
S.~Hanneke.
\newblock The optimal sample complexity of {PAC} learning.
\newblock {\em Journal of Machine Learning Research}, 17(38):1--15, 2016.
\newblock arXiv:1507.00473.

\bibitem{hardy&wright}
G.~H. Hardy and E.~M. Wright.
\newblock {\em An Introduction to the Theory of Numbers}.
\newblock Oxford University Press, New York, fifth edition, 1979.

\bibitem{hhl:lineq}
A.~Harrow, A.~Hassidim, and S.~Lloyd.
\newblock Quantum algorithm for solving linear systems of equations.
\newblock {\em Physical Review Letters}, 103(15):150502, 2009.
\newblock arXiv:0811.3171.

\bibitem{harvey&hoeven:fastermultj}
D.~Harvey and {J. van der} Hoeven.
\newblock Integer multiplication in time {$O(n\log n)$}.
\newblock {\em Annals of Mathematics}, 193(2):563--617, 2021.
\newblock Preprint hal-02070778 2019.

\bibitem{hastad:optimalj}
J.~H{\aa}stad.
\newblock Some optimal inapproximability results.
\newblock {\em Journal of the ACM}, 48(4):798--859, 2001.
\newblock Earlier version in STOC'97.

\bibitem{hansonea:loopholefreebell}
B.~Hensen, H.~Bernien, A.~E. Dr{\'e}au, A.~Reiserer, N.~Kalb, M.~S. Blok,
  J.~Ruitenberg, R.~F.~L. Vermeulen, R.~N. Schouten, C.~Abell{\'a}n, W.~Amaya,
  V.~Pruneri, M.~W. Mitchell, M.~Markham, D.~J. Twitchen, D.~Elkouss,
  S.~Wehner, T.~H. Taminiau, and R.~Hanson.
\newblock Loophole-free {B}ell inequality violation using electron spins
  separated by 1.3 kilometres.
\newblock {\em Nature}, 526, 29 October 2015.

\bibitem{holevo}
A.~S. Holevo.
\newblock Bounds for the quantity of information transmitted by a quantum
  communication channel.
\newblock {\em Problemy Peredachi Informatsii}, 9(3):3--11, 1973.
\newblock English translation in {\it Problems of Information Transmission},
  9:177--183, 1973.

\bibitem{hls:madv}
P.~H{\o}yer, T.~Lee, and R.~{\v{S}}palek.
\newblock Negative weights make adversaries stronger.
\newblock In {\em Proceedings of 39th ACM STOC}, pages 526--535, 2007.
\newblock quant-ph/0611054.

\bibitem{impagliazzo&wigderson:PisBPP}
R.~Impagliazzo and A.~Wigderson.
\newblock P = {BPP} if {E} requires exponential circuits: Derandomizing the
  {XOR} lemma.
\newblock In {\em Proceedings of 29th ACM STOC}, pages 220--229, 1997.

\bibitem{ims:nonAb}
G.~Ivanyos, F.~Magniez, and M.~Santha.
\newblock Efficient quantum algorithms for some instances of the non-{A}belian
  hidden subgroup problem.
\newblock {\em International Journal of Foundations of Computer Science},
  14(5):723--740, 2003.
\newblock Earlier version in SPAA'01. quant-ph/0102014.

\bibitem{iss:nil2groups}
G.~Ivanyos, L.~Sanselme, and M.~Santha.
\newblock An efficient quantum algorithm for the hidden subgroup problem in
  nil-2 groups.
\newblock {\em Algorithmica}, 62(1--2):480--498, 2012.
\newblock Earlier version in LATIN'08. arXiv:0707.1260.

\bibitem{JJUW10j}
R.~Jain, Z.~Ji, S.~Upadhyay, and J.~Watrous.
\newblock {QIP} = {PSPACE}.
\newblock {\em Journal of the ACM}, 58(6):30:1--30:27, 2011.
\newblock Earlier version in STOC'10. arXiv:0907.4737.

\bibitem{JWB:nonidQMA}
D.~Janzing, P.~Wocjan, and T.~Beth.
\newblock Non-identity check is {QMA}-complete.
\newblock {\em International Journal of Quantum Information}, 3(3):463--473,
  2005.
\newblock quant-ph/0305050.

\bibitem{jkm:nested}
S.~Jeffery, R.~Kothari, and F.~Magniez.
\newblock Nested quantum walks with quantum data structures.
\newblock In {\em Proceedings of 24th ACM-SIAM SODA}, pages 1474--1485, 2013.
\newblock arXiv:1210.1199.

\bibitem{jordan:gradient}
S.~Jordan.
\newblock Fast quantum algorithm for numerical gradient estimation.
\newblock {\em Physical Review Letters}, 95:050501, 2005.
\newblock quant-ph/0405146.

\bibitem{KLLN:breaking}
M.~Kaplan, G.~Leurent, A.~Leverrier, and M.~{Naya-Plasencia}.
\newblock Breaking symmetric cryptosystems using quantum period finding.
\newblock In {\em Proceedings of CRYPTO'16, Part II}, volume 9815 of {\em
  Lecture Notes in Computer Science}, pages 207--237, 2016.
\newblock arXiv:1602.05973.

\bibitem{katz&trevisan:ldc}
J.~Katz and L.~Trevisan.
\newblock On the efficiency of local decoding procedures for error-correcting
  codes.
\newblock In {\em Proceedings of 32nd ACM STOC}, pages 80--86, 2000.

\bibitem{KKR:localham}
J.~Kempe, A.~Yu. Kitaev, and O.~Regev.
\newblock The complexity of the local {H}amiltonian problem.
\newblock {\em SIAM Journal on Computing}, 35(5):1070--1097, 2006.
\newblock Earlier version in FSTTCS'04. quant-ph/0406180.

\bibitem{kruw:noisethresholdj}
J.~Kempe, O.~Regev, F.~Unger, and R.~{de} Wolf.
\newblock Upper bounds on the noise threshold for fault-tolerant quantum
  computing.
\newblock {\em Quantum Information and Computation}, 10(5--6):361--376, 2010.
\newblock Earlier version in ICALP'08. quant-ph/0802.1464.

\bibitem{KLLP:qmeans}
I.~Kerenidis, J.~Landman, A.~Luongo, and A.~Prakash.
\newblock q-means: A quantum algorithm for unsupervised machine learning.
\newblock In {\em Proceedings of the 33rd International Conference on Neural
  Information Processing Systems (NIPS'19), Paper 372}, page 4134–4144, 2019.
\newblock arXiv:1812.03584.

\bibitem{KP16}
I.~Kerenidis and A.~Prakash.
\newblock Quantum recommendation systems.
\newblock In {\em Proceedings of 8th Innovations in Theoretical Computer
  Science Conference}, volume~67 of {\em Leibniz International Proceedings in
  Informatics}, pages 49:1--49:21, 2017.
\newblock arXiv:1603.08675.

\bibitem{kerenidis&wolf:qldc}
I.~Kerenidis and R.~{de} Wolf.
\newblock Exponential lower bound for 2-query locally decodable codes via a
  quantum argument.
\newblock In {\em Proceedings of 35th ACM STOC}, pages 106--115, 2003.
\newblock quant-ph/0208062.

\bibitem{kerenidis&wolf:qldcj}
I.~Kerenidis and R.~{de} Wolf.
\newblock Exponential lower bound for 2-query locally decodable codes via a
  quantum argument.
\newblock {\em Journal of Computer and System Sciences}, 69(3):395--420, 2004.
\newblock Earlier version in STOC'03. quant-ph/0208062.

\bibitem{kimmelea:hamsim}
S.~Kimmel, G.~H.~Low C.~Lin, M.~Ozols, and T.~Yoder.
\newblock Hamiltonian simulation with optimal sample complexity.
\newblock {\em npj Quantum Information}, 3(13), 2017.
\newblock arXiv:1608.00281.

\bibitem{kitaev&watrous:qip}
A.~Kitaev and J.~Watrous.
\newblock Parallelization, amplification, and exponential time simulation of
  quantum interactive proof systems.
\newblock In {\em Proceedings of 32nd ACM STOC}, pages 608--617, 2000.

\bibitem{kitaev:stabilizer}
A.~Yu. Kitaev.
\newblock Quantum measurements and the {A}belian stabilizer problem.
\newblock quant-ph/9511026, 12 Nov 1995.

\bibitem{kitaev:qnp}
A.~Yu. Kitaev.
\newblock Quantum {NP}, January 1999.
\newblock Talk given at AQIP'99 conference, DePaul University, Chicago.

\bibitem{klartag&regev:lowerbound}
B.~Klartag and O.~Regev.
\newblock Quantum one-way communication is exponentially stronger than
  classical communication.
\newblock In {\em Proceedings of 43rd ACM STOC}, 2011.
\newblock arXiv:1009.3640.

\bibitem{klivans&melkebeek:GIsubexp}
A.~Klivans and D.~{van Melkebeek}.
\newblock Graph nonisomorphism has subexponential size proofs unless the
  polynomial-time hierarchy collapses.
\newblock {\em SIAM Journal on Computing}, 31(5):1501--1526, 2002.
\newblock Earlier version in STOC'99.

\bibitem{klz:treshold}
M.~Knill, R.~Laflamme, and W.~Zurek.
\newblock Threshold accuracy for quantum computation.
\newblock quant-ph/9610011, 15 Oct 1996.

\bibitem{knuth:acp2}
D.~E. Knuth.
\newblock {\em The Art of Computer Programming. Volume 2: Seminumerical
  Algorithms}.
\newblock Addison-Wesley, third edition, 1997.

\bibitem{KothariManohar:3LDC}
P.~Kothari and P.~Manohar.
\newblock An exponential lower bound for linear 3-query locally correctable
  codes.
\newblock In {\em Proceedings of 56th ACM STOC}, pages 776--787, 2024.

\bibitem{LC:qlearnTgate}
C-Y. Lai and H-C. Cheng.
\newblock Learning quantum circuits of some {$T$} gates.
\newblock {\em IEEE Transactions on Information Theory}, 68(6):3951--3964,
  2022.
\newblock arXiv:2106.12524.

\bibitem{legall:triangle}
F.~{Le Gall}.
\newblock Improved quantum algorithm for triangle finding via combinatorial
  arguments.
\newblock In {\em Proceedings of 55th IEEE FOCS}, pages 216--225, 2014.
\newblock arXiv:1407.0085.

\bibitem{lms:improvedtrianglej}
T.~Lee, F.~Magniez, and M.~Santha.
\newblock Improved quantum query algorithms for triangle finding and
  associativity testing.
\newblock {\em Algorithmica}, 77(2):459--486, 2017.
\newblock arXiv:1210.1014.

\bibitem{lmrss:stateconv}
T.~Lee, R.~Mittal, B.~Reichardt, R.~{\v{S}}palek, and M.~Szegedy.
\newblock Quantum query complexity of state conversion.
\newblock In {\em Proceedings of 52nd IEEE FOCS}, pages 344--353, 2011.
\newblock arXiv:1011.3020.

\bibitem{lenstras:sieve}
A.~K. Lenstra and H.~W. Lenstra, Jr.
\newblock {\em The Development of the Number Field Sieve}, volume 1554 of {\em
  Lecture Notes in Mathematics}.
\newblock Springer, 1993.

\bibitem{lenstra&pomerance:rigorous}
H.~W. Lenstra, Jr. and C.~Pomerance.
\newblock A rigorous time bound for factoring integers.
\newblock {\em Journal of the American Mathematical Society}, 5:483--516, 1992.

\bibitem{levin:np}
L.~Levin.
\newblock Universal search problems (translated from the {R}ussian).
\newblock {\em Problems of Information Transmission}, 9(3):115--–116, 1973.

\bibitem{linlin:qscomp}
L.~Lin.
\newblock Lecture notes on quantum algorithms for scientific computation, 2022.
\newblock arXiv:2201.08309.

\bibitem{linlin&tong:qls}
L.~Lin and Y.~Tong.
\newblock Optimal quantum eigenstate filtering with application to solving
  quantum linear systems.
\newblock arXiv:1910.14596, 31 Oct 2019.

\bibitem{liu:consistency}
Y.-K. Liu.
\newblock Consistency of local density matrices is {QMA}-complete.
\newblock In {\em Proceedings of 10th International Workshop on Randomization
  and Computation (RANDOM 2006)}, volume 4110 of {\em Lecture Notes in Computer
  Science}, pages 438--449, 2006.
\newblock quant-ph/0604166.

\bibitem{LCV:NrepQMA}
Y.-K. Liu, M.~Christandl, and F.~Verstraete.
\newblock Quantum computational complexity of the {N}-representability problem:
  {QMA} complete.
\newblock {\em Physical Review Letters}, 98(110503), 2007.
\newblock quant-ph/0609125.

\bibitem{lloyd:sim}
S.~Lloyd.
\newblock Universal quantum simulators.
\newblock {\em Science}, 273:1073--1078, 1996.

\bibitem{lmp:supunsup}
S.~Lloyd, M.~Mohseni, and P.~Rebentrost.
\newblock Quantum algorithms for supervised and unsupervised machine learning,
  1 Jul 2013.
\newblock arXiv:1307.0411.

\bibitem{lmr:pca}
S.~Lloyd, M.~Mohseni, and P.~Rebentrost.
\newblock Quantum principal component analysis.
\newblock {\em Nature Physics}, 10:631--633, 2013.
\newblock arXiv:1307.0401.

\bibitem{lo&chau:security}
H-K. Lo and H.~F. Chau.
\newblock Unconditional security of quantum key distribution over arbitrarily
  long distances.
\newblock {\em Science}, 283:2050--2056, 1999.
\newblock quant-ph/9803006.

\bibitem{LowChuang:Uniform}
G.~H. Low and I.~L. Chuang.
\newblock Hamiltonian simulation by uniform spectral amplification.
\newblock arXiv:1707.05391, 17 Jul 2017.

\bibitem{LowChuang:Qubitization}
G.~H. Low and I.~L. Chuang.
\newblock Hamiltonian simulation by qubitization.
\newblock arXiv:1610.06546, 20 Oct 2016.

\bibitem{LowChuang:OptHamSim}
G.~H. Low and I.~L. Chuang.
\newblock Optimal {H}amiltonian simulation by quantum signal processing.
\newblock {\em Physical Review Letters}, 118(1):010501, 2017.
\newblock arXiv:1606.02685.

\bibitem{LowYoderChuang:SignalProc}
G.~H. Low, T.~J. Yoder, and I.~L. Chuang.
\newblock Methodology of resonant equiangular composite quantum gates.
\newblock {\em Physical Review X}, 6(4):041067, 2016.
\newblock arXiv:1603.03996.

\bibitem{low:clifford}
R.~A. Low.
\newblock Learning and testing algorithms for the {C}lifford group.
\newblock {\em Physical Review A}, 80(052314), 2009.
\newblock arXiv:0907.2833.

\bibitem{LFKN:algebraicIP}
C.~Lund, L.~Fortnow, H.~Karloff, and N.~Nisan.
\newblock Algebraic methods for interactive proof systems.
\newblock {\em Journal of the ACM}, 39(4):859--868, 1992.
\newblock Earlier version in FOCS'90.

\bibitem{mnrs:searchwalk}
F.~Magniez, A.~Nayak, J.~Roland, and M.~Santha.
\newblock Search via quantum walk.
\newblock {\em SIAM Journal on Computing}, 40(1):142--164, 2011.
\newblock Earlier version in STOC'07. quant-ph/0608026.

\bibitem{mss:triangle}
F.~Magniez, M.~Santha, and M.~Szegedy.
\newblock Quantum algorithms for the triangle problem.
\newblock In {\em Proceedings of 16th ACM-SIAM SODA}, pages 1109--1117, 2005.
\newblock quant-ph/0310134.

\bibitem{mahadev:clasveri}
U.~Mahadev.
\newblock Classical verification of quantum computations.
\newblock In {\em Proceedings of 59th IEEE FOCS}, pages 259--267, 2018.
\newblock arXiv:1804.01082.

\bibitem{manin:radio}
Y.~Manin.
\newblock Vychislimoe i nevychislimoe (computable and noncomputable).
\newblock {\em Soviet Radio}, pages 13--15, 1980.
\newblock In Russian.

\bibitem{manin:qc}
Y.~Manin.
\newblock Classical computing, quantum computing, and {S}hor's factoring
  algorithm.
\newblock quant-ph/9903008, 2 Mar 1999.

\bibitem{marriott&watrous:qma}
C.~Marriott and J.~Watrous.
\newblock Quantum {Arthur-Merlin} games.
\newblock {\em Computational Complexity}, 14(2):122--152, 2005.
\newblock Earlier version in CCC'04. arXiv:cs/0506068.

\bibitem{mayers:security}
D.~Mayers.
\newblock Unconditional security in quantum cryptography.
\newblock quant-ph/9802025, 10 Feb 1998.

\bibitem{MRT:foundationsML}
M.~Mohri, A.~Rostamizadeh, and A.~Talwalkar.
\newblock {\em Foundations of Machine Learning}.
\newblock MIT Press, second edition, 2018.

\bibitem{mrr:genericQFT}
C.~Moore, D.~N. Rockmore, and A.~Russell.
\newblock Generic quantum {F}ourier transforms.
\newblock {\em ACM Transactions on Algorithms}, 2(4):707--723, 2006.
\newblock quant-ph/0304064.

\bibitem{mrs:defiesfourier}
C.~Moore, A.~Russell, and L.~Schulman.
\newblock The symmetric group defies strong {F}ourier sampling.
\newblock {\em SIAM Journal on Computing}, 37(6):1842--1864, 2008.
\newblock quant-ph/0501056+66. Earlier version in FOCS'05.

\bibitem{mosca&ekert:hidden}
M.~Mosca and A.~Ekert.
\newblock The hidden subgroup problem and eigenvalue estimation on a quantum
  computer.
\newblock In {\em Proceedings of 1st NASA QCQC conference}, volume 1509 of {\em
  Lecture Notes in Computer Science}, pages 174--188. Springer, 1998.
\newblock quant-ph/9903071.

\bibitem{nayak:qfa}
A.~Nayak.
\newblock Optimal lower bounds for quantum automata and random access codes.
\newblock In {\em Proceedings of 40th IEEE FOCS}, pages 369--376, 1999.
\newblock quant-ph/9904093.

\bibitem{newman:random}
I.~Newman.
\newblock Private vs.~common random bits in communication complexity.
\newblock {\em Information Processing Letters}, 39(2):67--71, 1991.

\bibitem{newman&szegedy:1round}
I.~Newman and M.~Szegedy.
\newblock Public vs.~private coin flips in one round communication games.
\newblock In {\em Proceedings of 28th ACM STOC}, pages 561--570, 1996.

\bibitem{nielsen&chuang:qc}
M.~A. Nielsen and I.~L. Chuang.
\newblock {\em Quantum Computation and Quantum Information}.
\newblock Cambridge University Press, 2000.

\bibitem{odonnell16}
R.~{O'Donnell} and J.~Wright.
\newblock Efficient quantum tomography.
\newblock In {\em Proceedings of 48th ACM STOC}, pages 899--912, 2016.
\newblock arXiv:1508.01907.

\bibitem{papadimitriou:cc}
C.~H. Papadimitriou.
\newblock {\em Computational Complexity}.
\newblock Addison-Wesley, 1994.

\bibitem{VQE}
A.~Peruzzo, J.~McClean, P.~Shadbolt, M-H. Yung, X-Q. Zhou, P.~Love,
  A.~Aspuru-Guzik, and J.~O’Brien.
\newblock A variational eigenvalue solver on a photonic quantum processor.
\newblock {\em Nature Communications}, 3:24, 2014.
\newblock arXiv:1304.3061.

\bibitem{preskill:faulttol}
J.~Preskill.
\newblock Fault-tolerant quantum computation.
\newblock In H-K. Lo, S.~Popescu, and T.~P. Spiller, editors, {\em Introduction
  to Quantum Computation}. World Scientific, 1998.
\newblock quant-ph/9712048.

\bibitem{preskill:qc40years}
J.~Preskill.
\newblock Quantum computing 40 years later.
\newblock In A.~Hey, editor, {\em Feynman Lectures on Computation}. Taylor \&\
  Francis Group, second edition, 2022.
\newblock arXiv:2106.10522.

\bibitem{QPG:improved}
H.~Qassim, H.~Pashayan, and D.~Gosset.
\newblock Improved upper bounds on the stabilizer rank of magic states.
\newblock {\em Quantum}, 5:606, 2021.
\newblock arXiv:2106.07740.

\bibitem{raussendorf&briegel:oneway1}
R.~Raussendorf and H.~J. Briegel.
\newblock Quantum computing via measurements only, 2000.
\newblock quant-ph/0010033.

\bibitem{raussendorf&briegel:oneway2}
R.~Raussendorf and H.~J. Briegel.
\newblock A one-way quantum computer.
\newblock {\em Physical Review Letters}, 86(22):5188--5191, 2001.
\newblock quant-ph/0510135.

\bibitem{razborov:qdisj}
A.~Razborov.
\newblock Quantum communication complexity of symmetric predicates.
\newblock {\em Izvestiya of the Russian Academy of Sciences, mathematics},
  67(1):159--176, 2003.
\newblock quant-ph/0204025.

\bibitem{rebentrost2014QSVM}
P.~Rebentrost, M.~Mohseni, and S.~Lloyd.
\newblock Quantum support vector machine for big data classification.
\newblock {\em Physical Review Letters}, 113(13):130503, 2014.
\newblock arXiv:1307.0471.

\bibitem{regev:lwe}
O.~Regev.
\newblock On lattices, learning with errors, random linear codes, and
  cryptography.
\newblock {\em Journal of the ACM}, 56(6):34:1--34:40, 2009.
\newblock Earlier version in STOC'13.

\bibitem{reichardt:tight}
B.~Reichardt.
\newblock Span programs and quantum query complexity: The general adversary
  bound is nearly tight for every {B}oolean function.
\newblock In {\em Proceedings of 50th IEEE FOCS}, pages 544--551, 2009.

\bibitem{reichardt:gametrees}
B.~Reichardt.
\newblock Faster quantum algorithm for evaluating game trees.
\newblock In {\em Proceedings of 22nd ACM-SIAM SODA}, pages 546--559, 2011.
\newblock arXiv:0907.1623.

\bibitem{reichardt:equiv}
B.~Reichardt.
\newblock Span programs are equivalent to quantum query algorithms.
\newblock {\em SIAM Journal on Computing}, 43(3):1206--1219, 2014.

\bibitem{reichardt&spalek:formula}
B.~Reichardt and R.~{\v{S}}palek.
\newblock Span-program-based quantum algorithm for evaluating formulas.
\newblock {\em Theory of Computing}, 8:291--319, 2012.
\newblock Earlier version in STOC'08. arXiv:0710.2630.

\bibitem{Riordan&Shannon:circuit}
J.~Riordan and C.~E. Shannon.
\newblock The number of two-terminal series-parallel networks.
\newblock {\em Journal of Mathematics and Physics}, 21:83--–93, 1942.

\bibitem{rsa}
R.~Rivest, A.~Shamir, and L.~Adleman.
\newblock A method for obtaining digital signatures and public key
  cryptosystems.
\newblock {\em Communications of the ACM}, 21:120--126, 1978.

\bibitem{rivest:crypto}
R.~L. Rivest.
\newblock Cryptography.
\newblock In van Leeuwen \cite{handbookA}, pages 717--755.

\bibitem{rivlin&cheney:approx}
T.~J. Rivlin and E.~W. Cheney.
\newblock A comparison of uniform approximations on an interval and a finite
  subset thereof.
\newblock {\em SIAM Journal on Numerical Analysis}, 3(2):311--320, 1966.

\bibitem{saeedi&arodz:qsvm}
S.~Saeedi and T.~Arodz.
\newblock Quantum sparse support vector machines, 2019.
\newblock arXiv:1902.01879.

\bibitem{SPA:semi}
S.~Saeedi, A.~Panahi, and T.~Arodz.
\newblock Quantum semi-supervised kernel learning.
\newblock {\em Quantum Machine Intelligence}, 3:24, 2021.

\bibitem{saks&wigderson:trees}
M.~Saks and A.~Wigderson.
\newblock Probabilistic {B}oolean decision trees and the complexity of
  evaluating game trees.
\newblock In {\em Proceedings of 27th IEEE FOCS}, pages 29--38, 1986.

\bibitem{santha:qrwsurvey}
M.~Santha.
\newblock Quantum walk based search algorithms.
\newblock In {\em Proceedings of 5th TAMC}, pages 31--46, 2008.
\newblock arXiv:0808.0059.

\bibitem{santoli&schaffner:simon}
T.~Santoli and C.~Schaffner.
\newblock Using {S}imon's algorithm to attack symmetric-key cryptographic
  primitives.
\newblock {\em Quantum Information and Computation}, 17(1\&\ 2):65--78, 2017.
\newblock arXiv:1603.07856.

\bibitem{schonhage&strassen}
A.~Sch{\"o}nhage and V.~Strassen.
\newblock Schnelle {M}ultiplikation grosser {Z}ahlen.
\newblock {\em Computing}, 7:281--292, 1971.

\bibitem{schoning:ksat}
U.~Sch{\"o}ning.
\newblock A probabilistic algorithm for {$k$-SAT} and constraint satisfaction
  problems.
\newblock In {\em Proceedings of 40th IEEE FOCS}, pages 410--414, 1999.

\bibitem{schuch&vertraete:dft}
N.~Schuch and F.~Verstraete.
\newblock Computational complexity of interacting electrons and fundamental
  limitations of density functional theory.
\newblock {\em Nature Physics}, 5:732–735, 2009.
\newblock arXiv:0712.0483.

\bibitem{schuld&killoran:feature}
M.~Schuld and N.~Killoran.
\newblock Quantum machine learning in feature {H}ilbert spaces.
\newblock {\em Physical Review Letters}, 122(13):040504, 2019.
\newblock arXiv:1803.07128.

\bibitem{schuld&petruccione:qmlbook}
M.~Schuld and F.~Petruccione.
\newblock {\em Machine Learning with Quantum Computers}.
\newblock Springer, second edition, 2021.

\bibitem{shwartz&david:learningbook}
S.~{Shalev-Shwartz} and S.~{Ben-David}.
\newblock {\em Understanding machine learning: From theory to algorithms}.
\newblock Cambridge University Press, 2014.

\bibitem{shamir:IPPSPACEj}
A.~Shamir.
\newblock {IP} = {PSPACE}.
\newblock {\em Journal of the ACM}, 39(4):869--877, 1992.
\newblock Earlier version in FOCS'90.

\bibitem{shen:IPPSPACEj}
A.~Shen.
\newblock {IP} = {PSPACE}: Simplified proof.
\newblock {\em Journal of the ACM}, 39(4):878--880, 1992.

\bibitem{sherstov:andor}
A.~Sherstov.
\newblock Approximating the {AND-OR} tree.
\newblock {\em Theory of Computing}, 9(20):653--663, 2013.

\bibitem{shor:scheme}
P.~W. Shor.
\newblock Scheme for reducing decoherence in quantum memory.
\newblock {\em Physical Review A}, 52:2493, 1995.

\bibitem{shor:factoring}
P.~W. Shor.
\newblock Polynomial-time algorithms for prime factorization and discrete
  logarithms on a quantum computer.
\newblock {\em SIAM Journal on Computing}, 26(5):1484--1509, 1997.
\newblock Earlier version in FOCS'94. quant-ph/9508027.

\bibitem{shoup:lowerbounddiscrlog}
V.~Shoup.
\newblock Lower bounds for discrete logarithms and related problems.
\newblock In {\em Proceedings of Eurocrypt'97}, volume 1233 of {\em Lecture
  Notes in Computer Science}, pages 256--266. Springer, 1997.
\newblock Revised version available at
  \url{http://www.shoup.net/papers/dlbounds1.pdf}.

\bibitem{simon:power}
D.~Simon.
\newblock On the power of quantum computation.
\newblock {\em SIAM Journal on Computing}, 26(5):1474--1483, 1997.
\newblock Earlier version in FOCS'94.

\bibitem{spalek&szegedy:adversaryj}
R.~{\v S}palek and M.~Szegedy.
\newblock All quantum adversary methods are equivalent.
\newblock {\em Theory of Computing}, 2(1):1--18, 2006.
\newblock Earlier version in ICALP'05, quant-ph/0409116.

\bibitem{steane:errcor}
A.~Steane.
\newblock Multiple particle interference and quantum error correction.
\newblock In {\em Proceedings of the Royal Society of London}, volume A452,
  pages 2551--2577, 1996.
\newblock quant-ph/9601029.

\bibitem{SZ:binarygcd}
D.~Stehl{\'{e}} and P.~Zimmermann.
\newblock A binary recursive gcd algorithm.
\newblock In {\em Algorithmic Number Theory (ANTS 2004)}, volume 3076 of {\em
  Lecture Notes in Computer Science}, pages 411--–425, 2004.

\bibitem{szegedy:qmarkov}
M.~Szegedy.
\newblock Quantum speed-up of {M}arkov chain based algorithms.
\newblock In {\em Proceedings of 45th IEEE FOCS}, pages 32--41, 2004.
\newblock quant-ph/0401053.

\bibitem{tang:RC}
E.~Tang.
\newblock A quantum-inspired classical algorithm for recommendation systems.
\newblock In {\em Proceedings of 51st ACM STOC}, pages 217--228, 2019.
\newblock arXiv:1807.04271.

\bibitem{terhal:qmemories}
B.~M. Terhal.
\newblock Quantum error correction for quantum memories.
\newblock {\em Reviews of Modern Physics}, 87:307, 2015.
\newblock arXiv:1302.3428.

\bibitem{trevisan:eccsurvey}
L.~Trevisan.
\newblock Some applications of coding theory in computational complexity.
\newblock {\em Quaderni di Matematica}, 13:347--424, 2004.

\bibitem{turing:compnumb}
A.~M. Turing.
\newblock On computable numbers, with an application to the
  {E}ntscheidungproblem.
\newblock In {\em Proceedings of the London Mathematical Society}, volume~42,
  pages 230--265, 1936.
\newblock Correction, ibidem (vol.~43), pages~544--546.

\bibitem{valiant:paclearning}
L.~Valiant.
\newblock A theory of the learnable.
\newblock {\em Communications of the ACM}, 27(11):1134--1142, 1984.

\bibitem{handbookA}
J.~van Leeuwen, editor.
\newblock {\em Handbook of Theoretical Computer Science. Volume~A: Algorithms
  and Complexity}.
\newblock MIT Press, Cambridge, MA, 1990.

\bibitem{nmr:period}
L.~Vandersypen, M.~Steffen, G.~Breyta, C.~Yannoni, R.~Cleve, and I.~Chuang.
\newblock Experimental realization of an order-finding algorithm with an {NMR}
  quantum computer.
\newblock {\em Physical Review Letters}, 85(25):5452--5455, 2000.
\newblock quant-ph/0007017.

\bibitem{vapnik:vcdimension}
V.~Vapnik and A.~Chervonenkis.
\newblock On the uniform convergence of relative frequencies of events to their
  probabilities.
\newblock {\em Theory of Probability \& Its Applications}, 16(2):264--280,
  1971.
\newblock English translation of 1968 Russian paper in \emph{Dokl.\ Akad.\
  Nauk.} 181(4).

\bibitem{vidick&watrous:qproofs}
T.~Vidick and J.~Watrous.
\newblock Quantum proofs.
\newblock {\em Foundations and Trends in Theoretical Computer Science},
  11(1--2):1–--215, 2015.

\bibitem{watrous:group}
J.~Watrous.
\newblock Succinct quantum proofs for properties of finite groups.
\newblock In {\em Proceedings of 41st IEEE FOCS}, pages 537--546, 2000.
\newblock quant-ph/0011023.

\bibitem{watrous:solvable}
J.~Watrous.
\newblock Quantum algorithms for solvable groups.
\newblock In {\em Proceedings of 33rd ACM STOC}, pages 60--67, 2001.

\bibitem{watrous:constantroundPSPACE}
J.~Watrous.
\newblock {PSPACE} has 2-round quantum interactive proof systems.
\newblock {\em Theoretical Computer Science}, 292(3):575--588, 2003.
\newblock Earlier version in STOC'99. arXiv:cs/9901015.

\bibitem{watrous:qcomplexitysurvey}
J.~Watrous.
\newblock Quantum computational complexity.
\newblock In {\em Encyclopedia of Complexity and Systems Science}. Springer,
  2009.
\newblock arXiv:0804.3401.

\bibitem{watrous:tqi}
J.~Watrous.
\newblock {\em The Theory of Quantum Information}.
\newblock Cambridge University Press, 2018.
\newblock Available at \url{https://cs.uwaterloo.ca/~watrous/TQI/}.

\bibitem{wehner&wolf:improvedldc}
S.~Wehner and R.~{de} Wolf.
\newblock Improved lower bounds for locally decodable codes and private
  information retrieval.
\newblock In {\em Proceedings of 32nd ICALP}, volume 3580 of {\em Lecture Notes
  in Computer Science}, pages 1424--1436, 2005.
\newblock quant-ph/0403140.

\bibitem{wolf:thesis}
R.~{de} Wolf.
\newblock {\em Quantum Computing and Communication Complexity}.
\newblock PhD thesis, University of Amsterdam, 2001.

\bibitem{woodruff:3querylower}
D.~Woodruff.
\newblock A quadratic lower bound for three-query linear locally decodable
  codes over any field.
\newblock In {\em Proceedings of 13th APPROX}, pages 766--779, 2010.

\bibitem{nocloning}
W.~K. Wootters and W.~H. Zurek.
\newblock A single quantum cannot be copied.
\newblock {\em Nature}, 299:802--803, 1982.

\bibitem{yao:distributive}
A.~C-C. Yao.
\newblock Some complexity questions related to distributive computing.
\newblock In {\em Proceedings of 11th ACM STOC}, pages 209--213, 1979.

\bibitem{yao:qcircuit}
A.~C-C. Yao.
\newblock Quantum circuit complexity.
\newblock In {\em Proceedings of 34th IEEE FOCS}, pages 352--360, 1993.

\bibitem{yekhanin:3ldcj}
S.~Yekhanin.
\newblock Towards 3-query locally decodable codes of subexponential length.
\newblock {\em Journal of the ACM}, 55(1), 2008.
\newblock Earlier version in STOC'07.

\end{thebibliography}

\end{document}